\documentclass[12pt]{report}
\usepackage[a4paper,width=160mm,top=25mm,bottom=25mm]{geometry}
\usepackage[utf8]{inputenc}
\usepackage{amsmath}
\usepackage{amsfonts}
\usepackage{amssymb}
\usepackage{amsthm}
\usepackage{caption}
\usepackage{graphicx}
\usepackage{tikz}
\usepackage{dcolumn}
\usepackage{hyperref}
\usepackage{verbatim}
\usepackage{color}
\usepackage{float}
\usepackage{environ}
\graphicspath{ {Images/} }
\usepackage{fancyhdr}
\usepackage[backend=biber,style=ieee]{biblatex}
\usepackage{blindtext}
\tikzset{Witten diagram/.style={execute at begin picture={%
\draw[blue,fill=white!20] circle[radius=\pgfkeysvalueof{/tikz/Witten/radius}];
}},vertex/.style={circle,fill,inner sep=1.5pt,node
contents={}},
Witten/.cd,radius/.initial=2cm}
\NewEnviron{wittendiagram}[1][]{\vcenter{\hbox{\begin{tikzpicture}[Witten diagram,#1]%
\BODY
\end{tikzpicture}}}}

\begin{document}

\newgeometry{width=220mm,top=20mm,bottom=20mm}

\begin{titlepage}

    \includegraphics[scale=0.07]{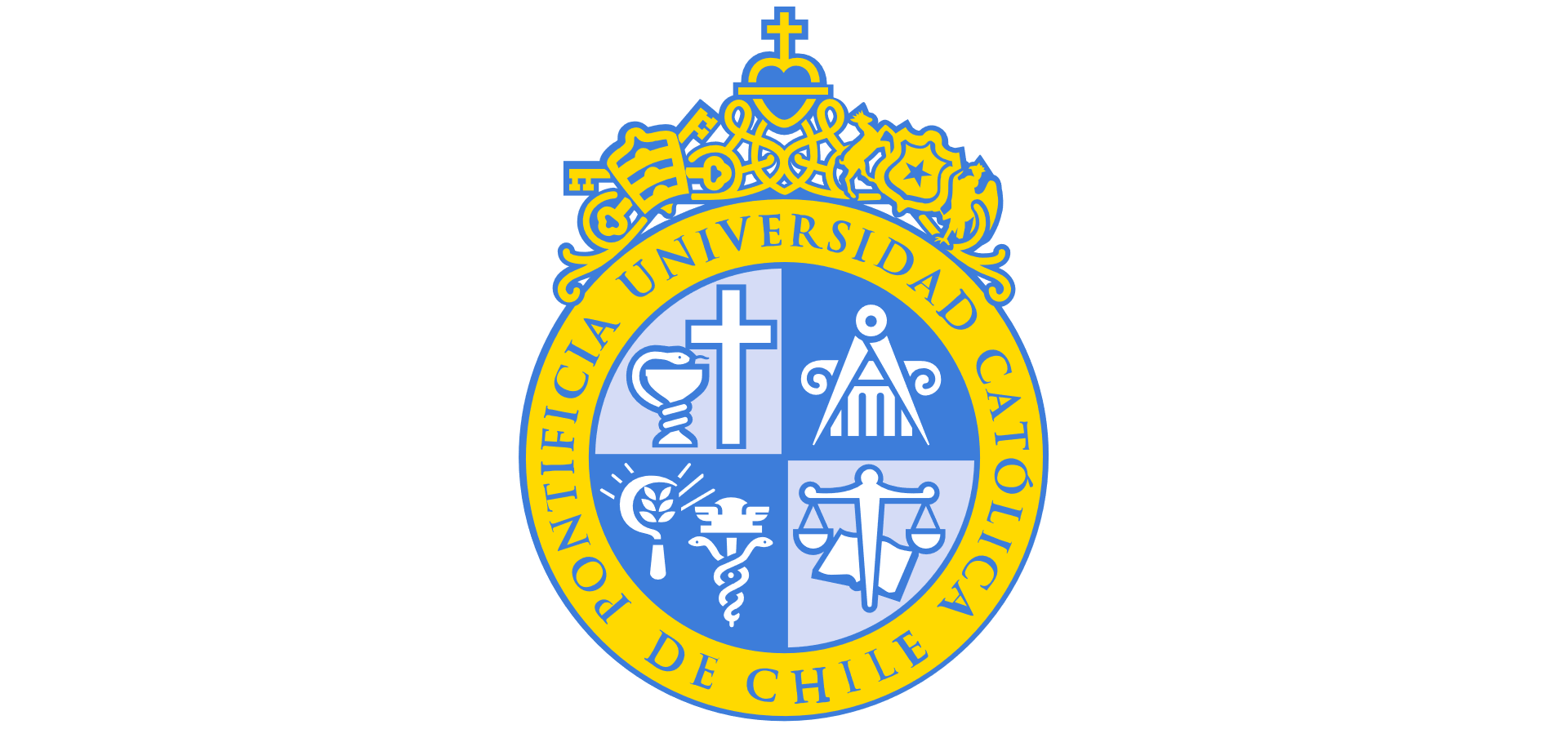}
    
    \vspace{-22mm}
    \small
    \hspace{35mm}Pontificia Universidad Católica de Chile
    
    \hspace{35mm}Facultad de Física
    
    \hspace{35mm}Instituto de Física
    
    \vspace{2.5cm}
    
    \begin{center}
        \large
        \textbf{QUANTUM SCALAR THEORIES IN THE\\
        ADS/CFT CORRESPONDENCE}
        
        \vspace{2cm}
        
        \large
        Ernesto Bianchi Palacios
        
        \vspace{2cm}
        
        \large
        Thesis submitted for the degree of\\
        Master of Science in Physics
        
        \vspace{2cm}
        
        Advisor:\\
        Máximo Bañados
        
        \vspace{0.5cm}
        Committee:\\
        Marco Aurelio Díaz\\
        Enrique Muñoz\\
        Rodrigo Soto
        
        \vspace{\fill}
        \normalsize
        Santiago, Chile\\
        July, 2021
    \end{center}
\end{titlepage}

\restoregeometry

\newpage


\pagenumbering{roman}

\chapter*{Abstract}
\addcontentsline{toc}{chapter}{Abstract}

This thesis contains original research on the conjectured AdS/CFT correspondence. Known are the holographic correlators resulting from this correspondence under its classical approximation, confirming the conjecture at this level. In this work we will explore its consequences beyond this approximation, considering the quantum corrections to these known dual correlators for the particular cases of scalar $\Phi^3$ and $\Phi^4$ theories on a fixed AdS$_{d+1}$ background. This will involve the development of a scheme that adds systematically order by order the quantum corrections to the correlators obtained classically together with the computation of these new contributions, introducing on the way sensible regularization and renormalization schemes for the different expected divergences. This process will show in a clear way the effect of these loop corrections through the main role played by the infrared and ultraviolet divergences, which is briefly summarized in: the quantum corrections to the holographic correlators produce anomalies in the resulting CFT in the form of an anomalous scaling dimension, confirming the validity of the conjecture beyond tree-level computations. All this study is carried out throughout this thesis for arbitrary values of the parameters of the theory in the bulk. With the intention to study the general ideas and results developed in this work, at the end of the thesis a particular case is analyzed.

\newpage


\chapter*{Acknowledgements}
\addcontentsline{toc}{chapter}{Acknowledgements}

I would like to thank the people who, at different levels, have been a part either directly or indirectly in my process of obtaining the Master's degree and in the work that this thesis has entailed.\par
Starting with my supervisor, Máximo Bañados, who not only introduced me to this exciting topic but also under his constant supervision and way of seeing and doing physics I have learned enormously, forming me on a professional but also on a personal level.\par
Part of our research group is also composed by Kostas Skenderis and Iván Muñoz. I thank you and Max, for our frequent meetings, whose valuable discussions have influenced and shaped an important part of this work.\par
Almost the entirety of this thesis was carried out under quarantine in these strange pandemic times, so I also wanted to thank my close friends and family for the company and support in the course of this work, especially to my parents Ernesto and Paula and to my siblings Isidora and Agustín, for their unconditional encouragement in my dreams and goals.\par
This work was funded in part by CONICYT FONDECYT Regular \#1201145.

\newpage


\pagenumbering{arabic}

\tableofcontents

\newpage


\chapter{Introduction}
Perhaps the most basic and fundamental search that physics tries to do is to identify which are the different agents (forces as Newton would have called them) existing in our reality that generate movements in the different physical objects that live in it. What has been extremely useful in this search is the mathematical language, allowing us to accurately describe the dynamics produced by these agents, identified so far by 4: electromagnetism, weak interaction, strong interaction and gravity. The different experiments that have been developed to study the dynamics of these 4 interactions suggest that their behaviors are dependent not only on the length scales at which they are tested but also on the energy scales, having to consider quantum and relativistic corrections in their mathematical descriptions. Despite this, the techniques developed to date have allowed us to satisfactorily describe the first 3 (e.m., weak and strong) of these 4 fundamental forces theoretically for all scales of interest. However, these same techniques that allow us to understand these 3 interactions at small distances applied to gravitation result in incurable contradictions, suggesting that the quantum nature of gravity is much more delicate.\par
The understanding of quantum gravity is perhaps the greatest theoretical challenge in contemporary physics, which is why different groups and programs have been fully dedicated to studying it. Unsurprisingly, the main objects of study of these programs have been black holes, since they concentrate a large amount of mass in an infinitesimally small space, where the quantum effects of gravity are expected to be important. The first remarkable results of these studies, consequence of general relativity and quantum mechanics, is that black holes would emit thermal radiation \cite{Hawking} and that also they would have an enormous entropy that scales not with their volume but with their area \cite{Bekenstein}. This seems to suggest that the microscopic nature of gravity is holographic, being able to encode the information that it contains in a space of one smaller dimension \cite{tHooft}\cite{Susskind}.\par
One of the most successful theories under development of quantum gravity that manages to explain this holography of gravity in a concrete and precise way is string theory. Particular cases of the holographic principle in this theory can be observed in the limit of low energy of certain string's dynamics, resulting in field theories on the product of $d+1$ dimensional anti-de Sitter space with a compact manifold. These theories would be included in the Hilbert space of certain large N gauge theories in $d$ dimensions that also occur to be invariant under the conformal group. The canonical example of this is the low energy limit of type IIB strings, where one obtains $\mathcal{N}=4$ super Yang-Mills \cite{JM1}. The modern dictionary of this holography that relates gravity theories in AdS$_{d+1}$ with particle theories in CFT$_d$ basically postulates that the partition function of the theory in AdS serves as a tool for the computation of the correlators of the corresponding CFT \cite{Witten}. However, the complexity of these calculations have reduced this study of holographic correlators to the approximation of the AdS path integral to its leading, on-shell version. Despite this, the resulting correlators of this approximation precisely correspond to those of a strongly coupled CFT. This remarkable property of gravity's holography is summarized in a strong/weak duality: classical computations on AdS correspond to quantum computations in the CFT.\par
The complexity that we mentioned previously lies in the development of a scheme that systematically incorporates the quantum corrections to the AdS path integral together with the analytical calculations of these corrections. However, the conjectured holography strongly restricts the form of these subleading contributions: they cannot break the conformal invariance of the holographic correlators. Since this AdS/CFT correspondence helps to elucidate the quantum nature of gravity, there are great reasons to verify that it is indeed true. This is precisely the objective of this work and this thesis, the complete study of the quantum corrections on the AdS side of the holographic correspondence and the resulting CFT correlators. It is worth mentioning that this is a highly unexplored area of research on the subject, so we expect that several of the results shown here will be of great interest.\par
The content of this thesis is organized as follows: as the name of the AdS/CFT correspondence suggests we cannot really talk about it and its consequences without a basic understanding of CFTs and AdS spaces, thus in Chapter 2 we review the very basic ingredients of CFTs and AdS spaces, which in turn at the end of the chapter will lead, hopefully, to an expected correspondence.\par
Then with the AdS/CFT correspondence explicitly stated, as a first exploration and familiarization into the topic in Chapter 3 we will proceed to study its consequences for what are perhaps the simplest cases, scalar field theories on a fixed AdS background. In here we develop the necessary tools that will allow us to obtain from these theories, and through the use of what will be understood as the classical approximation of the correspondence, the resulting approximated CFT n-point functions.\par Finally, having studied and gained some insight on the AdS/CFT conjecture through its classical approximation, in Chapter 4 we will proceed to study these same scalar theories but now embracing the full quantum nature of the correspondence. In here we develop a systematic scheme that adds order by order the respective quantum corrections to the previous approximated correlators, to then study and compute each one of these new contributions, process that will also force us to introduce sensitive regularization and renormalization schemes. This chapter will contain most of the original work done in this thesis.\par
We will end the discussion in this thesis with a concluding chapter, where we summarize the key theoretical and conceptual results of our work together with their implications.\par
Then comes the complementary material, which we present in the form of appendices: Appendix A, where we discuss the integrals found in our work that only involve the functions $K(x,\vec{y})$ (bulk-boundary propagator, introduced in section 3.1.4) and Appendix B, where we discuss those that also involve the functions $G(x,y)$ (bulk-bulk propagator, introduced in section 3.1.7). Most of these integrals are done separately from the main text as to not lose the focus of discussion.\par
Before diving into the work of this thesis, we clarify that the conventions used throughout it are $c=\hbar=1$.

\newpage


\chapter{CFTs, AdS Spaces and the Correspondence}
In this work we will study scalar field theories on a fixed anti-de Sitter background and how, using the conjectured AdS/CFT correspondence, we can obtain from these theories correlators for some dual conformal theory living in one less dimension.
However, as the name of the correspondence suggests, we cannot really talk about it and its consequences without a basic understanding of CFTs and AdS spaces. Therefore, by completeness, the objective of this chapter is to review the very basic ingredients of CFTs and AdS spaces, which in turn at the end of the chapter will lead, hopefully, to an expected correspondence.\par
Section 2.1 will cover everything we will need to know about conformal field theories, using as main references the canonical book by Di Francesco, Mathieu and Sénéchal \cite{DiFran} and the online lectures by Hugh Osborn \cite{Osborn1}. In particular, in section 2.1.1 we will introduce what are known as the conformal transformations, in section 2.1.2 we will study the group formed by these transformations, in section 2.1.3 the implication of conformal invariant field theories in their correlators and finally in section 2.1.4 the implication of these same invariant theories on the external sources that give rise to the CFT operators.\par
Section 2.2 in turn will cover everything we will need to know about anti-de Sitter spaces, using as the main reference the AdS section from the extensive review on the AdS/CFT correspondence by Aharony, Gubser, Maldacena, Ooguri and Oz \cite{AdSCFTBible}, complementing with various of the modern textbooks on general relativity. In particular, in section 2.2.1 we will introduce the AdS metric in convenient coordinates, in section 2.2.2 we will study the curvature equation that it satisfies, in section 2.2.3 the isometry transformations of AdS spaces and finally in section 2.2.4 the notion of boundaries that these spaces have in some limits.\par
Lastly, section 2.3 will be a brief review of the AdS/CFT correspondence itself, using as main references the core articles on the subject due to Maldacena \cite{JM1} and Witten \cite{Witten}.

\section{Conformal Field Theory}

\subsection{Conformal Transformations}

How is the structure of space and time related for different individuals living in it if each one of them is equipped with their own measuring sticks and clocks? Galileo would have responded that space is a rigid place and that time is universal, so every person would measure the same distance between two points and every clock would tick at the same rate:

\begin{equation}
     dx'^{\ 2} + dy'^{\ 2} + dz'^{\ 2} = dx^2 + dy^2 + dz^2,\hspace{1cm} dt' = dt
\end{equation}\

This statement implies that the different measurements done by the individuals can be related to each other simply through rotations and translations:

\begin{equation}
    \vec{x}\ ' = R\vec{x}+\vec{a},\hspace{1cm} t'=t
\end{equation}\

where $R$ is an orthogonal matrix and $\vec a$ is some vector. These are known as the Galilean transformations.\par
Einstein on the contrary would have argued that Galileo's point of view can't possible be true since it is ignoring a decisive piece of evidence, the fact that every person measures the same speed for a light ray. This inevitably connects space and time in a profound way, and the measurements that agree now are not necessarily the distances between two points in space nor the ticks from the clocks in time, but the distances between two points in spacetime as a whole:

\begin{equation}
    -dt'^{\ 2} + dx'^{\ 2} + dy'^{\ 2} + dz'^{\ 2} = -dt^2 + dx^2 + dy^2 + dz^2
\end{equation}\

The different measurements done by the individuals can be related to each other now through rotations, translations and boosts:

\begin{equation}
    x' = \Lambda x+a
\end{equation}\

where $\Lambda$ are the Lorentz matrices and $a$ is some constant vector. These are known as the Poincaré transformations and their key property, by construction, is that they are isometry transformations, i.e., they leave the spacetime metric invariant:

\begin{equation}\label{pt}
    x\rightarrow x' = \Lambda x+a,\hspace{1cm} \eta\rightarrow g'=\eta
\end{equation}\

where $\eta$ is the Minkowski metric $\eta=\text{diag}(-1,1,1,1)$. A natural exploration and generalization of these transformations are those which leave the metric invariant up to a factor, known as conformal transformations:

\begin{equation}\label{ct}
    x\rightarrow x'=x'(x),\hspace{1cm} \eta\rightarrow g'=\Omega^2(x)\eta
\end{equation}\

where $\Omega(x)$ is some function of the coordinates $x$. As we did for the Galilean and Poincaré transformations, we would like to know the explicit form of the coordinate transformations $x'(x)$ which satisfy this property, with their respective values of $\Omega(x)$. Usually, this is achieved by solving eq. (\ref{ct}) under infinitesimal transformations and from these solutions constructing their finite version. However, for the sake of simplicity we will follow here a more heuristic path.\par 
We already saw our first example of a conformal transformation, namely Poincaré transformations. Indeed, from eq. (\ref{pt}) we can directly see that these transformations satisfy eq. (\ref{ct}) with $\Omega(x)=1$. Another transformation which is also direct to see as conformal are dilations or rescaling of the coordinates $x\rightarrow x'=\lambda x$. Under these, the line element transform as:

\begin{equation}
    ds^2=\eta_{\mu\nu}dx^\mu dx^\nu\rightarrow ds'^{\ 2} = \lambda^2\eta_{\mu\nu}dx^\mu dx^\nu
\end{equation}\

therefore, the metric transform as $\eta\rightarrow g'=\lambda^2\eta$, resulting in a conformal transformation with $\Omega(x)=\lambda$.\par
There are two more transformations we can construct: inversions and special conformal transformations (SCT for short). Inversions are of the form $x^\mu\rightarrow x'^\mu=\frac{x^\mu}{x^2}$, where the differential $dx^\mu$ transform as:

\begin{equation}
    dx^\mu\rightarrow dx'^\mu = \frac{1}{x^2}I^\mu_{\ \nu}dx^\nu,\hspace{1cm} I^\mu_{\ \nu} = \delta^\mu_\nu-\frac{2x^\mu x_\nu}{x^2}
\end{equation}\

Under these, the line element transform as:

\begin{equation}
    ds^2=\eta_{\mu\nu}dx^\mu dx^\nu\rightarrow ds'^{\ 2} = \frac{1}{(x^2)^2}\eta_{\mu\nu}dx^\mu dx^\nu
\end{equation}\

where we used the fact that the matrices $I^\mu_{\ \nu}$ satisfy the easy to prove identity $\eta_{\mu\nu}I^\mu_{\ \alpha}I^\nu_{\ \beta}=\eta_{\alpha\beta}$. Therefore, the metric transform as $\eta\rightarrow g'=\frac{1}{(x^2)^2}\eta$, resulting in a conformal transformation with $\Omega(x)=\frac{1}{x^2}$.\par
Finally, the SCT are a more exotic type of transformation which can be intuitively understood as an inversion, followed by a translation, followed by another inversion of the original coordinates, resulting in:

\begin{equation}
    x^\mu \rightarrow x'^\mu = \frac{x^\mu+a^\mu x^2}{1+2ax+a^2x^2}
\end{equation}\

Under these, the line element transform as:

\begin{equation}
    ds^2=\eta_{\mu\nu}dx^\mu dx^\nu\rightarrow ds'^{\ 2} = \frac{1}{(1+2ax+a^2x^2)^2}\eta_{\mu\nu}dx^\nu dx^\nu
\end{equation}\

therefore, the metric transform as $\eta\rightarrow g'=\frac{1}{(1+2ax+a^2x^2)^2}\eta$, resulting in a conformal transformation with $\Omega(x)=\frac{1}{1+2ax+a^2x^2}$.\par
In summary, there is a natural generalization of Poincaré transformation called conformal transformations, which under a transformation of the coordinates $x\rightarrow x'=x'(x)$, the metric is left invariant up to an overall factor $\eta\rightarrow g'=\Omega^2(x)\eta$, and where the set of all transformations is given by:

\begin{align}\label{ct2}
    \text{Poincaré:}\hspace{0.5cm} &x'=\Lambda x + a,\hspace{2.1cm} \Omega(x)=1\nonumber\\
    \text{Rescaling:}\hspace{0.5cm} &x'=\lambda x,\hspace{2.9cm} \Omega(x)=\lambda\nonumber\\
    \text{Inversion:}\hspace{0.5cm} &x'^\mu=\frac{x^\mu}{x^2},\hspace{2.7cm} \Omega(x)=\frac{1}{x^2}\nonumber\\
    \text{SCT:}\hspace{0.5cm} &x'^\mu=\frac{x^\mu+a^\mu x^2}{1+2ax+a^2x^2},\hspace{0.5cm} \Omega(x)=\frac{1}{1+2ax+a^2x^2}
\end{align}\

where $\Lambda$ are the Lorentz matrices, $a$ is a constant vector and $\lambda$ is a constant scalar.

\subsection{Conformal Group}

What will be very enlightening when we study AdS spaces and their isometries, eventually leading to an expected correspondence, is the study of the group formed by the conformal transformations. The corresponding generators of the conformal algebra can be read from the transformations just derived (eq. (\ref{ct2}) just by looking at their infinitesimal version. Take for example infinitesimal translations, i.e., Poincaré transformations with $\Lambda = 1$ and $a$ small. The resulting transformation can be written as:

\begin{equation}
    x'^\mu \equiv (1+ia^\nu P_\nu)x^\mu
\end{equation}\

where we defined the translation generator $P_\nu=-i\partial_\nu$. In the same way for small rotations and boosts, small rescaling and small SCT, the resulting generators in each case are:

\begin{align}\label{cgen}
    \text{Translations:}&\hspace{0.5cm}P_\mu = -i\partial_\mu\nonumber\\
    \text{Rot. + Boosts:}&\hspace{0.5cm}L_{\mu\nu} = i(x_\mu\partial_\nu-x_\nu\partial_\mu)\nonumber\\
    \text{Rescaling:}&\hspace{0.5cm}D = -ix^\mu\partial_\mu\nonumber\\
    \text{SCT:}&\hspace{0.5cm}K_\mu = i(2x_\mu x^\nu\partial_\nu-x^2\partial_\mu)
\end{align}\

Notice that there is no generator for inversions. This is because inversions are discrete transformations, there isn't a parameter controlling the magnitude of these transformations that we can continuously vary to perform "small" inversions. With the generators at hand, the construction of the conformal algebra is straightforward, obtaining for every pair of commutators:

\begin{align}\label{calg}
    &[P_\mu,D] = -iP_\mu, \hspace{0.5cm} [P_\mu,L_{\alpha\beta}] = i(\eta_{\mu\alpha}P_\beta-\eta_{\mu\beta}P_\alpha), \hspace{0.5cm} [P_\mu,K_\nu] = 2i(\eta_{\mu\nu}D+L_{\mu\nu})\nonumber\\
    &\nonumber\\
    &[D,K_\mu] = -iK_\mu, \hspace{0.4cm} [L_{\mu\nu},K_\alpha] = i(\eta_{\nu\alpha}K_\mu-\eta_{\mu\alpha}K_\nu)\nonumber\\
    &\nonumber\\
    &[L_{\mu\nu},L_{\alpha\beta}] = i(\eta_{\nu\alpha}L_{\mu\beta} + \eta_{\mu\beta}L_{\nu\alpha} - \eta_{\mu\alpha}L_{\nu\beta} - \eta_{\nu\beta}L_{\mu\alpha})
\end{align}\

Any other commutator not listed here is simply zero. The conformal algebra written in this form is not very revealing so let us define a more interesting set of generators which will explicitly show the isomorphism of the conformal group to a more familiar one. First, let us quickly generalize the 4 spacetime dimensions we have been implicitly assuming to an arbitrary number $d$ ($\mu = 0,\dotsc,3 \rightarrow \mu=0,\dotsc,d-1$). Instead of the generators $P_\mu$, $L_{\mu\nu}$, $D$ and $K_\mu$ defined in eq. (\ref{cgen}) in $d$ dimensions, consider a new set $J_{ab}$ in $d+2$ dimensions defined in term of the previous generators as:

\begin{align}
    J_{\mu\nu} = L_{\mu\nu}, \hspace{0.5cm} J_{(d+1)d} = D, \hspace{0.5cm} J_{(d+1)\mu} = \frac{1}{2}(P_\mu+K_\mu), \hspace{0.5cm} J_{d\mu} = \frac{1}{2}(P_\mu-K_\mu)
\end{align}\

where $J_{ab}=-J_{ba}$, $a,b = 0,\dotsc,d+1$ and $\eta=\text{diag}(-1,1,\dotsc,1,-1)$. It is a nice exercise to check that the resulting conformal algebra (eq. (\ref{calg})) for the conformal generators written in terms of the new generators $J_{ab}$ simply reduces to:

\begin{equation}
    [J_{ab},J_{cd}] = i(\eta_{ad}J_{bc} + \eta_{bc}J_{ad} - \eta_{ac}J_{bd} - \eta_{bd}J_{ac})
\end{equation}\

But this is the same closed algebra followed by the generators of the Lorentz transformations $L_{\mu\nu}$! The conformal group in $d$ dimensions can be seen as the set of rotations and boosts, i.e. Lorentz transformations, in $d+2$ dimensions. In group terminology it is said that the conformal group in $d$ dimensions is isomorphic to $SO(d,2)$.

\subsection{Conformal Invariance in QFT}

Conformal transformations are not an isometry of flat space, since they add to the metric these $\Omega(x)$ factors which are not necessarily equal to 1. This would seem to suggest that any conformal study done in a non-gravitational system is a merely theoretical exploration, without much practical use. This is of course not the case. There are many interesting, real phenomena which in practice can be modeled as a conformal theory. Since these theories, by definition, are invariant under the full conformal group, in particular for rescaling of the coordinates, these are theories which usually look the same at every length scale. This observation gives a nice recipe for constructing conformal theories: simply don't consider any energy scale-dependent parameter, i.e., mass terms, dimensionful coupling constants, etc. Take for example a massless $\lambda\Phi^4$ theory in $d=4$:

\begin{equation}
    S = \int d^4x\ \Bigl(-\frac{1}{2}\eta^{\mu\nu}\partial_\mu\Phi\partial_\nu\Phi-\frac{\lambda}{4!}\Phi^4\Bigl)
\end{equation}\

Since we are working in units where $c=\hbar=1$, then $[S]=1$, $[x]=[E]^{-1}$, and therefore the units of the field and coupling constant are $[\Phi]=[E]$ and $[\lambda]=1$. The only parameter present in the action, $\lambda$, is dimensionless so the theory is conformal. One of the implications of field theories which are conformal is that under any conformal transformation of the coordinates, a simple redefinition of the field can bring back the action to its original form. Take for example the same action as before under a rescaling:

\begin{equation}
    x \rightarrow x'= cx,\hspace{0.5cm} S\rightarrow S'=\int d^4x\ \Bigl(-\frac{1}{2}\eta^{\mu\nu}\partial_\mu\bigl[c\Phi'(cx)\bigl]\partial_\nu\bigl[c\Phi'(cx)\bigl]-\frac{\lambda}{4!}\bigl[c\Phi'(cx)\bigl]^4\Bigl)
\end{equation}\

In this case we can recover the original form of the action by simply defining $c\Phi'(cx)\equiv\Phi(x)$. Notice that this implies a very specific rule of transformations for the fields under the conformal group. In general, a field is called primary if under a conformal transformation $x\rightarrow x'=x'(x)$, $\eta\rightarrow g'=\Omega^2(x)\eta$, it transforms as:

\begin{equation}\label{primtr}
    \Phi'(x') = \Omega^{-\Delta}(x)\Phi(x)
\end{equation}\

where the number $\Delta$ is called the scaling dimension of the field. In our previous example, since for rescaling $\Omega(x)=c$ we can rewrite the redefinition of the field as $\Phi'(x')=\Omega^{-1}(x)\Phi(x)$, therefore the massless $\lambda\Phi^4$ field in $d=4$ has a scaling dimension $\Delta = 1$.\par
In QFT one is usually interested in computing correlation functions, the most basic objects of a theory, since in these are contained the amplitudes of all the different processes that can occur. The functional form of these correlators is highly dependant on the particular symmetries of the theory under consideration. For instance, generally one is interested in relativistic theories which are invariant under Poincaré transformations, and this invariance is promoted to the correlators themselves: 2-point correlation functions must be a function of the distance between the 2 spacetime coordinates under consideration, and so on. This is where all the power and richness of CFTs come into play. Invariance under the conformal group is so restrictive that in some cases one can completely determine the functional form of the correlators! Take for example the n-point function of n primary scalar fields $O_{\Delta_i}(x_i)$ with scaling dimensions $\Delta_i$:

\begin{equation}\label{cftnpn}
    \langle O_{\Delta_1}(x_1)\dotsm O_{\Delta_n}(x_n)\rangle = \frac{\int DO\ O_{\Delta_1}(x_1)\dotsm O_{\Delta_n}(x_n)e^{iS}}{\int DO\ e^{iS}}
\end{equation}\

where $S=S[O_{\Delta_1},\dots,O_{\Delta_n}]$ is a conformally invariant action. Under a conformal transformation, the quantity $DO\ e^{iS}$ is invariant and the fields transform as eq. (\ref{primtr}), which translates into the transformation rule for the correlators:

\begin{equation}\label{corrt}
    \langle O_{\Delta_1}(x_1')\dotsm O_{\Delta_n}(x_n')\rangle = \Omega^{-\Delta_1}(x_1)\dotsm\Omega^{-\Delta_n}(x_n)\langle O_{\Delta_1}(x_1)\dotsm O_{\Delta_n}(x_n)\rangle
\end{equation}\

Let's see how in some cases eq. (\ref{corrt}) can fully determine the form of the n-point functions. Starting with $n=1$:

\begin{equation}\label{corrtn1}
    \langle O_{\Delta_1}(x_1')\rangle = \Omega^{-\Delta_1}(x_1)\langle O_{\Delta_1}(x_1)\rangle
\end{equation}\

For a translation $x_1' = x_1+a$, $\Omega(x_1)=1$, and rescaling $x_1'=\lambda x_1$, $\Omega(x_1)=\lambda$, eq. (\ref{corrtn1}) reduces in each case to:

\begin{equation}\label{transn1}
    \langle O_{\Delta_1}(x_1+a)\rangle = \langle O_{\Delta_1}(x_1)\rangle
\end{equation}

\begin{equation}\label{rescn1}
    \langle O_{\Delta_1}(\lambda x_1)\rangle = \lambda^{-\Delta_1}\langle O_{\Delta_1}(x_1)\rangle
\end{equation}\

Now, eq. (\ref{transn1}) alone implies that the 1-point function doesn't depend on the spacetime coordinate $x_1$, i.e., it is a constant $\langle O_{\Delta_1}(x_1)\rangle = c$, and this result in eq. (\ref{rescn1}) implies that $c = \lambda^{-\Delta_1}c$. Since this equality must hold for any values of $\lambda$ and $\Delta_1$, it must be that $c=0$. In other words:

\begin{equation}
    \langle O_{\Delta_1}(x_1)\rangle = 0
\end{equation}\

For $n=2$ the strategy to determine its functional form is the same. From eq. (\ref{corrt}) the 2-point function transform as:

\begin{equation}\label{corrtn2}
    \langle O_{\Delta_1}(x_1')O_{\Delta_2}(x_2')\rangle = \Omega^{-\Delta_1}(x_1)\Omega^{-\Delta_2}(x_2)\langle O_{\Delta_1}(x_1)O_{\Delta_2}(x_2)\rangle
\end{equation}\

For a Poincaré transformation $x_i'=\Lambda x_i+a$, $\Omega(x_i)=1$, and rescaling $x_i'=\lambda x_i$, $\Omega(x_i)=\lambda$, eq. (\ref{corrtn2}) reduces in each case to:

\begin{equation}\label{poincn2}
    \langle O_{\Delta_1}(\Lambda x_1+a)O_{\Delta_2}(\Lambda x_2+a)\rangle = \langle O_{\Delta_1}(x_1)O_{\Delta_2}(x_2)\rangle
\end{equation}

\begin{equation}\label{rescn2}
    \langle O_{\Delta_1}(\lambda x_1)O_{\Delta_2}(\lambda x_2)\rangle = \lambda^{-\Delta_1-\Delta_2}\langle O_{\Delta_1}(x_1)O_{\Delta_2}(x_2)\rangle
\end{equation}\

Now, eq. (\ref{poincn2}) alone implies that the 2-point function must be a function of the spacetime distance between $x_1$ and $x_2$, i.e., $\langle O_{\Delta_1}(x_1)O_{\Delta_2}(x_2)\rangle = f(\lvert x_1-x_2\rvert)$, and this result in eq. (\ref{rescn2}) implies that $f(\lambda\lvert x_1-x_2\rvert)=\lambda^{-\Delta_1-\Delta_2}f(\lvert x_1-x_2\rvert)$. Since this equality must hold for any values of $\lambda$ and $\Delta_i$, it must be that $f(\lvert x_1-x_2\rvert)=\frac{C_{\Delta_1,\Delta_2}}{\lvert x_1-x_2\rvert^{\Delta_1+\Delta_2}}$. In other words:

\begin{equation}\label{2pt1}
    \langle O_{\Delta_1}(x_1)O_{\Delta_2}(x_2)\rangle = \frac{C_{\Delta_1,\Delta_2}}{\lvert x_1-x_2\rvert^{\Delta_1+\Delta_2}}
\end{equation}\

where $C_{\Delta_1,\Delta_2}$ is some constant dependent on the scaling dimensions $\Delta_i$ but not on the spacetime coordinates $x_i$. Poincaré and rescaling transformations were sufficient to determine the functional form of the 2-point function, but actually there is one more piece of information that we can acquire through the exotic SCT. First, let us rewrite the transformation rule for the 2-point function (eq. (\ref{corrtn2})) in terms of eq. (\ref{2pt1}):

\begin{equation}\label{2pt2}
    \frac{1}{\lvert x_1'-x_2'\rvert^{\Delta_1+\Delta_2}} = \Omega^{-\Delta_1}(x_1)\Omega^{-\Delta_2}(x_2)\frac{1}{\lvert x_1-x_2\rvert^{\Delta_1+\Delta_2}}
\end{equation}\

It can be shown that under a conformal transformation, the distance between two points, $x_1$ and $x_2$, transform as:

\begin{equation}\label{distancetransf}
    \lvert x_1'-x_2'\rvert^2 = \Omega(x_1)\Omega(x_2)\lvert x_1-x_2\rvert^2
\end{equation}\

It is easy to check that this is indeed true case by case. Now, using this fact in eq. (\ref{2pt2}) it implies that:

\begin{equation}
    \frac{1}{\lvert x_1-x_2\rvert^{\Delta_1+\Delta_2}}\Omega^{-\frac{\Delta_1+\Delta_2}{2}}(x_1)\Omega^{-\frac{\Delta_1+\Delta_2}{2}}(x_2) = \Omega^{-\Delta_1}(x_1)\Omega^{-\Delta_2}(x_2)\frac{1}{\lvert x_1-x_2\rvert^{\Delta_1+\Delta_2}}
\end{equation}\

Then for 2-point functions under any conformal transformation it must be that:

\begin{equation}
    \Omega^{-\frac{\Delta_1+\Delta_2}{2}}(x_1)\Omega^{-\frac{\Delta_1+\Delta_2}{2}}(x_2) = \Omega^{-\Delta_1}(x_1)\Omega^{-\Delta_2}(x_2)
\end{equation}\

This is trivially satisfied by Poincaré and rescaling transformations where the quantities $\Omega(x_i)$ don't depend on the spacetime points $x_i$, but for SCT (and inversions) they do depend on $x_i$ and therefore the only possibility to satisfy this condition is if $\frac{\Delta_1+\Delta_2}{2} = \Delta_1$ and $\frac{\Delta_1+\Delta_2}{2} = \Delta_2$, i.e., if $\Delta_1=\Delta_2$. This further restricts the form of the 2-point function to:

\begin{equation}
    \langle O_{\Delta_1}(x_1)O_{\Delta_2}(x_2)\rangle = \frac{C_{\Delta_1,\Delta_2}\delta_{\Delta_1,\Delta_2}}{\lvert x_1-x_2\rvert^{\Delta_1+\Delta_2}}
\end{equation}\

where $\delta_{\Delta_1,\Delta_2}$ is the Kronecker delta.\par
For $n=3$, in exactly the same way as we did for the $n=2$ case, it can be shown that the functional form of the 3-point function can be determined to be:

\begin{equation}
    \langle O_{\Delta_1}(x_1)O_{\Delta_2}(x_2)O_{\Delta_3}(x_3)\rangle = \frac{C_{\Delta_1,\Delta_2,\Delta_3}}{\lvert x_1-x_2\rvert^{\Delta_1+\Delta_2-\Delta_3}\lvert x_2-x_3\rvert^{\Delta_2+\Delta_3-\Delta_1}\lvert x_3-x_1\rvert^{\Delta_3+\Delta_1-\Delta_2}}
\end{equation}\

where $C_{\Delta_1,\Delta_2,\Delta_3}$ is some constant dependent only on the scaling dimensions $\Delta_i$.\par
Now, for $n\geq4$ the study is more subtle. Nothing stops us from addressing the case $n=4$ in the same way as we did for $n=2$ and $n=3$, and if we did we would naively conclude that the functional form of the 4-point function must be:

\begin{equation}\label{4pt}
    \langle O_{\Delta_1}(x_1)O_{\Delta_2}(x_2)O_{\Delta_3}(x_3)O_{\Delta_4}(x_4)\rangle = \frac{C_{\Delta_1,\Delta_2,\Delta_3,\Delta_4}}{\prod_{i<j}\lvert x_i-x_j\rvert^{\Delta_i+\Delta_j-\frac{\sum_k\Delta_k}{3}}}
\end{equation}\

But actually for $n\geq4$, where we have 4 or more spacetime coordinates at disposal, we can start constructing conformal invariants, known as anharmonic or cross ratios, and the n-point functions can have an arbitrary dependence of these which are not determined by eq. (\ref{corrt}). In particular for $n=4$, where we have 4 coordinates to play with, we can construct two different cross ratios:

\begin{equation}\label{crossr}
    u = \frac{\lvert x_1-x_2\rvert^2\lvert x_3-x_4\rvert^2}{\lvert x_1-x_3\rvert^2\lvert x_2-x_4\rvert^2},\hspace{1cm} v = \frac{\lvert x_1-x_4\rvert^2\lvert x_2-x_3\rvert^2}{\lvert x_1-x_3\rvert^2\lvert x_2-x_4\rvert^2}
\end{equation}\

And the general form of the 4-point function is given by eq. (\ref{4pt}) up to some function of these invariants:

\begin{equation}
    \langle O_{\Delta_1}(x_1)O_{\Delta_2}(x_2)O_{\Delta_3}(x_3)O_{\Delta_4}(x_4)\rangle = \frac{f(u,v)}{\prod_{i<j}\lvert x_i-x_j\rvert^{\Delta_i+\Delta_j-\frac{\sum_k\Delta_k}{3}}}
\end{equation}\

In summary, CFTs are interesting theories with no dimensionful coupling constants in their actions and therefore they usually look the same at every length scale. Since the form of the correlators are sensible to the symmetries of the theory, and conformal symmetry is highly restrictive, in some cases one can completely determine their functional form. The resulting 1-, 2-, 3- and 4-point functions of primary scalar fields $O_{\Delta_i}(x_i)$ with scaling dimensions $\Delta_i$ are:

\begin{align}\label{summcorr}
    \text{1-pt fn:}&\hspace{0.5cm} \langle O_{\Delta_1}(x_1)\rangle =0\nonumber\\
    \text{2-pt fn:}&\hspace{0.5cm} \langle O_{\Delta_1}(x_1)O_{\Delta_2}(x_2)\rangle = \frac{C_{\Delta_1,\Delta_2}\delta_{\Delta_1,\Delta_2}}{\lvert x_1-x_2\rvert^{\Delta_1+\Delta_2}}\nonumber\\
    \text{3-pt fn:}&\hspace{0.5cm} \langle O_{\Delta_1}(x_1)O_{\Delta_2}(x_2)O_{\Delta_3}(x_3)\rangle = \frac{C_{\Delta_1,\Delta_2,\Delta_3}}{\lvert x_1-x_2\rvert^{\Delta_1+\Delta_2-\Delta_3}\lvert x_2-x_3\rvert^{\Delta_2+\Delta_3-\Delta_1}\lvert x_3-x_1\rvert^{\Delta_3+\Delta_1-\Delta_2}}\nonumber\\
    \text{4-pt fn:}&\hspace{0.5cm} \langle O_{\Delta_1}(x_1)O_{\Delta_2}(x_2)O_{\Delta_3}(x_3)O_{\Delta_4}(x_4)\rangle = \frac{f(u,v)}{\prod_{i<j}\lvert x_i-x_j\rvert^{\Delta_i+\Delta_j-\frac{\sum_k\Delta_k}{3}}}
\end{align}\

where $C_{\Delta_1,\Delta_2}$ and $C_{\Delta_1,\Delta_2,\Delta_3}$ are some constants dependent only on $\Delta_i$ and $f(u,v)$ is some function of the cross ratios $u$ and $v$ (eq. (\ref{crossr})).\par

In the particular case where one is interested in computing the n-point functions of the same primary scalar field of scaling dimension $\Delta$, they can be obtained from eq. (\ref{summcorr}) by simply taking $\Delta_i=\Delta$, resulting in:

\begin{align}\label{summcorr2}
    \text{1-pt fn:}&\hspace{0.5cm} \langle O_{\Delta}(x_1)\rangle =0\nonumber\\
    \text{2-pt fn:}&\hspace{0.5cm} \langle O_{\Delta}(x_1)O_{\Delta}(x_2)\rangle = \frac{C_\Delta}{\lvert x_1-x_2\rvert^{2\Delta}}\nonumber\\
    \text{3-pt fn:}&\hspace{0.5cm} \langle O_{\Delta}(x_1)O_{\Delta}(x_2)O_{\Delta}(x_3)\rangle = \frac{\tilde{C}_\Delta}{\lvert x_1-x_2\rvert^{\Delta}\lvert x_2-x_3\rvert^{\Delta}\lvert x_3-x_1\rvert^{\Delta}}\nonumber\\
    \text{4-pt fn:}&\hspace{0.5cm} \langle O_{\Delta}(x_1)O_{\Delta}(x_2)O_{\Delta}(x_3)O_{\Delta}(x_4)\rangle = \frac{f(u,v)}{\prod_{i<j}\lvert x_i-x_j\rvert^{\frac{2\Delta}{3}}}
\end{align}\

where $C_\Delta$ and $\tilde{C}_\Delta$ are some constants dependent only on $\Delta$.

\subsection{Conformal Source}

In practice, how does one compute CFT n-point functions of the form, say, eq. (\ref{cftnpn})? The strategy that one follows is the same as in regular QFT, one simply add to the generating functional $Z$ external sources $J_{\Delta_i}$ coupled to every operator $O_{\Delta_i}$ that one wishes to compute their correlators: 

\begin{equation}\label{gfcft}
    Z[J_{\Delta_1},\dotsc,J_{\Delta_n}] = \int DO\ e^{iS+i\int d^dx\ O_{\Delta_1}(x)J_{\Delta_1}(x)+\dotsb+i\int d^dx\ O_{\Delta_n}(x)J_{\Delta_n}(x)}
\end{equation}\

and then, performing the resulting integral on the operators $O$ with the use in most cases of perturbation theory, the computation of the n-point functions is reduced to the act of taking derivatives with respect to the external sources and then setting them to 0:

\begin{equation}
    \langle O_{\Delta_1}(x_1)\dotsm O_{\Delta_n}(x_n)\rangle = \frac{(-i)^n}{Z[0,\dotsc,0]}\frac{\delta^nZ[J_{\Delta_1},\dotsc,J_{\Delta_n}]}{\delta J_{\Delta_1}(x_1)\dotsm\delta J_{\Delta_n}(x_n)}\Bigl\rvert_{J_{\Delta_i}=0}
\end{equation}\

The idea we want to explore in this section is what are the consequences of the transformations rules for the primary operators $O_{\Delta_i}$ (eq. (\ref{primtr})) on the external sources $J_{\Delta_i}$ themselves. We know that for a conformal theory, the generating functional $Z$ is invariant under conformal transformations. Since both $DO$ and $e^{iS}$ in eq. (\ref{gfcft}) are separately invariant, this implies that every coupling between the operators $O_{\Delta_i}$ and their corresponding sources $J_{\Delta_i}$ must also be invariant:

\begin{equation}\label{ojcoup}
    \int d^dx'\ O'_{\Delta_i}(x')J'_{\Delta_i}(x') = \int d^dx\ O_{\Delta_i}(x)J_{\Delta_i}(x)
\end{equation}\

but we know how the operator and the measure transforms. The operator transform according to eq. (\ref{primtr}), $O'_{\Delta_i}(x')=\Omega^{-\Delta_i}O_{\Delta_i}(x)$, and the measure transform as $d^dx'=\sqrt{-\text{det}(g)}d^dx$, where $\sqrt{-\text{det}(g)}$ is the Jacobian of the transformation, but for a conformal transformation $\eta\rightarrow\Omega^2(x)\eta$, thus the measure transform as $d^dx'=\Omega^d(x)d^dx$. Therefore, eq. (\ref{ojcoup}) implies that:

\begin{equation}
    \int d^dx\ \Omega^{d-\Delta_i}(x)O_{\Delta_i}(x)J'_{\Delta_i}(x') = \int d^dx\ O_{\Delta_i}(x)J_{\Delta_i}(x)
\end{equation}\

For this equality to hold, the external sources must transform according to the rule:

\begin{equation}\label{confso}
    J'_{\Delta_i}(x') = \Omega^{-(d-\Delta_i)}(x)J_{\Delta_i}(x)
\end{equation}\

In other words, the corresponding sources $J_{\Delta_i}$ of the primary operators $O_{\Delta_i}$ of scaling dimensions $\Delta_i$ are also primary fields, with scaling dimensions $d-\Delta_i$. The realization of this fact will be very important in the last section of this chapter, where we will gather all the evidence to naturally state the AdS/CFT correspondence.


\section{Anti-de Sitter Space}

\subsection{Anti-de Sitter Geometry}

How can we describe the 2-dimensional surface of a sphere? There are of course many ways to do it, but one that is quite simple is acknowledging that it is the result of a sphere living in a flat space with one additional space-like dimension:

\begin{equation}
    ds^2 = dx^2+dy^2+dz^2,\hspace{1cm}x^2+y^2+z^2=R^2
\end{equation}\

where $R$ is the constant radius of the sphere. The spherical coordinates, by construction, not only satisfy the sphere equation but also it brings the line element $ds^2$ to the simple form:

\begin{equation}
    ds^2 = R^2(d\theta^2+\sin^2{\theta}d\phi^2)
\end{equation}\

which is the line element of a 2-dimensional surface of a sphere of radius R, as desired.\par
In exactly the same way, we can describe Anti-de Sitter space in $d+1$ spacetime dimensions as the embedding of an hyperboloid of radius $\ell$ in a flat, one higher time-like dimension spacetime:

\begin{equation}\label{hyperb}
    ds^2 = -dT_1^2-dT_2^2+d\vec{X_i}^2,\hspace{1cm} -T_1^2-T_2^2+\vec{X_i}^2=-\ell^2
\end{equation}\

where $i=1,\dotsc,d$. A nice set of coordinates which describe this hyperboloid are known as the Poincaré coordinates:

\begin{align}
    T_1 &= \frac{r}{\ell}t,\hspace{1cm} T_2 = \frac{\ell^2}{2r}\Bigl[1+\frac{r^2}{\ell^4}(\ell^2+\vec{x_i}^2-t^2)\Bigl]\nonumber\\
    X_i &= \frac{r}{\ell}x_i,\hspace{0.7cm} X_d = \frac{\ell^2}{2r}\Bigl[1-\frac{r^2}{\ell^4}(\ell^2-\vec{x_i}^2+t^2)\Bigl]
\end{align}\

where $x_i,t \in \mathbb{R}$, $r>0$ and now $i=1,\dotsc,d-1$. It is a nice exercise to check that these coordinates not only satisfy the hyperboloid equation but also they bring the metric to the simple form:

\begin{equation}
    ds^2 = \frac{\ell^2}{r^2}dr^2+\frac{r^2}{\ell^2}(-dt^2+d\vec{x_i}^2)
\end{equation}\

and if we further define $r=\frac{\ell^2}{x_0}$, $x_0>0$:

\begin{equation}
    ds^2 = \ell^2\Bigl(\frac{dx_0^2-dt^2+d\vec{x_i}^2}{x_0^2}\Bigl)
\end{equation}\

from where we can read the metric:

\begin{equation}\label{adsg}
    ds^2=g_{\mu\nu}dx^\mu dx^\nu,\hspace{1cm} g_{\mu\nu}=\frac{\ell^2}{x_0^2}\eta_{\mu\nu}
\end{equation}\

where $\mu,\nu=0,\dotsc,d$, $t\equiv x^{d}$ and $\eta=\text{diag}(1,\dotsc,1,-1)$. This is known as the AdS$_{d+1}$ metric in Poincaré coordinates.

\subsection{Einstein's Equations}

Great insight on AdS spaces can be gained by looking at the Einstein's equations satisfied by their metric (eq. \ref{adsg}). From this metric, it is a matter of direct calculation to compute the corresponding Christoffel symbols, obtaining:

\begin{equation}
    \Gamma^\rho_{\mu\nu} = \frac{1}{x_0}(\eta_{\mu\nu}\eta^{\rho 0}-\delta^\rho_\nu\delta^0_\mu-\delta^\rho_\mu\delta^0_\nu)
\end{equation}\

and from these directly follow the Riemann tensor:

\begin{equation}
    R^\rho_{\sigma\mu\nu} = \frac{1}{x_0^2}(\delta^\rho_\nu\eta_{\sigma\mu}-\delta^\rho_\mu\eta_{\sigma\nu})
\end{equation}\

the Ricci tensor:

\begin{equation}\label{ricci1}
    R_{\sigma\nu} = -\frac{d}{x_0^2}\eta_{\sigma\nu}
\end{equation}\

and the Ricci scalar:

\begin{equation}\label{ricci2}
    R = -\frac{d(d+1)}{\ell^2}
\end{equation}\

These quantities have the exact form expected for spaces that are maximally symmetric, i.e., spaces that are both homogeneous and isotropic. The Ricci scalar quantifies the intrinsic curvature at each point in space. Since in this case it is a negative constant, each point in AdS spaces is equally curved.\par
Maximally symmetric spaces are a solution to Einstein's equation in vacuum with the presence of a cosmological constant:

\begin{equation}
    R_{\mu\nu}-\frac{1}{2}Rg_{\mu\nu}+\Lambda g_{\mu\nu} = 0
\end{equation}\

From eqs. (\ref{ricci1}) and (\ref{ricci2}) it is direct to check that AdS spaces are indeed a solution to Einstein's equations with a cosmological constant value of:

\begin{equation}
    \Lambda = -\frac{1}{2}\frac{d(d-1)}{\ell^2}
\end{equation}

AdS spaces then can be understood as maximally symmetric solutions to Einstein's equations in vacuum with negative cosmological constant. Each point in space is equally curved, and for a given dimension, this curvature depends only on one parameter, the radius $\ell$ of the hyperboloid. As this radius gets larger, each region of AdS spaces becomes flatter.

\subsection{Anti-de Sitter Isometries}

What are the isometries of AdS spaces, i.e., the coordinate transformations that leave the AdS metric (eq. (\ref{adsg})) invariant? The answer to this question becomes very clear when looking at the hyperboloid equation (\ref{hyperb}). It is just the set of transformations that leave the hyperboloid equation invariant. Indeed, if we came up with a new set of coordinates that satisfy eq. (\ref{hyperb}), then following the same steps we did previously the resulting metric is, again, eq. (\ref{adsg}). But it is quite obvious what are these transformations. By definition, they are just the elements of the group $SO(d,2)$. But we already met this group before, it is the conformal group! Therefore, we expect that exactly the same set of transformations (Poincaré, rescaling, inversion and SCT, eq. (\ref{ct2})) to be isometry transformations for AdS spaces, with the condition that, since the AdS metric has explicit dependence on $x_0$, this coordinate doesn't get rotated nor translated. Therefore, the set of isometry transformations for AdS spaces are:

\begin{align}\label{adsiso}
    \text{Poincaré:}\hspace{1cm}& x'=\Lambda x+a\nonumber\\
    \text{Rescaling:}\hspace{1cm}& x' = \lambda x\nonumber\\
    \text{Inversion:}\hspace{1cm}& x'^\mu=\frac{x^\mu}{x^2}\nonumber\\
    \text{SCT:}\hspace{1cm}& x'^\mu = \frac{x^\mu+a^\mu x^2}{1+2ax+a^2x^2}
\end{align}\

where $\Lambda^0_\nu=\delta^0_\nu$ and $a^0=0$. It is easy to check that these transformations are isometries of AdS spaces. Indeed, since they are conformal transformations for flat spaces, they transform the flat part of the line element as $\eta_{\mu\nu}dx^\mu dx^\nu\rightarrow \Omega^2(x)\eta_{\mu\nu}dx^\mu dx^\nu$, which together with the resulting transformation rule for the $x_0$ coordinate:

\begin{equation}\label{x0transf}
    x_0'=\Omega(x)x_0
\end{equation}\

result in the same original AdS metric eq. (\ref{adsg}), as it should.

\subsection{Conformal Boundary}

Let us remember the AdS metric in Poincaré coordinates:

\begin{equation}
    ds^2 = \ell^2\Bigl(\frac{dx_0^2-dt^2+d\vec{x_i}^2}{x_0^2}\Bigl)
\end{equation}\

where $x_i,t\in\mathbb{R}$, and $x_0>0$. Notice that for fixed values of $x_0$ the resulting metric is, up to some overall constant, flat:

\begin{equation}
    ds^2\rvert_{x_0}=\frac{\ell^2}{x_0^2}(-dt^2+d\vec{x_i}^2)
\end{equation}\

There are two interesting limits of this metric, namely the two extreme values of the coordinate $x_0$. As we take larger values of $x_0$ all the spacetime starts collapsing into a point, and as we take smaller values the spacetime starts growing infinitely big. Now, these regions at $x_0=0$ and $x_0=\infty$ are not part of the domain of AdS spaces, but in some very sensitive notion we can not only include and patch these regions to AdS spaces, but even more we can define these regions as their boundaries.\par
The patching at $x_0=\infty$ is simple, in this region we just have to add a point. The patching at $x_0=0$ however is more subtle since in this region the metric diverges. Since it diverges as a second order pole, it doesn't yield a specific metric but a whole conformal structure instead. Consider a function $f(x)$ with a first order pole at $x_0=0$, then the metric:

\begin{equation}
    ds'^2=f^2(x)ds^2\lvert_{x_0=0}
\end{equation}\

is finite in this region but defined up to conformal transformations. Indeed, if $f(x)$ renders the AdS metric finite at $x_0=0$, then so does it too $f(x)\Omega(x)$ where $\Omega(x)$ is some function with no zeros or poles at $x_0=0$. In particular, if we take the function $f(x)$ to be $f(x)=\frac{x_0}{\ell}$, the resulting metric at $x_0=0$ is:

\begin{equation}\label{adscb}
    ds'^2=\Omega^2(x)(-dt^2+d\vec{x_i}^2)
\end{equation}\

which is finite, of one lower dimension and defined up to conformal transformations. AdS spaces then contain the notion of boundaries at the extreme values of the $x_0$ coordinate: a point at $x_0=\infty$ and a whole conformal structure of one lower dimension at $x_0=0$ \cite{Kostas}.

\section{The AdS/CFT Correspondence}

So far we have studied the very basics of conformal field theories and anti-de Sitter spaces. Now, the motivation for doing this has been twofold: to familiarize ourselves with the concepts and objects that we will use in the next chapters, but also to show that, although they are apparently very different and disconnected topics, they actually share important features in common. For instance, the set of conformal transformations in flat $d$ dimensional spacetime (eq. (\ref{ct2})) and the set of isometry transformations in $d+1$ dimensional AdS spaces (eq. (\ref{adsiso})) is the same, and also that AdS spaces have the notion of a boundary at $x_0=0$ where a whole conformal structure of one lower dimension resides (eq. (\ref{adscb})).\par
There seems to be some hints then on a possible correspondence between theories in CFT$_d$ and AdS$_{d+1}$. The final decisive piece of evidence that we will present which might render this correspondence not only possible but also expected, which will be derived in full detail in the next chapter, it is the fact that completely solving the dynamics for some field $\Phi(x)$ on a AdS$_{d+1}$ background only determines its functional form up to some arbitrary function $\varphi_0(\vec{x})$ not dependent on the $x_0$ coordinate, that also happens to be a primary field of scaling dimension $d-\Delta$, where $\Delta$ is some number dependent on the dimension of the AdS space and the mass parameter of the field $\Phi(x)$. Now, this fact is extremely remarkable for two reasons: first, it implies that the generating functional $Z_{\text{AdS}}$ of some field theory on a AdS$_{d+1}$ background will be a functional of this arbitrary function $\varphi_0(\vec{x})$, and second, the scaling dimension of this arbitrary function is exactly the expected for a conformal source of some operator $O_\Delta(\vec{x})$ of scaling dimension $\Delta$ (eq. (\ref{confso}))! \par
Putting all the evidence together, the resulting picture is very clear if not obvious: the generating functional $Z_{\text{AdS}}[\varphi_0]$ of a field $\Phi(x)$ on a AdS$_{d+1}$ background correspond to the generating functional $Z_{\text{CFT}}[\varphi_0]$ of a CFT$_d$ of some operator $O_\Delta(\vec{x})$ of scaling dimension $\Delta$ living at the conformal boundary $x_0=0$ of the AdS space, where $\varphi_0(\vec{x})$, the arbitrary function not determined by the dynamics on AdS, acts as a conformal source for the primary operators $O_\Delta(\vec{x})$ \cite{Witten}. In other words:

\begin{equation}\label{adscftcorresp}
    Z_{\text{CFT}}[\varphi_0] = \int DO\ e^{iS_{\text{CFT}}[O]+i\int d^dx\ O_\Delta(\vec{x})\varphi_0(\vec{x})} = \int_{\Phi[\varphi_0]} D\Phi\ e^{iS_{\text{AdS}}[\Phi]} = Z_{\text{AdS}}[\varphi_0]
\end{equation}\

where $S_{\text{CFT}}$ is some conformal invariant theory in $d$ dimensions and $S_{\text{AdS}}$ is some isometric AdS invariant theory in $d+1$ dimensions. This is the AdS/CFT correspondence. It implies that we can obtain correlators for some CFT$_d$ theory using as a starting point a field theory on AdS$_{d+1}$:

\begin{equation}
    \langle O_\Delta(\vec{x_1})\dotsm O_\Delta(\vec{x_n})\rangle_{\text{CFT}} = \frac{(-i)^n}{Z_{\text{AdS}}[\varphi_0=0]}\frac{\delta^nZ_{\text{AdS}}[\varphi_0]}{\delta\varphi_0(\vec{x_1})\dotsm\delta\varphi_0(\vec{x_n})}\Bigl\rvert_{\varphi_0=0}
\end{equation}
\begin{equation}
    \langle O_\Delta(\vec{x_1})\dotsm O_\Delta(\vec{x_n})\rangle_{\text{CFT,con}} = (-i)^n \frac{\delta^n\log{Z_{\text{AdS}}[\varphi_0]}}{\delta\varphi_0(\vec{x_1})\dotsm\delta\varphi_0(\vec{x_n})}\Bigl\rvert_{\varphi_0=0}
\end{equation}\

Notice how highly non-trivial this equivalence between both theories is. Indeed, it is saying that certain theories with gravity can be completely understood as some particle theory without gravity living in one less dimension! Moreover, this duality between both theories is expected to be strong/weak in the sense that classical computations on the AdS side are expected to be related through this correspondence with strongly quantum interacting phenomena on the CFT side. It must be taken into account however that this particular equivalence between both theories is a conjecture. Proving it would seem to require solving both sides of eq. (\ref{adscftcorresp}) independently, showing in this way that they are exactly the same. Needless to say, this is extremely difficult to do if not impossible. For this reason is that in this work we will limit ourselves in checking that the correlators obtained through this correspondence for certain scalar theories on AdS precisely correspond to those of a CFT.

\newpage


\chapter{Classical Scalar Theories in AdS/CFT}
With the AdS/CFT correspondence explicitly stated, as a first exploration and familiarization into the topic we will proceed to study its consequences for what are perhaps the simplest cases, these are, scalar field theories on a fixed AdS background. The objective of this chapter then is to develop the necessary tools that will allow us to obtain from these theories, and through the use of what will be understood as the classical approximation of the AdS/CFT correspondence, the resulting approximated CFT n-point functions.\par
Section 3.1 will cover all these tools needed for the simplest scalar case, the free field, using as the main reference the article by Skenderis \cite{Kostas}. In particular, in section 3.1.1 we will present the explicit AdS free field action to be studied, in section 3.1.2 its Euclidean rotated version which will be the signature used throughout this work, in section 3.1.3 as a first approximation to these calculations we will present the classical approximation of the AdS/CFT correspondence, in section 3.1.4 the classical solution to the bulk field, in section 3.1.5 the required renormalization scheme for the divergences present at the on-shell level of the AdS path integral, in section 3.1.6 through the use of the approximated correspondence the resulting CFT correlators dual to the free field on AdS and finally in section 3.1.7 how these same holographic correlators can be obtained from a more straightforward approach through what is known as holographic dictionary.\par
Section 3.2 in turn will cover the additional tools needed for interacting scalar theories, more concretely a $\Phi^3$ theory, using as the main reference the previous article by Skenderis. In particular, in section 3.2.1 we will present the explicit AdS $\Phi^3$ action to be studied under the classical approximation of the correspondence, in section 3.2.2 the classical solution to the bulk field, in section 3.2.3 through the use of the approximated correspondence the resulting CFT correlators dual to the $\Phi^3$ theory on AdS and finally in section 3.2.4 how these same holographic correlators can be obtained from the holographic dictionary.\par
Lastly, section 3.3 will cover the same tools developed as in section 3.2 now applied to a different interacting scalar theory, a $\Phi^4$ theory, using as the main reference the same article as before. In particular, in section 3.3.1 we will present the explicit AdS $\Phi^4$ action to be studied under the classical approximation of the correspondence, in section 3.3.2 the classical solution to the bulk field, in section 3.3.3 through the use of the approximated correspondence the resulting CFT correlators dual to the $\Phi^4$ theory on AdS and finally in section 3.3.4 how these same holographic correlators can be obtained from the holographic dictionary.

\section{Free Scalar Field}

\subsection{Free Field Action}

In the previous chapter we reviewed some very basics ingredients of CFTs and AdS spaces which, putting them together in the last section, naturally led to the realization of the AdS/CFT correspondence:

\begin{equation}\label{adscftc}
    Z_{\text{CFT}}[\varphi_0] = \int DO\ e^{iS_{\text{CFT}}[O]+i\int d^dx\ O_\Delta(\vec{x})\varphi_0(\vec{x})} = \int_{\Phi[\varphi_0]} D\Phi\ e^{iS_{\text{AdS}}[\Phi]} = Z_{\text{AdS}}[\varphi_0]
\end{equation}\

Among these ingredients was the fact that the set of conformal transformations in $d$ dimensions and the set of AdS isometry transformations in $d+1$ dimensions is the same. This is explicitly encoded in the correspondence where under this set of transformations the CFT side of the correspondence is invariant due to conformal invariance and the AdS side is invariant due to AdS isometry invariance.\par
In this section we want to explore what simple AdS isometry invariant actions we can construct to later perform a more concrete study of the correspondence. The simplest actions we can construct are, of course, for scalar theories where the corresponding fields transform trivially under diffeomorphisms. For these cases the invariant action terms we can write down are of the form:

\begin{equation}
    S_{\text{AdS}}[\Phi]=\int d^{d+1}x\sqrt{-\text{det}(g)}\ \bigl[c_1\Phi(x)+c_2\Phi^2(x)+\dotsb+c_3g^{\mu\nu}\partial_\mu\Phi(x)\partial_\nu\Phi(x)+\dotsb\bigl]
\end{equation}\

where $\Phi(x)$ is some scalar field, $g$ is the AdS$_{d+1}$ metric and $c_i$ are some constants. Let us take what would be the most simple non-trivial case, a free scalar field on a AdS$_{d+1}$ background:

\begin{equation}\label{freefield}
    S_{\text{AdS}}[\Phi]=\int d^{d+1}x\sqrt{-\text{det}(g)}\ \Bigl[-\frac{1}{2}g^{\mu\nu}\partial_\mu\Phi(x)\partial_\nu\Phi(x)-\frac{1}{2}m^2\Phi^2(x)\Bigl]
\end{equation}\

Indeed, this is nothing more than the natural generalization of a free scalar field in flat 4 dimensions to $4\rightarrow d+1$ dimensions on a $\eta\rightarrow g$ AdS background. This is the first action we are going to consider for a more concrete study of the correspondence.

\subsection{Euclidean Rotation}

So far every expression we have written down has been on Lorentzian signature of the metric where its time-like component is negative. But now that we will delve deeper into the calculations, the resulting equations will turn out to be more manageable in the Euclidean signature where the time-like component of the metric is positive. This change in the signature of the metric can be obtained by doing a Wick rotation of the time-like coordinate $t\rightarrow -it$. This rotation allows us to write the AdS metric in Euclidean Poincaré coordinates simply as:

\begin{equation}\label{eucg}
    ds^2=g_{\mu\nu}dx^\mu dx^\nu,\hspace{1cm} g_{\mu\nu}=\frac{\ell^2}{x_0^2}\delta_{\mu\nu}
\end{equation}\

where $\delta_{\mu\nu}$ is the Kronecker delta and the coordinates are understood to be $x=(x_0,\dotsc,x_d\equiv t)$. Not only the metric is redefined under this time transformation, but every quantity dependent on the time-like coordinate. For instance, the free scalar field action (eq. (\ref{freefield})) that we will be considering:

\begin{equation}\label{eucact}
    S_{\text{AdS}}\rightarrow i\int d^{d+1}x\sqrt{g}\ \Bigl[\frac{1}{2}g^{\mu\nu}\partial_\mu\Phi(x)\partial_\nu\Phi(x)+\frac{1}{2}m^2\Phi^2(x)\Bigl]\equiv iS_{\text{AdS}}
\end{equation}\

where we called the Jacobian $\sqrt{g}\equiv\sqrt{\text{det}(g)}$ and redefined the AdS action. The AdS/CFT correspondence also has a small tweak in its Euclidean version:

\begin{equation}\label{eucadscft}
    Z_{\text{CFT}}[\varphi_0] = \int DO\ e^{-S_{\text{CFT}}[O]+\int d^dx\ O_\Delta(\vec{x})\varphi_0(\vec{x})} = \int_{\Phi[\varphi_0]} D\Phi\ e^{-S_{\text{AdS}}[\Phi]} = Z_{\text{AdS}}[\varphi_0]
\end{equation}\

where we also assumed an Euclidean rotation for the CFT action $S_{\text{CFT}}\rightarrow iS_{\text{CFT}}$. And finally, the resulting correlators in the Euclidean version of the correspondence allow us to save writing the "i" factors compared to the Lorentzian signature:

\begin{equation}\label{euccorr1}
    \langle O_\Delta(\vec{x_1})\dotsm O_\Delta(\vec{x_n})\rangle_{\text{CFT}} = \frac{1}{Z_{\text{AdS}}[\varphi_0=0]}\frac{\delta^nZ_{\text{AdS}}[\varphi_0]}{\delta\varphi_0(\vec{x_1})\dotsm\delta\varphi_0(\vec{x_n})}\Bigl\rvert_{\varphi_0=0}
\end{equation}

\begin{equation}\label{euccorr2}
    \langle O_\Delta(\vec{x_1})\dotsm O_\Delta(\vec{x_n})\rangle_{\text{CFT,con}} = \frac{\delta^n\log{Z_{\text{AdS}}[\varphi_0]}}{\delta\varphi_0(\vec{x_1})\dotsm\delta\varphi_0(\vec{x_n})}\Bigl\rvert_{\varphi_0=0}
\end{equation}\

From now on we will work exclusively on the Euclidean signature unless stated otherwise.

\subsection{Semiclassical Approximation}

The main objects we will be interested in computing using the AdS/CFT correspondence are correlators for some conformal theory (eqs. (\ref{euccorr1}) and (\ref{euccorr2})). The impressive prediction of the correspondence is that these can be obtained from a very different starting point, a theory on AdS space. Let us see how this is achieved in what is possibly the simplest yet non-trivial example, the free scalar field. Consider the AdS part of the correspondence together with the free scalar field action on Euclidean signature:

\begin{equation}
    Z_{\text{AdS}} = \int D\Phi\ e^{-S_{\text{AdS}}[\Phi]},\hspace{0.5cm} S_{\text{AdS}}[\Phi] = \int d^{d+1}x\sqrt{g}\ \Bigl[\frac{1}{2}g^{\mu\nu}\partial_\mu\Phi(x)\partial_\nu\Phi(x)+\frac{1}{2}m^2\Phi^2(x)\Bigl]
\end{equation}\

The natural way to approach this path integral is by looking at quantum fluctuations around the classical solution of the AdS action. This is done by the change of variable $\Phi(x)=\phi(x)+h(x)$, where $\phi(x)$ is the on-shell field and $h(x)$ carries the quantum fluctuations around it. Under this division the AdS path integral takes the form:

\begin{equation}
    Z_{\text{AdS}} = e^{-S_{\text{AdS}}[\phi]}f[\phi]
\end{equation}\

where $S_{\text{AdS}}[\phi]$ is the same free field action as before and $f[\phi]$ is some functional of the classical field $\phi(x)$ coupled to the quantum field $h(x)$. This functional leftover of the change of variable will be precisely the responsible for the quantum corrections to the CFT correlators and will be the main object of study of the next chapter. However, in this chapter we will be interested only on the on-shell contributions to the correlators and therefore to this end, without loss of information, we will truncate\footnote{Since the holographic correlators are obtained from the AdS path integral by taking derivatives on it, thanks to the product rule any contribution coming from the functional $f[\phi]$ will be additive to those obtained from the on-shell part of the path integral. Therefore what we are achieving by truncating this functional is to ignore these additive contributions to the correlators which will be of the order of the quantum fluctuations $h(x)$.} for the moment the functional $f[\phi]$ of $Z_{\text{AdS}}$ resulting simply in:

\begin{equation}
    Z_{\text{AdS}} = e^{-S_{\text{AdS}}[\phi]}
\end{equation}\

This is the classical or saddle point approximation of the AdS/CFT correspondence.

\subsection{Classical Solution}

To continue advancing in the computation of the correlators we need the explicit form of the on-shell field $\phi(x)$. We know it is the classical solution of the AdS action, i.e., it satisfies the Euler-Lagrange equation:

\begin{equation}\label{eom0}
    (-\Box+m^2)\phi(x)=0
\end{equation}\

where we defined the Laplace operator in curvilinear coordinates $\Box\equiv\frac{1}{\sqrt{g}}\partial_\mu(\sqrt{g}g^{\mu\nu}\partial_\nu)$. This is the expected wave equation on a curved space. In Euclidean Poincaré coordinates it can be explicitly written as:

\begin{equation}\label{eom1}
    -x_0^2\partial_0^2\phi(x)-(1-d)x_0\partial_0\phi(x)-x_0^2\partial_i^2\phi(x)+m^2\phi(x)=0
\end{equation}\

where we set the AdS radius $\ell$ in the metric equal to 1. From now on we will keep this value, meaning that every length or energy scale will be measured in units of the AdS radius. We will explore a couple of ways to solve this equation, the first one being an asymptotic approach through a power series solution in the $x_0$ coordinate:

\begin{equation}\label{eomserie}
    \phi(x) = \sum_{n=0}^\infty x_0^{n+\Delta}\varphi_n(\vec{x})
\end{equation}\

where the number $\Delta$ and the functions $\varphi_n(\vec{x})$ will be highly restricted from the equation of motion. In fact, since the wave equation is of second order we expect that the infinite number of functions $\varphi_n(\vec{x})$ simply reduce to 2 linearly independent ones. The form of eq. (\ref{eom1}) suggests that we can fix the number $\Delta$ by quickly looking at a solution of the form $\phi(x)=x_0^\Delta$. By doing this one obtains the condition:

\begin{equation}\label{delta1}
    x_0^\Delta\bigl[\Delta(\Delta-d)-m^2\bigl] = 0 \implies \Delta(\Delta-d)=m^2
\end{equation}\

There are then two possible values for $\Delta$:

\begin{equation}\label{delta2}
    \Delta_\pm = \frac{d}{2}\pm\nu,\hspace{1cm}\nu\equiv\sqrt{\Bigl(\frac{d}{2}\Bigl)^2+m^2}
\end{equation}\

In our study we will focus exclusively on real and different values for $\Delta_+$ and $\Delta_-$, i.e., $\nu>0$. We will often refer to $\Delta_+$ simply as $\Delta$ and consequently to $\Delta_-$ as $d-\Delta$. Now, these two values generally give two different solutions for $\phi(x)$ of the form of eq. (\ref{eomserie}), one for each value of $\Delta$, but when $\Delta_+-\Delta_-(=2\nu)$ is an integer, one of the solutions will be contained in the other resulting in a linearly dependent set of solutions. If this is the case, in order two construct a second independent solution one considers a third term with a logarithmic function, resulting in a general solution to $\phi(x)$ of the form:

\begin{equation}
    \phi(x) = x_0^{d-\Delta}f(x)+x_0^{\Delta}g(x)+x_0^{\Delta}\ln{(x_0)}h(x)
\end{equation}\

where $2\nu\in\mathbb{N}$ and $f(x)$, $g(x)$ and $h(x)$ are some power series in $x_0$. As it will become clear in the next section, we will only need the field $\phi(x)$ up to order $x_0^\Delta$ in the computation of the holographic n-point functions. Therefore, the part of the solution we will be interested in can be written as:

\begin{equation}\
    \phi(x) = x_0^{d-\Delta}\sum_{n=0}^{2\nu-1}x_0^n\varphi_n(\vec{x})+x_0^\Delta\varphi_{2\nu}(\vec{x})+x_0^\Delta\ln{(x_0)}\psi(\vec{x})+\mathcal{O}(x_0^{\Delta<})
\end{equation}\

where most of the functions $\varphi_n(\vec{x})$, $\varphi_{2\nu}(\vec{x})$ and $\psi(\vec{x})$ are expected to be determined from eq. (\ref{eom1}). Indeed, calling the sum $\sum_{n=0}^{2\nu-1}x_0^n\varphi_n(\vec{x})\equiv\varphi(x)$ and plugin the solution in the equation of motion one finds the condition:

\begin{equation}\label{eomcon}
    (2\nu-1)x_0^{d-\Delta+1}\partial_0\varphi(x)-x_0^{d-\Delta+2}\partial_0^2\varphi(x)-x_0^{d-\Delta+2}\partial_i^2\varphi(x)-2\nu x_0^\Delta \psi(\vec{x}) + \mathcal{O}(x_0^{\Delta<}) = 0
\end{equation}\

Since we are considering that $2\nu$ is a positive integer, $\nu$ itself will be a positive integer or positive half-integer ($\nu=\frac{1}{2},1,\frac{3}{2},\dotsc$). If $\nu=\frac{1}{2}$, eq. (\ref{eomcon}) reduces to:

\begin{equation}
    -x_0^\Delta\psi(\vec{x}) + \mathcal{O}(x_0^{\Delta<}) = 0\implies \psi(\vec{x}) = 0
\end{equation}\

and the general solution of the wave equation, up to order $x_0^\Delta$, is given by:

\begin{equation}
    \phi(x) = x_0^{\Delta-1}\varphi_0(\vec{x})+x_0^\Delta\varphi_1(\vec{x}) + \mathcal{O}(x_0^{\Delta<})
\end{equation}\

where $\varphi_0(\vec{x})$ and $\varphi_1(\vec{x})$ are some arbitrary functions not determined by the equation of motion. If $\nu=1$, using the power series representation of $\varphi(x)$, eq. (\ref{eomcon}) reduces to:

\begin{equation}
    x_0^{\Delta-1}\varphi_1(\vec{x})-x_0^\Delta\bigl[\partial_i^2\varphi_0(\vec{x})+2\psi(\vec{x})\bigl] + \mathcal{O}(x_0^{\Delta<}) = 0\implies \varphi_1(\vec{x})=0,\hspace{0.5cm} \psi(\vec{x}) = -\frac{1}{2}\partial_i^2\varphi_0(\vec{x})
\end{equation}\

and the general solution of the wave equation, up to order $x_0^\Delta$, is given by:

\begin{equation}
    \phi(x) = x_0^{\Delta-2}\varphi_0(\vec{x})+x_0^\Delta\varphi_2(\vec{x}) - \frac{1}{2}x_0^\Delta\ln{(x_0)}\partial_i^2\varphi_0(\vec{x})+ \mathcal{O}(x_0^{\Delta<})
\end{equation}\

where, again, $\varphi_0(\vec{x})$ and $\varphi_1(\vec{x})$ are some arbitrary functions. Finally, if $\nu>1$, using the power series representation of $\varphi(x)$, eq. (\ref{eomcon}) reduces to:

\begin{align}
    &(2\nu-1)x_0^{d-\Delta+1}\varphi_1(\vec{x})+\sum_{n=0}^{2\nu-3}x_0^{d-\Delta+2+n}\bigl[(n+2)(2\nu-n-2)\varphi_{n+2}(\vec{x})-\partial_i^2\varphi_n(\vec{x})\bigl]\nonumber\\
    &-x_0^\Delta\bigl[\partial_i^2\varphi_{2\nu-2}(\vec{x})+2\nu\psi(\vec{x})\bigl]+\mathcal{O}(x_0^{\Delta<}) = 0
\end{align}\

from where we can read the conditions:

\begin{equation}
    \varphi_1(\vec{x}) = 0,\hspace{0.5cm} \varphi_{n+2}(\vec{x}) = \frac{\partial_i^2\varphi_n(\vec{x})}{(n+2)(2\nu-n-2)},\hspace{0.5cm} \psi(\vec{x}) = -\frac{1}{2\nu}\partial_i^2\varphi_{2\nu-2}(\vec{x})
\end{equation}\

The first two conditions imply that:

\begin{equation}
    \varphi_{2n+1}(\vec{x})=0,\hspace{1cm} \varphi_{2n}(\vec{x}) = \frac{\Gamma(\nu-n)}{4^nn!\Gamma(\nu)}(\partial_i^2)^n\varphi_0(\vec{x})
\end{equation}\

which replacing in the last condition results in:

\begin{equation}
    \psi(\vec{x}) =
    \begin{cases}
    0,\hspace{5cm} \nu\in\mathbb{N}-\frac{1}{2}\\
    -\frac{1}{2}\frac{1}{4^{\nu-1}\Gamma(\nu)\Gamma(\nu+1)}(\partial_i^2)^\nu\varphi_0(\vec{x}),\hspace{0.5cm} \nu\in\mathbb{N}
    \end{cases}
\end{equation}\

Therefore, when $\nu$ is a positive integer the general solution of the wave equation (eq. (\ref{eom0})), up to order $x_0^\Delta$, is given by:

\begin{align}\label{onshell1}
    \phi(x) = &x_0^{d-\Delta}\sum_{n=0}^{\nu-1}x_0^{2n}\frac{\Gamma(\nu-n)}{4^nn!\Gamma(\nu)}(\partial_i^2)^n\varphi_0(\vec{x})+x_0^\Delta\varphi_{2\nu}(\vec{x})\nonumber\\
    &-\frac{1}{2}\frac{1}{4^{\nu-1}\Gamma(\nu)\Gamma(\nu+1)}x_0^\Delta\ln{(x_0)}(\partial_i^2)^\nu\varphi_0(\vec{x}) + \mathcal{O}(x_0^{\Delta<})
\end{align}\

and when $\nu$ is a positive half-integer the general solution, up to this same order, is instead:

\begin{equation}\label{onshell2}
    \phi(x) = x_0^{d-\Delta}\sum_{n=0}^{\nu-\frac{1}{2}}x_0^{2n}\frac{\Gamma(\nu-n)}{4^nn!\Gamma(\nu)}(\partial_i^2)^n\varphi_0(\vec{x})+x_0^\Delta\varphi_{2\nu}(\vec{x})+\mathcal{O}(x_0^{\Delta<})
\end{equation}\

where $\varphi_0(\vec{x})$ and $\varphi_{2\nu}(\vec{x})$ are some arbitrary functions not determined by the equation of motion. Notice that, for any value of $\nu$, as the on-shell field $\phi(x)$ approaches the conformal boundary of the AdS space at $x_0=0$, it has an asymptotic expansion of the form:

\begin{equation}\label{expphi}
    \phi(x) = x_0^{d-\Delta}\varphi_0(\vec{x})+\dotsb+x_0^\Delta\varphi_{2\nu}(\vec{x})+\dotsb+\mathcal{O}(x_0^{\Delta<})
\end{equation}\

As we anticipated in the last section of the previous chapter where we formulated the AdS/CFT correspondence, the function $\varphi_0(\vec{x})$ is precisely the conformal source for some scalar operator $O_\Delta(\vec{x})$ living at the boundary of AdS, and as we will see in the next section the exact form for its correlators will be given by the function $\varphi_{2\nu}(\vec{x})$. Up to now, where we have only solved the equation of motion asymptotically, these functions are completely arbitrary and disconnected between them, but in fact fully solving eq. (\ref{eom1}) for a regular field in the interior of AdS forces $\varphi_{2\nu}(\vec{x})$ to be a functional of $\varphi_0(\vec{x})$.\par
To fully solve the equation of motion we will look for a solution in terms of a Green's function:

\begin{equation}
    \phi(x) = \int d^dy\ K(x,\vec{y})\varphi_0(\vec{y})
\end{equation}\

where the Green's function $K(x,\vec{y})$ is known as the bulk-boundary propagator, since it propagates the field from the boundary of the AdS space to a point in the bulk. Replacing this form of the solution into the equation of motion implies that the bulk-boundary propagator itself must satisfy the wave equation:

\begin{equation}
    (-\Box+m^2)K(x,\vec{y}) = 0
\end{equation}\

There is a quite clever way to solve it which can be seen by remembering from eqs. (\ref{delta1}) and (\ref{delta2}) that $x_0^\Delta$ itself is a solution of the wave equation if $\Delta=\Delta_\pm$, where $\Delta_\pm=\frac{d}{2}\pm\nu$. Then, a more general solution to it can be written as:

\begin{equation}
    (-\Box+m^2)\bigl(c_{\Delta_+}x_0^{\Delta_+}+c_{\Delta_-}x_0^{\Delta_-}\bigl) = 0
\end{equation}\

Now, since the metric is invariant under AdS isometries, in particular for inversions and translations of the form $x^\mu\rightarrow\frac{x^\mu}{x^2}$ and $x\rightarrow x-\vec{y}$ respectively, then the wave equation is invariant as well and thus under these transformations it must also hold that:

\begin{equation}
    (-\Box+m^2)\Bigl(c_{\Delta_+}\Bigl[\frac{x_0}{(x-\vec{y})^2}\Bigl]^{\Delta_+}+c_{\Delta_-}\Bigl[\frac{x_0}{(x-\vec{y})^2}\Bigl]^{\Delta_-}\Bigl) = 0
\end{equation}\

Therefore, by arguments of existence and uniqueness of the solution, the general form of the bulk-boundary propagator is given by:

\begin{equation}\label{generalk}
    K(x,\vec{y}) = c_{\Delta}\Bigl[\frac{x_0}{(x-\vec{y})^2}\Bigl]^{\Delta}+c_{d-\Delta}\Bigl[\frac{x_0}{(x-\vec{y})^2}\Bigl]^{d-\Delta}
\end{equation}\

where, again, we simply called $\Delta_+\equiv\Delta$ and consequently $\Delta_-\equiv d-\Delta$. The coefficients $c_\Delta$ and $c_{d-\Delta}$ can be determined with the imposition of appropriate boundary conditions. In particular we will be interested in two boundary conditions, that the resulting fields are regular in the interior of the AdS space, that is to say, fields that vanish fast enough as $x_i\rightarrow\pm\infty$ but also as $x_0\rightarrow\infty$, and that their asymptotic expansion behave like eq. (\ref{expphi}) as they approach the conformal boundary at $x_0\rightarrow0$. Since in most cases of interest the second term in eq. (\ref{generalk}) will be the responsible for divergences as we study the field in the limit $x_0\rightarrow\infty$, the first condition sets the value of the coefficient $c_{d-\Delta}$ to 0. This reduces the general form of the regular fields in the interior of AdS to be of the form:

\begin{equation}\label{k1}
    \phi(x) = \int d^dy\ K(x,\vec{y})\varphi_0(\vec{y}),\hspace{1cm} K(x,\vec{y}) = c_{\Delta}\Bigl[\frac{x_0}{(x-\vec{y})^2}\Bigl]^{\Delta}
\end{equation}\

where the coefficient $c_\Delta$ is yet to be determined. The second condition, that the asymptotic expansion of the field should agree with the one found previously, implies that in the limit $x_0\rightarrow0$ the leading terms in the expansion of the bulk-boundary propagator must be:

\begin{equation}\label{kexp1}
    K(x,\vec{y}) \underset{x_0\rightarrow0}{=} x_0^{d-\Delta}\delta^d(\vec{x}-\vec{y})+\dotsb+x_0^\Delta f(\vec{x},\vec{y})+\dotsb+\mathcal{O}(x_0^{\Delta<})
\end{equation}\

where $f(\vec{x},\vec{y})$ is some function and $\delta^d(\vec{x}-\vec{y})$ is the Dirac delta. Indeed, replacing this behavior in $\phi(x)$ one recovers that $\phi(x)=x_0^{d-\Delta}\varphi_0(\vec{x})+\dotsb+x_0^\Delta\varphi_{2\nu}(\vec{x})+\dotsb+\mathcal{O}(x_0^{\Delta<})$, for some $\varphi_{2\nu}(\vec{x})$ in terms of $f(\vec{x},\vec{y})$ and $\varphi_0(\vec{y})$, which is the correct expansion (eq. (\ref{expphi})). Let us see then that eq. (\ref{kexp1}) is precisely the expansion of the bulk-boundary propagator from eq. (\ref{k1}), finding explicitly the forms of $c_\Delta$ and $f(\vec{x},\vec{y})$ in the process. In the case that $\vec{x}\neq\vec{y}$, a simple Taylor expansion near the point $x_0 = 0$ suggests that:

\begin{equation}
    K(x,\vec{y}) \underset{\vec{x}\neq\vec{y}}{\underset{x_0\rightarrow0}{=}} x_0^\Delta\frac{c_\Delta}{\lvert\vec{x}-\vec{y}\rvert^{2\Delta}} + \mathcal{O}(x_0^{\Delta<})\implies f(\vec{x},\vec{y}) = \frac{c_\Delta}{\lvert\vec{x}-\vec{y}\rvert^{2\Delta}}
\end{equation}\

completely fixing the functional form of the function $f(\vec{x},\vec{y})$ up to the constant $c_\Delta$. The resulting expansion however when not only $x_0\rightarrow0$ but also $\vec{x}\rightarrow\vec{y}$ is more interesting. Eq. (\ref{kexp1}) is telling us that in these limits the leading order in the expansion of the bulk-boundary propagator must be a Dirac delta with support at $\vec{x}=\vec{y}$, of order $x_0^{d-\Delta}$. This fact further implies that:

\begin{align}\label{cond2k}
    \lim_{x_0\rightarrow0} x_0^{\Delta-d}K(x,\vec{y}) = \delta^d(\vec{x}-\vec{y}) \implies \lim_{x_0\rightarrow0}\int d^dy\ x_0^{\Delta-d}K(x,\vec{y}) &= \int d^dy\ \delta^d(\vec{x}-\vec{y})=1
\end{align}\

In our case, the explicit form of $K(x,\vec{y})$ found in eq. (\ref{k1}) normally goes to 0 as $x_0\rightarrow0$, but as $\vec{x}\rightarrow\vec{y}$ also, it no longer vanishes, in fact it diverges. This is the expected behavior of a quantity behaving as a Dirac delta with support at $\vec{x}=\vec{y}$. Moreover, when integrated in the $\vec{y}$ coordinate in the limit of eq. (\ref{cond2k})\footnote{The integral can be easily done after a translation $\vec{y}\rightarrow\vec{y}+\vec{x}$ and rescaling $\vec{y}\rightarrow x_0\vec{y}$ using spherical coordinates.}:

\begin{equation}
    \lim_{x_0\rightarrow0}\int d^dy\ x_0^{\Delta-d}K(x,\vec{y}) = c_\Delta\frac{\pi^{\frac{d}{2}}\Gamma(\nu)}{\Gamma(\Delta)}
\end{equation}\

Therefore eq. (\ref{cond2k}), which contains the explicit behavior as a Dirac delta function that the bulk-boundary propagator must have in the limits $x_0\rightarrow0$ and $\vec{x}\rightarrow\vec{y}$, is satisfied for the $K(x,\vec{y})$ found in eq. (\ref{k1}) if the coefficient $c_\Delta$ is chosen such that:

\begin{equation}
    c_\Delta\frac{\pi^{\frac{d}{2}}\Gamma(\nu)}{\Gamma(\Delta)} = 1\implies c_\Delta = \frac{\Gamma(\Delta)}{\pi^{\frac{d}{2}}\Gamma(\nu)}
\end{equation}\

In summary, the complete solution of the wave equation $(-\Box+m^2)\phi(x)=0$ which is regular in the interior of the AdS space with appropriate boundary behavior as it approaches the conformal boundary of AdS at $x_0=0$ can be written in terms of a Green's function $K(x,\vec{y})$, the bulk-boundary propagator which also satisfies the wave equation $(-\Box+m^2)K(x,\vec{y})=0$, as:

\begin{equation}\label{ffsumm}
    \phi(x) = \int d^dy\ K(x,\vec{y})\varphi_0(\vec{y}),\hspace{0.5cm} K(x,\vec{y}) = c_\Delta\Bigl[\frac{x_0}{(x-\vec{y})^2}\Bigl]^\Delta,\hspace{0.5cm} c_\Delta=\frac{\Gamma(\Delta)}{\pi^{\frac{d}{2}}\Gamma(\nu)}
\end{equation}\

where the asymptotic expansion of the bulk-boundary propagator $K(x,\vec{y})$ and bulk field $\phi(x)$ as $x_0\rightarrow0$ are given by:

\begin{equation}\label{expbubop}
    K(x,\vec{y}) = x_0^{d-\Delta}\delta^d(\vec{x}-\vec{y})+\dotsb+x_0^\Delta\frac{c_\Delta}{\lvert\vec{x}-\vec{y}\rvert^{2\Delta}}+\dotsb+\mathcal{O}(x_0^{\Delta<})
\end{equation}

\begin{equation}\label{phi2nu}
    \phi(x) = x_0^{d-\Delta}\varphi_0(\vec{x})+\dotsb+x_0^\Delta\varphi_{2\nu}(\vec{x})+\dotsb+\mathcal{O}(x_0^{\Delta<}),\hspace{0.5cm} \varphi_{2\nu}(\vec{x}) = \int d^dy\ \frac{c_\Delta}{\lvert\vec{x}-\vec{y}\rvert^{2\Delta}}\varphi_0(\vec{y})
\end{equation}\

This form of the solution not only shows our promised statement that the function $\varphi_{2\nu}(\vec{x})$ present in the expansion of the field (function that as we will see will be crucial in the values of the correlators) becomes a functional of $\varphi_0(\vec{y})$, but also that the transformation rules for $\varphi_0(\vec{y})$ are precisely those of a conformal source to some $O_\Delta(\vec{y})$ operator of scaling dimension $\Delta$, realization which was key in the formulation of the AdS/CFT correspondence at the end of the last chapter. Indeed, this can easily be seen from the trivial transformation rule for the field $\phi(x)$ itself. Under AdS isometries, using eqs. (\ref{distancetransf}) and (\ref{x0transf}), the bulk-boundary propagator transforms as:

\begin{equation}\label{bubotr}
    K(x',\vec{y'}) = \Omega^{-\Delta}(\vec{y})K(x,\vec{y})
\end{equation}\

Since $\phi(x)$ is a scalar field, it transforms trivially under diffeomorphisms $\phi'(x')=\phi(x)$. This fact implies that:

\begin{equation}
    \phi'(x') = \int d^dy\ K(x,\vec{y})\Omega^{d-\Delta}(\vec{y})\varphi'_0(\vec{y'})\overset{!}{=} \int d^dy\ K(x,\vec{y})\varphi_0(\vec{y})= \phi(x)
\end{equation}\

where we used the transformation rules for $K(x,\vec{y})$ and the measure $d^dy' = \Omega^d(\vec{y})d^dy$. In other words, the function $\varphi_0(\vec{y})$ itself must transform as:

\begin{equation}
    \varphi'_0(\vec{y'}) = \Omega^{-(d-\Delta)}(\vec{y})\varphi_0(\vec{y})
\end{equation}\

This is exactly the transformation rule expected for a conformal source of a primary scalar operator $O_\Delta(\vec{y})$ of scaling dimension $\Delta$ (eq. (\ref{confso})), as we anticipated.

\subsection{Holographic Renormalization}

We are one step away from being able to compute our first correlation functions using the AdS/CFT correspondence for what is the simplest case, a free scalar field. Now that we know how the on-shell field $\phi(x)$ depends on the dual source $\varphi_0(\vec{x})$, it is a matter of direct calculation to compute the functional derivatives of $Z_{\text{AdS}}[\varphi_0]=e^{-S_{\text{AdS}}[\phi]}$ with respect to $\varphi_0(\vec{x})$ with the intention to compute the CFT correlators eqs. (\ref{euccorr1}) and (\ref{euccorr2}). With this objective in mind then, varying the AdS path integral once:

\begin{equation}
    \delta Z_{\text{AdS}}[\varphi_0] =-Z_{\text{AdS}}[\varphi_0]\int d^{d+1}x\ \Bigl(\partial_\mu\bigl[\sqrt{g}g^{\mu\nu}\partial_\nu\phi(x)\delta\phi(x)\bigl]+\sqrt{g}(-\Box+m^2)\phi(x)\delta\phi(x)\Bigl)
\end{equation}\

where integration by parts was done. Now, the second term vanishes due to the Euler-Lagrange equation (eq. (\ref{eom0})) and the first term is a boundary term evaluated at the boundaries of every coordinate. As we mentioned in the last section, we are interested in fields which are regular in the interior of the AdS space, that is to say, fields that vanish fast enough as $x_i\rightarrow\pm\infty$ but also as $x_0\rightarrow\infty$. All the interesting behavior of the bulk fields occur as they approach the conformal boundary of the AdS space at $x_0=0$, which reduces the variation of the AdS action to:

\begin{equation}\label{varz}
    \delta Z_{\text{AdS}}[\varphi_0] =Z_{\text{AdS}}[\varphi_0]\int d^dx\ x_0^{1-d}\partial_0\phi(x)\delta\phi(x)\Bigl\rvert_{x_0=0}
\end{equation}\

where we used the explicit form of the AdS metric. From this expression it is clear why we only need the expansion of the field $\phi(x)$ up to order $x_0^\Delta$. Considering higher order terms will result in contributions of the form $x_0^{0<}$ which will vanish when we evaluate the integrand at $x_0=0$, not playing any role in the computation of the correlators. But even more interesting, the contributions from the leading terms in the expansion of $\phi(x)$ will be of the form $x_0^{<0}$ which will diverge when evaluated at $x_0=0$, making the variation of the AdS path integral $\delta Z_{\text{AdS}}$ ill-defined. This is not surprising at all when one takes into account the fact that the AdS/CFT correspondence is a weak/strong duality meaning that weakly coupled (classical) theories on the AdS side are related to strongly coupled (quantum) theories on the CFT side. Since we are considering some classical theory approximation on AdS, we are computing the CFT correlators for some quantum theory which are usually UV-divergent due to loops contributions, divergence which must be present somehow on the AdS side due to the AdS/CFT dictionary. In this case, the divergence is of the type IR and it is because the combination found in eq. (\ref{varz}) of the AdS metric together with the bulk field diverges as it approaches the conformal boundary.\par
If we are able to write the IR-divergences present in eq. (\ref{varz}) as the total variation of some quantity which transforms correctly under AdS isometries, we can always add a boundary term to the AdS action with exactly this same value but opposite sign which not only it will not modify the Euler-Lagrange equations but also it will preserve the AdS invariance of the action, property that is extremely important for the validity of the correspondence. Adding such boundary term to the action would exactly cancel all divergences, rendering a finite predictive action.
This process of extracting the sensitive information from the ill-defined variation of the AdS path integral is known as holographic renormalization and it consists of the following way: to not manipulate divergent terms let us first regularize $\delta Z_{\text{AdS}}$ with a small IR-regulator $\varepsilon$ where instead of evaluating its integrand at the conformal boundary $x_0=0$ we evaluate it at some small distance from it $x_0=\varepsilon$:

\begin{equation}\label{regvarz}
    \delta Z_{\text{AdS}}[\varphi_0] =\lim_{\varepsilon\rightarrow0}Z_{\text{AdS}}[\varphi_0]\int d^dx\ x_0^{1-d}\partial_0\phi(x)\delta\phi(x)\Bigl\rvert_{x_0=\varepsilon}
\end{equation}\

Before replacing the explicit form of the field $\phi(x)$, since we only need it up to order $x_0^\Delta$, its asymptotic expansions given by eqs. (\ref{onshell1}) and (\ref{onshell2}) will be sufficient for this study. To address both cases at the same time we will compactly write these expansions as:

\begin{equation}\label{compact}
    \phi(x) = x_0^{d-\Delta}\varphi(x) + x_0^\Delta\varphi_{2\nu}(\vec{x}) + x_0^\Delta\ln{(x_0)}\psi(\vec{x})+\mathcal{O}(x_0^{\Delta<})
\end{equation}\

where we called:

\begin{equation}\label{varphi2}
    \varphi(x) =
    \begin{cases}
    \sum_{n=0}^{\nu-\frac{1}{2}}x_0^{2n}\frac{\Gamma(\nu-n)}{4^nn!\Gamma(\nu)}(\partial_i^2)^n\varphi_0(\vec{x}),\hspace{0.5cm} \nu\in\mathbb{N}-\frac{1}{2}\\
    \sum_{n=0}^{\nu-1}x_0^{2n}\frac{\Gamma(\nu-n)}{4^nn!\Gamma(\nu)}(\partial_i^2)^n\varphi_0(\vec{x}),\hspace{0.5cm} \nu\in\mathbb{N}
    \end{cases}
\end{equation}

\begin{equation}\label{psi2}
    \psi(\vec{x}) =
    \begin{cases}
    0,\hspace{5cm} \nu\in\mathbb{N}-\frac{1}{2}\\
    -\frac{1}{2}\frac{1}{4^{\nu-1}\Gamma(\nu)\Gamma(\nu+1)}(\partial_i^2)^\nu\varphi_0(\vec{x}),\hspace{0.5cm} \nu\in\mathbb{N}
    \end{cases}
\end{equation}\

Replacing then eq. (\ref{compact}) into eq. (\ref{regvarz}) one finds that:

\begin{align}
    \delta Z_{\text{AdS}}[\varphi_0] =\lim_{\varepsilon\rightarrow0}Z_{\text{AdS}}[\varphi_0]\int d^dx\ \Bigl(&(d-\Delta)\bigl[\varepsilon^{-2\nu}\varphi(x)\delta\varphi(x)+\varphi(x)\delta\varphi_{2\nu}(\vec{x})+\ln{(\varepsilon)}\varphi(x)\delta\psi(\vec{x})\bigl]\nonumber\\
    &+\varepsilon^{-2\nu+1}\partial_0\varphi(x)\delta\varphi(x)+\Delta\varphi_{2\nu}(\vec{x})\delta\varphi(x)+\Delta\ln{(\varepsilon)}\psi(\vec{x})\delta\varphi(x)\nonumber\\
    &+\psi(\vec{x})\delta\varphi(x)+\mathcal{O}(\varepsilon^{0<})\Bigl)
\end{align}\

where the bulk coordinates $x$ are understood to be $x=(\varepsilon,\vec{x})$. It is clear that many of these terms will diverge as we take the limit $\varepsilon\rightarrow0$. Indeed, take for example the first term for some particular simple case, say $\nu=1$. It reduces to $\varepsilon^{-2\nu}\varphi(x)\delta\varphi(x)\underset{\nu=1}{=}\varepsilon^{-2}\varphi_0(\vec{x})\delta\varphi_0(\vec{x})$, which is clearly divergent for $\varepsilon=0$. As we explained before, if we can write these divergences as total variations of some AdS invariant quantities we will be able to renormalize them through a suitable boundary action term. Following this line then, let us write the variations on the square bracket as $\varphi(x)\delta\varphi(x) = \delta\bigl[\frac{1}{2}\varphi^2(x)\bigl]$, $\varphi(x)\delta\varphi_{2\nu}(\vec{x}) = \delta\bigl[\varphi(x)\varphi_{2\nu}(\vec{x})\bigl]-\varphi_{2\nu}(\vec{x})\delta\varphi(x)$ and $\varphi(x)\delta\psi(\vec{x}) = \delta\bigl[\varphi(x)\psi(\vec{x})\bigl]-\psi(\vec{x})\delta\varphi(x)$, resulting in:

\begin{align}\label{varz5}
    \delta Z_{\text{AdS}}[\varphi_0] =\lim_{\varepsilon\rightarrow0}Z_{\text{AdS}}[\varphi_0]\int d^dx\ \Bigl(&\frac{1}{2}(d-\Delta)\delta\bigl[\varepsilon^{-2\nu}\varphi^2(x)+2\varphi(x)\varphi_{2\nu}(\vec{x})+2\ln{(\varepsilon)}\varphi(x)\psi(\vec{x})\bigl]\nonumber\\
    &+\varepsilon^{-2\nu+1}\partial_0\varphi(x)\delta\varphi(x)+2\nu\varphi_{2\nu}(\vec{x})\delta\varphi(x)+2\nu\ln{(\varepsilon)}\psi(\vec{x})\delta\varphi(x)\nonumber\\
    &+\psi(\vec{x})\delta\varphi(x)+\mathcal{O}(\varepsilon^{0<})\Bigl)
\end{align}\

The terms on the square bracket, up to order $\varepsilon^0$, are nothing but the square of the field $\phi(x)$ times the corresponding Jacobian:

\begin{equation}
     \varepsilon^{-2\nu}\varphi^2(x)+2\varphi(x)\varphi_{2\nu}(\vec{x})+2\ln{(\varepsilon)}\varphi(x)\psi(\vec{x}) = \sqrt{\gamma}\phi^2(x) + \mathcal{O}(\varepsilon^{0<})
\end{equation}\

where we defined the induced metric on the boundary $\gamma_{ij}=\frac{1}{\varepsilon^2}\delta_{ij}$. The first term on the second line can also be written as a total variation in terms of the field and induced metric:

\begin{equation}
    \varepsilon^{-2\nu+1}\partial_0\varphi(x)\delta\varphi(x) = \delta\Bigl[\sqrt{\gamma}\frac{1}{4(\nu-1)}\phi(x)\Box_\gamma\phi(x)+\dotsb\Bigl]
\end{equation}\

where we defined the Laplace operator $\Box_\gamma\equiv\frac{1}{\sqrt{\gamma}}\partial_i(\sqrt{\gamma}\gamma^{ij}\partial_j)=\varepsilon^2\partial_i^2$, used that $\Box_\gamma\phi(x)\delta\phi(x)=\delta\bigl[\frac{1}{2}\phi(x)\Box_\gamma\phi(x)\bigl]$, and where the triple dots represent terms with higher order derivatives. In exactly the same way we can also write the third term on the second line and the term on the third line of eq. (\ref{varz5}) as total variations in terms of the field and induced metric using that:

\begin{equation}
    \psi(\vec{x})\delta\varphi(x) = \delta\Bigl[\sqrt{\gamma}\frac{1}{2}C_\nu \phi(x)\Box_\gamma^\nu\phi(x)\Bigl] + \mathcal{O}(\varepsilon^{0<})
\end{equation}\

where $C_\nu$ summarizes the constants in eq. (\ref{psi2}):

\begin{equation}\label{cnu}
    C_\nu =
    \begin{cases}
    0,\hspace{3.1cm} \nu\in\mathbb{N}-\frac{1}{2}\\
    -\frac{1}{2}\frac{1}{4^{\nu-1}\Gamma(\nu)\Gamma(\nu+1)},\hspace{0.5cm} \nu\in\mathbb{N}
    \end{cases}
\end{equation}\

All this allows us to group the divergent terms in eq. (\ref{varz5}) as a total variation of some quantity which transforms correctly under diffeomorphisms:

\begin{align}\label{varz10}
    \delta Z_{\text{AdS}}[\varphi_0] =\lim_{\varepsilon\rightarrow0}Z_{\text{AdS}}[\varphi_0]\int d^dx\ \Bigl(&\delta\Bigl\{\sqrt{\gamma}\Bigl[\frac{1}{2}(d-\Delta)\phi^2(x)+\nu\ln{(\varepsilon)}C_\nu \phi(x)\Box_\gamma^\nu\phi(x)\nonumber\\
    &\hspace{1.2cm}+\frac{1}{2}C_\nu \phi(x)\Box_\gamma^\nu\phi(x)+\frac{1}{4(\nu-1)}\phi(x)\Box_\gamma\phi(x)+\dotsb\Bigl]\Bigl\}\nonumber\\
    &+2\nu\varphi_{2\nu}(\vec{x})\delta\varphi(x)+\mathcal{O}(\varepsilon^{0<})\Bigl)
\end{align}\

The only term that cannot be written as a total variation is $2\nu\varphi_{2\nu}(\vec{x})\delta\varphi(x)$ and this will be precisely the term that will give rise to the values of the correlators. Now, continuing with the renormalization process, we can always add a boundary term to the AdS action of the form:

\begin{equation}\label{actionren}
    S_\text{AdS}[\phi]\rightarrow S_\text{AdS}[\phi] + \int d^dx\sqrt{\gamma}\ B\bigl(\phi(x)\bigl)\Bigl\rvert_{x_0=0}
\end{equation}\

where $B\bigl(\phi(x)\bigl)$ is some AdS invariant function of the fields $\phi(x)$. The addition of this boundary term modifies the variation of the AdS path integral eq. (\ref{varz10}) to:

\begin{align}
    \delta Z_{\text{AdS}}[\varphi_0] =\lim_{\varepsilon\rightarrow0}Z_{\text{AdS}}[\varphi_0]\int d^dx\ \Bigl(&\delta\Bigl\{\sqrt{\gamma}\Bigl[\frac{1}{2}(d-\Delta)\phi^2(x)+\nu\ln{(\varepsilon)}C_\nu \phi(x)\Box_\gamma^\nu\phi(x)+\nonumber\\
    &\hspace{1.3cm}\frac{1}{2}C_\nu \phi(x)\Box_\gamma^\nu\phi(x)+\frac{1}{4(\nu-1)}\phi(x)\Box_\gamma\phi(x)+\dotsb\nonumber\\
    &\hspace{1.3cm}-B\bigl(\phi(x)\bigl)\Bigl]\Bigl\}+2\nu\varphi_{2\nu}(\vec{x})\delta\varphi(x)+\mathcal{O}(\varepsilon^{0<})\Bigl)
\end{align}\

Then simply choosing this boundary term to be\footnote{For $\nu<\frac{3}{2}$ the terms $\frac{1}{4(\nu-1)}\phi(x)\Box_\gamma\phi(x)+\dotsb$ are not present, while for $\nu=\frac{3}{2}$ only the higher order derivative terms represented by the triple dots are absent.}:

\begin{align}\label{boundaryterm}
    B\bigl(\phi(x)\bigl) = &\frac{1}{2}(d-\Delta)\phi^2(x)+\nu\ln{(\varepsilon)}C_\nu \phi(x)\Box_\gamma^\nu\phi(x)+\frac{1}{2}C_\nu \phi(x)\Box_\gamma^\nu\phi(x)+\nonumber\\
    &\frac{1}{4(\nu-1)}\phi(x)\Box_\gamma\phi(x)+\dotsb
\end{align}\

renders the variation of the AdS path integral finite, allowing us to safely approach the conformal boundary, i.e., take the IR-regulator $\varepsilon$ equal to 0:

\begin{equation}\label{renvarz}
    \delta Z_{\text{AdS}}[\varphi_0] =Z_{\text{AdS}}[\varphi_0]\int d^dx\ 2\nu\varphi_{2\nu}(\vec{x})\delta\varphi_0(\vec{x})
\end{equation}\

This is the holographic renormalization procedure.

\subsection{Correlation Functions}

The whole study of the obtention of CFT correlators starting from an AdS path integral in the AdS/CFT correspondence is reduced to the computation of the finite, renormalized variation $\delta Z_{\text{AdS}}$ as a functional of the corresponding conformal sources. Once the explicit form of this variation is found, the obtention of every n-point function is just a simple exercise of taking derivatives. In the particular case of a free scalar field in AdS, this study led us to eq. (\ref{renvarz}) where the explicit form of $\varphi_{2\nu}(\vec{x})$ is given by eq. (\ref{phi2nu}):

\begin{equation}
    \delta Z_{\text{AdS}}[\varphi_0] =Z_{\text{AdS}}[\varphi_0]\int\int d^dxd^dy\ \frac{2\nu c_\Delta}{\lvert\vec{x}-\vec{y}\rvert^{2\Delta}}\varphi_0(\vec{y})\delta\varphi_0(\vec{x})
\end{equation}\

From this expression it is just a matter of direct calculation to obtain the corresponding correlators eqs. (\ref{euccorr1}) and (\ref{euccorr2}). The resulting 1-, 2-, 3- and 4-point functions for some primary scalar operator $O_\Delta(\vec{x})$ of scaling dimension $\Delta$ dual to a free scalar field $\Phi(x)$ in AdS$_{d+1}$ are given by:

\begin{align}\label{ffcorr}
    \text{1-pt fn:}\hspace{0.5cm}&\langle O_\Delta(\vec{y_1})\rangle_{\text{CFT}} = \langle O_\Delta(\vec{y_1})\rangle_{\text{CFT,con}} = 0\nonumber\\
    \text{2-pt fn:}\hspace{0.5cm}&\langle O_\Delta(\vec{y_1})O_\Delta(\vec{y_2})\rangle_{\text{CFT}} = \langle O_\Delta(\vec{y_1})O_\Delta(\vec{y_2})\rangle_{\text{CFT,con}} = \frac{2\nu c_\Delta}{\lvert\vec{y_1}-\vec{y_2}\rvert^{2\Delta}}\nonumber\\
    \text{3-pt fn:}\hspace{0.5cm}&\langle O_\Delta(\vec{y_1})O_\Delta(\vec{y_2})O_\Delta(\vec{y_3})\rangle_{\text{CFT}} = \langle O_\Delta(\vec{y_1})O_\Delta(\vec{y_2})O_\Delta(\vec{y_3})\rangle_{\text{CFT,con}} = 0\nonumber\\
    \text{4-pt fn:}\hspace{0.5cm}&\langle O_\Delta(\vec{y_1})O_\Delta(\vec{y_2})O_\Delta(\vec{y_3})O_\Delta(\vec{y_4})\rangle_{\text{CFT}} = \frac{2\nu c_\Delta}{\lvert\vec{y_1}-\vec{y_2}\rvert^{2\Delta}}\frac{2\nu c_\Delta}{\lvert\vec{y_3}-\vec{y_4}\rvert^{2\Delta}}+(\vec{y_2}\leftrightarrow\vec{y_3})+(\vec{y_2}\leftrightarrow\vec{y_4})\nonumber\\
    &\langle O_\Delta(\vec{y_1})O_\Delta(\vec{y_2})O_\Delta(\vec{y_3})O_\Delta(\vec{y_4})\rangle_{\text{CFT,con}} = 0
\end{align}\

\begin{figure}[h]
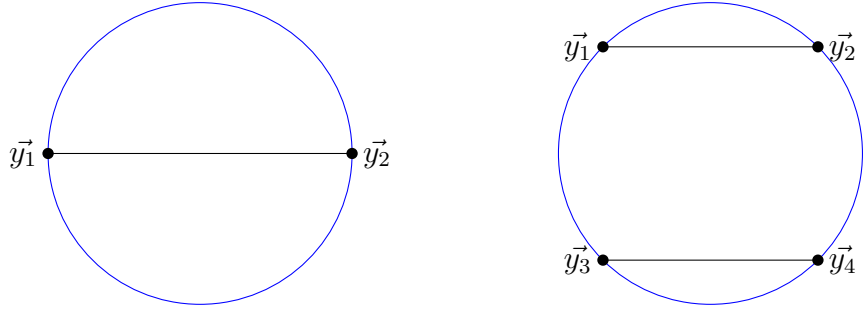

    \[\begin{wittendiagram}
    \draw (-2,0) node[vertex] -- (2,0)
    (2,0) node[vertex]
    
    (-2,0) node[left]{$\vec{y_1}$}
    (2,0) node[right]{$\vec{y_2}$}
    
    ;
  \end{wittendiagram}\hspace{2cm}
  \begin{wittendiagram}
    \draw (-1.4142,1.4142) node[vertex] -- (1.4142,1.4142)
    (-1.4142,-1.4142) node[vertex] -- (1.4142,-1.4142)
    (1.4142,1.4142) node[vertex]
    (1.4142,-1.4142) node[vertex]
    
    (-1.4142,1.4142) node[left]{$\vec{y_1}$}
    (-1.4142,-1.4142) node[left]{$\vec{y_3}$}
    (1.4142,1.4142) node[right]{$\vec{y_2}$}
    (1.4142,-1.4142) node[right]{$\vec{y_4}$}
    
    ;
  \end{wittendiagram}\]
  \caption{Pictorial representation of the contributions to the 2- and 4-point functions.}
\end{figure}

Notice how the holographic 1-, 2- and 3-point functions follow the expected form for a conformal invariant theory as it was derived in eq. (\ref{summcorr2}). It is not obvious however given its current form that this is also true for the holographic 4-point function just obtained. What we can assert at the moment is that this correlator depends of the product of 2-point functions evaluated at different boundary points along with its permutations which, pictorially, correspond exactly to the disconnected tree-level diagrams expected from a free field theory. Fortunately, the conformal form of the 4-point function can be exposed directly through simple manipulations of the boundary points. Indeed, adding a convenient factor of 1 to the first term of the correlator we can rewrite it as:

\begin{equation}\label{1term4pf}
    \frac{2\nu c_\Delta}{\lvert\vec{y_1}-\vec{y_2}\rvert^{2\Delta}}\frac{2\nu c_\Delta}{\lvert\vec{y_3}-\vec{y_4}\rvert^{2\Delta}} = \frac{(2\nu c_\Delta)^2}{\prod_{i<j}\lvert\vec{y_{ij}}\rvert^{\frac{2\Delta}{3}}}\Bigl(\frac{\lvert\vec{y_{13}}\rvert\lvert\vec{y_{14}}\rvert\lvert\vec{y_{23}}\rvert\lvert\vec{y_{24}}\rvert}{\lvert\vec{y_{12}}\rvert^2\lvert\vec{y_{34}}\rvert^2}\Bigl)^{\frac{2\Delta}{3}}
\end{equation}\

where we defined $\lvert\vec{y_{ij}}\rvert\equiv\lvert\vec{y_i}-\vec{y_j}\rvert$. The resulting fraction of the boundary points inside the parentheses can be written in terms of the conformal invariant cross ratios $u$ and $v$ defined in eq. (\ref{crossr}) simply as $u^{-1}v^{\frac{1}{2}}$, which allows us to express eq. (\ref{1term4pf}) as:

\begin{equation}
    \frac{2\nu c_\Delta}{\lvert\vec{y_1}-\vec{y_2}\rvert^{2\Delta}}\frac{2\nu c_\Delta}{\lvert\vec{y_3}-\vec{y_4}\rvert^{2\Delta}} = \frac{(2\nu c_\Delta)^2}{\prod_{i<j}\lvert\vec{y_{ij}}\rvert^{\frac{2\Delta}{3}}}u^{-\frac{2\Delta}{3}}v^{\frac{\Delta}{3}}
\end{equation}\

This result let us rewrite the holographic 4-point function into the form:

\begin{equation}
    \langle O_\Delta(\vec{y_1})O_\Delta(\vec{y_2})O_\Delta(\vec{y_3})O_\Delta(\vec{y_4})\rangle_{\text{CFT}} = \frac{(2\nu c_\Delta)^2}{\prod_{i<j}\lvert\vec{y_{ij}}\rvert^{\frac{2\Delta}{3}}}u^{-\frac{2\Delta}{3}}v^{\frac{\Delta}{3}}+(\vec{y_2}\leftrightarrow\vec{y_3})+(\vec{y_2}\leftrightarrow\vec{y_4})
\end{equation}\

The objective of these manipulations was precisely to be able to write the first term of the correlator in its desired conformal form, as in eq. (\ref{summcorr2}). But what about its permutations? It is a nice exercise to check that under the interchange of $\vec{y_2}$ with $\vec{y_3}$ the multiplication $\prod_{i<j}\lvert\vec{y_{ij}}\rvert^{\frac{2\Delta}{3}}$ remains invariant while the cross ratios $u$ and $v$ are mapped to $\frac{1}{u}$ and $\frac{v}{u}$ respectively, and equivalently under the interchange of $\vec{y_2}$ with $\vec{y_4}$ the multiplication $\prod_{i<j}\lvert\vec{y_{ij}}\rvert^{\frac{2\Delta}{3}}$ also remains invariant while the cross ratios $u$ and $v$ are mapped now to $v$ and $u$ respectively. Therefore, since the only quantity that changes when interchanging the boundary points is the particular function on the cross ratios, the permutations of the first term of the correlator also have the expected conformal form. Finally then, these facts allow us to write the holographic 4-point function, exposing explicitly its conformal form, as:

\begin{equation}\label{conformalform4pf}
    \langle O_\Delta(\vec{y_1})O_\Delta(\vec{y_2})O_\Delta(\vec{y_3})O_\Delta(\vec{y_4})\rangle_{\text{CFT}} = \frac{(2\nu c_\Delta)^2}{\prod_{i<j}\lvert\vec{y_{ij}}\rvert^{\frac{2\Delta}{3}}}u^{-\frac{2\Delta}{3}}v^{\frac{\Delta}{3}}+\Bigl(u,v\rightarrow\frac{1}{u},\frac{v}{u}\Bigl)+(u,v\rightarrow v,u)
\end{equation}\

In summary, the resulting 1-, 2-, 3- and 4-point functions for some primary scalar operator $O_\Delta(\vec{x})$ of scaling dimension $\Delta$ dual to a free scalar field $\Phi(x)$ in AdS$_{d+1}$ are given by:

\begin{align}\label{ffcorr2}
    \text{1-pt fn:}\hspace{0.5cm}&\langle O_\Delta(\vec{y_1})\rangle_{\text{CFT}} = \langle O_\Delta(\vec{y_1})\rangle_{\text{CFT,con}} = 0\nonumber\\
    \text{2-pt fn:}\hspace{0.5cm}&\langle O_\Delta(\vec{y_1})O_\Delta(\vec{y_2})\rangle_{\text{CFT}} = \langle O_\Delta(\vec{y_1})O_\Delta(\vec{y_2})\rangle_{\text{CFT,con}} = \frac{2\nu c_\Delta}{\lvert\vec{y_1}-\vec{y_2}\rvert^{2\Delta}}\nonumber\\
    \text{3-pt fn:}\hspace{0.5cm}&\langle O_\Delta(\vec{y_1})O_\Delta(\vec{y_2})O_\Delta(\vec{y_3})\rangle_{\text{CFT}} = \langle O_\Delta(\vec{y_1})O_\Delta(\vec{y_2})O_\Delta(\vec{y_3})\rangle_{\text{CFT,con}} = 0\nonumber\\
    \text{4-pt fn:}\hspace{0.5cm}&\langle O_\Delta(\vec{y_1})O_\Delta(\vec{y_2})O_\Delta(\vec{y_3})O_\Delta(\vec{y_4})\rangle_{\text{CFT}} =\frac{(2\nu c_\Delta)^2}{\prod_{i<j}\lvert\vec{y_{ij}}\rvert^{\frac{2\Delta}{3}}}u^{-\frac{2\Delta}{3}}v^{\frac{\Delta}{3}}+\Bigl(u,v\rightarrow\frac{1}{u},\frac{v}{u}\Bigl)\nonumber\\
    &\hspace{6.5cm}+(u,v\rightarrow v,u)\nonumber\\
    &\langle O_\Delta(\vec{y_1})O_\Delta(\vec{y_2})O_\Delta(\vec{y_3})O_\Delta(\vec{y_4})\rangle_{\text{CFT,con}} = 0
\end{align}\

The form of these correlators are exactly those dictated by eq. (\ref{summcorr2}), expected for a conformal theory. These results greatly motivate and contribute to the belief of the validity of the AdS/CFT conjecture.

\subsection{Holographic Dictionary}

The n-point functions just obtained were the result of a careful study of a free scalar field theory on AdS with delicate boundary behavior as it approaches its conformal boundary through the use of the AdS/CFT correspondence, however the relative complexity of this procedure makes us wonder if there is any other more direct approach. With this concern in mind, we want to check if there is any relation between these desired boundary correlators for the operator $O_\Delta(\vec{x})$ with the simple and known bulk correlators for the field $\Phi(x)$. Consider the n-point functions of the free scalar field $\Phi(x)$ on a AdS background obtained from the usual Feynman rules:

\begin{align}\label{bulkcorrff}
    \text{1-pt fn:}\hspace{0.5cm}&\langle \Phi(y_1)\rangle = \langle \Phi(y_1)\rangle_{\text{con}} = 0\nonumber\\
    \text{2-pt fn:}\hspace{0.5cm}&\langle \Phi(y_1)\Phi(y_2)\rangle = \langle \Phi(y_1)\Phi(y_2)\rangle_{\text{con}} = G(y_1,y_2)\nonumber\\
    \text{3-pt fn:}\hspace{0.5cm}&\langle \Phi(y_1)\Phi(y_2)\Phi(y_3)\rangle = \langle \Phi(y_1)\Phi(y_2)\Phi(y_3)\rangle_{\text{con}} = 0\nonumber\\
    \text{4-pt fn:}\hspace{0.5cm}&\langle \Phi(y_1)\Phi(y_2)\Phi(y_3)\Phi(y_4)\rangle = G(y_1,y_2)G(y_3,y_4)+(y_2\leftrightarrow y_3)+(y_2\leftrightarrow y_4)\nonumber\\
    &\langle \Phi(y_1)\Phi(y_2)\Phi(y_3)\Phi(y_4)\rangle_{\text{con}} = 0
\end{align}\

where $G(x,z)$ is the Green's function of the wave operator:

\begin{equation}\label{bbpropeq}
    (-\Box+m^2)G(x,z) = \frac{1}{\sqrt{g}}\delta^{d+1}(x-z)
\end{equation}\

whose solution in terms of the AdS invariant variable $\xi$ is given by\footnote{The invariance of $\xi$ can be seen directly from its representation in terms of the chordal distance $u$: $\xi=\frac{1}{1+u}$, where $u$ is given by: $u=\frac{(x-z)^2}{2x_0z_0}$. Under an AdS isometry transformation, the Euclidean distance $(x-z)^2$ transforms according to eq. (\ref{distancetransf}) and the radial coordinates $x_0$ and $z_0$ according to eq. (\ref{x0transf}), which implies the transformation rule for $u$: $u(x',z')=\frac{(x'-z')^2}{2x'_0z'_0}=\frac{\Omega(x)\Omega(z)(x-z)^2}{2\Omega(x)x_0\Omega(z)z_0}=\frac{(x-z)^2}{2x_0z_0}=u(x,z)$. Since $u$ is invariant, and $\xi$ is a function of $u$, then $\xi$ is also invariant.} \cite{Freedman3}:

\begin{equation}\label{G}
    G(x,z) = \frac{2^{-\Delta}c_\Delta}{2\nu}\xi^\Delta\ _2 F_1\Bigl(\frac{\Delta}{2},\frac{\Delta+1}{2};\nu+1;\xi^2\Bigl),\hspace{0.5cm} \xi = \frac{2x_0z_0}{x_0^2+z_0^2+(\vec{x}-\vec{z})^2}
\end{equation}\

where the function $_2F_1$ is the Gauss' hypergeometric function. It is a nice exercise to check that this solution indeed satisfies the Green's equation. $G(x,z)$ is also known as the bulk-bulk propagator since it propagates the field between two points in the bulk of the AdS space. When one of its points is taken to the conformal boundary it has an expansion of the form:

\begin{align}\label{expbubup1}
    G(x,z) &\underset{x_0\rightarrow0}{=} \frac{x_0^\Delta}{2\nu}K(z,\vec{x}) + \mathcal{O}(x_0^{\Delta<})
\end{align}\

where we used that $_2F_1(\dotsc;0)=1$ and identified the explicit form of the bulk-boundary propagator. When both points are taken to the boundary however, using the known expansion for the bulk-boundary propagator eq. (\ref{expbubop}), it has the form:

\begin{equation}\label{expbubup2}
    G(x,z) \underset{x_0,z_0\rightarrow0}{=} (\text{contact terms}) + \frac{x_0^\Delta}{2\nu}\frac{z_0^\Delta}{2\nu}\frac{2\nu c_\Delta}{\lvert\vec{x}-\vec{z}\rvert^{2\Delta}}+(\text{subleading terms})
\end{equation}\

These nice expansions for the bulk-bulk propagator allow us to easily relate the boundary n-point functions with the bulk n-point functions simply as the former being the extension of the internal points of the latter to the conformal boundary of the AdS space \cite{FreedmanBook}! Indeed, by just applying these behaviors into the bulk correlators eq. (\ref{bulkcorrff}), the resulting quantities (up to contact terms which can always be renormalized with appropriate local counterterms) are:

\begin{align}
    \text{1-pt fn:}\hspace{0.5cm}&\lim_{y_{1,0}\rightarrow0}\frac{2\nu}{y_{1,0}^\Delta}\langle\Phi(y_1)\rangle = \lim_{y_{1,0}\rightarrow0}\frac{2\nu}{y_{1,0}^\Delta}\langle \Phi(y_1)\rangle_{\text{con}} = 0\nonumber\\
    \text{2-pt fn:}\hspace{0.5cm}&\lim_{y_{1,0},y_{2,0}\rightarrow0}\frac{2\nu}{y_{1,0}^\Delta}\frac{2\nu}{y_{2,0}^\Delta}\langle \Phi(y_1)\Phi(y_2)\rangle = \lim_{y_{1,0},y_{2,0}\rightarrow0}\frac{2\nu}{y_{1,0}^\Delta}\frac{2\nu}{y_{2,0}^\Delta}\langle \Phi(y_1)\Phi(y_2)\rangle_{\text{con}} = \frac{2\nu c_\Delta}{\lvert\vec{y_1}-\vec{y_2}\rvert^{2\Delta}}\nonumber\\
    \text{3-pt fn:}\hspace{0.5cm}&\lim_{y_{1,0},y_{2,0},y_{3,0}\rightarrow0}\frac{2\nu}{y_{1,0}^\Delta}\frac{2\nu}{y_{2,0}^\Delta}\frac{2\nu}{y_{3,0}^\Delta}\langle \Phi(y_1)\Phi(y_2)\Phi(y_3)\rangle\nonumber\\
    &\hspace{3cm}=
    \lim_{y_{1,0},y_{2,0},y_{3,0}\rightarrow0}\frac{2\nu}{y_{1,0}^\Delta}\frac{2\nu}{y_{2,0}^\Delta}\frac{2\nu}{y_{3,0}^\Delta}\langle \Phi(y_1)\Phi(y_2)\Phi(y_3)\rangle_{\text{con}}=0\nonumber\\
    \text{4-pt fn:}\hspace{0.5cm}&\lim_{y_{1,0},y_{2,0},y_{3,0},y_{4,0}\rightarrow0}\frac{2\nu}{y_{1,0}^\Delta}\frac{2\nu}{y_{2,0}^\Delta}\frac{2\nu}{y_{3,0}^\Delta}\frac{2\nu}{y_{4,0}^\Delta}\langle \Phi(y_1)\Phi(y_2)\Phi(y_3)\Phi(y_4)\rangle =\nonumber\\
    &\hspace{6cm}\frac{2\nu c_\Delta}{\lvert\vec{y_1}-\vec{y_2}\rvert^{2\Delta}}\frac{2\nu c_\Delta}{\lvert\vec{y_3}-\vec{y_4}\rvert^{2\Delta}}+(\vec{y_2}\leftrightarrow\vec{y_3})+(\vec{y_2}\leftrightarrow\vec{y_4})\nonumber\\
    &\lim_{y_{1,0},y_{2,0},y_{3,0},y_{4,0}\rightarrow0}\frac{2\nu}{y_{1,0}^\Delta}\frac{2\nu}{y_{2,0}^\Delta}\frac{2\nu}{y_{3,0}^\Delta}\frac{2\nu}{y_{4,0}^\Delta}\langle \Phi(y_1)\Phi(y_2)\Phi(y_3)\Phi(y_4)\rangle_{\text{con}} = 0
\end{align}\

but these values are nothing but the boundary n-point functions just obtained through the AdS/CFT procedure for the operator $O_\Delta(\vec{x})$ dual to the field $\Phi(x)$, eq. (\ref{ffcorr})! This fact suggests the boundary/bulk n-point functions equivalence:

\begin{equation}\label{dict1}
    \langle O_\Delta(\vec{y_1})\dotsm O_\Delta(\vec{y_n})\rangle_{\text{CFT}} \equiv \lim_{y_{1,0},\dotsc,y_{n,0}\rightarrow0} \frac{2\nu}{y_{1,0}^\Delta}\dotsm\frac{2\nu}{y_{n,0}^\Delta}\langle\Phi(y_1)\dotsm\Phi(y_n)\rangle
\end{equation}

\begin{equation}\label{dict2}
    \langle O_\Delta(\vec{y_1})\dotsm O_\Delta(\vec{y_n})\rangle_{\text{CFT,con}} \equiv \lim_{y_{1,0},\dotsc,y_{n,0}\rightarrow0} \frac{2\nu}{y_{1,0}^\Delta}\dotsm\frac{2\nu}{y_{n,0}^\Delta}\langle\Phi(y_1)\dotsm\Phi(y_n)\rangle_{\text{con}}
\end{equation}\

This is known as the holographic dictionary. Here we gave an heuristic derivation of this equivalence for the very particular case of a free scalar field on the AdS side in the classical approximation of the AdS/CFT correspondence, but note however that this dictionary as it is presented in eqs. (\ref{dict1}) and (\ref{dict2}) it is conjectured to hold for any theory on AdS and at the full quantum level of the AdS/CFT correspondence. The validity of this claim will be put to the test throughout our work as we consider more complicated and interesting theories.

\section{\texorpdfstring{$\Phi^3$}{} Theory}

\subsection{Semiclassical Approximation}

Up to now we have reviewed in detail how starting from a concrete field theory on a AdS background we can obtain the corresponding dual CFT correlators, in the particular case of a free scalar field through the use of the classical or saddle point approximation of the AdS/CFT correspondence. Now, of course free fields are not the only theories we can consider on AdS, and the correspondence is conjectured to be true not only in the classical approximation but at the full quantum level. The quantum corrections to the correlators will be the main topic of the next chapter. In what remains of the current chapter however, we will study how we can complicate the current picture by considering now self-interacting terms in the AdS action with the intention to not only start constructing more interesting theories but also to further test the validity of the AdS/CFT correspondence.\par
The most natural step in difficulty from the current picture is, of course, adding a $\Phi^3(x)$ self-interacting term to the free scalar field AdS action, from which we expect new contributions on top of the recently found correlators (eq. (\ref{ffcorr})). Therefore, the theory that we will have under study in this section is:

\begin{equation}\label{phi3theory}
    Z_{\text{AdS}} = \int D\Phi\ e^{-S_{\text{AdS}}[\Phi]},\hspace{0.2cm} S_{\text{AdS}}[\Phi] = \int d^{d+1}x\sqrt{g}\ \Bigl[\frac{1}{2}g^{\mu\nu}\partial_\mu\Phi(x)\partial_\nu\Phi(x)+\frac{1}{2}m^2\Phi^2(x)+\frac{\lambda}{3!}\Phi^3(x)\Bigl]
\end{equation}\

This is known as a $\Phi^3$ theory on a Euclidean AdS background. Just like we did for the free field case, the natural way to approach this path integral is by looking at quantum fluctuations $h(x)$ around the classical solution $\phi(x)$ of the AdS action through the change of variable $\Phi(x) = \phi(x) + h(x)$, resulting in:

\begin{equation}
    Z_{\text{AdS}} = e^{-S_{\text{AdS}}[\phi]}f[\phi]
\end{equation}\

where $S_{\text{AdS}}[\phi]$ is the same $\phi^3$ action  and $f[\phi]$, as we mentioned before, is some functional of the classical field $\phi(x)$ coupled to the quantum field $h(x)$ responsible for the quantum corrections to the dual CFT correlators. Then, adopting the classical or saddle point approximation of the AdS/CFT correspondence, we will truncate for the moment this functional from $Z_{\text{AdS}}$, focusing only on the on-shell contributions to the correlators:

\begin{equation}
    Z_{\text{AdS}} = e^{-S_{\text{AdS}}[\phi]}
\end{equation}\

This is the classical AdS path integral in which we will work on.

\subsection{Classical Solution}

Following the same steps as in the free field case, to continue advancing in the computation of the correlators we need the explicit dependence of the on-shell field $\phi(x)$ in terms of the dual source $\varphi_0(\vec{x})$. We know it is the classical solution of the AdS action, i.e., it satisfies the Euler-Lagrange equation:

\begin{equation}\label{eqphi3}
    (-\Box+m^2)\phi(x) = -\frac{\lambda}{2}\phi^2(x)
\end{equation}\

Now, exactly solving this equation is very hard. However, if we can think of the parameter $\lambda$ (which mediates the strength of the self-interaction) as being in some sense "small", we can resort to perturbation theory, easily solving for the form of the field $\phi(x)$ as an expansion of this parameter. With this objective in mind then, we will look for a solution to eq. (\ref{eqphi3}) of the form\footnote{Considering the expansion of the field up to order $\lambda^2$ will be sufficient to completely compute the correlation functions up to the 4-point functions.}:

\begin{equation}\label{expphi3}
    \phi(x) = \phi_{(0)}(x) + \lambda\phi_{(1)}(x) + \lambda^2\phi_{(2)}(x) + \mathcal{O}(\lambda^3)
\end{equation}\

where the functions $\phi_{(i)}(x)$ are to be determined. Therefore, replacing the expansion eq. (\ref{expphi3}) into the equation of motion eq. (\ref{eqphi3}), we find at each order in $\lambda$:

\begin{align}
    \text{Order } \lambda^0:\ & (-\Box+m^2)\phi_{(0)}(x) = 0\nonumber\\
    \text{Order } \lambda^1:\ & (-\Box+m^2)\phi_{(1)}(x) = -\frac{1}{2}\phi_{(0)}^2(x) \nonumber\\
    \text{Order } \lambda^2:\ & (-\Box+m^2)\phi_{(2)}(x) = -\phi_{(0)}(x)\phi_{(1)}(x)
\end{align}\

We already solved the resulting homogeneous equation at order $\lambda^0$. This is nothing but the equation of motion of a free field, eq. (\ref{eom0}). The solution we found which is regular in the interior of AdS with appropriate boundary behavior is given by:

\begin{equation}
    \phi_{(0)}(x) = \int d^dy\ K(x,\vec{y})\varphi_0(\vec{y})
\end{equation}\

where $K(x,\vec{y})$ is the bulk-boundary propagator given by eq. (\ref{ffsumm}). The resulting inhomogeneous equations at orders $\lambda^1$ and $\lambda^2$ can be easily solved in terms of the Green's function $G(x,z)$ of the wave operator eq. (\ref{bbpropeq}), also known as the bulk-bulk propagator whose form is given by eq. (\ref{G}). Any inhomogeneous wave equation can be directly solved in terms of this propagator simply as:

\begin{equation}\label{bubuproperty}
    (-\Box+m^2)f(x) = g(x) \implies f(x) = \int d^{d+1}z\sqrt{g}\ G(x,z)g(z)
\end{equation}\

Indeed, by just applying the wave operator on both sides of this solution and using the definition of the Green's function eq. (\ref{bbpropeq}) one recovers the original equation, as expected. Then, using this property we can easily solve the resulting equation at order $\lambda^1$ for $\phi_{(1)}(x)$:

\begin{equation}
    \phi_{(1)}(x) = -\frac{1}{2}\int d^{d+1}x_1\sqrt{g}\ G(x,x_1)\Bigl[\int d^dy\ K(x_1,\vec{y})\varphi_0(\vec{y})\Bigl]^2
\end{equation}\

and consequently, in exactly the same way we can solve the resulting equation at order $\lambda^2$ for $\phi_{(2)}(x)$:

\begin{align}
    \phi_{(2)}(x) &= \frac{1}{2}\int d^{d+1}x_1\sqrt{g}\ G(x,x_1)\Bigl[\int d^dy\ K(x_1,\vec{y})\varphi_0(\vec{y})\Bigl]\nonumber\\\
    &\hspace{3cm}\times\int d^{d+1}x_2\sqrt{g}\ G(x_1,x_2)\Bigl[\int d^dy\ K(x_2,\vec{y})\varphi_0(\vec{y})\Bigl]^2
\end{align}\

Therefore, the explicit form of the on-shell field $\phi(x)$ as a functional of the dual source $\varphi_0(\vec{y})$, up to order $\lambda^2$ in the self-interacting coupling constant, is:

\begin{align}\label{solphi3}
    \phi(x) = &\int d^dy\ K(x,\vec{y})\varphi_0(\vec{y}) - \frac{\lambda}{2}\int d^{d+1}x_1\sqrt{g}\ G(x,x_1)\Bigl[\int d^dy\ K(x_1,\vec{y})\varphi_0(\vec{y})\Bigl]^2\nonumber\\
    &+\frac{\lambda^2}{2}\int d^{d+1}x_1\sqrt{g}\ G(x,x_1)\Bigl[\int d^dy\ K(x_1,\vec{y})\varphi_0(\vec{y})\Bigl]\nonumber\\
    &\hspace{3cm}\times\int d^{d+1}x_2\sqrt{g}\ G(x_1,x_2)\Bigl[\int d^dy\ K(x_2,\vec{y})\varphi_0(\vec{y})\Bigl]^2+\mathcal{O}(\lambda^3)
\end{align}\

The next step in the computation of the CFT correlators is to replace this solution for $\phi(x)$ in the classical AdS path integral, variate it with respect to the dual source $\varphi_0(\vec{x})$ and then extract from the resulting ill-defined variation the sensitive information which will give rise to the particular finite values of the n-point functions, process which is known as holographic renormalization. Now, as we saw in detail for the case of a free field, this process only requires knowing the on-shell field $\phi(x)$ up to order $x_0^\Delta$ since the resulting quantities in this procedure coming from higher order terms will simply vanish, not making any contribution to the correlators. To this end, to know the field $\phi(x)$ just obtained (eq. (\ref{solphi3})) up to order $x_0^\Delta$, we need the expansion of both propagators eqs. (\ref{expbubop}) and (\ref{expbubup1}). Since the leading term in the expansion of $G(x,z)$ is already of the order of $x_0^\Delta$, the resulting form of the field $\phi(x)$ up to this order after using the expansion of both propagators is exactly the same as the one found for the free field (eq. (\ref{compact}))! The only difference with the current case is what we understand by the function $\varphi_{2\nu}(\vec{x})$, where now for a $\phi^3$ theory it will receive self-interacting contributions of the order of the coupling constant $\lambda$ and its powers. Indeed, expanding the propagators in eq. (\ref{solphi3}) we easily recover the same asymptotic form of the free field:

\begin{equation}
    \phi(x) = x_0^{d-\Delta}\varphi(x) + x_0^\Delta\varphi_{2\nu}(\vec{x}) + x_0^\Delta\ln{(x_0)}\psi(\vec{x})+\mathcal{O}(x_0^{\Delta<})
\end{equation}\

where the functions $\varphi(x)$ and $\psi(\vec{x})$ are given by eqs. (\ref{varphi2}) and (\ref{psi2}) respectively, and where now the function $\varphi_{2\nu}(\vec{x})$, i.e., the terms of the order $x_0^\Delta$ in the expansion of the field, is given by:

\begin{align}\label{phi2nu3}
    \varphi_{2\nu}(\vec{x}) = &\int d^dy\ \frac{c_\Delta}{\lvert\vec{x}-\vec{y}\rvert^{2\Delta}}\varphi_0(\vec{y}) - \frac{\lambda}{4\nu}\int d^{d+1}x_1\sqrt{g}\ K(x_1,\vec{x})\Bigl[\int d^dy\ K(x_1,\vec{y})\varphi_0(\vec{y})\Bigl]^2\nonumber\\
    &+\frac{\lambda^2}{4\nu}\int d^{d+1}x_1\sqrt{g}\ K(x_1,\vec{x})\Bigl[\int d^dy\ K(x_1,\vec{y})\varphi_0(\vec{y})\Bigl]\nonumber\\
    &\hspace{2.5cm}\times\int d^{d+1}x_2\sqrt{g}\ G(x_1,x_2)\Bigl[\int d^dy\ K(x_2,\vec{y})\varphi_0(\vec{y})\Bigl]^2+\mathcal{O}(\lambda^3)
\end{align}\

Of course for $\lambda = 0$ it reduces to the value obtained for the free field case. Having the expansion of the field at hand, the next step is to holographic renormalize the ill-defined variation of the AdS path integral, in order to construct a finite, predictive quantity. Notice however that for the free field this process didn't require the explicit form of $\varphi_{2\nu}(\vec{x})$. Since the only difference with the current case is what we understand by this function, this implies that we can renormalize the variation of the path integral in exactly the same way as we did before! This is, adding the boundary term eq. (\ref{boundaryterm}) to the AdS action of the form of eq. (\ref{actionren}) renders the variation of the AdS path integral finite, given by:

\begin{align}\label{varzphi3}
    \delta Z_{\text{AdS}}[\varphi_0] =Z_{\text{AdS}}[\varphi_0]\int d^dx\ 2\nu\varphi_{2\nu}(\vec{x})\delta\varphi_0(\vec{x})
\end{align}\

where now with the presence of a $\phi^3$ self-interacting term, the function $\varphi_{2\nu}(\vec{x})$ is given by eq. (\ref{phi2nu3}).

\subsection{Correlation Functions}

The finite, renormalized variation $\delta Z_{\text{AdS}}$ as a functional of the corresponding conformal sources is the main object to be computed in the study of the dual correlators since once its explicit form is found, every n-point function can be directly obtained from it through a simple exercise of just taking derivatives. For a $\phi^3$ theory on AdS, this study led us to eq. (\ref{varzphi3}) where the quantity $\varphi_{2\nu}(\vec{x})$ is given by eq. (\ref{phi2nu3}):

\begin{align}\label{varzphi3explicit}
    \delta Z_{\text{AdS}}[\varphi_0] =Z_{\text{AdS}}[\varphi_0]\int d^dx\Bigl\{&\int d^dy\ \frac{2\nu c_\Delta}{\lvert\vec{x}-\vec{y}\rvert^{2\Delta}}\varphi_0(\vec{y})\nonumber\\
    &-\frac{\lambda}{2}\int d^{d+1}x_1\sqrt{g}\ K(x_1,\vec{x})\Bigl[\int d^dy\ K(x_1,\vec{y})\varphi_0(\vec{y})\Bigl]^2\nonumber\\
    &+\frac{\lambda^2}{2}\int d^{d+1}x_1\sqrt{g}\ K(x_1,\vec{x})\Bigl[\int d^dy\ K(x_1,\vec{y})\varphi_0(\vec{y})\Bigl]\nonumber\\
    &\hspace{1cm}\times\int d^{d+1}x_2\sqrt{g}\ G(x_1,x_2)\Bigl[\int d^dy\ K(x_2,\vec{y})\varphi_0(\vec{y})\Bigl]^2\nonumber\\
    &+\mathcal{O}(\lambda^3)\Bigl\}\delta\varphi_0(\vec{x})
\end{align}\

From this expression it is just a matter of direct calculation to obtain the corresponding correlators eqs. (\ref{euccorr1}) and (\ref{euccorr2}). The resulting 1-, 2-, 3- and 4-point functions for some primary scalar operator $O_\Delta(\vec{x})$ of scaling dimension $\Delta$ dual to a $\Phi^3$ self-interacting scalar field in AdS$_{d+1}$ are given by:

\begin{align}\label{phi3corr}
    \text{1-pt fn:}\hspace{0.5cm}&\langle O_\Delta(\vec{y_1})\rangle_{\text{CFT}} = \langle O_\Delta(\vec{y_1})\rangle_{\text{CFT,con}} = 0\nonumber\\
    \text{2-pt fn:}\hspace{0.5cm}&\langle O_\Delta(\vec{y_1})O_\Delta(\vec{y_2})\rangle_{\text{CFT}} = \langle O_\Delta(\vec{y_1})O_\Delta(\vec{y_2})\rangle_{\text{CFT,con}} = \frac{2\nu c_\Delta}{\lvert\vec{y_1}-\vec{y_2}\rvert^{2\Delta}}\nonumber\\
    \text{3-pt fn:}\hspace{0.5cm}&\langle O_\Delta(\vec{y_1})O_\Delta(\vec{y_2})O_\Delta(\vec{y_3})\rangle_{\text{CFT}} = \langle O_\Delta(\vec{y_1})O_\Delta(\vec{y_2})O_\Delta(\vec{y_3})\rangle_{\text{CFT,con}}\nonumber\\
    &\hspace{4.8cm}= -\lambda\int K(x_1,\vec{y_1})K(x_1,\vec{y_2})K(x_1,\vec{y_3})\nonumber\\
    \text{4-pt fn:}\hspace{0.5cm}&\langle O_\Delta(\vec{y_1})O_\Delta(\vec{y_2})O_\Delta(\vec{y_3})O_\Delta(\vec{y_4})\rangle_{\text{CFT}} =\frac{2\nu c_\Delta}{\lvert\vec{y_1}-\vec{y_2}\rvert^{2\Delta}}\frac{2\nu c_\Delta}{\lvert\vec{y_3}-\vec{y_4}\rvert^{2\Delta}}+(\vec{y_2}\leftrightarrow\vec{y_3})+(\vec{y_2}\leftrightarrow\vec{y_4})\nonumber\\
    &\hspace{3.5cm}+\lambda^2\int\int K(x_1,\vec{y_1})K(x_1,\vec{y_2})G(x_1,x_2)K(x_2,\vec{y_3})K(x_2,\vec{y_4})\times 3\nonumber\\
    &\langle O_\Delta(\vec{y_1})O_\Delta(\vec{y_2})O_\Delta(\vec{y_3})O_\Delta(\vec{y_4})\rangle_{\text{CFT,con}} =\nonumber\\
    &\hspace{4cm}\lambda^2\int\int K(x_1,\vec{y_1})K(x_1,\vec{y_2})G(x_1,x_2)K(x_2,\vec{y_3})K(x_2,\vec{y_4})\times 3
\end{align}\

where, with the intention to keep the notation short, we defined\\ $\int\int\dotsm\equiv\int d^{d+1}x_1\sqrt{g}\int d^{d+1}x_2\sqrt{g}\dotsm$ and represented the different permutations of the integrals as a multiplicative factor at the end of each. Notice the new contributions of the order of the self-interacting coupling constant $\lambda$ to the 3- and 4-point functions, compared to the free field case. Pictorially, these new contributions correspond exactly to the connected tree-level diagrams expected from a $\Phi^3$ self-interacting theory. Since these integrals are contributing to specific correlators which are conjectured to be of the form dictated by eq. (\ref{summcorr2}), the functional form of their results is strongly conditioned purely from conformal symmetry arguments. We will proceed then to study these quantities in detail through their explicit computation.

\subsubsection{3-Point Function}

In the holographic CFT 3-point function dual to a $\Phi^3$ self-interacting theory on AdS we then find a contribution of the form:

\begin{equation}\label{phi3classint1}
   I(\vec{y_1},\vec{y_2},\vec{y_3}) = -\lambda\int d^{d+1}x_1\sqrt{g}\ K(x_1,\vec{y_1})K(x_1,\vec{y_2})K(x_1,\vec{y_3})
\end{equation}\

\begin{figure}[h]
    \[\begin{wittendiagram}
    \draw (-1.732,1) node[vertex] -- (0,0)
    (1.732,1) node[vertex] -- (0,0)
    (0,-2) node[vertex] -- (0,0)
    (0,0) node[vertex]
    
     (-1.732,1) node[left]{$\vec{y_1}$}
    (1.732,1) node[right]{$\vec{y_2}$}
    (0,-2) node[below]{$\vec{y_3}$}
    
    ;
  \end{wittendiagram}\]
  \caption{Pictorial representation of the contribution to the 3-point function.}
\end{figure}

These integrals involving only the bulk-boundary propagator are common objects in the study of holographic correlators, which in the AdS/CFT literature can be found under the name of D-functions. The computation of these quantities are in principle straightforward but tedious, so with the intention to not lose the focus of discussion we will give them a separate treatment from the main text, dedicating the entire Appendix A to their delicate study. What will matter to us right now is that their definition in eq. (\ref{dfunc}) allows us to write the integral that we are interested in computing eq. (\ref{phi3classint1}) in the form of:

\begin{equation}
   I(\vec{y_1},\vec{y_2},\vec{y_3}) = -\lambda c^3_\Delta D_{\Delta\Delta\Delta}(\vec{y_1},\vec{y_2},\vec{y_3})
\end{equation}\

where $\tilde{K}^\Delta(x_1,\vec{y_i})$ is the unnormalized bulk-boundary propagator, eq. (\ref{unbubop}). The complete study of this particular D-function can be found in section A.2 of Appendix A, concluding in its value in eq. (\ref{3dfunc}). Using this value then in our present case we find that the final result of the integral eq. (\ref{phi3classint1}) is given by:

\begin{equation}\label{i3phi3}
    I(\vec{y_1},\vec{y_2},\vec{y_3}) = -\lambda c^3_\Delta \frac{\pi^\frac{d}{2}}{2}\frac{\Gamma(\frac{3\Delta-d}{2})}{\Gamma(\Delta)^3}\frac{\Gamma(\frac{\Delta}{2})^3}{\lvert\vec{y_1}-\vec{y_2}\rvert^\Delta \lvert\vec{y_2}-\vec{y_3}\rvert^\Delta \lvert\vec{y_3}-\vec{y_1}\rvert^\Delta}
\end{equation}\

respecting, of course, the functional form expected for contributions to CFT 3-point functions derived in eq. (\ref{summcorr2}).

\subsubsection{4-Point Function}

In the holographic CFT 4-point function dual to a $\Phi^3$ self-interacting theory on AdS we find a contribution of the form:

\begin{align}\label{phi34pfclass1}
   I(\vec{y_1},\vec{y_2},\vec{y_3},\vec{y_4}) = &\lambda^2\int d^{d+1}x_1\sqrt{g}\int d^{d+1}x_2\sqrt{g}\ K(x_1,\vec{y_1})K(x_1,\vec{y_2})G(x_1,x_2)K(x_2,\vec{y_3})K(x_2,\vec{y_4})\nonumber\\
   &+(\vec{y_2}\leftrightarrow\vec{y_3})+(\vec{y_2}\leftrightarrow\vec{y_4})
\end{align}\

\begin{figure}[h]
    \[\begin{wittendiagram}
    \draw (-1.4142,1.4142) node[vertex] -- (-1,0)
    (-1.4142,-1.4142) node[vertex] -- (-1,0)
    (1.4142,1.4142) node[vertex] -- (1,0)
    (1.4142,-1.4142) node[vertex] -- (1,0)
    (-1,0) node[vertex] -- (1,0)
    (1,0) node[vertex]
    
    (-1.4142,1.4142) node[left]{$\vec{y_1}$}
    (-1.4142,-1.4142) node[left]{$\vec{y_2}$}
    (1.4142,1.4142) node[right]{$\vec{y_3}$}
    (1.4142,-1.4142) node[right]{$\vec{y_4}$}
    
    ;
  \end{wittendiagram}\]
  \caption{Pictorial representation of the contribution to the 4-point function.}
\end{figure}

which in terms of the unnormalized bulk-boundary propagator eq. (\ref{unbubop}) we can write it as:

\begin{align}\label{phi34pfclass2}
   I(\vec{y_1},\vec{y_2},\vec{y_3},\vec{y_4}) = &\lambda^2c^4_\Delta\int d^{d+1}x_1\sqrt{g}\ \tilde{K}^\Delta(x_1,\vec{y_1})\tilde{K}^\Delta(x_1,\vec{y_2})\nonumber\\
   &\times\int d^{d+1}x_2\sqrt{g}\ G(x_1,x_2)\tilde{K}^\Delta(x_2,\vec{y_3})\tilde{K}^\Delta(x_2,\vec{y_4})+(\vec{y_2}\leftrightarrow\vec{y_3})+(\vec{y_2}\leftrightarrow\vec{y_4})
\end{align}\

The solving strategy for this quantity will be brute force. We will start by first computing one of the integrals, say the $x_2$ integral, hoping that the remaining integral in $x_1$ will be familiar to us, which as we will see will indeed be the case. Notice however that unlike the D-functions, the $x_2$ integral in this case not only contains the bulk-boundary propagator but also the bulk-bulk propagator, which is of course expected for diagrams with internal lines on AdS. These integrals involving both propagators are common objects in the study of holographic correlators, specially at their quantum corrections as we will see in the next chapter. The computation of these integrals are in principle straightforward but tedious, so just like we did for the D-functions, with the intention to not lose the focus of discussion we will give them a separate treatment from the main text, dedicating the entire Appendix B to their delicate study. What will matter to us right now is the discussion that takes place at the beginning of this appendix, where we prove that the argument of the bulk-bulk propagator ranges between 0 and 1 in the entire region of integration, allowing us to express it in its convergent power series representation eq. (\ref{Gseries}), which in turn it further allows us to write the $x_2$ integral in eq. (\ref{phi34pfclass2}) in the form of:

\begin{align}
    \int d^{d+1}x_2\sqrt{g}\ G(x_1,x_2)\tilde{K}^\Delta(x_2,\vec{y_3})\tilde{K}^\Delta(x_2,\vec{y_4}) = &\frac{2^{-\Delta}c_\Delta}{2\nu}\sum_{k=0}^\infty\frac{(\frac{\Delta}{2})_k(\frac{\Delta+1}{2})_k}{(\nu+1)_k\ k!}\nonumber\\
    &\times\int d^{d+1}x_2\sqrt{g}\ \xi^{\Delta+2k}\tilde{K}^\Delta(x_2,\vec{y_3})\tilde{K}^\Delta(x_2,\vec{y_4})
\end{align}\

The complete study of this type of integrals can be found in section B.4 of Appendix B, concluding in its value in eq. (\ref{xikkform}). Using this formula then for the particular values $\Delta_1=\Delta+2k$, $\Delta_2=\Delta$ and $\Delta_3=\Delta$, we find that the result of the $x_2$ integral is given by:

\begin{align}\label{phi34pfclass3}
    \int d^{d+1}x_2\sqrt{g}\ G(x_1,x_2)\tilde{K}^\Delta(x_2,\vec{y_3})\tilde{K}^\Delta(x_2,\vec{y_4}) = &\tilde{K}^{\Delta}(x_1,\vec{y_3})\tilde{K}^{\Delta}(x_1,\vec{y_4})\pi^{\frac{d+1}{2}}\frac{2^{-\Delta}c_\Delta}{2\nu}\frac{\Gamma(\frac{3\Delta-d}{2})\Gamma(\frac{\Delta}{2})}{\Gamma(\frac{\Delta+1}{2})\Gamma(\frac{3\Delta}{2})}\nonumber\\
    &\hspace{-5cm}\times\sum_{k=0}^\infty\frac{(\frac{\Delta}{2})_k(\frac{\Delta}{2})_k(\frac{3\Delta-d}{2})_k}{(\nu+1)_k(\frac{3\Delta}{2})_k\ k!}\ _2F_1\Bigl(\Delta,\Delta;\frac{3\Delta}{2}+k;1-\tilde{K}(x_1,\vec{y_3})\tilde{K}(x_1,\vec{y_4})\lvert\vec{y_{34}}\rvert^2\Bigl)
\end{align}\

where we wrote the $k$-dependent Gamma functions in terms of their Pochhammer symbols $\Gamma(a+k)=\Gamma(a)(a)_k$ and moved all the $k$-independent terms out of the sum. A nice clue on how to proceed with the calculations is to note that if we are able to write this hypergeometric function being summed as a power series in $\tilde{K}(x_1,\vec{y_3})\tilde{K}(x_1,\vec{y_4})\lvert\vec{y_{34}}\rvert^2$ the value of the $x_2$ integral would consist of a sum of 2 bulk-boundary propagators of different scaling dimensions, value which when replaced back into the original integral eq. (\ref{phi34pfclass2}) would result in a sum of integrals of 4 bulk-boundary propagators in the $x_1$ variable. But we already discussed these integrals involving only bulk-boundary propagators when we studied the contribution to the 3-point function, these integrals are precisely the D-functions which are reviewed in detail in Appendix A. Therefore, if we are able to write the hypergeometric function in eq. (\ref{phi34pfclass3}) as a series in the bulk-boundary propagators, we can solve for eq. (\ref{phi34pfclass1}) in terms of D-functions just like we did for the 3-point function. But this can be easily achieved using the known linear transformation of the hypergeometric function \cite{HypFunc2}:

\begin{align}
   _2F_1(a,b;c;z) = &\frac{\Gamma(c)\Gamma(c-a-b)}{\Gamma(c-a)\Gamma(c-b)}\ _2F_1(a,b;a+b+1-c;1-z)\nonumber\\
   &+\frac{\Gamma(c)\Gamma(a+b-c)}{\Gamma(a)\Gamma(b)}(1-z)^{c-a-b}\ _2F_1(c-a,c-b;1+c-a-b;1-z) 
\end{align}\

which for the particular values $a=\Delta$, $b=\Delta$, $c=\frac{3\Delta}{2}+k$ and $z=1-\tilde{K}(x_1,\vec{y_3})\tilde{K}(x_1,\vec{y_4})\lvert\vec{y_{34}}\rvert^2$ let us rewrite eq. (\ref{phi34pfclass3}) as:

\begin{align}\label{phi34pfclass4}
    \int d^{d+1}x_2\sqrt{g}\ G(x_1,x_2)\tilde{K}^\Delta(x_2,\vec{y_3})\tilde{K}^\Delta(x_2,\vec{y_4})&=\tilde{K}^\Delta(x_1,\vec{y_3})\tilde{K}^\Delta(x_1,\vec{y_4})\pi^{\frac{d+1}{2}}\frac{2^{-\Delta}c_\Delta}{2\nu}\frac{\Gamma(\frac{3\Delta-d}{2})\Gamma(\frac{\Delta}{2})}{\Gamma(\frac{\Delta+1}{2})}\nonumber\\
    &\hspace{-5.2cm}\times\Bigl\{\frac{\Gamma(-\frac{\Delta}{2})}{\Gamma(\frac{\Delta}{2})^2}\sum_{k=0}^\infty\frac{(\frac{3\Delta-d}{2})_k(-\frac{\Delta}{2})_k}{(\nu+1)_k\ k!}\ _2F_1\Bigl(\Delta,\Delta;\frac{\Delta}{2}+1-k;\tilde{K}(x_1,\vec{y_3})\tilde{K}(x_1,\vec{y_4})\lvert\vec{y_{34}}\rvert^2\Bigl)\nonumber\\
    &\hspace{-4.4cm}+\frac{\Gamma(\frac{\Delta}{2})}{\Gamma(\Delta)^2}\sum_{k=0}^\infty\frac{(\frac{\Delta}{2})_k(\frac{\Delta}{2})_k(\frac{3\Delta-d}{2})_k(\frac{\Delta}{2})_{-k}}{(\nu+1)_k\ k!}\Bigl[\tilde{K}(x_1,\vec{y_3})\tilde{K}(x_1,\vec{y_4})\lvert\vec{y_{34}}\rvert^2\Bigl]^{-\frac{\Delta}{2}+k}\nonumber\\
    &\hspace{-2.2cm}\times\ _2F_1\Bigl(\frac{\Delta}{2}+k,\frac{\Delta}{2}+k;1-\frac{\Delta}{2}+k;\tilde{K}(x_1,\vec{y_3})\tilde{K}(x_1,\vec{y_4})\lvert\vec{y_{34}}\rvert^2\Bigl)\Bigl\}
\end{align}\

It turns out that these 2 resulting sums can be solved in closed form using known properties of sums, Pochhammer symbols and hypergeometric functions. The first sum can be computed to give:

\begin{align}\label{phi3classsum1}
    &\sum_{k=0}^\infty\frac{(\frac{3\Delta-d}{2})_k(-\frac{\Delta}{2})_k}{(\nu+1)_k\ k!}\ _2F_1\Bigl(\Delta,\Delta;\frac{\Delta}{2}+1-k;\tilde{K}(x_1,\vec{y_3})\tilde{K}(x_1,\vec{y_4})\lvert\vec{y_{34}}\rvert^2\Bigl)\nonumber\\
    &\hspace{1.5cm}=\frac{\Gamma(\nu+1)}{\Gamma(1-\frac{\Delta}{2})\Gamma(\frac{3\Delta-d}{2}+1)}\ _3F_2\Bigl(\Delta,\Delta,1;\frac{\Delta}{2}+1,\frac{3\Delta-d}{2}+1;\tilde{K}(x_1,\vec{y_3})\tilde{K}(x_1,\vec{y_4})\lvert\vec{y_{34}}\rvert^2\Bigl)
\end{align}\

where we expressed the hypergeometric function in its power series representation, used that $(a)_{-k}=\frac{(-1)^k}{(1-a)_k}$, solved the resulting sum in $k$ in terms of the $_2F_1$ function of unit argument where $_2F_1(a,b;c;1)=\frac{\Gamma(c)\Gamma(c-a-b)}{\Gamma(c-a)\Gamma(c-b)}$ and finally identified the representation of the generalized hypergeometric function $_3F_2$. Similarly, the second sum in eq. (\ref{phi34pfclass4}) result in the value:

\begin{align}\label{phi3classsum2}
    &\sum_{k=0}^\infty\frac{(\frac{\Delta}{2})_k(\frac{\Delta}{2})_k(\frac{3\Delta-d}{2})_k(\frac{\Delta}{2})_{-k}}{(\nu+1)_k\ k!}\Bigl[\tilde{K}(x_1,\vec{y_3})\tilde{K}(x_1,\vec{y_4})\lvert\vec{y_{34}}\rvert^2\Bigl]^{-\frac{\Delta}{2}+k}\nonumber\\
    &\hspace{0.7cm}\times\ _2F_1\Bigl(\frac{\Delta}{2}+k,\frac{\Delta}{2}+k;1-\frac{\Delta}{2}+k;\tilde{K}(x_1,\vec{y_3})\tilde{K}(x_1,\vec{y_4})\lvert\vec{y_{34}}\rvert^2\Bigl)\nonumber\\
    &\hspace{2.5cm}=\Bigl[\tilde{K}(x_1,\vec{y_3})\tilde{K}(x_1,\vec{y_4})\lvert\vec{y_{34}}\rvert^2\Bigl]^{-\frac{\Delta}{2}}\ _2F_1\Bigl(\frac{\Delta}{2},\frac{\Delta}{2};\nu+1;\tilde{K}(x_1,\vec{y_3})\tilde{K}(x_1,\vec{y_4})\lvert\vec{y_{34}}\rvert^2\Bigl)
\end{align}\

where we expressed the hypergeometric function in its power series representation, used that $(a)_{-k}=\frac{(-1)^k}{(1-a)_k}$ and $(a)_k(a+k)_l=(a)_{k+l}$, also used that $\sum_{l=0}^\infty\sum_{k=0}^\infty a_{l,k}=\sum_{l=0}^\infty\sum_{k=0}^l a_{l-k,k}$, solved the resulting sum in $k$ in terms of the terminating $_2F_1$ function of unit argument where $_2F_1(a,b;c;1)=\frac{\Gamma(c)\Gamma(c-a-b)}{\Gamma(c-a)\Gamma(c-b)}$ and finally identified the representation of the $_2F_1$ function. Eq. (\ref{phi3classsum1}) together with eq. (\ref{phi3classsum2}) allow us to express the result for the $\int GKK$ integral eq. (\ref{phi34pfclass4}) in the nice closed form:

\begin{align}
    &\int d^{d+1}x_2\sqrt{g}\ G(x_1,x_2)\tilde{K}^\Delta(x_2,\vec{y_3})\tilde{K}^\Delta(x_2,\vec{y_4})=\tilde{K}^\Delta(x_1,\vec{y_3})\tilde{K}^\Delta(x_1,\vec{y_4})\pi^{\frac{d+1}{2}}\frac{2^{-\Delta}c_\Delta}{2\nu}\frac{\Gamma(\frac{3\Delta-d}{2})\Gamma(\frac{\Delta}{2})}{\Gamma(\frac{\Delta+1}{2})}\nonumber\\
    &\hspace{0.5cm}\times\Bigl\{-\frac{\Gamma(\nu+1)}{\Gamma(\frac{\Delta}{2})\Gamma(\frac{\Delta}{2}+1)\Gamma(\frac{3\Delta-d}{2}+1)}\ _3F_2\Bigl(\Delta,\Delta,1;\frac{\Delta}{2}+1,\frac{3\Delta-d}{2}+1;\tilde{K}(x_1,\vec{y_3})\tilde{K}(x_1,\vec{y_4})\lvert\vec{y_{34}}\rvert^2\Bigl)\nonumber\\
    &\hspace{1.2cm}+\frac{\Gamma(\frac{\Delta}{2})}{\Gamma(\Delta)^2}\Bigl[\tilde{K}(x_1,\vec{y_3})\tilde{K}(x_1,\vec{y_4})\lvert\vec{y_{34}}\rvert^2\Bigl]^{-\frac{\Delta}{2}}\ _2F_1\Bigl(\frac{\Delta}{2},\frac{\Delta}{2};\nu+1;\tilde{K}(x_1,\vec{y_3})\tilde{K}(x_1,\vec{y_4})\lvert\vec{y_{34}}\rvert^2\Bigl)\Bigl\}
\end{align}\

where we used that $\frac{\Gamma(-\frac{\Delta}{2})}{\Gamma(\frac{\Delta}{2})\Gamma(1-\frac{\Delta}{2})}=-\frac{1}{\Gamma(\frac{\Delta}{2}+1)}$. Remember that we are trying to compute the contribution to the holographic 4-point function coming from the $\Phi^3$ self-interaction of the bulk field on AdS, eq. (\ref{phi34pfclass2}). Replacing then the result for the $x_2$ integral just found back into the quantity we are trying to compute, we find that it reduces to:

\begin{align}\label{phi34pfclass5}
   I(\vec{y_1},\vec{y_2},\vec{y_3},\vec{y_4}) = &\lambda^2c^4_\Delta\pi^{\frac{d+1}{2}}\frac{2^{-\Delta}c_\Delta}{2\nu}\frac{\Gamma(\frac{3\Delta-d}{2})\Gamma(\frac{\Delta}{2})}{\Gamma(\frac{\Delta+1}{2})}\nonumber\\
   &\hspace{-2cm}\times\int d^{d+1}x_1\sqrt{g}\ \tilde{K}^\Delta(x_1,\vec{y_1})\tilde{K}^\Delta(x_1,\vec{y_2})\tilde{K}^\Delta(x_1,\vec{y_3})\tilde{K}^\Delta(x_1,\vec{y_4})\nonumber\\
   &\hspace{-2cm}\times\Bigl\{\frac{\Gamma(\frac{\Delta}{2})}{\Gamma(\Delta)^2}\Bigl[\tilde{K}(x_1,\vec{y_3})\tilde{K}(x_1,\vec{y_4})\lvert\vec{y_{34}}\rvert^2\Bigl]^{-\frac{\Delta}{2}}\ _2F_1\Bigl(\frac{\Delta}{2},\frac{\Delta}{2};\nu+1;\tilde{K}(x_1,\vec{y_3})\tilde{K}(x_1,\vec{y_4})\lvert\vec{y_{34}}\rvert^2\Bigl)\nonumber\\
   &\hspace{-3cm}-\frac{\Gamma(\nu+1)}{\Gamma(\frac{\Delta}{2})\Gamma(\frac{\Delta}{2}+1)\Gamma(\frac{3\Delta-d}{2}+1)}\ _3F_2\Bigl(\Delta,\Delta,1;\frac{\Delta}{2}+1,\frac{3\Delta-d}{2}+1;\tilde{K}(x_1,\vec{y_3})\tilde{K}(x_1,\vec{y_4})\lvert\vec{y_{34}}\rvert^2\Bigl)\Bigl\}\nonumber\\
   &+(\vec{y_2}\leftrightarrow\vec{y_3})+(\vec{y_2}\leftrightarrow\vec{y_4})
\end{align}\

As we commented before, we can proceed with the calculations by simply writing these hypergeometric functions as a power series in their argument, resulting for eq. (\ref{phi34pfclass5}) in a sum of integrals of bulk-boundary propagators, each one solvable in terms of the D-functions eq. (\ref{dfunc}). Therefore following this path, the definition of the D-functions allows us to write this integral in the form of:

\begin{align}\label{phi34pfclass6}
   I(\vec{y_1},\vec{y_2},\vec{y_3},\vec{y_4}) = &\lambda^2c^4_\Delta\pi^{\frac{d+1}{2}}\frac{2^{-\Delta}c_\Delta}{2\nu}\frac{\Gamma(\frac{3\Delta-d}{2})\Gamma(\frac{\Delta}{2})}{\Gamma(\frac{\Delta+1}{2})}\Bigl[\frac{\Gamma(\frac{\Delta}{2})}{\Gamma(\Delta)^2}\sum_{k=0}^\infty\frac{(\frac{\Delta}{2})_k(\frac{\Delta}{2})_k}{(\nu+1)_k\ k!}D_{\Delta\Delta\frac{\Delta}{2}+k\frac{\Delta}{2}+k}\lvert\vec{y_{34}}\rvert^{-\Delta+2k}\nonumber\\
   &-\frac{\Gamma(\nu+1)}{\Gamma(\frac{\Delta}{2})\Gamma(\frac{\Delta}{2}+1)\Gamma(\frac{3\Delta-d}{2}+1)}\sum_{k=0}^\infty\frac{(\Delta)_k(\Delta)_k(1)_k}{(\frac{\Delta}{2}+1)_k(\frac{3\Delta-d}{2}+1)_k\ k!}D_{\Delta\Delta\Delta+k\Delta+k}\lvert\vec{y_{34}}\rvert^{2k}\Bigl]\nonumber\\
   &+(\vec{y_2}\leftrightarrow\vec{y_3})+(\vec{y_2}\leftrightarrow\vec{y_4})
\end{align}\

This result is in perfect agreement with what was found by D'Hoker, Freedman and Rastelli in \cite{Freedman4} and also independently by Dolan and Osborn in \cite{Osborn3}. The complete study of these type of D-functions can be found in section A.3 of Appendix A, concluding in its value in eq. (\ref{4dfuncgeneral}). Using this formula then for the particular values $\Delta_1=\Delta_2=\Delta$ and $\Delta_3=\Delta_4=\frac{\Delta}{2}+k$, we find that the first D-function in eq. (\ref{phi34pfclass6}) can be written as:

\begin{equation}
    D_{\Delta\Delta\frac{\Delta}{2}+k\frac{\Delta}{2}+k}\lvert\vec{y_{34}}\rvert^{-\Delta+2k} = \frac{\pi^\frac{d}{2}}{2}\frac{\Gamma(\frac{3\Delta-d}{2}+k)}{\Gamma(\Delta)^2\Gamma(\frac{\Delta}{2}+k)^2}\frac{u^\frac{\Delta}{3}v^\frac{\Delta}{3}}{\prod_{i<j}\lvert\vec{y_{ij}}\rvert^{\frac{2\Delta}{3}}}H\Bigl(\Delta,\Delta,\frac{\Delta}{2}+1-k,2\Delta;u,v\Bigl)
\end{equation}\

where the function $H(\dotsc;u,v)$ represents a series expansion on both cross ratios as discussed in Appendix A. Using the same formula eq. (\ref{4dfuncgeneral}) for the particular values $\Delta_1=\Delta_2=\Delta$ and $\Delta_3=\Delta_4=\Delta+k$, we find that the second D-function in eq. (\ref{phi34pfclass6}) can be written as:

\begin{equation}
    D_{\Delta\Delta\Delta+k\Delta+k}\lvert\vec{y_{34}}\rvert^{2k} = \frac{\pi^\frac{d}{2}}{2}\frac{\Gamma(2\Delta-\frac{d}{2}+k)}{\Gamma(\Delta)^2\Gamma(\Delta+k)^2}\frac{u^{\frac{\Delta}{3}}v^{\frac{\Delta}{3}}}{\prod_{i<j}\lvert\vec{y_{ij}}\rvert^{\frac{2\Delta}{3}}}H(\Delta,\Delta,1-k,2\Delta;u,v)
\end{equation}\

Replacing the values of both D-functions back into eq. (\ref{phi34pfclass6}), simplifying common terms and writing the permutations in terms of the cross ratios $u$ and $v$ as done in eq. (\ref{conformalform4pf}), we find that the final result of the contribution eq. (\ref{phi34pfclass1}) can be written in the form of:

\begin{align}\label{i4phi3}
   I(\vec{y_1},\vec{y_2},\vec{y_3},\vec{y_4}) = &\lambda^2c^4_\Delta\frac{\pi^\frac{d}{2}}{8}\frac{\Gamma(\frac{3\Delta-d}{2})\Gamma(\frac{\Delta}{2})}{\Gamma(\Delta)^4}\frac{u^\frac{\Delta}{3}v^\frac{\Delta}{3}}{\prod_{i<j}\lvert\vec{y_{ij}}\rvert^{\frac{2\Delta}{3}}}\nonumber\\
   &\times\Bigl[\sum_{k=0}^\infty\frac{\Gamma(\frac{3\Delta-d}{2}+k)}{\Gamma(\nu+1+k)\ k!}H\Bigl(\Delta,\Delta,\frac{\Delta}{2}+1-k,2\Delta;u,v\Bigl)\nonumber\\
   &\hspace{0.7cm}-\sum_{k=0}^\infty\frac{\Gamma(2\Delta-\frac{d}{2}+k)}{\Gamma(\frac{\Delta}{2}+1+k)\Gamma(\frac{3\Delta-d}{2}+1+k)}H(\Delta,\Delta,1-k,2\Delta;u,v)\Bigl]\nonumber\\
   &+\Bigl(u,v\rightarrow\frac{1}{u},\frac{v}{u}\Bigl) + (u,v\rightarrow v,u)
\end{align}\

respecting, of course, the functional form expected for contributions to CFT 4-point functions derived in eq. (\ref{summcorr2}).

\subsubsection{Final Correlators}

Finally then, replacing the results for the integrals eqs. (\ref{i3phi3}) and (\ref{i4phi3}) back into eq. (\ref{phi3corr}) and writing the disconnected part of the 4-point correlator in its conformal form eq. (\ref{conformalform4pf}), the resulting 1-, 2-, 3- and 4-point functions for some primary scalar operator $O_\Delta(\vec{x})$ of scaling dimension $\Delta$ dual to a $\Phi^3$ self-interacting scalar field in AdS$_{d+1}$ are given by:

\begin{align}\label{phi3corr2}
    \text{1-pt fn:}\hspace{0.5cm}&\langle O_\Delta(\vec{y_1})\rangle_{\text{CFT}} = \langle O_\Delta(\vec{y_1})\rangle_{\text{CFT,con}} = 0\nonumber\\
    \text{2-pt fn:}\hspace{0.5cm}&\langle O_\Delta(\vec{y_1})O_\Delta(\vec{y_2})\rangle_{\text{CFT}} = \langle O_\Delta(\vec{y_1})O_\Delta(\vec{y_2})\rangle_{\text{CFT,con}} = \frac{2\nu c_\Delta}{\lvert\vec{y_1}-\vec{y_2}\rvert^{2\Delta}}\nonumber\\
    \text{3-pt fn:}\hspace{0.5cm}&\langle O_\Delta(\vec{y_1})O_\Delta(\vec{y_2})O_\Delta(\vec{y_3})\rangle_{\text{CFT}} = \langle O_\Delta(\vec{y_1})O_\Delta(\vec{y_2})O_\Delta(\vec{y_3})\rangle_{\text{CFT,con}}\nonumber\\
    &\hspace{4.8cm}= -\lambda c^3_\Delta \frac{\pi^\frac{d}{2}}{2}\frac{\Gamma(\frac{3\Delta-d}{2})}{\Gamma(\Delta)^3}\frac{\Gamma(\frac{\Delta}{2})^3}{\lvert\vec{y_1}-\vec{y_2}\rvert^\Delta \lvert\vec{y_2}-\vec{y_3}\rvert^\Delta \lvert\vec{y_3}-\vec{y_1}\rvert^\Delta}\nonumber\\
    \text{4-pt fn:}\hspace{0.5cm}&\langle O_\Delta(\vec{y_1})O_\Delta(\vec{y_2})O_\Delta(\vec{y_3})O_\Delta(\vec{y_4})\rangle_{\text{CFT}} =\frac{(2\nu c_\Delta)^2}{\prod_{i<j}\lvert\vec{y_{ij}}\rvert^{\frac{2\Delta}{3}}}u^{-\frac{2\Delta}{3}}v^{\frac{\Delta}{3}}+\Bigl(u,v\rightarrow\frac{1}{u},\frac{v}{u}\Bigl)\nonumber\\
    &\hspace{2cm}+(u,v\rightarrow v,u)+\lambda^2c^4_\Delta\frac{\pi^\frac{d}{2}}{8}\frac{\Gamma(\frac{3\Delta-d}{2})\Gamma(\frac{\Delta}{2})}{\Gamma(\Delta)^4}\frac{u^\frac{\Delta}{3}v^\frac{\Delta}{3}}{\prod_{i<j}\lvert\vec{y_{ij}}\rvert^{\frac{2\Delta}{3}}}\nonumber\\
    &\hspace{2cm}\times\Bigl[\sum_{k=0}^\infty\frac{\Gamma(\frac{3\Delta-d}{2}+k)}{\Gamma(\nu+1+k)\ k!}H\Bigl(\Delta,\Delta,\frac{\Delta}{2}+1-k,2\Delta;u,v\Bigl)\nonumber\\
   &\hspace{2.7cm}-\sum_{k=0}^\infty\frac{\Gamma(2\Delta-\frac{d}{2}+k)}{\Gamma(\frac{\Delta}{2}+1+k)\Gamma(\frac{3\Delta-d}{2}+1+k)}H(\Delta,\Delta,1-k,2\Delta;u,v)\Bigl]\nonumber\\
   &\hspace{2cm}+\Bigl(u,v\rightarrow\frac{1}{u},\frac{v}{u}\Bigl) + (u,v\rightarrow v,u)\nonumber\\
    &\langle O_\Delta(\vec{y_1})O_\Delta(\vec{y_2})O_\Delta(\vec{y_3})O_\Delta(\vec{y_4})\rangle_{\text{CFT,con}} =\lambda^2c^4_\Delta\frac{\pi^\frac{d}{2}}{8}\frac{\Gamma(\frac{3\Delta-d}{2})\Gamma(\frac{\Delta}{2})}{\Gamma(\Delta)^4}\frac{u^\frac{\Delta}{3}v^\frac{\Delta}{3}}{\prod_{i<j}\lvert\vec{y_{ij}}\rvert^{\frac{2\Delta}{3}}}\nonumber\\
    &\hspace{2cm}\times\Bigl[\sum_{k=0}^\infty\frac{\Gamma(\frac{3\Delta-d}{2}+k)}{\Gamma(\nu+1+k)\ k!}H\Bigl(\Delta,\Delta,\frac{\Delta}{2}+1-k,2\Delta;u,v\Bigl)\nonumber\\
   &\hspace{2.7cm}-\sum_{k=0}^\infty\frac{\Gamma(2\Delta-\frac{d}{2}+k)}{\Gamma(\frac{\Delta}{2}+1+k)\Gamma(\frac{3\Delta-d}{2}+1+k)}H(\Delta,\Delta,1-k,2\Delta;u,v)\Bigl]\nonumber\\
   &\hspace{2cm}+\Bigl(u,v\rightarrow\frac{1}{u},\frac{v}{u}\Bigl) + (u,v\rightarrow v,u)
\end{align}\

The form of these correlators are exactly those dictated by eq. (\ref{summcorr2}), expected for a conformal theory. These results greatly motivate and contribute to the belief of the validity of the AdS/CFT conjecture.

\subsection{Holographic Dictionary}

The n-point functions just obtained were the result of a careful study of a $\Phi^3$ self-interacting scalar field theory on AdS with delicate boundary behavior as it approaches its conformal boundary through the use of the AdS/CFT correspondence, however the relative complexity of this procedure makes us wonder if there is any other more direct approach. For the free scalar field case this concern led us to relate the desired boundary correlators for the operator $O_\Delta(\vec{x})$ with the simple and known bulk correlators for the field $\Phi(x)$ defining what is known as the holographic dictionary eqs. (\ref{dict1}) and (\ref{dict2}). These boundary/bulk correlators equivalence certainly hold for free fields on AdS but we want to verify if it remains true for less trivial theories, in particular for $\Phi^3$ theories. Consider the tree-level n-point functions of the $\Phi^3$ self-interacting scalar field $\Phi(x)$ on a AdS background obtained from the usual Feynman rules:

\begin{align}\label{bulkcorrphi3}
    \text{1-pt fn:}\hspace{0.5cm}&\langle \Phi(y_1)\rangle = \langle \Phi(y_1)\rangle_{\text{con}} = 0\nonumber\\
    \text{2-pt fn:}\hspace{0.5cm}&\langle \Phi(y_1)\Phi(y_2)\rangle = \langle \Phi(y_1)\Phi(y_2)\rangle_{\text{con}} = G(y_1,y_2)\nonumber\\
    \text{3-pt fn:}\hspace{0.5cm}&\langle \Phi(y_1)\Phi(y_2)\Phi(y_3)\rangle = \langle \Phi(y_1)\Phi(y_2)\Phi(y_3)\rangle_{\text{con}} = -\lambda\int G(x_1,y_1)G(x_1,y_2)G(x_1,y_3)\nonumber\\
    \text{4-pt fn:}\hspace{0.5cm}&\langle \Phi(y_1)\Phi(y_2)\Phi(y_3)\Phi(y_4)\rangle = G(y_1,y_2)G(y_3,y_4)+(y_2\leftrightarrow y_3)+(y_2\leftrightarrow y_4)\nonumber\\
    &\hspace{3cm}+\lambda^2\int\int G(x_1,y_1)G(x_1,y_2)G(x_1,x_2)G(x_2,y_3)G(x_2,y_4)\times3\nonumber\\
    &\langle \Phi(y_1)\Phi(y_2)\Phi(y_3)\Phi(y_4)\rangle_{\text{con}} =\nonumber\\
    &\hspace{3.5cm}\lambda^2\int\int G(x_1,y_1)G(x_1,y_2)G(x_1,x_2)G(x_2,y_3)G(x_2,y_4)\times3
\end{align}\

where $\int\int\dotsm\equiv\int d^{d+1}x_1\sqrt{g}\int d^{d+1}x_2\sqrt{g}\dotsm$. The nice expansions for the bulk-bulk propagator eqs. (\ref{expbubup1}) and (\ref{expbubup2}) allow us to easily confirm the equivalence between the boundary n-point functions with the bulk n-point functions simply as the former being the extension of the internal points of the latter to the conformal boundary of the AdS space. Indeed, by just applying these behaviors into the bulk correlators eq. (\ref{bulkcorrphi3}), the resulting quantities (up to contact terms which can always be renormalized with appropriate local counterterms) are:

\begin{align}\label{hdphi3class}
    \text{1-pt fn:}\hspace{0.5cm}&\lim_{y_{1,0}\rightarrow0}\frac{2\nu}{y_{1,0}^\Delta}\langle\Phi(y_1)\rangle = \lim_{y_{1,0}\rightarrow0}\frac{2\nu}{y_{1,0}^\Delta}\langle \Phi(y_1)\rangle_{\text{con}} = 0\nonumber\\
    \text{2-pt fn:}\hspace{0.5cm}&\lim_{y_{1,0},y_{2,0}\rightarrow0}\frac{2\nu}{y_{1,0}^\Delta}\frac{2\nu}{y_{2,0}^\Delta}\langle \Phi(y_1)\Phi(y_2)\rangle = \lim_{y_{1,0},y_{2,0}\rightarrow0}\frac{2\nu}{y_{1,0}^\Delta}\frac{2\nu}{y_{2,0}^\Delta}\langle \Phi(y_1)\Phi(y_2)\rangle_{\text{con}} = \frac{2\nu c_\Delta}{\lvert\vec{y_1}-\vec{y_2}\rvert^{2\Delta}}\nonumber\\
    \text{3-pt fn:}\hspace{0.5cm}&\lim_{y_{1,0},y_{2,0},y_{3,0}\rightarrow0}\frac{2\nu}{y_{1,0}^\Delta}\frac{2\nu}{y_{2,0}^\Delta}\frac{2\nu}{y_{3,0}^\Delta}\langle \Phi(y_1)\Phi(y_2)\Phi(y_3)\rangle =\nonumber\\
    &\lim_{y_{1,0},y_{2,0},y_{3,0}\rightarrow0}\frac{2\nu}{y_{1,0}^\Delta}\frac{2\nu}{y_{2,0}^\Delta}\frac{2\nu}{y_{3,0}^\Delta}\langle \Phi(y_1)\Phi(y_2)\Phi(y_3)\rangle_{\text{con}}=-\lambda\int K(x_1,\vec{y_1})K(x_1,\vec{y_2})K(x_1,\vec{y_3})\nonumber\\
    \text{4-pt fn:}\hspace{0.5cm}&\lim_{y_{1,0},y_{2,0},y_{3,0},y_{4,0}\rightarrow0}\frac{2\nu}{y_{1,0}^\Delta}\frac{2\nu}{y_{2,0}^\Delta}\frac{2\nu}{y_{3,0}^\Delta}\frac{2\nu}{y_{4,0}^\Delta}\langle \Phi(y_1)\Phi(y_2)\Phi(y_3)\Phi(y_4)\rangle =\nonumber\\
    &\hspace{3.3cm}\frac{2\nu c_\Delta}{\lvert\vec{y_1}-\vec{y_2}\rvert^{2\Delta}}\frac{2\nu c_\Delta}{\lvert\vec{y_3}-\vec{y_4}\rvert^{2\Delta}}+(\vec{y_2}\leftrightarrow\vec{y_3})+(\vec{y_2}\leftrightarrow\vec{y_4})\nonumber\\
    &\hspace{3.3cm}+\lambda^2\int\int K(x_1,\vec{y_1})K(x_1,\vec{y_2})G(x_1,x_2)K(x_2,\vec{y_3})K(x_2,\vec{y_4})\times3\nonumber\\
    &\lim_{y_{1,0},y_{2,0},y_{3,0},y_{4,0}\rightarrow0}\frac{2\nu}{y_{1,0}^\Delta}\frac{2\nu}{y_{2,0}^\Delta}\frac{2\nu}{y_{3,0}^\Delta}\frac{2\nu}{y_{4,0}^\Delta}\langle \Phi(y_1)\Phi(y_2)\Phi(y_3)\Phi(y_4)\rangle_{\text{con}} =\nonumber\\
    &\hspace{3.8cm}\lambda^2\int\int K(x_1,\vec{y_1})K(x_1,\vec{y_2})G(x_1,x_2)K(x_2,\vec{y_3})K(x_2,\vec{y_4})\times3
\end{align}\

but these values are nothing but the boundary n-point functions just obtained through the AdS/CFT procedure for the operator $O_\Delta(\vec{x})$ dual to the field $\Phi(x)$, eq. (\ref{phi3corr})! This fact confirms the boundary/bulk n-point functions equivalence eqs. (\ref{dict1}) and (\ref{dict2}), known as holographic dictionary, for a $\Phi^3$ theory on AdS in the classical approximation of the AdS/CFT correspondence. The validity of this dictionary for a $\Phi^3$ theory at the full quantum level of the AdS/CFT correspondence will be put to the test in the next chapter.

\section{\texorpdfstring{$\Phi^4$}{} Theory}

\subsection{Semiclassical Approximation}

We have reviewed in detail how starting from a concrete field theory on a AdS background we can obtain the corresponding dual CFT correlators, in the particular cases of a free scalar field and a $\Phi^3$ self-interacting scalar field through the use of the classical or saddle point approximation of the AdS/CFT correspondence. Continuing in the line of complicating the current picture with the intention to not only construct more interesting theories but also to further test the validity of the AdS/CFT correspondence, we will study now another self-interacting scalar theory on the AdS side. Instead of adding a $\Phi^3(x)$ self-interacting term to the free scalar field AdS action as in the previous case, we will consider now a $\Phi^4(x)$ term of the form:

\begin{equation}\label{phi4theory}
    Z_{\text{AdS}} = \int D\Phi\ e^{-S_{\text{AdS}}[\Phi]},\hspace{0.5cm} S_{\text{AdS}}[\Phi] = \int d^{d+1}x\sqrt{g}\ \Bigl[\frac{1}{2}g^{\mu\nu}\partial_\mu\Phi(x)\partial_\nu\Phi(x)+\frac{1}{2}m^2\Phi^2(x)+\frac{\lambda}{4!}\Phi^4(x)\Bigl]
\end{equation}\

This is known as a $\Phi^4$ theory on a Euclidean AdS background. Just like we did in the previous cases, the natural way to approach this path integral is by looking at quantum fluctuations $h(x)$ around the classical solution $\phi(x)$ of the AdS action through the change of variable $\Phi(x)=\phi(x)+h(x)$, resulting in:

\begin{equation}
    Z_{\text{AdS}} = e^{-S_{\text{AdS}}[\phi]}f[\phi]
\end{equation}\

where $S_{\text{AdS}}[\phi]$ is the same $\phi^4$ action and $f[\phi]$ is the responsible for the quantum corrections to the dual CFT correlators. Then, adopting the classical or saddle point approximation of the AdS/CFT correspondence, we will truncate for the moment this functional from $Z_{\text{AdS}}$, focusing only on the on-shell contributions to the correlators:

\begin{equation}
    Z_{\text{AdS}} = e^{-S_{\text{AdS}}[\phi]}
\end{equation}\

This is the classical AdS path integral in which we will work in.

\subsection{Classical Solution}

Following the same steps as in the previous cases, to continue advancing in the computation of the correlators we need the explicit dependence of the on-shell field $\phi(x)$ in terms of the dual source $\varphi_0(\vec{x})$. We know it is the classical solution of the AdS action, i.e., it satisfies the Euler-Lagrange equation:

\begin{equation}\label{eleqphi4}
    (-\Box+m^2)\phi(x) = -\frac{\lambda}{6}\phi^3(x)
\end{equation}\

Exactly solving this equation is very hard, but if we can think of the parameter $\lambda$ as being in some sense "small", we can resort to perturbation theory, easily solving for the form of the field $\phi(x)$ as an expansion of this parameter. With this objective in mind then, we will look for a solution to eq. (\ref{eleqphi4}) of the form\footnote{Considering the expansion of the field up to order $\lambda$ will be sufficient to completely compute the correlation functions up to the 4-point functions.}:

\begin{equation}\label{expphi4}
    \phi(x) = \phi_{(0)}(x)+\lambda\phi_{(1)}(x)+\mathcal{O}(\lambda^2)
\end{equation}\

where the functions $\phi_{(i)}(x)$ are to be determined. Therefore, replacing the expansion eq. (\ref{expphi4}) into the equation of motion eq. (\ref{eleqphi4}), we find at each order in $\lambda$:

\begin{align}
    \text{Order } \lambda^0:\ & (-\Box+m^2)\phi_{(0)}(x) = 0\nonumber\\
    \text{Order } \lambda^1:\ & (-\Box+m^2)\phi_{(1)}(x) = -\frac{1}{6}\phi_{(0)}^3(x)
\end{align}\

We already solved the resulting homogeneous equation at order $\lambda^0$. This is nothing but the equation of motion of a free field, eq. (\ref{eom0}). The solution we found which is regular in the interior of AdS with appropriate boundary behavior is given by:

\begin{equation}
    \phi_{(0)}(x) = \int d^dy\ K(x,\vec{y})\varphi_0(\vec{y})
\end{equation}\

where $K(x,\vec{y})$ is the bulk-boundary propagator given by eq. (\ref{ffsumm}). The resulting inhomogeneous equation at order $\lambda^1$ can be easily solved in terms of the Green's function $G(x,z)$ of the wave operator eq. (\ref{bbpropeq}), also known as the bulk-bulk propagator whose form is given by eq. (\ref{G}). Any inhomogeneous wave equation can be directly solved in terms of this propagator using the property eq. (\ref{bubuproperty}), in particular for $\phi_{(1)}(x)$:

\begin{equation}
    \phi_{(1)}(x) = -\frac{1}{6}\int d^{d+1}x_1\sqrt{g}\ G(x,x_1)\Bigl[\int d^dy\ K(x_1,\vec{y})\varphi_0(\vec{y})\Bigl]^3
\end{equation}\

Therefore, the explicit form of the on-shell field $\phi(x)$ as a functional of the dual source $\varphi_0(\vec{y})$, up to order $\lambda$ in the self-interacting coupling constant, is:

\begin{equation}\label{solphi4}
    \phi(x) = \int d^dy\ K(x,\vec{y})\varphi_0(\vec{y})-\frac{\lambda}{6}\int d^{d+1}x_1\sqrt{g}\ G(x,x_1)\Bigl[\int d^dy\ K(x_1,\vec{y})\varphi_0(\vec{y})\Bigl]^3+\mathcal{O}(\lambda^2)
\end{equation}\

The next step in the computation of the CFT correlators is to replace this solution for $\phi(x)$ in the classical AdS path integral and to holographic renormalize its resulting ill-defined variation with respect to the dual source, process which will give rise to the particular finite values of the n-point functions. As we have already seen though, this procedure only requires knowing the on-shell field $\phi(x)$ up to order $x_0^\Delta$, since the resulting quantities coming from higher order terms will simply vanish, not making any contribution to the correlators. To this end then, expanding the propagators in eq. (\ref{solphi4}), as in the $\Phi^3$ case, we easily recover the same asymptotic form of the free field:

\begin{equation}
    \phi(x) = x_0^{d-\Delta}\varphi(x) + x_0^\Delta\varphi_{2\nu}(\vec{x}) + x_0^\Delta\ln{(x_0)}\psi(\vec{x})+\mathcal{O}(x_0^{\Delta<})
\end{equation}\

where the functions $\varphi(x)$ and $\psi(\vec{x})$ are given by eqs. (\ref{varphi2}) and (\ref{psi2}) respectively, and where now the function $\varphi_{2\nu}(\vec{x})$, i.e., the terms of the order $x_0^\Delta$ in the expansion of the field, is given by:

\begin{equation}\label{phi2nu4}
    \varphi_{2\nu}(\vec{x}) = \int d^dy\ \frac{c_\Delta}{\lvert\vec{x}-\vec{y}\rvert^{2\Delta}}\varphi_0(\vec{y}) - \frac{\lambda}{12\nu}\int d^{d+1}x_1\sqrt{g}\ K(x_1,\vec{x})\Bigl[\int d^dy\ K(x_1,\vec{y})\varphi_0(\vec{y})\Bigl]^3+\mathcal{O}(\lambda^2)
\end{equation}\

For $\lambda=0$ it reduces to the value obtained for the free field, as expected. Moreover, since the expansion of the field has exactly the same form as in the free field case, the only difference with the current case being what we understand by $\varphi_{2\nu}(\vec{x})$, and for the free field the holographic renormalization procedure didn't require the explicit form of this function, this implies that we can renormalize the current theory in exactly the same way as we did before. This is, adding the boundary term eq. (\ref{boundaryterm}) to the AdS action of the form of eq. (\ref{actionren}) renders the variation of the AdS path integral finite, given by:

\begin{align}\label{varzphi4}
    \delta Z_{\text{AdS}}[\varphi_0] =Z_{\text{AdS}}[\varphi_0]\int d^dx\ 2\nu\varphi_{2\nu}(\vec{x})\delta\varphi_0(\vec{x})
\end{align}\

where now with the presence of a $\phi^4$ self-interacting term, the function $\varphi_{2\nu}(\vec{x})$ is given by eq. (\ref{phi2nu4}).

\subsection{Correlation Functions}

The finite, renormalized variation $\delta Z_{\text{AdS}}$ as a functional of the corresponding conformal sources is the main object to be computed in the study of the dual correlators since once its explicit form is found, every n-point function can be directly obtained from it through a simple exercise of just taking derivatives. For a $\phi^4$ theory on AdS, this study led us to eq. (\ref{varzphi4}) where the quantity $\varphi_{2\nu}(\vec{x})$ is given by eq. (\ref{phi2nu4}):

\begin{align}\label{varzphi4explicit}
    \delta Z_{\text{AdS}}[\varphi_0] =Z_{\text{AdS}}[\varphi_0]\int d^dx\Bigl\{&\int d^dy\ \frac{2\nu c_\Delta}{\lvert\vec{x}-\vec{y}\rvert^{2\Delta}}\varphi_0(\vec{y})\nonumber\\
    &-\frac{\lambda}{6}\int d^{d+1}x_1\sqrt{g}\ K(x_1,\vec{x})\Bigl[\int d^dy\ K(x_1,\vec{y})\varphi_0(\vec{y})\Bigl]^3\nonumber\\
    &+\mathcal{O}(\lambda^2)\Bigl\}\delta\varphi_0(\vec{x})
\end{align}\

From this expression it is just a matter of direct calculation to obtain the corresponding correlators eqs. (\ref{euccorr1}) and (\ref{euccorr2}). The resulting 1-, 2-, 3- and 4-point functions for some primary scalar operator $O_\Delta(\vec{x})$ of scaling dimension $\Delta$ dual to a $\Phi^4$ self-interacting scalar field in AdS$_{d+1}$ are given by:

\begin{align}\label{phi4corr}
    \text{1-pt fn:}\hspace{0.5cm}&\langle O_\Delta(\vec{y_1})\rangle_{\text{CFT}} = \langle O_\Delta(\vec{y_1})\rangle_{\text{CFT,con}} = 0\nonumber\\
    \text{2-pt fn:}\hspace{0.5cm}&\langle O_\Delta(\vec{y_1})O_\Delta(\vec{y_2})\rangle_{\text{CFT}} = \langle O_\Delta(\vec{y_1})O_\Delta(\vec{y_2})\rangle_{\text{CFT,con}} = \frac{2\nu c_\Delta}{\lvert\vec{y_1}-\vec{y_2}\rvert^{2\Delta}}\nonumber\\
    \text{3-pt fn:}\hspace{0.5cm}&\langle O_\Delta(\vec{y_1})O_\Delta(\vec{y_2})O_\Delta(\vec{y_3})\rangle_{\text{CFT}} = \langle O_\Delta(\vec{y_1})O_\Delta(\vec{y_2})O_\Delta(\vec{y_3})\rangle_{\text{CFT,con}} = 0\nonumber\\
    \text{4-pt fn:}\hspace{0.5cm}&\langle O_\Delta(\vec{y_1})O_\Delta(\vec{y_2})O_\Delta(\vec{y_3})O_\Delta(\vec{y_4})\rangle_{\text{CFT}} =\frac{2\nu c_\Delta}{\lvert\vec{y_1}-\vec{y_2}\rvert^{2\Delta}}\frac{2\nu c_\Delta}{\lvert\vec{y_3}-\vec{y_4}\rvert^{2\Delta}}+(\vec{y_2}\leftrightarrow\vec{y_3})+(\vec{y_2}\leftrightarrow\vec{y_4})\nonumber\\
    &\hspace{6.7cm}-\lambda\int K(x_1,\vec{y_1})K(x_1,\vec{y_2})K(x_1,\vec{y_3})K(x_1,\vec{y_4})\nonumber\\
    &\langle O_\Delta(\vec{y_1})O_\Delta(\vec{y_2})O_\Delta(\vec{y_3})O_\Delta(\vec{y_4})\rangle_{\text{CFT,con}} = -\lambda \int K(x_1,\vec{y_1})K(x_1,\vec{y_2})K(x_1,\vec{y_3})K(x_1,\vec{y_4})
\end{align}\

where, with the intention to keep the notation short, we defined $\int\equiv\int d^{d+1}x_1\sqrt{g}$. Notice the new contribution of the order of the self-interacting coupling constant $\lambda$ to the 4-point function, compared to the free field case. Pictorially, this new contribution correspond exactly to the connected tree-level diagram expected from a $\Phi^4$ self-interacting theory. Since this integral is contributing to a specific correlator which is conjectured to be of the form dictated by eq. (\ref{summcorr2}), the functional form of its result is strongly conditioned purely from conformal symmetry arguments. We will proceed then to study this quantity in detail through its explicit computation.

\subsubsection{4-Point Function}

In the holographic CFT 4-point function dual to a $\Phi^4$ self-interacting theory on AdS we then find a contribution of the form:

\begin{equation}\label{phi4classint1}
    I(\vec{y_1},\vec{y_2},\vec{y_3},\vec{y_4}) = -\lambda \int d^{d+1}x_1\sqrt{g}\ K(x_1,\vec{y_1})K(x_1,\vec{y_2})K(x_1,\vec{y_3})K(x_1,\vec{y_4})
\end{equation}\

\begin{figure}[h]
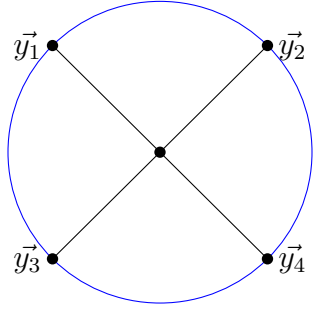

    \[\begin{wittendiagram}
    \draw (-1.4142,1.4142) node[vertex] -- (1.4142,-1.4142)
    (-1.4142,-1.4142) node[vertex] -- (1.4142,1.4142)
    (1.4142,1.4142) node[vertex]
    (1.4142,-1.4142) node[vertex]
    (0,0) node[vertex]
    
    (-1.4142,1.4142) node[left]{$\vec{y_1}$}
    (-1.4142,-1.4142) node[left]{$\vec{y_3}$}
    (1.4142,1.4142) node[right]{$\vec{y_2}$}
    (1.4142,-1.4142) node[right]{$\vec{y_4}$}
    
    ;
  \end{wittendiagram}\]
  \caption{Pictorial representation of the contribution to the 4-point function.}
\end{figure}

As we already mentioned previously in the study of the $\Phi^3$ theory, these integrals involving only the bulk-boundary propagator, the D-functions as they are known in the AdS/CFT literature, are common objects in the study of holographic correlators. Since their computation are in principle straightforward but tedious, with the intention to not lose the focus of discussion we give them a separate treatment from the main text, dedicating the entire Appendix A to their delicate study. What will matter to us right now is that their definition in eq. (\ref{dfunc}) allows us to write the integral that we are interested in computing eq. (\ref{phi4classint1}) in the form of:

\begin{equation}
    I(\vec{y_1},\vec{y_2},\vec{y_3},\vec{y_4}) = -\lambda c^4_\Delta D_{\Delta\Delta\Delta\Delta}(\vec{y_1},\vec{y_2},\vec{y_3},\vec{y_4})
\end{equation}\

where $\tilde{K}^\Delta(x_1,\vec{y_i})$ is the unnormalized bulk-boundary propagator, eq. (\ref{unbubop}). The complete study of this particular D-function can be found in section A.3 of Appendix A, concluding in its value in eq. (\ref{4dfunc}). Using this value then in our present case we find that the final result of the integral eq. (\ref{phi4classint1}) is given by:

\begin{equation}\label{i4phi4}
    I(\vec{y_1},\vec{y_2},\vec{y_3},\vec{y_4}) = -\lambda c^4_\Delta \frac{\pi^{\frac{d}{2}}}{2}\frac{\Gamma(2\Delta-\frac{d}{2})}{\Gamma(\Delta)^4}\frac{u^{\frac{\Delta}{3}}v^{\frac{\Delta}{3}}}{\prod_{i<j}\lvert\vec{y_i}-\vec{y_j}\rvert^{\frac{2\Delta}{3}}}H(\Delta,\Delta,1,2\Delta;u,v)
\end{equation}\

where the function $H(\dotsc;u,v)$ represents a series expansion on both cross ratios as discussed in Appendix A. This result respects, of course, the functional form expected for contributions to CFT 4-point functions derived in eq. (\ref{summcorr2}).

\subsubsection{Final Correlators}

Finally then, replacing the result for the integral eq. (\ref{i4phi4}) back into eq. (\ref{phi4corr}) and writing the disconnected part of the 4-point correlator in its conformal form eq. (\ref{conformalform4pf}), the resulting 1-, 2-, 3- and 4-point functions for some primary scalar operator $O_\Delta(\vec{x})$ of scaling dimension $\Delta$ dual to a $\Phi^4$ self-interacting scalar field in AdS$_{d+1}$ are given by:

\begin{align}\label{phi4corr2}
    \text{1-pt fn:}\hspace{0.5cm}&\langle O_\Delta(\vec{y_1})\rangle_{\text{CFT}} = \langle O_\Delta(\vec{y_1})\rangle_{\text{CFT,con}} = 0\nonumber\\
    \text{2-pt fn:}\hspace{0.5cm}&\langle O_\Delta(\vec{y_1})O_\Delta(\vec{y_2})\rangle_{\text{CFT}} = \langle O_\Delta(\vec{y_1})O_\Delta(\vec{y_2})\rangle_{\text{CFT,con}} = \frac{2\nu c_\Delta}{\lvert\vec{y_1}-\vec{y_2}\rvert^{2\Delta}}\nonumber\\
    \text{3-pt fn:}\hspace{0.5cm}&\langle O_\Delta(\vec{y_1})O_\Delta(\vec{y_2})O_\Delta(\vec{y_3})\rangle_{\text{CFT}} = \langle O_\Delta(\vec{y_1})O_\Delta(\vec{y_2})O_\Delta(\vec{y_3})\rangle_{\text{CFT,con}} = 0\nonumber\\
    \text{4-pt fn:}\hspace{0.5cm}&\langle O_\Delta(\vec{y_1})O_\Delta(\vec{y_2})O_\Delta(\vec{y_3})O_\Delta(\vec{y_4})\rangle_{\text{CFT}} =\frac{(2\nu c_\Delta)^2}{\prod_{i<j}\lvert\vec{y_{ij}}\rvert^{\frac{2\Delta}{3}}}u^{-\frac{2\Delta}{3}}v^{\frac{\Delta}{3}}+\Bigl(u,v\rightarrow\frac{1}{u},\frac{v}{u}\Bigl)\nonumber\\
    &\hspace{1cm}+(u,v\rightarrow v,u)-\lambda c^4_\Delta \frac{\pi^{\frac{d}{2}}}{2}\frac{\Gamma(2\Delta-\frac{d}{2})}{\Gamma(\Delta)^4}\frac{u^{\frac{\Delta}{3}}v^{\frac{\Delta}{3}}}{\prod_{i<j}\lvert\vec{y_i}-\vec{y_j}\rvert^{\frac{2\Delta}{3}}}H(\Delta,\Delta,1,2\Delta;u,v)\nonumber\\
    &\langle O_\Delta(\vec{y_1})O_\Delta(\vec{y_2})O_\Delta(\vec{y_3})O_\Delta(\vec{y_4})\rangle_{\text{CFT,con}} =\nonumber\\
    &\hspace{3.8cm}-\lambda c^4_\Delta \frac{\pi^{\frac{d}{2}}}{2}\frac{\Gamma(2\Delta-\frac{d}{2})}{\Gamma(\Delta)^4}\frac{u^{\frac{\Delta}{3}}v^{\frac{\Delta}{3}}}{\prod_{i<j}\lvert\vec{y_i}-\vec{y_j}\rvert^{\frac{2\Delta}{3}}}H(\Delta,\Delta,1,2\Delta;u,v)
\end{align}\

The form of these correlators are exactly those dictated by eq. (\ref{summcorr2}), expected for a conformal theory. These results greatly motivate and contribute to the belief of the validity of the AdS/CFT conjecture.

\subsection{Holographic Dictionary}

The n-point functions just obtained were the result of a careful study of a $\Phi^4$ self-interacting scalar field theory on AdS with delicate boundary behavior as it approaches its conformal boundary through the use of the AdS/CFT correspondence, however the relative complexity of this procedure makes us wonder if there is any other more direct approach. For the free scalar field case this concern led us to relate the desired boundary correlators for the operator $O_\Delta(\vec{x})$ with the simple and known bulk correlators for the field $\Phi(x)$ defining what is known as the holographic dictionary eqs. (\ref{dict1}) and (\ref{dict2}). These boundary/bulk correlators equivalence turned out to be true not only for free fields on AdS but also for $\Phi^3$ self-interacting fields. Now we want to verify if this relation still holds for $\Phi^4$ theories. Consider the tree-level n-point functions of the $\Phi^4$ self-interacting scalar field $\Phi(x)$ on a AdS background obtained from the usual Feynman rules:

\begin{align}\label{bulkcorrphi4}
    \text{1-pt fn:}\hspace{0.5cm}&\langle \Phi(y_1)\rangle = \langle \Phi(y_1)\rangle_{\text{con}} = 0\nonumber\\
    \text{2-pt fn:}\hspace{0.5cm}&\langle \Phi(y_1)\Phi(y_2)\rangle = \langle \Phi(y_1)\Phi(y_2)\rangle_{\text{con}} = G(y_1,y_2)\nonumber\\
    \text{3-pt fn:}\hspace{0.5cm}&\langle \Phi(y_1)\Phi(y_2)\Phi(y_3)\rangle = \langle \Phi(y_1)\Phi(y_2)\Phi(y_3)\rangle_{\text{con}} = 0\nonumber\\
    \text{4-pt fn:}\hspace{0.5cm}&\langle \Phi(y_1)\Phi(y_2)\Phi(y_3)\Phi(y_4)\rangle = G(y_1,y_2)G(y_3,y_4)+(y_2\leftrightarrow y_3)+(y_2\leftrightarrow y_4)\nonumber\\
    &\hspace{4.8cm}-\lambda\int G(x_1,y_1)G(x_1,y_2)G(x_1,y_3)G(x_1,y_4)\nonumber\\
    &\langle \Phi(y_1)\Phi(y_2)\Phi(y_3)\Phi(y_4)\rangle_{\text{con}} = -\lambda\int G(x_1,y_1)G(x_1,y_2)G(x_1,y_3)G(x_1,y_4)
\end{align}\

where $\int\equiv\int d^{d+1}x_1\sqrt{g}$. The nice expansions for the bulk-bulk propagator eqs. (\ref{expbubup1}) and (\ref{expbubup2}) allow us to easily confirm the equivalence between the boundary n-point functions with the bulk n-point functions simply as the former being the extension of the internal points of the latter to the conformal boundary of the AdS space. Indeed, by just applying these behaviors into the bulk correlators eq. (\ref{bulkcorrphi4}), the resulting quantities (up to contact terms which can always be renormalized with appropriate local counterterms) are:

\begin{align}\label{hdphi4class}
    \text{1-pt fn:}\hspace{0.5cm}&\lim_{y_{1,0}\rightarrow0}\frac{2\nu}{y_{1,0}^\Delta}\langle\Phi(y_1)\rangle = \lim_{y_{1,0}\rightarrow0}\frac{2\nu}{y_{1,0}^\Delta}\langle \Phi(y_1)\rangle_{\text{con}} = 0\nonumber\\
    \text{2-pt fn:}\hspace{0.5cm}&\lim_{y_{1,0},y_{2,0}\rightarrow0}\frac{2\nu}{y_{1,0}^\Delta}\frac{2\nu}{y_{2,0}^\Delta}\langle \Phi(y_1)\Phi(y_2)\rangle = \lim_{y_{1,0},y_{2,0}\rightarrow0}\frac{2\nu}{y_{1,0}^\Delta}\frac{2\nu}{y_{2,0}^\Delta}\langle \Phi(y_1)\Phi(y_2)\rangle_{\text{con}} = \frac{2\nu c_\Delta}{\lvert\vec{y_1}-\vec{y_2}\rvert^{2\Delta}}\nonumber\\
    \text{3-pt fn:}\hspace{0.5cm}&\lim_{y_{1,0},y_{2,0},y_{3,0}\rightarrow0}\frac{2\nu}{y_{1,0}^\Delta}\frac{2\nu}{y_{2,0}^\Delta}\frac{2\nu}{y_{3,0}^\Delta}\langle \Phi(y_1)\Phi(y_2)\Phi(y_3)\rangle \nonumber\\
    &\hspace{2cm}=
    \lim_{y_{1,0},y_{2,0},y_{3,0}\rightarrow0}\frac{2\nu}{y_{1,0}^\Delta}\frac{2\nu}{y_{2,0}^\Delta}\frac{2\nu}{y_{3,0}^\Delta}\langle \Phi(y_1)\Phi(y_2)\Phi(y_3)\rangle_{\text{con}}=0\nonumber\\
    \text{4-pt fn:}\hspace{0.5cm}&\lim_{y_{1,0},y_{2,0},y_{3,0},y_{4,0}\rightarrow0}\frac{2\nu}{y_{1,0}^\Delta}\frac{2\nu}{y_{2,0}^\Delta}\frac{2\nu}{y_{3,0}^\Delta}\frac{2\nu}{y_{4,0}^\Delta}\langle \Phi(y_1)\Phi(y_2)\Phi(y_3)\Phi(y_4)\rangle =\nonumber\\
    &\hspace{6cm}\frac{2\nu c_\Delta}{\lvert\vec{y_1}-\vec{y_2}\rvert^{2\Delta}}\frac{2\nu c_\Delta}{\lvert\vec{y_3}-\vec{y_4}\rvert^{2\Delta}}+(\vec{y_2}\leftrightarrow\vec{y_3})+(\vec{y_2}\leftrightarrow\vec{y_4})\nonumber\\
    &\hspace{6cm}-\lambda\int K(x_1,\vec{y_1})K(x_1,\vec{y_2})K(x_1,\vec{y_3})K(x_1,\vec{y_4})\nonumber\\
    &\lim_{y_{1,0},y_{2,0},y_{3,0},y_{4,0}\rightarrow0}\frac{2\nu}{y_{1,0}^\Delta}\frac{2\nu}{y_{2,0}^\Delta}\frac{2\nu}{y_{3,0}^\Delta}\frac{2\nu}{y_{4,0}^\Delta}\langle \Phi(y_1)\Phi(y_2)\Phi(y_3)\Phi(y_4)\rangle_{\text{con}} =\nonumber\\
    &\hspace{6cm}-\lambda\int K(x_1,\vec{y_1})K(x_1,\vec{y_2})K(x_1,\vec{y_3})K(x_1,\vec{y_4})
\end{align}\

but these values are nothing but the boundary n-point functions just obtained through the AdS/CFT procedure for the operator $O_\Delta(\vec{x})$ dual to the field $\Phi(x)$, eq. (\ref{phi4corr})! This fact confirms the boundary/bulk n-point functions equivalence eqs. (\ref{dict1}) and (\ref{dict2}), known as holographic dictionary, for a $\Phi^4$ theory on AdS in the classical approximation of the AdS/CFT correspondence. The validity of this dictionary for a $\Phi^4$ theory at the full quantum level of the AdS/CFT correspondence will be put to the test in the next chapter.

\newpage


\chapter{Quantum Scalar Theories in AdS/CFT}
Having studied and gained some insight on the AdS/CFT conjecture through its classical approximation for interacting scalar field theories on AdS, we will proceed to study these same theories but now embracing the full quantum nature of the correspondence. The objective of this chapter then is to develop a systematic scheme that adds order by order the respective quantum corrections to the previous holographic correlators obtained through the approximated correspondence, to then study and compute each one of these new contributions that will have the form of loop integrals in the bulk, process that will also force us to introduce sensitive regularization and renormalization schemes. This chapter will contain most of the original work done in this thesis.\par
Section 4.1 will cover the complete study of a $\Phi^3$ theory on AdS and its renormalized, quantum corrected holographic CFT correlators. In particular, in section 4.1.1 we will present the scheme that adds order by order in $\lambda$ the quantum corrections to the AdS path integral, in section 3.1.2 the resulting holographic n-point functions obtained from this corrected quantity, in section 3.1.3 how these same holographic correlators can be obtained from the holographic dictionary, in section 3.1.4 how in the properly normalized $\Phi^3$ theory all the tadpole contributions to the correlators are canceled, in section 3.1.5 the introduction of the corresponding IR and UV regularization schemes to be used on the loop integrals throughout this work, in sections 4.1.6, 4.1.7 and 4.1.8 the study and computation of these regularized integrals present in the 2-, 3- and 4-point functions respectively together with their proper renormalization, process which is summarized in the final section 4.1.9.\par
Lastly, section 4.2 will cover the same study but now for an interacting $\Phi^4$ theory on AdS and its resulting renormalized, quantum corrected CFT correlators. In particular, in section 4.2.1 we will present the scheme that adds order by order in $\lambda$ the quantum corrections to its AdS path integral, in section 4.2.2 the resulting holographic n-point functions obtained from this corrected quantity, in section 4.2.3 how these same holographic correlators can be obtained from the holographic dictionary, in section 4.2.4 the application of the same IR and UV regularization schemes introduced before to the loop integrals, in sections 4.2.5 and 4.2.6 the study and computation of these regularized integrals present in the 2- and 4-point functions respectively together with their proper renormalization, process which is summarized in the section 4.2.7. Finally, in section 4.2.8 we study a concrete example of a $\Phi^4$ theory on AdS and its resulting CFT correlators, applying the general ideas developed throughout the chapter.

\section{\texorpdfstring{$\Phi^3$}{} Theory}

\subsection{Semiclassical Approximation}

In the previous chapter we studied in detail how to obtain the corresponding CFT$_d$ correlators for the cases of a free scalar field, a $\Phi^3$ scalar theory and a $\Phi^4$ scalar theory on a AdS$_{d+1}$ background using the classical or saddle point approximation of the AdS/CFT correspondence, where we looked for solutions to the bulk fields $\Phi(x)$ as quantum fluctuations $h(x)$ around the classical solution $\phi(x)$ of the AdS actions of the form of $\Phi(x) = \phi(x) + h(x)$, and kept only the resulting on-shell contributions in the AdS path integrals, truncating the remaining off-shell terms. Of course under this approximation we only got a part of the full answer, yet the pieces we obtained in every n-point function were completely consistent with the expected for conformal theories. The objective that we will set ourselves in this chapter is to compute the remaining parts of the n-point functions obtained previously for the self-interacting theories, fully embracing the quantum nature of the path integrals. This study will not only give a more complete and satisfactory answer to the holographic correlators for these theories, showing in the process the concrete role of the quantum corrections coming from the off-shell part of the path integrals, but also a much stronger verification of the validity of the AdS/CFT correspondence.\par
We will start this study in the same order as before, tackling the $\Phi^3$ self-interacting scalar theory first and then, following the same steps and using the same ideas developed here, leaving the $\Phi^4$ case for last. The starting point will be the $\Phi^3$ theory defined in eq. (\ref{phi3theory}), together with its holographic renormalization eq. (\ref{actionren}):

\begin{equation}\
    Z_{\text{AdS}}[\varphi_0] = \int D\Phi\ e^{-S_{\text{AdS}}[\Phi]-\int d^dx\sqrt{\gamma}\ B(\Phi(x))\rvert_{x_0=0}}
\end{equation}\

where $S_{\text{AdS}}[\Phi]$ is the $\Phi^3$ action:

\begin{equation}
    S_{\text{AdS}}[\Phi] = \int d^{d+1}x\sqrt{g}\ \Bigl[\frac{1}{2}g^{\mu\nu}\partial_\mu\Phi(x)\partial_\nu\Phi(x)+\frac{1}{2}m^2\Phi^2(x)+\frac{\lambda}{3!}\Phi^3(x)\Bigl]
\end{equation}\

and where $B(\Phi(x))$ is the boundary term eq. (\ref{boundaryterm}), counterterm responsible for the renormalization of the infrared divergences coming from the on-shell part of the path integral:

\begin{align}
    B\bigl(\Phi(x)\bigl) = &\frac{1}{2}(d-\Delta)\Phi^2(x)+\nu\ln{(x_0)}C_\nu \Phi(x)\Box_\gamma^\nu\Phi(x)+\frac{1}{2}C_\nu \Phi(x)\Box_\gamma^\nu\Phi(x)\nonumber\\
    &+\frac{1}{4(\nu-1)}\Phi(x)\Box_\gamma\Phi(x)+\dotsb
\end{align}\

where the triple dots represent higher order derivative terms. We will proceed then in the same way as we did before for the saddle point approximation of the correspondence, looking at quantum fluctuations $h(x)$ around the classical solution $\phi(x)$ of the AdS action through the change of variable $\Phi(x)=\phi(x)+h(x)$, but now keeping track of every quantity resulting from this separation. Under this change of variable the AdS path integral transforms as:

\begin{equation}\label{phi3pisep1}
    Z_{\text{AdS}}[\varphi_0] = e^{-\int d^dx\sqrt{\gamma}\ B(\phi(x))\rvert_{x_0=0}}\int Dh\ e^{-S_{\text{AdS}}[\phi+h]}
\end{equation}\

where we used that the on-shell field is functionally fixed, i.e., $D\phi=0$, and postulated that the quantum fluctuations of the bulk field are only contained in the interior of the AdS space, vanishing sufficiently fast at its boundaries. In other words, all the non-normalizable behavior of $\Phi(x)$ is contained in $\phi(x)$. This assumption for the quantum fluctuations further allows us to write the resulting AdS action as:

\begin{equation}\label{phi3pisep2}
    S_{\text{AdS}}[\phi+h] = S_{\text{AdS}}[\phi] + S_{\text{AdS}}[h] + \frac{\lambda}{2}\int d^{d+1}x\sqrt{g}\ \phi(x)h^2(x)
\end{equation}\

where we integrated by parts dropping the quantum fluctuations $h(x)$ being evaluated at the boundaries of the AdS space, used the classical equation satisfied by $\phi(x)$ and finally identified the original form of the AdS action now for the different fields. We see that the $\Phi^3$ AdS action does not act as a linear functional under the field's change of variable ($S_{\text{AdS}}[\phi+h]\neq S_{\text{AdS}}[\phi] + S_{\text{AdS}}[h]$) due to the presence of the last term in eq. (\ref{phi3pisep2}) coming from the self-interaction. This quantity, where the on-shell part of the bulk field is directly coupled to its quantum fluctuations, can be seen as a deformation to the linearity of the action and it will be precisely the responsible for the quantum corrections to the classical correlators found in the previous chapter. Indeed, replacing eq. (\ref{phi3pisep2}) back into eq. (\ref{phi3pisep1}), the resulting $\Phi^3$ AdS path integral under the change of variable $\Phi(x)=\phi(x)+h(x)$ is:

\begin{equation}
    Z_{\text{AdS}}[\varphi_0] = e^{-S_{\text{AdS}}[\phi]-\int d^dx\sqrt{\gamma}\ B(\phi(x))\bigl\rvert_{x_0=0}}\int Dh\ e^{-S_\text{AdS}[h]-\frac{\lambda}{2}\int d^{d+1}x\sqrt{g}\ \phi(x)h^2(x)}
\end{equation}\

and if we think of the parameter $\lambda$, which mediates the strength of the self-interaction, as being in some sense "small", we can resort to perturbation theory solving for the form of the AdS path integral as an expansion in this parameter:

\begin{align}\label{phi3pisep3}
    Z_{\text{AdS}}[\varphi_0] = &e^{-S_{\text{AdS}}[\phi]-\int d^dx\sqrt{\gamma}\ B(\phi(x))\bigl\rvert_{x_0=0}}\Bigl[\int Dh\ e^{-S_\text{AdS}[h]}-\frac{\lambda}{2}\int\phi(x_1)\int Dh\ h^2(x_1)e^{-S_\text{AdS}[h]}\nonumber\\
    &+\frac{\lambda^2}{8}\int\int\phi(x_1)\phi(x_2)\int Dh\ h^2(x_1)h^2(x_2)e^{-S_\text{AdS}[h]}\nonumber\\
    &-\frac{\lambda^3}{48}\int\int\int\phi(x_1)\phi(x_2)\phi(x_3)\int Dh\ h^2(x_1)h^2(x_2)h^2(x_3)e^{-S_\text{AdS}[h]}\Bigl]+\mathcal{O}(\lambda^4)
\end{align}\

where we expanded the deformation term up to order $\lambda^3$ and separated the integral in $h$ into the 4 resulting terms, defining in the process $\int\int\dotsm\equiv\int d^{d+1}x_1\sqrt{g}\int d^{d+1}x_2\sqrt{g}\dotsm$ to keep the notation short. Note that in this expansion we come across path integrals in the field $h(x)$ of the form $\int Dh\ h(x_1)\dotsm h(x_n)e^{-S_{\text{AdS}}[h]}$. Since this field $h(x)$, unlike the complete bulk field $\Phi(x)$, is thought to be perfectly regular at the boundaries of the current space under consideration, these resulting path integrals can be solved in exactly the same way as the ones encountered in ordinary quantum field theories. This realization motivates us to define the ordinary n-point functions in the bulk:

\begin{equation}\label{npphi3}
    G_n(x_1,\dotsc,x_n) \equiv \frac{\int Dh\ h(x_1)\dotsm h(x_n)e^{-S_{\text{AdS}}[h]}}{\int Dh\ e^{-S_{\text{AdS}}[h]}}
\end{equation}\

where these quantities are expected to be solved, again, as an expansion in $\lambda$ with each resulting term involving only the bulk-bulk propagator. In terms of these functions each path integral in the field $h(x)$ present in eq. (\ref{phi3pisep3}) can be solved directly obtaining, up to order $\lambda^3$ in the coupling constant, the normalized AdS path integral:

\begin{align}\label{phi3pisep4}
    \frac{Z_{\text{AdS}}[\varphi_0]}{Z_{\text{AdS}}[\varphi_0=0]} = &e^{-S_{\text{AdS}}[\phi]-\int d^dx\sqrt{\gamma}\ B(\phi(x))\bigl\rvert_{x_0=0}}\Bigl[1-\frac{\lambda}{2}\int\phi(x_1)G_2(x_1,x_1)\nonumber\\
    &+\frac{\lambda^2}{8}\int\int\phi(x_1)\phi(x_2)G_4(x_1,x_1,x_2,x_2)\nonumber\\
    &-\frac{\lambda^3}{48}\int\int\int\phi(x_1)\phi(x_2)\phi(x_3)G_6(x_1,x_1,x_2,x_2,x_3,x_3)\Bigl]+\mathcal{O}(\lambda^4)
\end{align}\

where, since $\phi(x)\bigl\rvert_{\varphi_0=0}=0$ (eq. (\ref{solphi3})), we used that $Z_{\text{AdS}}[\varphi_0=0]=\int Dh\ e^{-S_{\text{AdS}}[h]}$. Now, as we just said, each one of these n-point functions in the bulk $G_n(x_1,\dotsc,x_n)$ can be solved in exactly the same way as in ordinary QFT, this is, using the same regular methods of adding an external source coupled to the field in the generating functional, then performing the resulting integral with the use again of perturbation theory and finally computing the desired n-point function through the corresponding derivatives of the source, which at the end of the calculation are set to 0. This process for the particular n-point functions present in eq. (\ref{phi3pisep4}) results in, first for $G_2(x_1,x_1)$ up to order $\lambda^2$:

\begin{align}
    G_2(x_1,x_1) = &G(x_1,x_1) + \frac{\lambda^2}{2}\int\int G(x_1,x_2)G(x_1,x_3)G^2(x_2,x_3)\nonumber\\
    &+\frac{\lambda^2}{4}\int\int G(x_1,x_2)G(x_1,x_3)G(x_2,x_2)G(x_3,x_3)\nonumber\\
    &+\frac{\lambda^2}{2}\int\int G^2(x_1,x_2)G(x_2,x_3)G(x_3,x_3) + \mathcal{O}(\lambda^3)
\end{align}\

where $\int\int\equiv\int d^{d+1}x_2\sqrt{g}\int d^{d+1}x_3\sqrt{g}$, then for $G_4(x_1,x_1,x_2,x_2)$ up to order $\lambda^1$:

\begin{equation}
    G_4(x_1,x_1,x_2,x_2) = 2G^2(x_1,x_2)+G(x_1,x_1)G(x_2,x_2)+\mathcal{O}(\lambda^2)
\end{equation}\

and finally for $G_6(x_1,x_1,x_2,x_2,x_3,x_3)$ up to order $\lambda^0$:

\begin{align}
    G_6(x_1,x_1,x_2,x_2,x_3,x_3) = &8G(x_1,x_2)G(x_2,x_3)G(x_3,x_1)+G(x_1,x_1)G(x_2,x_2)G(x_3,x_3)\nonumber\\
    &+\bigl[2G(x_1,x_1)G^2(x_2,x_3)+(x_1\leftrightarrow x_2)+(x_1\leftrightarrow x_3)\bigl]+\mathcal{O}(\lambda)
\end{align}\

where the quantities $G(x,z)$ are the usual bulk-bulk propagators. These results allow us to finally express the complete expansion of the normalized AdS path integral up to order $\lambda^3$ in the self-interacting coupling constant as:

\begin{align}\label{fullexpzphi3}
    \frac{Z_{\text{AdS}}[\varphi_0]}{Z_{\text{AdS}}[\varphi_0=0]} =& e^{-S_{\text{AdS}}[\phi]-\int d^dx\sqrt{\gamma}\ B(\phi(x))\bigl\rvert_{x_0=0}}\Bigl[1-\frac{\lambda}{2}\int\phi(x_1)G(x_1,x_1)\nonumber\\
    &-\frac{\lambda^3}{4}\int\int\int\phi(x_1)G(x_1,x_2)G(x_1,x_3)G^2(x_2,x_3)\nonumber\\
    &-\frac{\lambda^3}{8}\int\int\int\phi(x_1)G(x_1,x_2)G(x_1,x_3)G(x_2,x_2)G(x_3,x_3)\nonumber\\
    &-\frac{\lambda^3}{4}\int\int\int\phi(x_1)G^2(x_1,x_2)G(x_2,x_3)G(x_3,x_3)\nonumber\\
    &+\frac{\lambda^2}{4}\int\int\phi(x_1)\phi(x_2)G^2(x_1,x_2)\nonumber\\
    &+\frac{\lambda^2}{8}\int\int\phi(x_1)\phi(x_2)G(x_1,x_1)G(x_2,x_2)\nonumber\\
    &-\frac{\lambda^3}{6}\int\int\int\phi(x_1)\phi(x_2)\phi(x_3)G(x_1,x_2)G(x_2,x_3)G(x_3,x_1)\nonumber\\
    &-\frac{\lambda^3}{48}\int\int\int\phi(x_1)\phi(x_2)\phi(x_3)G(x_1,x_1)G(x_2,x_2)G(x_3,x_3)\nonumber\\
    &-\frac{\lambda^3}{8}\int\int\int\phi(x_1)\phi(x_2)\phi(x_3)G(x_1,x_1)G^2(x_2,x_3)\Bigl]+\mathcal{O}(\lambda^4)
\end{align}\

remembering that the on-shell field $\phi(x)$ in this same $\lambda$ expansion is given by eq. (\ref{solphi3}):

\begin{align}\label{solphi3ch4}
    \phi(x) = &\int d^dy\ K(x,\vec{y})\varphi_0(\vec{y}) - \frac{\lambda}{2}\int d^{d+1}x_1\sqrt{g}\ G(x,x_1)\Bigl[\int d^dy\ K(x_1,\vec{y})\varphi_0(\vec{y})\Bigl]^2\nonumber\\
    &+\frac{\lambda^2}{2}\int d^{d+1}x_1\sqrt{g}\ G(x,x_1)\Bigl[\int d^dy\ K(x_1,\vec{y})\varphi_0(\vec{y})\Bigl]\nonumber\\
    &\hspace{3cm}\times\int d^{d+1}x_2\sqrt{g}\ G(x_1,x_2)\Bigl[\int d^dy\ K(x_2,\vec{y})\varphi_0(\vec{y})\Bigl]^2+\mathcal{O}(\lambda^3)
\end{align}\

Eq. (\ref{fullexpzphi3}) together with eq. (\ref{solphi3ch4}) give us the concrete and explicit dependence of the $\Phi^3$ AdS path integral as a functional of the dual source $\varphi_0(\vec{y})$, ready to be differentiated with the intention to compute the quantum corrected CFT holographic correlators.

\subsection{Correlation Functions}

Plugging the explicit form of the on-shell field $\phi(x)$ (eq. (\ref{solphi3ch4})) into the normalized AdS path integral $\frac{Z_{\text{AdS}}[\varphi_0]}{Z_{\text{AdS}}[\varphi_0=0]}$ (eq. (\ref{fullexpzphi3})) and keeping terms of order $\lambda^3$, the obtention of the holographic correlators eqs. (\ref{euccorr1}) and (\ref{euccorr2}) up to this order in the coupling constant is reduced to a simple exercise of taking derivatives, where since the on-shell part of the path integral is holographic renormalized the variation of this part is understood to be given by eq. (\ref{varzphi3explicit}). The resulting quantum corrected holographic correlators from this process for some primary scalar operator $O_\Delta(\vec{x})$ of scaling dimension $\Delta$ dual to a $\Phi^3$ self-interacting scalar field in AdS$_{d+1}$ are given by the 1-point functions:

\begin{align}\label{1pfphi3quantum}
    \langle O_\Delta(\vec{y_1})\rangle_{\text{CFT}} = \langle O_\Delta(\vec{y_1})\rangle_{\text{CFT,con}} = &-\frac{\lambda}{2}\int K(x_1,\vec{y_1})G(x_1,x_1)\nonumber\\
    &\hspace{-1cm}-\frac{\lambda^3}{4}\int\int\int K(x_1,\vec{y_1})G(x_1,x_2)G(x_1,x_3)G^2(x_2,x_3)\nonumber\\
    &\hspace{-1cm}-\frac{\lambda^3}{8}\int\int\int K(x_1,\vec{y_1})G(x_1,x_2)G(x_1,x_3)G(x_2,x_2)G(x_3,x_3)\nonumber\\
    &\hspace{-1cm}-\frac{\lambda^3}{4}\int\int\int K(x_1,\vec{y_1})G^2(x_1,x_2)G(x_2,x_3)G(x_3,x_3)+\mathcal{O}(\lambda^4)
\end{align}\

the 2-point functions:

\begin{align}\label{2pfphi3quantum}
    \langle O_\Delta(\vec{y_1})O_\Delta(\vec{y_2})\rangle_{\text{CFT}} = &\frac{2\nu c_\Delta}{\lvert\vec{y_1}-\vec{y_2}\rvert^{2\Delta}}+\frac{\lambda^2}{2}\int\int K(x_1,\vec{y_1})K(x_1,\vec{y_2})G(x_1,x_2)G(x_2,x_2)\nonumber\\
    &+\frac{\lambda^2}{4}\int\int K(x_1,\vec{y_1})K(x_2,\vec{y_2})G(x_1,x_1)G(x_2,x_2)\nonumber\\
    &+\frac{\lambda^2}{2}\int\int K(x_1,\vec{y_1})K(x_2,\vec{y_2})G^2(x_1,x_2)+\mathcal{O}(\lambda^4)\nonumber\\
    \langle O_\Delta(\vec{y_1})O_\Delta(\vec{y_2})\rangle_{\text{CFT,con}} = &\frac{2\nu c_\Delta}{\lvert\vec{y_1}-\vec{y_2}\rvert^{2\Delta}}+\frac{\lambda^2}{2}\int\int K(x_1,\vec{y_1})K(x_1,\vec{y_2})G(x_1,x_2)G(x_2,x_2)\nonumber\\
    &+\frac{\lambda^2}{2}\int\int K(x_1,\vec{y_1})K(x_2,\vec{y_2})G^2(x_1,x_2)+\mathcal{O}(\lambda^4)
\end{align}\

the 3-point functions:

\begin{align}\label{3pfphi3quantum}
    \langle O_\Delta(\vec{y_1})O_\Delta(\vec{y_2})O_\Delta(\vec{y_3})\rangle_{\text{CFT}} = &-\lambda\int K(x_1,\vec{y_1})K(x_1,\vec{y_2})K(x_1,\vec{y_3})\nonumber\\
    &\hspace{-3.5cm}-\frac{\lambda^3}{2}\int\int\int K(x_1,\vec{y_1})K(x_2,\vec{y_2})K(x_2,\vec{y_3})G(x_1,x_2)G(x_1,x_3)G(x_3,x_3)\times 3\nonumber\\
    &\hspace{-3.5cm}-\frac{\lambda^3}{4}\int\int\int K(x_1,\vec{y_1})K(x_2,\vec{y_2})K(x_2,\vec{y_3})G(x_1,x_1)G(x_2,x_3)G(x_3,x_3)\times 3\nonumber\\
    &\hspace{-3.5cm}-\frac{\lambda^3}{2}\int\int\int K(x_1,\vec{y_1})K(x_2,\vec{y_2})K(x_2,\vec{y_3})G^2(x_1,x_3)G(x_3,x_2)\times 3\nonumber\\
    &\hspace{-3.5cm}-\lambda^3\int\int\int K(x_1,\vec{y_1})K(x_2,\vec{y_2})K(x_3,\vec{y_3})G(x_1,x_2)G(x_2,x_3)G(x_3,x_1)\nonumber\\
    &\hspace{-3.5cm}-\frac{\lambda^3}{8}\int\int\int K(x_1,\vec{y_1})K(x_2,\vec{y_2})K(x_3,\vec{y_3})G(x_1,x_1)G(x_2,x_2)G(x_3,x_3)\nonumber\\
    &\hspace{-3.5cm}-\frac{\lambda^3}{4}\int\int\int K(x_1,\vec{y_1})K(x_2,\vec{y_2})K(x_3,\vec{y_3})G(x_1,x_1)G^2(x_2,x_3)\times 3\nonumber\\
    &\hspace{-3.5cm}+\frac{2\nu c_\Delta}{\lvert\vec{y_1}-\vec{y_2}\rvert^{2\Delta}}\Bigl[-\frac{\lambda}{2}\int K(x_1,\vec{y_3})G(x_1,x_1)\nonumber\\
    &\hspace{-0.9cm}-\frac{\lambda^3}{4}\int\int\int K(x_1,\vec{y_3})G(x_1,x_2)G(x_1,x_3)G^2(x_2,x_3)\nonumber\\
    &\hspace{-0.9cm}-\frac{\lambda^3}{8}\int\int\int K(x_1,\vec{y_3})G(x_1,x_2)G(x_1,x_3)G(x_2,x_2)G(x_3,x_3)\nonumber\\
    &\hspace{-0.9cm}-\frac{\lambda^3}{4}\int\int\int K(x_1,\vec{y_3})G^2(x_1,x_2)G(x_2,x_3)G(x_3,x_3)\Bigl]\times 3 + \mathcal{O}(\lambda^4)\nonumber\\
    \langle O_\Delta(\vec{y_1})O_\Delta(\vec{y_2})O_\Delta(\vec{y_3})\rangle_{\text{CFT,con}} = &-\lambda\int K(x_1,\vec{y_1})K(x_1,\vec{y_2})K(x_1,\vec{y_3})\nonumber\\
    &\hspace{-3.5cm}-\frac{\lambda^3}{2}\int\int\int K(x_1,\vec{y_1})K(x_2,\vec{y_2})K(x_2,\vec{y_3})G(x_1,x_2)G(x_1,x_3)G(x_3,x_3)\times 3\nonumber\\
    &\hspace{-3.5cm}-\frac{\lambda^3}{2}\int\int\int K(x_1,\vec{y_1})K(x_2,\vec{y_2})K(x_2,\vec{y_3})G^2(x_1,x_3)G(x_3,x_2)\times 3\nonumber\\
    &\hspace{-3.5cm}-\lambda^3\int\int\int K(x_1,\vec{y_1})K(x_2,\vec{y_2})K(x_3,\vec{y_3})G(x_1,x_2)G(x_2,x_3)G(x_3,x_1)+\mathcal{O}(\lambda^4)
\end{align}\

and finally by the 4-point functions:

\begin{align}\label{4pfphi3quantum}
    \langle O_\Delta(\vec{y_1})O_\Delta(\vec{y_2})O_\Delta(\vec{y_3})O_\Delta(\vec{y_4})\rangle_{\text{CFT}} =&\frac{2\nu c_\Delta}{\lvert\vec{y_1}-\vec{y_2}\rvert^{2\Delta}}\frac{2\nu c_\Delta}{\lvert\vec{y_3}-\vec{y_4}\rvert^{2\Delta}}+(\vec{y_2}\leftrightarrow\vec{y_3})+(\vec{y_2}\leftrightarrow\vec{y_4})\nonumber\\
    &\hspace{-2cm}+\lambda^2\int\int K(x_1,\vec{y_1})K(x_1,\vec{y_2})G(x_1,x_2)K(x_2,\vec{y_3})K(x_2,\vec{y_4})\times3\nonumber\\
    &\hspace{-2cm}+\frac{\lambda^2}{2}\int\int K(x_1,\vec{y_1})G(x_1,x_1)K(x_2,\vec{y_2})K(x_2,\vec{y_3})K(x_2,\vec{y_4})\times4\nonumber\\
    &\hspace{-2cm}+\frac{2\nu c_\Delta}{\lvert\vec{y_1}-\vec{y_2}\rvert^{2\Delta}}\Bigl[\frac{\lambda^2}{2}\int\int K(x_1,\vec{y_3})K(x_1,\vec{y_4})G(x_1,x_2)G(x_2,x_2)\nonumber\\
    &\hspace{0.6cm}+\frac{\lambda^2}{4}\int\int K(x_1,\vec{y_3})K(x_2,\vec{y_4})G(x_1,x_1)G(x_2,x_2)\nonumber\\
    &\hspace{0.6cm}+\frac{\lambda^2}{2}\int\int K(x_1,\vec{y_3})K(x_2,\vec{y_4})G^2(x_1,x_2)\Bigl]+\mathcal{O}(\lambda^4)\nonumber\\
    \langle O_\Delta(\vec{y_1})O_\Delta(\vec{y_2})O_\Delta(\vec{y_3})O_\Delta(\vec{y_4})\rangle_{\text{CFT,con}} =&\nonumber\\
    &\hspace{-2.5cm}\lambda^2\int\int K(x_1,\vec{y_1})K(x_1,\vec{y_2})G(x_1,x_2)K(x_2,\vec{y_3})K(x_2,\vec{y_4})\times3+\mathcal{O}(\lambda^4)
\end{align}\

where, with the intention to keep the notation short, we defined\\ $\int\int\dotsm\equiv\int d^{d+1}x_1\sqrt{g}\int d^{d+1}x_2\sqrt{g}\dotsm$ and represented the different permutations of the integrals as a multiplicative factor at the end of each. Notice the new contributions to the correlators in comparison with those obtained under the classical approximation of the AdS/CFT correspondence, eq. (\ref{phi3corr}). Pictorially, these new terms correspond exactly to the loops diagrams expected from a regular QFT $\Phi^3$ self-interacting theory resulting from a perturbative expansion in the self-interacting coupling constant, even agreeing with the same coefficients! Since these new integrals are contributing to specific correlators which are conjectured to be of the form dictated by eq. (\ref{summcorr2}), the functional form of their results is strongly conditioned purely from conformal symmetry arguments. We will proceed then to study these quantities in detail through their explicit computation.

\subsection{Holographic Dictionary}

Before jumping straight into the calculations of the new contributions to the holographic correlators it will be useful to check if the holographic dictionary defined in eqs. (\ref{dict1}) and (\ref{dict2}), which relates the desired boundary correlators for the operator $O_\Delta(\vec{x})$ with the simple and known bulk correlators for the field $\Phi(x)$, is still valid for the recently obtained quantum corrected 1-, 2-, 3- and 4-point functions since if this is the case it will allow us to use nice (and even desired, as we will discuss shortly) properties of $\Phi^3$ theories which will result in a simplification to the explicit form of these correlators.\par
To keep the discussion clean and short, the validity of the holographic dictionary for the present case can be argued to hold without doing any computation through the understanding of why it holds for its classical counterpart in the first place. In that case, under the appropriate limits the values of the bulk tree-level n-point functions obtained from the usual Feynman rules of scalar $\Phi^3$ theories (eq. (\ref{hdphi3class})) matched exactly the boundary correlators obtained through the classical approximation of the AdS/CFT correspondence (eq. (\ref{phi3corr})) mainly because the latter essentially contained the same type of diagrams expected from Feynman rules as the former, even with the same coefficients, with the exception that the external legs of said diagrams had been replaced with bulk-boundary propagators. But this replacement precisely matched the effective dictionary coming from the nice expansions of the bulk-bulk propagators eqs. (\ref{expbubup1}) and (\ref{expbubup2}) which for the bulk correlators implied the simple recipe "replace external $G$'s with $K$'s", inevitably resulting in exactly the same correlators as those found through the classical approximation of the AdS/CFT correspondence. These facts unequivocally led us to relate the boundary n-point functions with the bulk n-point functions simply as the former being the extension of the internal points of the latter to the conformal boundary of the AdS space.\par
Having said this, it is straightforward to see that exactly the same is happening to the now quantum corrected boundary correlators for the operator $O_\Delta(\vec{x})$ just found, eqs. (\ref{1pfphi3quantum}), (\ref{2pfphi3quantum}), (\ref{3pfphi3quantum}) and (\ref{4pfphi3quantum}). Indeed, these correlators correspond exactly to the ones obtained from Feynman rules where the external bulk-bulk propagators have been replaced with bulk-boundary propagators, but this is precisely the effective action of the holographic dictionary, implying that we can always formulate these boundary correlators as the appropriate limit of some bulk correlators which follow the Feynman rules of a $\Phi^3$ theory. Unsurprisingly, these bulk correlators are nothing but the correlators for the bulk field $\Phi$, confirming in this way for the present case the boundary/bulk n-point functions equivalence.

\subsection{Tadpole Renormalization}

So far the discussion of the holographic dictionary has been a mere curiosity, which allowed us to relate the correlators living at the conformal boundary of the AdS space to those living in its interior. However, as we will see in this segment, this curiosity taken seriously has some profound implications for the boundary correlators we are interested in computing. These implications come from the fact that these desired correlators can be thought of simply as the limit of some ordinary and well-known $\Phi^3$ theory, being able then to use the familiar machinery known for these theories. One of the many features of $\Phi^3$ theories which will be extremely useful and revealing for our current purpose of computing eqs. (\ref{1pfphi3quantum}), (\ref{2pfphi3quantum}), (\ref{3pfphi3quantum}) and (\ref{4pfphi3quantum}), is the liberty to redefine the value of the vacuum expectation (the VEV) of the field $\Phi(x)$ through the simple addition of a counterterm linear in the field of the form $Y(\lambda)\Phi(x)$, where the parameter $Y(\lambda)$ that helps fixing the desired value for the VEV of the field can be determined perturbatively in the coupling constant. In general one is interested in normalizing the VEV of the field to 0, which for $\Phi^3$ theories has the remarkable result of renormalizing not only the tadpole diagrams present in the 1-point function but also those present in higher point functions \cite{Srednicki}! But notice precisely how the tadpole contributions present in the holographic 1-point function just derived (eq. (\ref{1pfphi3quantum})) are spoiling the conformal structure of this correlator, expected to be of the form dictated by eq. (\ref{summcorr2}). With the recent discussion we can already understand why this is the case: the bulk field $\Phi$ is simply not well normalized.\par
Considering then the properly normalized theory\\ $S_{\text{AdS}}[\Phi]\rightarrow S_{\text{AdS}}[\Phi]+\int d^{d+1}x\sqrt{g}\ Y(\lambda)\Phi(x)$, where the parameter $Y(\lambda)$ is chosen accordingly, the holographic dictionary assures us that the resulting boundary 1-, 2, 3-, and 4-point functions for some operator $O_\Delta(\vec{x})$ will be the same as those obtained previously, with the exception that the tadpole contributions have been renormalized from every n-point function:

\begin{align}\label{phi3corrtadren}
    \text{1-pt fn:}\hspace{0.25cm}&\langle O_\Delta(\vec{y_1})\rangle_{\text{CFT}} = \langle O_\Delta(\vec{y_1})\rangle_{\text{CFT,con}} = 0\nonumber\\
    \text{2-pt fn:}\hspace{0.25cm}&\langle O_\Delta(\vec{y_1})O_\Delta(\vec{y_2})\rangle_{\text{CFT}} = \langle O_\Delta(\vec{y_1})O_\Delta(\vec{y_2})\rangle_{\text{CFT,con}} = \frac{2\nu c_\Delta}{\lvert\vec{y_1}-\vec{y_2}\rvert^{2\Delta}}\nonumber\\
    &\hspace{4cm}+\frac{\lambda^2}{2}\int\int K(x_1,\vec{y_1})K(x_2,\vec{y_2})G^2(x_1,x_2)+\mathcal{O}(\lambda^4)\nonumber\\
    \text{3-pt fn:}\hspace{0.25cm}&\langle O_\Delta(\vec{y_1})O_\Delta(\vec{y_2})O_\Delta(\vec{y_3})\rangle_{\text{CFT}} = \langle O_\Delta(\vec{y_1})O_\Delta(\vec{y_2})O_\Delta(\vec{y_3})\rangle_{\text{CFT,con}} = \nonumber\\
    &\hspace{1cm}-\lambda\int K(x_1,\vec{y_1})K(x_1,\vec{y_2})K(x_1,\vec{y_3})\nonumber\\
    &\hspace{1cm}-\frac{\lambda^3}{2}\int\int\int K(x_1,\vec{y_1})K(x_2,\vec{y_2})K(x_2,\vec{y_3})G^2(x_1,x_3)G(x_3,x_2)\times 3\nonumber\\
    &\hspace{1cm}-\lambda^3\int\int\int K(x_1,\vec{y_1})K(x_2,\vec{y_2})K(x_3,\vec{y_3})G(x_1,x_2)G(x_2,x_3)G(x_3,x_1)+\mathcal{O}(\lambda^4)\nonumber\\
    \text{4-pt fn:}\hspace{0.25cm}&\langle O_\Delta(\vec{y_1})O_\Delta(\vec{y_2})O_\Delta(\vec{y_3})O_\Delta(\vec{y_4})\rangle_{\text{CFT}} =\frac{2\nu c_\Delta}{\lvert\vec{y_1}-\vec{y_2}\rvert^{2\Delta}}\frac{2\nu c_\Delta}{\lvert\vec{y_3}-\vec{y_4}\rvert^{2\Delta}}+(\vec{y_2}\leftrightarrow\vec{y_3})+(\vec{y_2}\leftrightarrow\vec{y_4})\nonumber\\
    &\hspace{2.4cm}+\lambda^2\int\int K(x_1,\vec{y_1})K(x_1,\vec{y_2})G(x_1,x_2)K(x_2,\vec{y_3})K(x_2,\vec{y_4})\times3\nonumber\\
    &\hspace{2.4cm}+\frac{2\nu c_\Delta}{\lvert\vec{y_1}-\vec{y_2}\rvert^{2\Delta}}\frac{\lambda^2}{2}\int\int K(x_1,\vec{y_3})K(x_2,\vec{y_4})G^2(x_1,x_2)\times6 + \mathcal{O}(\lambda^4)\nonumber\\
    &\langle O_\Delta(\vec{y_1})O_\Delta(\vec{y_2})O_\Delta(\vec{y_3})O_\Delta(\vec{y_4})\rangle_{\text{CFT,con}} =\nonumber\\
    &\hspace{2.5cm}\lambda^2\int\int K(x_1,\vec{y_1})K(x_1,\vec{y_2})G(x_1,x_2)K(x_2,\vec{y_3})K(x_2,\vec{y_4})\times3+\mathcal{O}(\lambda^4)
\end{align}\

Since this properly normalized theory is the one that delivers conformal correlators, these are the ones that will be of interest to us and that therefore we will proceed to calculate.

\subsection{Regularization Schemes}

We are one step away from fully diving into the computation of the quantum corrections to the holographic correlators dual to a $\Phi^3$ theory on AdS. The last matter that we will see before this, with the intention to keep these computations organized and clean, is the introduction of the appropriate regularization schemes that we will use in this study, necessary for the handling of the different divergent quantities (as we will see in detail) present in the n-point functions, eq. (\ref{phi3corrtadren}).

\subsubsection*{IR Regularization}

The first type of divergences that we will deal with are those of the type IR (infrared) coming from the different loops integrals in eq. (\ref{phi3corrtadren}) as their internal points $x_i$ being integrated approach the conformal boundary of the AdS space at $x_{i,0}=0$. But we have already discussed this type of divergence, it was precisely the divergence we encountered when we were trying to compute the tree-level contributions to the holographic correlators, motivating the entire holographic renormalization program. As discussed in there (section 3.1.5) this infrared divergence was completely expected due to the weak/strong duality that the AdS/CFT correspondence implies, since we are in fact computing the correlators for some strongly coupled quantum theory, correlators which are usually UV-divergent due to loops contributions. Since the quantum contributions to the holographic correlators are out of the scope of the holographic renormalization, procedure which only renormalizes the on-shell contributions coming from the AdS path integral, these new terms in eq. (\ref{phi3corrtadren}) given the reasons just said are expected to be IR-divergent as well. The renormalization of these infinities will be identical in spirit with those found in ordinary QFTs as we will see later but with regards to their regularization, it will be extremely satisfactory to find that the same regularization scheme introduced in the holographic renormalization procedure which captured the correct structure of the classical contributions to the holographic correlators, also does it for their quantum contributions. This scheme consisted in manipulating the IR-divergent quantities not in the conformal boundary itself at $x_0=0$ but instead at some small distance from it at $x_0=\varepsilon$, where the limit $\varepsilon\rightarrow0$ is understood. In our present case this scheme gives us a natural, and rather obvious, approach on how to treat these possible IR-divergent integrals: simply integrate them up to some small distance $\varepsilon$! The resulting correlators from this scheme can be simply put into the form:

\begin{align}\label{phi3corrtadren2}
    \text{1-pt fn:}\hspace{0.25cm}&\langle O_\Delta(\vec{y_1})\rangle_{\text{CFT}} = \langle O_\Delta(\vec{y_1})\rangle_{\text{CFT,con}} = 0\nonumber\\
    \text{2-pt fn:}\hspace{0.25cm}&\langle O_\Delta(\vec{y_1})O_\Delta(\vec{y_2})\rangle_{\text{CFT}} = \langle O_\Delta(\vec{y_1})O_\Delta(\vec{y_2})\rangle_{\text{CFT,con}} = \frac{2\nu c_\Delta}{\lvert\vec{y_1}-\vec{y_2}\rvert^{2\Delta}}\nonumber\\
    &\hspace{4cm}+\frac{\lambda^2}{2}\int_\varepsilon\int_\varepsilon K(x_1,\vec{y_1})K(x_2,\vec{y_2})G^2(x_1,x_2)+\mathcal{O}(\lambda^4)\nonumber\\
    \text{3-pt fn:}\hspace{0.25cm}&\langle O_\Delta(\vec{y_1})O_\Delta(\vec{y_2})O_\Delta(\vec{y_3})\rangle_{\text{CFT}} = \langle O_\Delta(\vec{y_1})O_\Delta(\vec{y_2})O_\Delta(\vec{y_3})\rangle_{\text{CFT,con}} = \nonumber\\
    &\hspace{1cm}-\lambda\int K(x_1,\vec{y_1})K(x_1,\vec{y_2})K(x_1,\vec{y_3})\nonumber\\
    &\hspace{1cm}-\frac{\lambda^3}{2}\int_\varepsilon\int_\varepsilon\int_\varepsilon K(x_1,\vec{y_1})K(x_2,\vec{y_2})K(x_2,\vec{y_3})G^2(x_1,x_3)G(x_3,x_2)\times 3\nonumber\\
    &\hspace{1cm}-\lambda^3\int_\varepsilon\int_\varepsilon\int_\varepsilon K(x_1,\vec{y_1})K(x_2,\vec{y_2})K(x_3,\vec{y_3})G(x_1,x_2)G(x_2,x_3)G(x_3,x_1)+\mathcal{O}(\lambda^4)\nonumber\\
    \text{4-pt fn:}\hspace{0.25cm}&\langle O_\Delta(\vec{y_1})O_\Delta(\vec{y_2})O_\Delta(\vec{y_3})O_\Delta(\vec{y_4})\rangle_{\text{CFT}} =\frac{2\nu c_\Delta}{\lvert\vec{y_1}-\vec{y_2}\rvert^{2\Delta}}\frac{2\nu c_\Delta}{\lvert\vec{y_3}-\vec{y_4}\rvert^{2\Delta}}+(\vec{y_2}\leftrightarrow\vec{y_3})+(\vec{y_2}\leftrightarrow\vec{y_4})\nonumber\\
    &\hspace{2.4cm}+\lambda^2\int\int K(x_1,\vec{y_1})K(x_1,\vec{y_2})G(x_1,x_2)K(x_2,\vec{y_3})K(x_2,\vec{y_4})\times3\nonumber\\
    &\hspace{2.4cm}+\frac{2\nu c_\Delta}{\lvert\vec{y_1}-\vec{y_2}\rvert^{2\Delta}}\frac{\lambda^2}{2}\int_\varepsilon\int_\varepsilon K(x_1,\vec{y_3})K(x_2,\vec{y_4})G^2(x_1,x_2)\times6 + \mathcal{O}(\lambda^4)\nonumber\\
    &\langle O_\Delta(\vec{y_1})O_\Delta(\vec{y_2})O_\Delta(\vec{y_3})O_\Delta(\vec{y_4})\rangle_{\text{CFT,con}} =\nonumber\\
    &\hspace{2.5cm}\lambda^2\int\int K(x_1,\vec{y_1})K(x_1,\vec{y_2})G(x_1,x_2)K(x_2,\vec{y_3})K(x_2,\vec{y_4})\times3+\mathcal{O}(\lambda^4)
\end{align}\

where we regularized the quantum contributions in eq. (\ref{phi3corrtadren}), defining in the process the IR-regularized integrals $\int_\varepsilon\int_\varepsilon\dotsm\equiv\int_{x_{1,0}=\varepsilon}d^{d+1}x_1\sqrt{g}\int_{x_{2,0}=\varepsilon}d^{d+1}x_2\sqrt{g}\dotsm$. When we need to keep these regulators or when can we safely take them as 0 will be an interesting study which we will leave pending for a brief moment when we finally start computing these integrals. But before this, we still need to introduce one more regulator to able to compute the delicate value of these integrals, which we will proceed to do next.

\subsubsection*{UV Regularization}

The second type of divergences that we will deal with are those of the type UV (ultraviolet) coming from the different loops integrals in eq. (\ref{phi3corrtadren2}) as the bulk-bulk propagators contained in them get integrated at more closer and closer points. Indeed, take for example the loop integral found in the 2-point functions. In this case, both $x_1$ and $x_2$ integrals are carried out in the entire region of the AdS space, thus they will contain contributions coming from when both points coincide with each other at $x_1=x_2$. But how does its integrand behave in this region, in particular its bulk-bulk propagator? For tree-level computations this question did not concern us and a naive approach to the resulting integrals involving the bulk-bulk propagator fortunately resulted in finite and convergent values. However in general this will no longer be true for loops computations, forcing us to take action on the matter. An easy way to see that this study of integrals involving $G(x_1,x_2)$ becomes delicate is to observe how this propagator behaves in this conflictive region where $x_1=x_2$. From the definition of the bulk-bulk propagator in terms of the parameter $\xi$, eq. (\ref{G}), it is direct to check that in this region of coincident points $\xi$ takes the simple value of 1, which reduces the form of the propagator into:

\begin{equation}
    G(x_1,x_1)=\frac{2^{-\Delta}c_\Delta}{2\nu}\ _2F_1\Bigl(\frac{\Delta}{2},\frac{\Delta+1}{2};\nu+1;1\Bigl)
\end{equation}\

Generalized hypergeometric functions of real parameters and unit argument of the form $_{p+1}F_p(a_1,\dotsc,a_{p+1};b_1,\dotsc,b_p;1)$ are convergent if $\sum b_i-\sum a_j >0$ \cite{GenHypFuncConv}. For the hypergeometric function in $G(x_1,x_1)$ this criteria implies the convergence condition $\nu+1-\frac{\Delta}{2}-\frac{\Delta+1}{2}=\frac{1-d}{2}>0$, i.e., $d<1$. In fact, the explicit divergence of the bulk-bulk propagator can be extracted out from its hypergeometric function representation using the known Euler's transformation for hypergeometric functions eq. (\ref{eultrans}), which puts the propagator in the equivalent form:

\begin{equation}\label{Grepdiv}
    G(x_1,x_2) = \frac{2^{-\Delta}c_\Delta}{2\nu}\frac{\xi^\Delta}{(1-\xi^2)^{\frac{d-1}{2}}}\ _2F_1\Bigl(\frac{\Delta-d}{2}+1,\frac{\Delta-d+1}{2};\nu+1;\xi^2\Bigl)
\end{equation}\

The convergence criteria applied to this new form for the hypergeometric function is summarized in $d>1$, therefore for dimensions greater than 1 all the divergence of the bulk-bulk propagator as its internal points $x_1$ and $x_2$ gets closer one to another, i.e., as $\xi\rightarrow1$, is contained in its factor $\frac{1}{(1-\xi^2)^{\frac{d-1}{2}}}$. Of course we will be interested in studying theories on the boundary of the AdS space of dimensions greater than 1 as well, meaning that we will be usually carrying ultraviolet divergences coming from the loops computations present on the AdS side.\par
Since we do not want to reduce the range of our study to the small region of $d<1$, we will have to find a way to regularize these ultraviolet infinities coming from the bulk-bulk propagator hoping that eventually the theory itself will have the capacity to absorb them, which as we will see will lead to interesting renormalizability conditions for the resulting dual theories coming from scalar theories on AdS. There are many sensitive ways in which we can regularize these infinities occurring from the physics at small distances, but the scheme we finally opted for is highly satisfactory not only because it has a clear picture and interpretation of the mechanism behind it, but also because it keeps intact the AdS symmetry of every quantity, which translates into the conservation of the CFT symmetry in the resulting holographic correlators. This scheme can be understood simply as a point-splitting approach whose motivation comes from observing how $\xi$, the parameter that contains the combination of how $G(x_1,x_2)$ depends on $x_1$ and $x_2$, behaves as $x_2$ approaches $x_1$ while keeping a small proper AdS distance towards it, i.e., for $x_2=x_1+dx_1$ where $dx_1$ satisfies $ds^2=g_{\mu\nu}dx_1^\mu dx_1^\nu$, $g_{\mu\nu}$ being the AdS metric $g_{\mu\nu}=\frac{\delta_{\mu\nu}}{x_{1,0}^2}$. In this case it is easy to check that $\xi$ behaves as:

\begin{equation}
    \xi(x_1,x_1+dx_1) = 1-\frac{ds^2}{2}+\mathcal{O}(dx_1^{2<})
\end{equation}\

in which calling $\kappa\equiv\frac{ds^2}{2}$, we can rewrite it up to this order in $\kappa$ simply as:

\begin{equation}\label{xisep}
    \xi(x_1,x_1+dx_1) = \frac{1}{1+\kappa}
\end{equation}\

Here $\kappa$ constitutes the UV-regulator that we will use in our study, which since $ds^2$ is a small positive number, then so is $\kappa$, implying that keeping separated the points $x_1$ and $x_2$ by a small proper AdS distance results in a value for $\xi$ arbitrarily close to 1 depending on the specific value of $\kappa$ but not 1, which is exactly what we are interested in achieving. This feature for the point-splitting approach is nice because if we are able to redefine $\xi$ such that for the case $x_1\neq x_2$ it reduces to its ordinary value, while for the case $x_1=x_2$ it reduces to its regularized form eq. (\ref{xisep}), we would have solved the problem of ultraviolet regularization since simply replacing this new parameter into $G(\xi)$ we will have cured the divergences coming from its integration at coincident points. This realization motivates us to define the regularized $\xi$ (same regularization scheme used by Bertan, Sachs and Skvortsov in \cite{Sachs}):

\begin{equation}
    \xi_\kappa\equiv\frac{\xi}{1+\kappa}
\end{equation}\

which in turn allows us to define the regularized version of the bulk-bulk propagator:

\begin{equation}
    G_\kappa(\xi)\equiv G(\xi_\kappa) = G\Bigl(\frac{\xi}{1+\kappa}\Bigl)
\end{equation}\

Since under AdS isometry transformations the parameter $\xi$ is invariant, it is trivial to check that this regulator preserves the AdS symmetry of the propagator. Moreover, for $x_1\neq x_2$ we can always take $\kappa=0$ recovering the original form of the propagator, while for $x_1=x_2$ the regulator $\kappa$ precisely handles the ultraviolet divergences coming from it. The resulting holographic correlators from this regularization scheme can be simply stated then as:

\begin{align}\label{phi3corrtadren3}
    \text{1-pt fn:}\hspace{0.25cm}&\langle O_\Delta(\vec{y_1})\rangle_{\text{CFT}} = \langle O_\Delta(\vec{y_1})\rangle_{\text{CFT,con}} = 0\nonumber\\
    \text{2-pt fn:}\hspace{0.25cm}&\langle O_\Delta(\vec{y_1})O_\Delta(\vec{y_2})\rangle_{\text{CFT}} = \langle O_\Delta(\vec{y_1})O_\Delta(\vec{y_2})\rangle_{\text{CFT,con}} = \frac{2\nu c_\Delta}{\lvert\vec{y_1}-\vec{y_2}\rvert^{2\Delta}}\nonumber\\
    &\hspace{4cm}+\frac{\lambda^2}{2}\int_\varepsilon\int_\varepsilon K(x_1,\vec{y_1})K(x_2,\vec{y_2})G^2_\kappa(x_1,x_2)+\mathcal{O}(\lambda^4)\nonumber\\
    \text{3-pt fn:}\hspace{0.25cm}&\langle O_\Delta(\vec{y_1})O_\Delta(\vec{y_2})O_\Delta(\vec{y_3})\rangle_{\text{CFT}} = \langle O_\Delta(\vec{y_1})O_\Delta(\vec{y_2})O_\Delta(\vec{y_3})\rangle_{\text{CFT,con}} = \nonumber\\
    &\hspace{1cm}-\lambda\int K(x_1,\vec{y_1})K(x_1,\vec{y_2})K(x_1,\vec{y_3})\nonumber\\
    &\hspace{1cm}-\frac{\lambda^3}{2}\int_\varepsilon\int_\varepsilon\int_\varepsilon K(x_1,\vec{y_1})K(x_2,\vec{y_2})K(x_2,\vec{y_3})G^2_\kappa(x_1,x_3)G_\kappa(x_3,x_2)\times 3\nonumber\\
    &\hspace{1cm}-\lambda^3\int_\varepsilon\int_\varepsilon\int_\varepsilon K(x_1,\vec{y_1})K(x_2,\vec{y_2})K(x_3,\vec{y_3})G_\kappa(x_1,x_2)G_\kappa(x_2,x_3)G_\kappa(x_3,x_1)+\mathcal{O}(\lambda^4)\nonumber\\
    \text{4-pt fn:}\hspace{0.25cm}&\langle O_\Delta(\vec{y_1})O_\Delta(\vec{y_2})O_\Delta(\vec{y_3})O_\Delta(\vec{y_4})\rangle_{\text{CFT}} =\frac{2\nu c_\Delta}{\lvert\vec{y_1}-\vec{y_2}\rvert^{2\Delta}}\frac{2\nu c_\Delta}{\lvert\vec{y_3}-\vec{y_4}\rvert^{2\Delta}}+(\vec{y_2}\leftrightarrow\vec{y_3})+(\vec{y_2}\leftrightarrow\vec{y_4})\nonumber\\
    &\hspace{2.4cm}+\lambda^2\int\int K(x_1,\vec{y_1})K(x_1,\vec{y_2})G(x_1,x_2)K(x_2,\vec{y_3})K(x_2,\vec{y_4})\times3\nonumber\\
    &\hspace{2.4cm}+\frac{2\nu c_\Delta}{\lvert\vec{y_1}-\vec{y_2}\rvert^{2\Delta}}\frac{\lambda^2}{2}\int_\varepsilon\int_\varepsilon K(x_1,\vec{y_3})K(x_2,\vec{y_4})G^2_\kappa(x_1,x_2)\times6 + \mathcal{O}(\lambda^4)\nonumber\\
    &\langle O_\Delta(\vec{y_1})O_\Delta(\vec{y_2})O_\Delta(\vec{y_3})O_\Delta(\vec{y_4})\rangle_{\text{CFT,con}} =\nonumber\\
    &\hspace{2.5cm}\lambda^2\int\int K(x_1,\vec{y_1})K(x_1,\vec{y_2})G(x_1,x_2)K(x_2,\vec{y_3})K(x_2,\vec{y_4})\times3+\mathcal{O}(\lambda^4)
\end{align}\

where we UV-regularized the quantum contributions in eq. (\ref{phi3corrtadren2}). These are the integrals that we will finally compute, process which we will proceed to do next.

\subsection{2-Point Function}

The regularized holographic 2-point functions dual to a $\Phi^3$ self-interacting theory on AdS are given by:

\begin{align}
    \langle O_\Delta(\vec{y_1})O_\Delta(\vec{y_2})\rangle_{\text{CFT}} &= \langle O_\Delta(\vec{y_1})O_\Delta(\vec{y_2})\rangle_{\text{CFT,con}}\nonumber\\
    &=\frac{2\nu c_\Delta}{\lvert\vec{y_1}-\vec{y_2}\rvert^{2\Delta}}+\frac{\lambda^2}{2}\int_\varepsilon\int_\varepsilon K(x_1,\vec{y_1})K(x_2,\vec{y_2})G^2_\kappa(x_1,x_2)+\mathcal{O}(\lambda^4)
\end{align}\

from where we see the quantum correction they receive coming from the loop integral:

\begin{equation}\label{2pfphi3regquantum0}
    I(\vec{y_1},\vec{y_2}) = \frac{\lambda^2}{2}\int_{x_{1,0}=\varepsilon}d^{d+1}x_1\sqrt{g}\int_{x_{2,0}=\varepsilon}d^{d+1}x_2\sqrt{g}\ K(x_1,\vec{y_1})K(x_2,\vec{y_2})G^2_\kappa(x_1,x_2)
\end{equation}\

In order to compute the complete 2-point function up to this order in the expansion of $\lambda$, we will proceed then to compute this quantity.

\subsubsection{The "Eye" Diagram}

\begin{figure}[h]
    \[\begin{wittendiagram}
    \draw (-2,0) node[vertex] -- (-1,0)
    (-1,0) node[vertex]
    (1,0) node[vertex]
    (2,0) node[vertex] -- (1,0)
    
     (-2,0) node[left]{$\vec{y_1}$}
    (2,0) node[right]{$\vec{y_2}$}
    
    (0,0) circle (1) ;
  \end{wittendiagram}\]
  \caption{Pictorial representation of the "eye" diagram.}
\end{figure}

In terms of the unnormalized bulk-boundary propagator we can rewrite the loop integral eq. (\ref{2pfphi3regquantum0}) (which we will refer to it as the "eye" diagram) as:

\begin{equation}\label{2pfphi3regquantum1}
    I(\vec{y_1},\vec{y_2}) = \frac{\lambda^2c^2_\Delta}{2}\int_{x_{1,0}=\varepsilon}d^{d+1}x_1\sqrt{g}\ \tilde{K}^\Delta(x_1,\vec{y_1})\int_{x_{2,0}=\varepsilon}d^{d+1}x_2\sqrt{g}\ G^2_\kappa(x_1,x_2)\tilde{K}^\Delta(x_2,\vec{y_2})
\end{equation}\

As we have done for past integrals, the solving strategy for this quantity will be brute force. We will start by first computing one of the integrals, say the $x_2$ integral, hoping that the remaining integral in $x_1$ will be familiar to us, which as we will see will indeed be the case. The first question that we will be interested in answering is when in this $x_2$ integral we can safely take the IR-regulator $\varepsilon$ equal to 0. The infrared convergence region of this integral can be seen directly by studying how its integrand behaves as it approaches the boundary of AdS. In this case, using the explicit form of the metric and the known expansion of both propagators, we obtain that:

\begin{equation}\label{g2kirstudy}
    \sqrt{g}\ G^2_\kappa(x_1,x_2)\tilde{K}^\Delta(x_2,\vec{y_2}) \underset{x_{2,0}\rightarrow0}{\sim} x_{2,0}^{-d-1}x_{2,0}^{2\Delta}x_{2,0}^{d-\Delta}= x_{2,0}^{-1+\Delta}
\end{equation}\

From here we conclude that the $x_2$ integral will be IR-convergent as long as $\Delta$ is a positive number. Since in this work we are considering the cases where $\Delta>\frac{d}{2}$ (as discussed in section 3.1.4), where $d$ is the dimension of the CFT theory living on the boundary of the AdS space, $\Delta$ will be always positive, implying that we can always take in this integral $\varepsilon=0$.\par
As we will see for every integral contributing to the holographic correlators, a remarkable feature of them is that their general structure can be derived by simply using AdS isometries transformations as change of variables. For the case of this $x_2$ integral performing the sequence of translation, inversion, translation and rescaling, allows us to extract all the external dependence from the integral:

\begin{align}\label{g2kstructure}
    \int d^{d+1}x_2\sqrt{g}\ G^2_\kappa(x_1,x_2)\tilde{K}^\Delta(x_2,\vec{y_2}) &= \int d^{d+1}x_2\sqrt{g}\ G^2_\kappa(x_1',x_2)\tilde{K}^\Delta(x_2,\vec{0})\nonumber\\
    &= \int d^{d+1}x_2\sqrt{g}\ G^2_\kappa(x_1'',x_2)x_{2,0}^\Delta\nonumber\\
    &= \int d^{d+1}x_2\sqrt{g}\ G^2_\kappa\bigl((x_{1,0}'',\vec{0}),x_2\bigl)x_{2,0}^\Delta\nonumber\\
    &= x_{1,0}''^\Delta\int d^{d+1}x_2\sqrt{g}\ G^2_\kappa\bigl((1,\vec{0}),x_2\bigl)x_{2,0}^\Delta\nonumber\\
    &\equiv C_{G^2K}(\kappa)\tilde{K}^\Delta(x_1,\vec{y_2})
\end{align}\

where, using the invariance of the AdS measure and the bulk-bulk propagator and the transformation rules of the bulk-boundary propagator, in the first equality we performed the translation $x_2\rightarrow x_2+\vec{y_2}$ and defined $x_1'\equiv x_1-\vec{y_2}$, in the second equality we performed the inversion $x^\mu_2\rightarrow \frac{x^\mu_2}{x^2_2}$ and defined $x'^\mu_1\equiv \frac{x''^\mu_1}{x''^2_1}$, in the third equality we performed the translation $x_2\rightarrow x_2+\vec{x_1}''$, in the fourth equality the rescaling $x_2\rightarrow x''_{1,0}x_2$ and in the final equality we remembered that $x''^\mu_1=\frac{x'^\mu_1}{x'^2_1}$ where $x'_1=x_1-\vec{y_2}$, also noticing that the remaining integral of this sequence of change of variables is just a function of the UV-regulator $\kappa$ which we simply called $C_{G^2K}(\kappa)$. This result is noteworthy, it is telling us that the integral $\int G^2K$ is proportional to $K$ where all the possible ultraviolet divergence coming from the bulk-bulk propagator being evaluated at coincident points is contained in the proportionality constant. In fact, notice that the power of $G$ did not play any role in this demonstration, which implies that this statement is true whenever the integral is IR-convergent. Doing the same infrared convergence study as in eq. (\ref{g2kirstudy}) now for a general power $n$ of the bulk-bulk propagator it is easy to verify that the integral will be IR-convergent for $n>1$. This implies then the more general result:

\begin{equation}\label{intgnkpropk}
    \int d^{d+1}x_2\sqrt{g}\ G^n_\kappa(x_1,x_2)\tilde{K}^\Delta(x_2,\vec{y_2}) = C_{G^nK}(\kappa)\tilde{K}^\Delta(x_1,\vec{y_2}),\hspace{1cm} \text{for }n>1
\end{equation}\

where $C_{G^nK}(\kappa)$ contains all the possible UV-divergence coming from the integral. The explicit form of this quantity is obtainable through the brute force calculation of the integral, process which will also show its ultraviolet convergence region. Let us then proceed to do this study for the particular integral that we are interested in computing, that is, $\int G^2K$. For this calculation, it will turn out to be useful to use the representation eq. (\ref{Grepdiv}) of the bulk-bulk propagator, where all its UV-divergence has been extracted out from the hypergeometric function. In terms of this representation then, we can write its regularized version squared simply as:

\begin{equation}
    G^2_\kappa(x_1,x_2) = \Bigl(\frac{2^{-\Delta}c_\Delta}{2\nu}\Bigl)^2\frac{\xi_\kappa^{2\Delta}}{(1-\xi_\kappa^2)^{d-1}}\Bigl[\ _2F_1\Bigl(\frac{\Delta-d}{2}+1,\frac{\Delta-d+1}{2};\nu+1;\xi_\kappa^2\Bigl)\Bigl]^2
\end{equation}\

where $\xi_\kappa=\frac{\xi}{1+\kappa}$. Using the double sum property $\sum_{k=0}^\infty\sum_{l=0}^\infty a_{k,l}=\sum_{k=0}^\infty\sum_{l=0}^k a_{k-l,l}$, the square of the hypergeometric function can be written as a single sum in the regularized parameter $\xi_\kappa$:

\begin{equation}\label{2f1square}
    \Bigl[\ _2F_1\Bigl(\frac{\Delta-d}{2}+1,\frac{\Delta-d+1}{2};\nu+1;\xi_\kappa^2\Bigl)\Bigl]^2 =\sum_{k=0}^\infty a^{(2)}_k\xi_\kappa^{2k}
\end{equation}\

where we defined the coefficient:

\begin{equation}\label{a2kcoef}
    a^{(2)}_k\equiv\sum_{l=0}^k\frac{(\frac{\Delta-d}{2}+1)_{k-l}(\frac{\Delta-d+1}{2})_{k-l}(\frac{\Delta-d}{2}+1)_l(\frac{\Delta-d+1}{2})_l}{(\nu+1)_{k-l}(\nu+1)_l\ (k-l)!\ l!}
\end{equation}\

This nice form for the square of the hypergeometric function further allows us to express the square of the regularized propagator in the form of:

\begin{equation}\label{reggsquare}
    G^2_\kappa(x_1,x_2) = \Bigl(\frac{2^{-\Delta}c_\Delta}{2\nu}\Bigl)^2\sum_{k=0}^\infty a^{(2)}_k\sum_{l=0}^\infty\frac{(d-1)_l}{l!}\Bigl(\frac{\xi}{1+\kappa}\Bigl)^{2\Delta+2k+2l}
\end{equation}\

where we used that $\frac{1}{(1-\xi_\kappa^2)^{d-1}}=\ _1F_0(d-1;\xi_\kappa^2)=\sum_{l=0}^\infty\frac{(d-1)_l}{l!}\xi_\kappa^{2l}$ and the explicit form of $\xi_\kappa$. Therefore, the integral of $\int G^2K$ with the bulk-bulk propagator written in this form can be expressed as:

\begin{align}
    \int d^{d+1}x_2\sqrt{g}\ G^2_\kappa(x_1,x_2)\tilde{K}^\Delta(x_2,\vec{y_2}) = &\Bigl(\frac{2^{-\Delta}c_\Delta}{2\nu}\Bigl)^2\sum_{k=0}^\infty a^{(2)}_k\sum_{l=0}^\infty\frac{(d-1)_l}{l!}\Bigl(\frac{1}{1+\kappa}\Bigl)^{2\Delta+2k+2l}\nonumber\\
    &\hspace{3cm}\times\int d^{d+1}x_2\sqrt{g}\ \xi^{2\Delta+2k+2l}\tilde{K}^\Delta(x_2,\vec{y_2})
\end{align}\

The complete study of this type of integrals can be found in section B.3 of Appendix B, concluding in its value in eq. (\ref{xikform}). Using this formula then for the particular values $\Delta_1=2\Delta+2k+2l$ and $\Delta_2=\Delta$, we find that the result of the $x_2$ integral is given by:

\begin{align}\label{intg2kvalue}
    \int d^{d+1}x_2\sqrt{g}\ G^2_\kappa(x_1,x_2)\tilde{K}^\Delta(x_2,\vec{y_2}) = &\tilde{K}^\Delta(x_1,\vec{y_2})\pi^\frac{d+1}{2}\Bigl(\frac{2^{-\Delta}c_\Delta}{2\nu}\Bigl)^2\sum_{k=0}^\infty a^{(2)}_k\frac{\Gamma(\frac{3\Delta-d}{2}+k)\Gamma(\frac{\Delta}{2}+k)}{\Gamma(\Delta+k)\Gamma(\Delta+\frac{1}{2}+k)}\nonumber\\
    &\hspace{-4cm}\times\Bigl(\frac{1}{1+\kappa}\Bigl)^{2\Delta+2k}\ _3F_2\Bigl(d-1,\frac{3\Delta-d}{2}+k,\frac{\Delta}{2}+k;\Delta+k,\Delta+\frac{1}{2}+k;\Bigl(\frac{1}{1+\kappa}\Bigl)^2\Bigl)\nonumber\\
    \equiv &\ C_{G^2K}(\kappa)\tilde{K}^\Delta(x_1,\vec{y_2})
\end{align}\

where we identified the representation of the generalized hypergeometric function $_3F_2$ and denoted all the terms not dependent on the external points simply by $C_{G^2K}(\kappa)$. This result for the $\int G^2K$ integral obtained from its explicit computation has precisely the structure expected from AdS isometry arguments eq. (\ref{intgnkpropk}), where in this case the value of the constant $C_{G^2K}(\kappa)$ is found to be:

\begin{align}
    C_{G^2K}(\kappa) = \pi^\frac{d+1}{2}\Bigl(\frac{2^{-\Delta}c_\Delta}{2\nu}\Bigl)^2\sum_{k=0}^\infty &a^{(2)}_k\frac{\Gamma(\frac{3\Delta-d}{2}+k)\Gamma(\frac{\Delta}{2}+k)}{\Gamma(\Delta+k)\Gamma(\Delta+\frac{1}{2}+k)}\Bigl(\frac{1}{1+\kappa}\Bigl)^{2\Delta+2k}\nonumber\\
    &\hspace{-1cm}\times\ _3F_2\Bigl(d-1,\frac{3\Delta-d}{2}+k,\frac{\Delta}{2}+k;\Delta+k,\Delta+\frac{1}{2}+k;\Bigl(\frac{1}{1+\kappa}\Bigl)^2\Bigl)
\end{align}\

where in turn the coefficient $a^{(2)}_k$ was defined in eq. (\ref{a2kcoef}). Notice how the ultraviolet convergence region of this quantity can be read directly from its generalized hypergeometric function. Indeed, as $\kappa\rightarrow0$ the argument of this function goes to 1, which applying the corresponding convergence criteria introduced in the last section implies the convergence condition $\Delta+k+\Delta+\frac{1}{2}+k-(d-1)-(\frac{3\Delta-d}{2}+k)-(\frac{\Delta}{2}+k)=\frac{3-d}{2}>0$, that is, $d<3$. In other words, the $\int G^2K$ integral is UV-divergent for values of the dimension $d$ equal or greater than 3, and UV-convergent otherwise, being able to safely take $\kappa=0$ in this case.\par

Remember that we are trying to compute the loop integral contributing to the holographic 2-point function dual to a $\Phi^3$ theory on AdS, eq. (\ref{2pfphi3regquantum1}). Replacing then the nice result just found for the $x_2$ integral back into the quantity we are trying to compute it reduces to:

\begin{equation}
    I(\vec{y_1},\vec{y_2}) = \frac{\lambda^2C_{G^2K}(\kappa)c^2_\Delta}{2}\int_{x_{1,0}=\varepsilon}d^{d+1}x_1\sqrt{g}\ \tilde{K}^\Delta(x_1,\vec{y_1})\tilde{K}^\Delta(x_1,\vec{y_2})
\end{equation}\

If we were to put the normalization factors $c_\Delta$ back to the unnormalized bulk-boundary propagators $\tilde{K}^\Delta$ it would result in the integral of 2 normalized bulk-boundary propagators times some number. As we will see during the calculations of these loops integral, it will be practical to name this overall number resulting from this process. Let us define then what eventually will be understood as the one-particle irreducible (or simply 1PI) contributions to the correlators:

\begin{equation}\label{1piphi3}
    \Pi(\kappa) \equiv \frac{\lambda^2C_{G^2K}(\kappa)}{2}
\end{equation}\

In terms of this quantity, the integral we are trying to compute is written as:

\begin{equation}
    I(\vec{y_1},\vec{y_2}) = \Pi(\kappa)c_\Delta^2\int_{x_{1,0}=\varepsilon}d^{d+1}x_1\sqrt{g}\ \tilde{K}^\Delta(x_1,\vec{y_1})\tilde{K}^\Delta(x_1,\vec{y_2})
\end{equation}\

The remaining integral in $x_1$, if it were not for the IR-regulator $\varepsilon$, should remind you of the D-functions defined in eq. (\ref{dfunc}) of Appendix A. Indeed, it is an integral of just bulk-boundary propagators. However, as it is discussed in detail in section A.4 of this appendix, a simple power counting of this integral suggests that it is logarithmically divergent in the lower limit of integration of the radial coordinate $x_{1,0}$. As we have already discussed previously, these infinities coming from quantities being evaluated at the conformal boundary of AdS spaces are no surprise since we are actually computing CFT correlators which are expected to be divergent. This realization was precisely what motivated the introduction of the IR-regulator $\varepsilon$ in the first place, which in turn also motivates the definition of the regularized version of this particular D-function in eq. (\ref{d2funcreg1}). Therefore, in terms of this function we can express the integral we are interested in computing in the form of:

\begin{equation}
    I(\vec{y_1},\vec{y_2}) = \Pi(\kappa)c_\Delta^2D^{(\varepsilon)}_{\Delta\Delta}(\vec{y_1},\vec{y_2})
\end{equation}\

The complete study of this particular D-function can be found in section A.4 of Appendix A, concluding in its value in eq. (\ref{d2funcreg2}). Using this value then in our present case we find that the final result of the integral eq. (\ref{2pfphi3regquantum0}) is given by:

\begin{equation}\label{2pfphi3regquantum5}
    I(\vec{y_1},\vec{y_2}) = -\frac{2\nu c_\Delta}{\lvert\vec{y_1}-\vec{y_2}\rvert^{2\Delta}}\frac{\Pi(\kappa)}{\nu}\ln{\Bigl(\frac{\varepsilon}{\lvert\vec{y_1}-\vec{y_2}\rvert}\Bigl)}
\end{equation}\

where we have written it conveniently for the upcoming study. The presence of the logarithm in this result seems to break the conformal structure expected for contributions to the 2-point function of a CFT as it was derived in eq. (\ref{summcorr2}), however as we will see next when we consider the complete correlator, we will realize that the result just found corresponds exactly to the expansion of a conformal anomaly up to this same order in the self-interacting coupling constant, realization that will also provide us with a natural renormalization scheme of both IR and UV divergences equivalent to those used in ordinary QFTs.

\subsubsection{Correlator Renormalization}

Replacing the result just found then for the "eye" diagram back into the holographic 2-point functions, we find that they can be factorized into the form:

\begin{align}
    \langle O_\Delta(\vec{y_1})O_\Delta(\vec{y_2})\rangle_{\text{CFT}} &= \langle O_\Delta(\vec{y_1})O_\Delta(\vec{y_2})\rangle_{\text{CFT,con}}\nonumber\\
    &=\frac{2\nu c_\Delta}{\lvert\vec{y_1}-\vec{y_2}\rvert^{2\Delta}}\Bigl[1-\frac{\Pi(\kappa)}{\nu}\ln{\Bigl(\frac{\varepsilon}{\lvert\vec{y_1}-\vec{y_2}\rvert}\Bigl)}\Bigl]+\mathcal{O}(\lambda^4)
\end{align}\

The terms inside the square bracket, up to this same order in $\lambda$, correspond to the known Taylor series of an exponent:

\begin{equation}
    1-\frac{\Pi(\kappa)}{\nu}\ln{\Bigl(\frac{\varepsilon}{\lvert\vec{y_1}-\vec{y_2}\rvert}\Bigl)} = \Bigl(\frac{\varepsilon}{\lvert\vec{y_1}-\vec{y_2}\rvert}\Bigl)^{-\frac{\Pi(\kappa)}{\nu}}+\mathcal{O}(\lambda^4)
\end{equation}\

This fact allows us to express the regularized holographic 2-point functions in the nice compact form:

\begin{equation}\label{reg2pfgphi3qm}
    \langle O_\Delta(\vec{y_1})O_\Delta(\vec{y_2})\rangle_{\text{CFT}} = \langle O_\Delta(\vec{y_1})O_\Delta(\vec{y_2})\rangle_{\text{CFT,con}}
    =\varepsilon^{-\frac{\Pi(\kappa)}{\nu}}\frac{2\nu c_\Delta}{\lvert\vec{y_1}-\vec{y_2}\rvert^{2\Delta-\frac{\Pi(\kappa)}{\nu}}}+\mathcal{O}(\lambda^4)
\end{equation}\

With the correlators written in this form, it is direct to see what are the effects of the quantum corrections coming from the off-shell part of the AdS path integral to the 2-point function found previously under the classical approximation of the AdS/CFT correspondence. Indeed, they contribute with an overall rescaling to the correlator along with a shift in its scaling dimension! Now, of course as we take the understood limits $\varepsilon\rightarrow0$ and $\kappa\rightarrow0$ this correlator becomes divergent so it is necessary the introduction of a delicate renormalization scheme in order to absorb the respective infinities. Fortunately, the nice form of eq. (\ref{reg2pfgphi3qm}) allows us to renormalize it in exactly the same spirit as it is done for ordinary QFTs, this is, by understanding the parameters of the theory not as physical constants but bare quantities, opening the possibility of a renormalization scheme through their definition. Take for example the anomalous dimension of the correlators. Notice that their exponent can be written as:

\begin{equation}
    2\Delta-\frac{\Pi(\kappa)}{\nu} = 2\Bigl[\frac{d}{2}+\sqrt{\Bigl(\frac{d}{2}\Bigl)^2+m^2-\Pi(\kappa)}\Bigl]+\mathcal{O}(\lambda^4)
\end{equation}\

where we used that $\Delta=\frac{d}{2}+\nu$, $\nu=\sqrt{(\frac{d}{2})^2+m^2}$ and the known Taylor series of the square root. The exponent written in this form strongly suggests the renormalization of the UV-divergences coming from $\Pi(\kappa)$ (that is, whenever $d>2$) through a redefinition of the bulk's mass parameter $m^2$. Indeed, redefining this parameter in the AdS bulk action simply as:

\begin{equation}
    m^2\rightarrow m^2+\delta m^2
\end{equation}\

where the counterterm is expected to be of order $\delta m^2=\mathcal{O}(\lambda^2)$ adds, up to order $\lambda^3$, a new counterterm interaction to the holographic 2-point function eq. (\ref{reg2pfgphi3qm}) of the form:

\begin{align}
    \langle O_\Delta(\vec{y_1})O_\Delta(\vec{y_2})\rangle_{\text{CFT}} &= \langle O_\Delta(\vec{y_1})O_\Delta(\vec{y_2})\rangle_{\text{CFT,con}}\nonumber\\
    &=\varepsilon^{-\frac{\Pi(\kappa)}{\nu}}\frac{2\nu c_\Delta}{\lvert\vec{y_1}-\vec{y_2}\rvert^{2\Delta-\frac{\Pi(\kappa)}{\nu}}}-\delta m^2\int_{x_0=\varepsilon}d^{d+1}x\sqrt{g}\ K(x,\vec{y_1})K(x,\vec{y_2})+\mathcal{O}(\lambda^4)
\end{align}\

This new contribution to the correlator can be solved in terms of the D-function eq. (\ref{d2funcreg2}):

\begin{equation}
    -\delta m^2\int_{x_0=\varepsilon}d^{d+1}x\sqrt{g}\ K(x,\vec{y_1})K(x,\vec{y_2}) = \frac{2\nu c_\Delta}{\lvert\vec{y_1}-\vec{y_2}\rvert^{2\Delta}}\frac{\delta m^2}{\nu}\ln{\Bigl(\frac{\varepsilon}{\lvert\vec{y_1}-\vec{y_2}\rvert}\Bigl)}
\end{equation}\

which ultimately results in:

\begin{align}
    \langle O_\Delta(\vec{y_1})O_\Delta(\vec{y_2})\rangle_{\text{CFT}} &= \langle O_\Delta(\vec{y_1})O_\Delta(\vec{y_2})\rangle_{\text{CFT,con}}\nonumber\\
    &=\varepsilon^{\frac{\delta m^2-\Pi(\kappa)}{\nu}}\frac{2\nu c_\Delta}{\lvert\vec{y_1}-\vec{y_2}\rvert^{2\Delta+\frac{\delta m^2-\Pi(\kappa)}{\nu}}}+\mathcal{O}(\lambda^4)
\end{align}\

Therefore, denoting the 1PI contributions $\Pi(\kappa)$ as $\Pi(\kappa)=\Pi_\infty(\kappa)+\Pi_0(\kappa)$, where all its UV-divergent terms are contained in $\Pi_\infty(\kappa)$, the infinities present in the correlators coming from the ultraviolet divergences of the loops integrals can be renormalized away through the convenient choice of the counterterm $\delta m^2$ as:

\begin{equation}
    \delta m^2 = \Pi_\infty(\kappa)
\end{equation}\

resulting in the UV-renormalized holographic 2-point functions:

\begin{equation}
    \langle O_\Delta(\vec{y_1})O_\Delta(\vec{y_2})\rangle_{\text{CFT}} = \langle O_\Delta(\vec{y_1})O_\Delta(\vec{y_2})\rangle_{\text{CFT,con}}=\varepsilon^{-\frac{\Pi_0(0)}{\nu}}\frac{2\nu c_\Delta}{\lvert\vec{y_1}-\vec{y_2}\rvert^{2\Delta-\frac{\Pi_0(0)}{\nu}}}+\mathcal{O}(\lambda^4)
\end{equation}\

where we safely took the limit $\kappa=0$. We are still half way in the renormalization process as we still have to deal with the infrared divergence of the correlators. However, noticing that it acts simply as an overall factor, this strongly suggests its renormalization through a redefinition of the bulk field $\Phi(x)$. Indeed, redefining it in the AdS bulk action simply as:

\begin{equation}
    \Phi(x) \rightarrow \sqrt{Z(\lambda)}\Phi(x)
\end{equation}\

adds a new factor to the holographic 2-point function of the form:

\begin{equation}
    \langle O_\Delta(\vec{y_1})O_\Delta(\vec{y_2})\rangle_{\text{CFT}} = \langle O_\Delta(\vec{y_1})O_\Delta(\vec{y_2})\rangle_{\text{CFT,con}}=\frac{\varepsilon^{-\frac{\Pi_0(0)}{\nu}}}{Z(\lambda)}\frac{2\nu c_\Delta}{\lvert\vec{y_1}-\vec{y_2}\rvert^{2\Delta-\frac{\Pi_0(0)}{\nu}}}+\mathcal{O}(\lambda^4)
\end{equation}\

Therefore, the infinities present in the correlators coming from the infrared divergences of the loops integrals can be renormalized away through the convenient choice of the counterterm $Z(\lambda)$ as:

\begin{equation}
    Z(\lambda) = \varepsilon^{-\frac{\Pi_0(0)}{\nu}} = 1-\frac{\Pi_0(0)}{\nu}\ln{(\varepsilon)}+\mathcal{O}(\lambda^4)
\end{equation}\

resulting in both IR and UV renormalized holographic 2-point functions:

\begin{equation}
    \langle O_\Delta(\vec{y_1})O_\Delta(\vec{y_2})\rangle_{\text{CFT}} = \langle O_\Delta(\vec{y_1})O_\Delta(\vec{y_2})\rangle_{\text{CFT,con}}=\frac{2\nu c_\Delta}{\lvert\vec{y_1}-\vec{y_2}\rvert^{2\Delta-\frac{\Pi_0(0)}{\nu}}}+\mathcal{O}(\lambda^4)
\end{equation}\

where the limits $\varepsilon=\kappa=0$ have been taken and where $\Pi_0(0)$ denote the UV-finite part of the 1PI contributions $\Pi(\kappa)$. Since for $d<3$ the quantity $\Pi(\kappa)$ is already UV-finite, this implies that in these cases $\Pi_0(\kappa)=\Pi(\kappa)$.

\subsection{3-Point Function}

The regularized holographic 3-point functions dual to a $\Phi^3$ self-interacting theory on AdS are given by:

\begin{align}
    \langle O_\Delta(\vec{y_1})O_\Delta(\vec{y_2})O_\Delta(\vec{y_3})\rangle_{\text{CFT}} = &\langle O_\Delta(\vec{y_1})O_\Delta(\vec{y_2})O_\Delta(\vec{y_3})\rangle_{\text{CFT,con}}= \nonumber\\
    &\hspace{-2cm}-\lambda\int K(x_1,\vec{y_1})K(x_1,\vec{y_2})K(x_1,\vec{y_3})\nonumber\\
    &\hspace{-2cm}-\frac{\lambda^3}{2}\int_\varepsilon\int_\varepsilon\int_\varepsilon K(x_1,\vec{y_1})K(x_2,\vec{y_2})K(x_2,\vec{y_3})G^2_\kappa(x_1,x_3)G_\kappa(x_3,x_2)\times 3\nonumber\\
    &\hspace{-2cm}-\lambda^3\int_\varepsilon\int_\varepsilon\int_\varepsilon K(x_1,\vec{y_1})K(x_2,\vec{y_2})K(x_3,\vec{y_3})G_\kappa(x_1,x_2)G_\kappa(x_2,x_3)G_\kappa(x_3,x_1)\nonumber\\
    &\hspace{-2cm}+\mathcal{O}(\lambda^4)
\end{align}\

from where we see the quantum corrections they receive coming from the loop integrals:

\begin{align}
    I_1(\vec{y_1},\vec{y_2},\vec{y_3}) = &-\frac{\lambda^3}{2}\int_\varepsilon\int_\varepsilon\int_\varepsilon K(x_1,\vec{y_1})K(x_2,\vec{y_2})K(x_2,\vec{y_3})G^2_\kappa(x_1,x_3)G_\kappa(x_3,x_2)\nonumber\\
    &+(\vec{y_1}\leftrightarrow\vec{y_2})+(\vec{y_1}\leftrightarrow\vec{y_3})
\end{align}

\begin{equation}
    I_2(\vec{y_1},\vec{y_2},\vec{y_3}) = -\lambda^3\int_\varepsilon\int_\varepsilon\int_\varepsilon K(x_1,\vec{y_1})K(x_2,\vec{y_2})K(x_3,\vec{y_3})G_\kappa(x_1,x_2)G_\kappa(x_2,x_3)G_\kappa(x_3,x_1)
\end{equation}\

In order to compute the complete 3-point function up to this order in the expansion of $\lambda$, we will proceed then to compute these quantities. We will start by studying $I_1$, leaving $I_2$ for last.

\subsubsection{The Reducible "Eye" Diagram}

\begin{figure}[h]
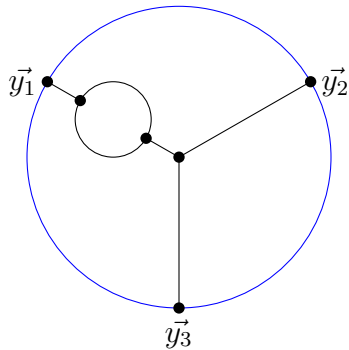

    \[\begin{wittendiagram}
    \draw (-1.732,1) node[vertex] -- (-1.299,0.75)
    (1.732,1) node[vertex] -- (0,0)
    (0,-2) node[vertex] -- (0,0)
    (0,0) node[vertex] -- (-0.433,0.25)
    (-0.433,0.25) node[vertex]
    (-1.299,0.75) node[vertex]
    
     (-1.732,1) node[left]{$\vec{y_1}$}
    (1.732,1) node[right]{$\vec{y_2}$}
    (0,-2) node[below]{$\vec{y_3}$}
    
    (-0.866,0.5) circle (0.5);
  \end{wittendiagram}\]
  \caption{Pictorial representation of the reducible "eye" diagram.}
\end{figure}

In terms of the unnormalized bulk-boundary propagator we can rewrite the loop integral $I_1$ (which we will refer to it as the reducible "eye" diagram) as:

\begin{align}\label{3pfphi3regquantum11}
    I_1(\vec{y_1},\vec{y_2},\vec{y_3}) = &-\frac{\lambda^3c_\Delta^3}{2}\int_{x_{2,0}=\varepsilon}d^{d+1}x_2\sqrt{g}\ \tilde{K}^\Delta(x_2,\vec{y_2})\tilde{K}^\Delta(x_2,\vec{y_3})\int_{x_{3,0}=\varepsilon}d^{d+1}x_3\sqrt{g}\ G_\kappa(x_3,x_2)\nonumber\\
    &\hspace{1cm}\times\int_{x_{1,0}=\varepsilon}d^{d+1}x_1\sqrt{g}\ G^2_\kappa(x_3,x_1)\tilde{K}^\Delta(x_1,\vec{y_1})+(\vec{y_1}\leftrightarrow\vec{y_2})+(\vec{y_1}\leftrightarrow\vec{y_3})
\end{align}\

Continuing in the same line we have followed so far for the integrals that we have stumble upon, we will try to solve these that are present in eq. (\ref{3pfphi3regquantum11}) one by one hoping that every resulting integral from this iteration will be familiar to us, which as we will see will indeed be the case. We will start this study by noticing that the $x_1$ integral is nothing but the IR-convergent $x_2$ integral that we just faced for the 2-point function, eq. (\ref{intg2kvalue}). Therefore, safely taking $\varepsilon=0$ in this integral and replacing its known result, we find that $I_1$ reduces to:

\begin{align}
    I_1(\vec{y_1},\vec{y_2},\vec{y_3}) = &-\frac{\lambda^3c_\Delta^3C_{G^2K}(\kappa)}{2}\int_{x_{2,0}=\varepsilon}d^{d+1}x_2\sqrt{g}\ \tilde{K}^\Delta(x_2,\vec{y_2})\tilde{K}^\Delta(x_2,\vec{y_3})\nonumber\\
    &\hspace{2cm}\times \int_{x_{3,0}=\varepsilon}d^{d+1}x_3\sqrt{g}\ G_\kappa(x_2,x_3)\tilde{K}^\Delta(x_3,\vec{y_1})+(\vec{y_1}\leftrightarrow\vec{y_2})+(\vec{y_1}\leftrightarrow\vec{y_3})
\end{align}\

The resulting integral in $x_3$, as it was discussed in the last section for the "eye" diagram of the 2-point function and also as it is discussed in detail in section B.5 of Appendix B, a simple power counting suggests that, if it were not for the $\varepsilon$-regulator, it would be logarithmically divergent in the lower limit of integration of the radial coordinate $x_{3,0}$. Precisely the role of this regulator is not only to tame this divergence but also to capture the correct behavior of the integral hidden in it. The complete study of the $x_3$ integral can be found in the section of the appendix mentioned above, concluding in its value in eq. (\ref{scintgk}). Using this value then in our present case we find that $I_1$ further reduces to:

\begin{align}
    I_1(\vec{y_1},\vec{y_2},\vec{y_3}) = &\lambda c_\Delta^3\frac{\Pi(\kappa)}{2\nu}\int d^{d+1}x_1\sqrt{g}\ \tilde{K}^\Delta(x_1,\vec{y_1})\ln{\bigl(\varepsilon\tilde{K}(x_1,\vec{y_1})\bigl)}\tilde{K}^\Delta(x_1,\vec{y_2})\tilde{K}^\Delta(x_1,\vec{y_3})\nonumber\\
    &+(\vec{y_1}\leftrightarrow\vec{y_2})+(\vec{y_1}\leftrightarrow\vec{y_3})
\end{align}\

where we wrote the constant factors in terms of $\Pi(\kappa)$ (eq. (\ref{1piphi3})), called the integrated variable $x_2\rightarrow x_1$, and performed a simple power counting in the radial coordinate $x_{1,0}$ as it approaches the boundaries, realizing that the integral is IR-convergent and therefore allowing us to simply take the regulator $\varepsilon$ equal to 0. As we will see shortly, instead of trying to compute this integral it will turn out to be much simpler to keep it in this form for now, since later when grouping it with the other contributions to the 3-point function it will result in a natural factorization into a known integral. Let us proceed then to study the second loop contribution, $I_2$.

\subsubsection{The "O" Diagram}

\begin{figure}[h]
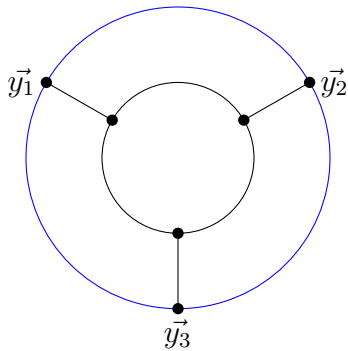

    \[\begin{wittendiagram}
    \draw (-1.732,1) node[vertex]
    (1.732,1) node[vertex]
    (0,-2) node[vertex]
    (-0.866,0.5) node[vertex] -- (-1.732,1)
    (0.866,0.5) node[vertex] -- (1.732,1)
    (0,-1) node[vertex] -- (0,-2)
    
     (-1.732,1) node[left]{$\vec{y_1}$}
    (1.732,1) node[right]{$\vec{y_2}$}
    (0,-2) node[below]{$\vec{y_3}$}
    
    (0,0) circle (1);
  \end{wittendiagram}\]
  \caption{Pictorial representation of the "O" diagram.}
\end{figure}

In terms of the unnormalized bulk-boundary propagator we can rewrite the loop integral $I_2$ (which we will refer to it as the "O" diagram) as:

\begin{equation}
    I_2(\vec{y_1},\vec{y_2},\vec{y_3}) = -\lambda^3c^3_\Delta\int_\varepsilon\int_\varepsilon\int_\varepsilon \tilde{K}^\Delta(x_1,\vec{y_1})\tilde{K}^\Delta(x_2,\vec{y_2})\tilde{K}^\Delta(x_3,\vec{y_3})G_\kappa(x_1,x_2)G_\kappa(x_2,x_3)G_\kappa(x_3,x_1)
\end{equation}\

Unfortunately, the integration formulas developed in this work do not allow us to compute integrals involving more than one bulk-bulk propagator being evaluated at different points, however this does not stop us from studying the structure of this quantity through AdS isometries transformations as we did for the integral $\int G^2K$ in eq. (\ref{g2kstructure}). Indeed, this can be achieved by first noticing that a simple power counting in the radial coordinates suggests that the "O" diagram is IR-convergent allowing us to compute the integrals up to $x_{i,0}=0$, fact which further allows us to extract all the external dependence from the integrals through the sequence of translation, inversion, translation and rescaling:

\begin{align}
    &\int\int\int \tilde{K}^\Delta(x_1,\vec{y_1})\tilde{K}^\Delta(x_2,\vec{y_2})\tilde{K}^\Delta(x_3,\vec{y_3})G_\kappa(x_1,x_2)G_\kappa(x_2,x_3)G_\kappa(x_3,x_1)\nonumber\\
    &\hspace{1cm}= \int\int\int \tilde{K}^\Delta(x_1,\vec{y_{13}})\tilde{K}^\Delta(x_2,\vec{y_{23}})\tilde{K}^\Delta(x_3,\vec{0})G_\kappa(x_1,x_2)G_\kappa(x_2,x_3)G_\kappa(x_3,x_1)\nonumber\\
    &\hspace{1cm}= \lvert\vec{y_{13}}'\rvert^{2\Delta}\lvert\vec{y_{23}}'\rvert^{2\Delta}\int\int\int \tilde{K}^\Delta(x_1,\vec{y_{13}}')\tilde{K}^\Delta(x_2,\vec{y_{23}}')x^\Delta_{3,0}G_\kappa(x_1,x_2)G_\kappa(x_2,x_3)G_\kappa(x_3,x_1)\nonumber\\
    &\hspace{1cm}= \lvert\vec{y_{13}}'\rvert^{2\Delta}\lvert\vec{y_{23}}'\rvert^{2\Delta}\int\int\int \tilde{K}^\Delta(x_1,\vec{y_{13}}'-\vec{y_{23}}')\tilde{K}^\Delta(x_2,\vec{0})x^\Delta_{3,0}G_\kappa(x_1,x_2)G_\kappa(x_2,x_3)G_\kappa(x_3,x_1)\nonumber\\
    &\hspace{1cm}= \frac{\lvert\vec{y_{13}}'\rvert^{2\Delta}\lvert\vec{y_{23}}'\rvert^{2\Delta}}{\lvert\vec{y_{13}}'-\vec{y_{23}}'\rvert^\Delta}\int\int\int \tilde{K}^\Delta(x_1,\hat{n})\tilde{K}^\Delta(x_2,\vec{0})x^\Delta_{3,0}G_\kappa(x_1,x_2)G_\kappa(x_2,x_3)G_\kappa(x_3,x_1)\nonumber\\
    &\hspace{1cm}= \frac{1}{\lvert\vec{y_{12}}\rvert^\Delta\lvert\vec{y_{23}}\rvert^\Delta\lvert\vec{y_{31}}\rvert^\Delta}\int\int\int \tilde{K}^\Delta(x_1,\hat{n})\tilde{K}^\Delta(x_2,\vec{0})x^\Delta_{3,0}G_\kappa(x_1,x_2)G_\kappa(x_2,x_3)G_\kappa(x_3,x_1)
\end{align}\

where, using the invariance of the AdS measure and the bulk-bulk propagator and the transformation rules of the bulk-boundary propagator, in the first equality we performed the translations $x_i\rightarrow x_i+\vec{y_3}$, in the second equality we performed the inversions $x_i^\mu\rightarrow \frac{x_i^\mu}{x_i^2}$ and defined $\vec{y_{ij}}'=\frac{\vec{y_{ij}}}{\lvert\vec{y_{ij}}\rvert^2}$, in the third equality we performed the translations $x_i\rightarrow x_i+\vec{y_{23}}'$, in the fourth equality we performed the rescaling $x_i\rightarrow\lvert\vec{y_{13}}'-\vec{y_{23}}'\rvert x_i$ and defined the unit vector $\hat{n}\equiv\frac{\vec{y_{13}}'-\vec{y_{23}}'}{\lvert\vec{y_{13}}'-\vec{y_{23}}'\rvert}$, and in the final equality we wrote the external points in terms of the originals. This result is noteworthy, it is telling us that the "O" diagram has exactly the conformal form expected for contributions to the 3-point function of a CFT, where all its possible UV-divergence coming from the integrals is contained in its overall factor. In fact, for the upcoming calculations it will be useful to write the conformal dependence of the diagram in terms of the D-function eq. (\ref{3dfunc}):

\begin{equation}
    \frac{1}{\lvert\vec{y_{12}}\rvert^\Delta\lvert\vec{y_{23}}\rvert^\Delta\lvert\vec{y_{31}}\rvert^\Delta} = \frac{2}{\pi^\frac{d}{2}}\frac{\Gamma(\Delta)^3}{\Gamma(\frac{3\Delta-d}{2})\Gamma(\frac{\Delta}{2})^3}D_{\Delta\Delta\Delta}(\vec{y_1},\vec{y_2},\vec{y_3})
\end{equation}\

which allows us to write the loop integral more compactly:

\begin{align}
    &\int\int\int \tilde{K}^\Delta(x_1,\vec{y_1})\tilde{K}^\Delta(x_2,\vec{y_2})\tilde{K}^\Delta(x_3,\vec{y_3})G_\kappa(x_1,x_2)G_\kappa(x_2,x_3)G_\kappa(x_3,x_1)\nonumber\\
    &\hspace{0.5cm}= \frac{2}{\pi^\frac{d}{2}}\frac{\Gamma(\Delta)^3}{\Gamma(\frac{3\Delta-d}{2})\Gamma(\frac{\Delta}{2})^3}D_{\Delta\Delta\Delta}\int\int\int \tilde{K}^\Delta(x_1,\hat{n})\tilde{K}^\Delta(x_2,\vec{0})x^\Delta_{3,0}G_\kappa(x_1,x_2)G_\kappa(x_2,x_3)G_\kappa(x_3,x_1)\nonumber\\
    &\hspace{0.5cm}\equiv C_{GGG}(\kappa)D_{\Delta\Delta\Delta}(\vec{y_1},\vec{y_2},\vec{y_3})
\end{align}\

where in the second line we denoted all the constant factors simply as $C_{GGG}(\kappa)$. Finally then, with the result of the integrals written in this form, the complete "O" diagram $I_2$ can be expressed as:

\begin{equation}
    I_2(\vec{y_1},\vec{y_2},\vec{y_3}) = -\lambda^3c^3_\Delta C_{GGG}(\kappa)D_{\Delta\Delta\Delta}(\vec{y_1},\vec{y_2},\vec{y_3})
\end{equation}\

Next we will see how these loops contributions to the 3-point function, $I_1$ and $I_2$, can be understood as the expansion of not any conformal anomaly, but to the exact same anomaly dictated by the 2-point function, where in the present case the effective coupling constant also receives a correction. This realization will provide us with a natural and consistent renormalization scheme of both IR and UV divergences for all the holographic n-point functions, equivalent to those schemes used in ordinary QFTs.

\subsubsection{Correlator Renormalization}

Replacing the results just found then for the reducible "eye" diagram and the "O" diagram back into the holographic 3-point functions, we find that they can be written as:

\begin{align}
    \langle O_\Delta(\vec{y_1})O_\Delta(\vec{y_2})O_\Delta(\vec{y_3})\rangle_{\text{CFT}} &= \langle O_\Delta(\vec{y_1})O_\Delta(\vec{y_2})O_\Delta(\vec{y_3})\rangle_{\text{CFT,con}} \nonumber\\
    &\hspace{-2cm}=-\lambda c_\Delta^3\int d^{d+1}x_1\sqrt{g}\ \tilde{K}^\Delta(x_1,\vec{y_1})\tilde{K}^\Delta(x_1,\vec{y_2})\tilde{K}^\Delta(x_1,\vec{y_3})\nonumber\\
    &\hspace{-1.5cm}+\lambda c_\Delta^3\frac{\Pi(\kappa)}{2\nu}\int d^{d+1}x_1\sqrt{g}\ \tilde{K}^\Delta(x_1,\vec{y_1})\ln{\bigl(\varepsilon\tilde{K}(x_1,\vec{y_1})\bigl)}\tilde{K}^\Delta(x_1,\vec{y_2})\tilde{K}^\Delta(x_1,\vec{y_3})\nonumber\\
    &\hspace{-1.5cm}+(\vec{y_1}\leftrightarrow\vec{y_2})+(\vec{y_1}\leftrightarrow\vec{y_3})-\lambda^3c^3_\Delta C_{GGG}(\kappa)D_{\Delta\Delta\Delta}(\vec{y_1},\vec{y_2},\vec{y_3})+\mathcal{O}(\lambda^4)
\end{align}\

Now, up to this same order in the coupling constant, it is easy to see that the tree-level term together with the reducible "eye" diagrams can be factorized into a known integral by realizing that they can be written in the form:

\begin{align}
    &-\lambda c_\Delta^3\int d^{d+1}x_1\sqrt{g}\ \tilde{K}^\Delta(x_1,\vec{y_1})\tilde{K}^\Delta(x_1,\vec{y_2})\tilde{K}^\Delta(x_1,\vec{y_3})\nonumber\\
    &+\lambda c_\Delta^3\frac{\Pi(\kappa)}{2\nu}\int d^{d+1}x_1\sqrt{g}\ \tilde{K}^\Delta(x_1,\vec{y_1})\ln{\bigl(\varepsilon\tilde{K}(x_1,\vec{y_1})\bigl)}\tilde{K}^\Delta(x_1,\vec{y_2})\tilde{K}^\Delta(x_1,\vec{y_3})\nonumber\\
    &+(\vec{y_1}\leftrightarrow\vec{y_2})+(\vec{y_1}\leftrightarrow\vec{y_3})\nonumber\\
    &\hspace{2cm}=\varepsilon^{-\frac{3\Pi(\kappa)}{2\nu}}\times-\lambda c_\Delta^3D_{\Delta-\frac{\Pi(\kappa)}{2\nu}\Delta-\frac{\Pi(\kappa)}{2\nu}\Delta-\frac{\Pi(\kappa)}{2\nu}}(\vec{y_1},\vec{y_2},\vec{y_3})+\mathcal{O}(\lambda^4)
\end{align}\

where we wrote the resulting integral in terms of the D-function. This result allows us to express the 3-point correlators in terms of these functions as:

\begin{align}\label{regiruv3pf}
    \langle O_\Delta(\vec{y_1})O_\Delta(\vec{y_2})O_\Delta(\vec{y_3})\rangle_{\text{CFT}} =& \langle O_\Delta(\vec{y_1})O_\Delta(\vec{y_2})O_\Delta(\vec{y_3})\rangle_{\text{CFT,con}} \nonumber\\
    =&\varepsilon^{-\frac{3\Pi(\kappa)}{2\nu}}\times-\lambda c_\Delta^3D_{\Delta-\frac{\Pi(\kappa)}{2\nu}\Delta-\frac{\Pi(\kappa)}{2\nu}\Delta-\frac{\Pi(\kappa)}{2\nu}}(\vec{y_1},\vec{y_2},\vec{y_3})\nonumber\\
    &-\lambda^3c^3_\Delta C_{GGG}(\kappa)D_{\Delta\Delta\Delta}(\vec{y_1},\vec{y_2},\vec{y_3})+\mathcal{O}(\lambda^4)
\end{align}\

With the correlators written in this form, it is direct to see what are the effects of the quantum corrections coming from the off-shell part of the AdS path integral to the 3-point functions found previously under the classical approximation of the AdS/CFT correspondence. Indeed, up to order $\lambda^4$ in the coupling constant, they contribute with an overall rescaling to the correlator along with a shift in its scaling dimension, just like for the 2-point functions, with the difference that in the current case the effective self-interacting coupling constant between the bulk fields also receives a correction coming from the "O" diagram. What is remarkable however about eq. (\ref{regiruv3pf}) is that the resulting rescaling and anomalous dimension of the 3-point functions are exactly the same as those dictated by the 2-point function! This fact implies that the very same redefinitions of the bulk's parameters done for the 2-point function not only have the effect of renormalizing the divergences present there, but also for the divergences present in the 3-point function, where now a redefinition of the coupling constant $\lambda$ is also needed. To see this, consider a redefinition of the bulk's self-interacting coupling constant $\lambda$ in the AdS bulk action of the form:

\begin{equation}
    \lambda\rightarrow\lambda+\delta\lambda
\end{equation}\

where the counterterm is expected to be of order $\delta\lambda=\mathcal{O}(\lambda^3)$. This redefinition of $\lambda$ adds, up to order $\lambda^3$, a new counterterm interaction to the holographic 3-point function eq. (\ref{regiruv3pf}) of the form:

\begin{align}
    \langle O_\Delta(\vec{y_1})O_\Delta(\vec{y_2})O_\Delta(\vec{y_3})\rangle_{\text{CFT}} =& \langle O_\Delta(\vec{y_1})O_\Delta(\vec{y_2})O_\Delta(\vec{y_3})\rangle_{\text{CFT,con}} \nonumber\\
    =&\varepsilon^{-\frac{3\Pi(\kappa)}{2\nu}}\times-\lambda c_\Delta^3D_{\Delta-\frac{\Pi(\kappa)}{2\nu}\Delta-\frac{\Pi(\kappa)}{2\nu}\Delta-\frac{\Pi(\kappa)}{2\nu}}(\vec{y_1},\vec{y_2},\vec{y_3})\nonumber\\
    &-\bigl[\delta\lambda+\lambda^3C_{GGG}(\kappa)\bigl]c_\Delta^3D_{\Delta\Delta\Delta}(\vec{y_1},\vec{y_2},\vec{y_3})+\mathcal{O}(\lambda^4)
\end{align}\

Therefore, denoting the constant $C_{GGG}(\kappa)$ as $C_{GGG}(\kappa) = C^\infty_{GGG}(\kappa) + C^0_{GGG}(\kappa)$, where all its UV-divergent terms are contained in $C^\infty_{GGG}(\kappa)$, the infinities present in the correlators coming from the ultraviolet divergence of this quantity can be renormalized away through the convenient choice of the counterterm $\delta\lambda$ as:

\begin{equation}
    \delta\lambda = -\lambda^3C^\infty_{GGG}(\kappa)
\end{equation}\

resulting in the partially renormalized holographic 3-point functions:

\begin{align}\label{regiruv3pf2}
    \langle O_\Delta(\vec{y_1})O_\Delta(\vec{y_2})O_\Delta(\vec{y_3})\rangle_{\text{CFT}} =& \langle O_\Delta(\vec{y_1})O_\Delta(\vec{y_2})O_\Delta(\vec{y_3})\rangle_{\text{CFT,con}} \nonumber\\
    =&\varepsilon^{-\frac{3\Pi(\kappa)}{2\nu}}\times-\lambda c_\Delta^3D_{\Delta-\frac{\Pi(\kappa)}{2\nu}\Delta-\frac{\Pi(\kappa)}{2\nu}\Delta-\frac{\Pi(\kappa)}{2\nu}}(\vec{y_1},\vec{y_2},\vec{y_3})\nonumber\\
    &-\lambda^3c^3_\Delta C^0_{GGG}(\kappa)D_{\Delta\Delta\Delta}(\vec{y_1},\vec{y_2},\vec{y_3})+\mathcal{O}(\lambda^4)
\end{align}\

We are not done with the renormalization process as we still have to deal with the other divergences of the correlators. However, as we already anticipated, the very same redefinitions for the bulk's parameters introduced in the study of the 2-point function exactly renormalize the divergences present in the current case for the 3-point function. Take for example the divergent anomalous dimension. The redefinition of the bulk's mass parameter $m^2$ as $m^2+\delta m^2$ in the AdS bulk action (where $\delta m^2=\mathcal{O}(\lambda^2)$) adds, up to order $\lambda^3$, new counterterms interactions to the holographic 3-point function of the form:

\begin{align}
    &\langle O_\Delta(\vec{y_1})O_\Delta(\vec{y_2})O_\Delta(\vec{y_3})\rangle_{\text{CFT}},\ \langle O_\Delta(\vec{y_1})O_\Delta(\vec{y_2})O_\Delta(\vec{y_3})\rangle_{\text{CFT,con}}\nonumber\\
    &\hspace{3cm}\rightarrow\langle O_\Delta(\vec{y_1})O_\Delta(\vec{y_2})O_\Delta(\vec{y_3})\rangle_{\text{CFT}},\ \langle O_\Delta(\vec{y_1})O_\Delta(\vec{y_2})O_\Delta(\vec{y_3})\rangle_{\text{CFT,con}}\nonumber\\
    &\hspace{3.6cm}+\lambda\delta m^2\int_\varepsilon\int_\varepsilon K(x_1,\vec{y_1})G(x_1,x_2)K(x_2,\vec{y_2})K(x_2,\vec{y_3})\nonumber\\
    &\hspace{3.6cm}+(\vec{y_1}\leftrightarrow\vec{y_2})+(\vec{y_1}\leftrightarrow\vec{y_3})+\mathcal{O}(\lambda^4)
\end{align}\

But notice that these new contributions to the correlators have exactly the same form as the reducible "eye" diagrams, where the coefficient $\Pi(\kappa)$ has been replaced by $-\delta m^2$. This implies that considering such counterterm interactions coming from the redefinition of $m^2$ in a earlier step in the computation of the 3-point function, both contributions can be exactly factorized resulting for the current expressions in the replacement of $\Pi(\kappa)\rightarrow\Pi(\kappa)-\delta m^2$. Therefore, the exact same choice for the counterterm $\delta m^2$ as $\delta m^2=\Pi_\infty(\kappa)$ made in the renormalization of the 2-point function also renormalizes the UV-divergences of the anomalous dimension present in the 3-point function, resulting in the UV-renormalized correlators:

\begin{align}
    \langle O_\Delta(\vec{y_1})O_\Delta(\vec{y_2})O_\Delta(\vec{y_3})\rangle_{\text{CFT}} =& \langle O_\Delta(\vec{y_1})O_\Delta(\vec{y_2})O_\Delta(\vec{y_3})\rangle_{\text{CFT,con}} \nonumber\\
    =&\varepsilon^{-\frac{3\Pi_0(0)}{2\nu}}\times-\lambda c_\Delta^3D_{\Delta-\frac{\Pi_0(0)}{2\nu}\Delta-\frac{\Pi_0(0)}{2\nu}\Delta-\frac{\Pi_0(0)}{2\nu}}(\vec{y_1},\vec{y_2},\vec{y_3})\nonumber\\
    &-\lambda^3c^3_\Delta C^0_{GGG}(0)D_{\Delta\Delta\Delta}(\vec{y_1},\vec{y_2},\vec{y_3})+\mathcal{O}(\lambda^4)
\end{align}\

where we safely took $\kappa=0$. Finally, the redefinition of the bulk field $\Phi(x)$ as $\Phi(x)\rightarrow\sqrt{Z(\lambda)}\Phi(x)$ in the AdS bulk action adds a new factor to the holographic 3-point function of the form:

\begin{align}
    \langle O_\Delta(\vec{y_1})O_\Delta(\vec{y_2})O_\Delta(\vec{y_3})\rangle_{\text{CFT}} =& \langle O_\Delta(\vec{y_1})O_\Delta(\vec{y_2})O_\Delta(\vec{y_3})\rangle_{\text{CFT,con}} \nonumber\\
    =&\frac{\varepsilon^{-\frac{3\Pi_0(0)}{2\nu}}}{Z(\lambda)^{\frac{3}{2}}}\times-\lambda c_\Delta^3D_{\Delta-\frac{\Pi_0(0)}{2\nu}\Delta-\frac{\Pi_0(0)}{2\nu}\Delta-\frac{\Pi_0(0)}{2\nu}}(\vec{y_1},\vec{y_2},\vec{y_3})\nonumber\\
    &-\lambda^3c^3_\Delta C^0_{GGG}(0)D_{\Delta\Delta\Delta}(\vec{y_1},\vec{y_2},\vec{y_3})+\mathcal{O}(\lambda^4)
\end{align}\

where in the last term we used that $Z(\lambda)=1+\mathcal{O}(\lambda^2)$, ignoring contributions of the order $\lambda^4$. With the correlator written in this form, it is direct to see that the exact same choice for the counterterm $Z(\lambda)$ as $Z(\lambda)=\varepsilon^{-\frac{\Pi_0(0)}{\nu}}$ made in the renormalization of the 2-point functions also renormalizes the IR-divergence of the overall rescaling of the 3-point function, resulting in both IR and UV renormalized holographic 3-point functions:

\begin{align}
    \langle O_\Delta(\vec{y_1})O_\Delta(\vec{y_2})O_\Delta(\vec{y_3})\rangle_{\text{CFT}} =& \langle O_\Delta(\vec{y_1})O_\Delta(\vec{y_2})O_\Delta(\vec{y_3})\rangle_{\text{CFT,con}} \nonumber\\
    =&-\lambda c_\Delta^3D_{\Delta-\frac{\Pi_0(0)}{2\nu}\Delta-\frac{\Pi_0(0)}{2\nu}\Delta-\frac{\Pi_0(0)}{2\nu}}(\vec{y_1},\vec{y_2},\vec{y_3})\nonumber\\
    &-\lambda^3c^3_\Delta C^0_{GGG}(0)D_{\Delta\Delta\Delta}(\vec{y_1},\vec{y_2},\vec{y_3})+\mathcal{O}(\lambda^4)
\end{align}\

where the limits $\varepsilon=\kappa=0$ have been taken. The complete study of these type of D-functions can be found in section A.2 of Appendix A, concluding in its value in eq. (\ref{3dfunc}). Using this formula then in our present case we find that the explicit form of the renormalized 3-point functions can be expressed as:

\begin{align}
    \langle O_\Delta(\vec{y_1})O_\Delta(\vec{y_2})O_\Delta(\vec{y_3})\rangle_{\text{CFT}} =& \langle O_\Delta(\vec{y_1})O_\Delta(\vec{y_2})O_\Delta(\vec{y_3})\rangle_{\text{CFT,con}} \nonumber\\
    &\hspace{-2.5cm}=-\lambda c_\Delta^3\frac{\pi^\frac{d}{2}}{2}\frac{\Gamma(\frac{3\Delta-d}{2}-\frac{3\Pi_0(0)}{4\nu})}{\Gamma(\Delta-\frac{\Pi_0(0)}{2\nu})^3}\frac{\Gamma(\frac{\Delta}{2}-\frac{\Pi_0(0)}{4\nu})^3}{\lvert\vec{y_1}-\vec{y_2}\rvert^{\Delta-\frac{\Pi_0(0)}{2\nu}} \lvert\vec{y_2}-\vec{y_3}\rvert^{\Delta-\frac{\Pi_0(0)}{2\nu}} \lvert\vec{y_3}-\vec{y_1}\rvert^{\Delta-\frac{\Pi_0(0)}{2\nu}}}\nonumber\\
    &\hspace{-2cm}-\lambda^3c^3_\Delta C^0_{GGG}(0)\frac{\pi^\frac{d}{2}}{2}\frac{\Gamma(\frac{3\Delta-d}{2})}{\Gamma(\Delta)^3}\frac{\Gamma(\frac{\Delta}{2})^3}{\lvert\vec{y_1}-\vec{y_2}\rvert^\Delta \lvert\vec{y_2}-\vec{y_3}\rvert^\Delta \lvert\vec{y_3}-\vec{y_1}\rvert^\Delta}+\mathcal{O}(\lambda^4)
\end{align}\

where the coefficients $\Pi_0(0)$ and $C^0_{GGG}(0)$ denote the UV-finite parts of the 1PI contributions $\Pi(\kappa)$ and the "O" diagram, respectively.

\subsection{4-Point Function}

The regularized holographic 4-point functions dual to a $\Phi^3$ self-interacting theory on AdS are given by:

\begin{align}
    \langle O_\Delta(\vec{y_1})O_\Delta(\vec{y_2})O_\Delta(\vec{y_3})O_\Delta(\vec{y_4})\rangle_{\text{CFT}} =&\frac{2\nu c_\Delta}{\lvert\vec{y_1}-\vec{y_2}\rvert^{2\Delta}}\frac{2\nu c_\Delta}{\lvert\vec{y_3}-\vec{y_4}\rvert^{2\Delta}}+(\vec{y_2}\leftrightarrow\vec{y_3})+(\vec{y_2}\leftrightarrow\vec{y_4})\nonumber\\
    &\hspace{-2.5cm}+\lambda^2\int\int K(x_1,\vec{y_1})K(x_1,\vec{y_2})G(x_1,x_2)K(x_2,\vec{y_3})K(x_2,\vec{y_4})\times3\nonumber\\
    &\hspace{-2.5cm}+\frac{2\nu c_\Delta}{\lvert\vec{y_1}-\vec{y_2}\rvert^{2\Delta}}\frac{\lambda^2}{2}\int_\varepsilon\int_\varepsilon K(x_1,\vec{y_3})K(x_2,\vec{y_4})G^2_\kappa(x_1,x_2)\times6+\mathcal{O}(\lambda^4)\nonumber\\
    \langle O_\Delta(\vec{y_1})O_\Delta(\vec{y_2})O_\Delta(\vec{y_3})O_\Delta(\vec{y_4})\rangle_{\text{CFT,con}} =&\nonumber\\
    &\hspace{-2.5cm}\lambda^2\int\int K(x_1,\vec{y_1})K(x_1,\vec{y_2})G(x_1,x_2)K(x_2,\vec{y_3})K(x_2,\vec{y_4})\times3+\mathcal{O}(\lambda^4)
\end{align}\

from where we see the quantum corrections they receive coming from the loop integral:

\begin{align}\label{4pfphi3regquantum0}
    I(\vec{y_1},\vec{y_2},\vec{y_3},\vec{y_4}) = &\frac{2\nu c_\Delta}{\lvert\vec{y_1}-\vec{y_2}\rvert^{2\Delta}}\frac{\lambda^2}{2}\int_{x_{1,0}=\varepsilon}d^{d+1}x_1\sqrt{g}\int_{x_{2,0}=\varepsilon}d^{d+1}x_2\sqrt{g}\ K(x_1,\vec{y_3})K(x_2,\vec{y_4})G^2_\kappa(x_1,x_2)\nonumber\\
    &+(\vec{y_1}\leftrightarrow\vec{y_3})+(\vec{y_1}\leftrightarrow\vec{y_4})+(\vec{y_2}\leftrightarrow\vec{y_3})+(\vec{y_2}\leftrightarrow\vec{y_4})+(\vec{y_1}\leftrightarrow\vec{y_3},\vec{y_2}\leftrightarrow\vec{y_4})
\end{align}\

In order to compute the complete 4-point function up to this order in the expansion of $\lambda$, we will proceed then to compute this quantity.

\subsubsection{The Disconnected "Eye" Diagram}

\begin{figure}[h]
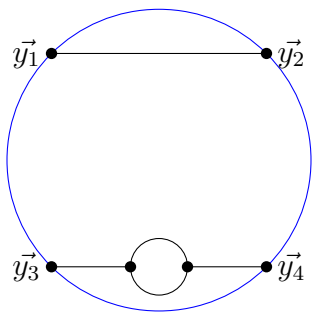

    \[\begin{wittendiagram}
    \draw (-1.4142,1.4142) node[vertex] -- (1.4142,1.4142)
    (-1.4142,-1.4142) node[vertex] -- (-0.3535,-1.4142)
    (1.4142,1.4142) node[vertex]
    (1.4142,-1.4142) node[vertex]
    (-0.375,-1.4142) node[vertex]
    (0.375,-1.4142) node[vertex] -- (1.4142,-1.4142)
    
    (-1.4142,1.4142) node[left]{$\vec{y_1}$}
    (-1.4142,-1.4142) node[left]{$\vec{y_3}$}
    (1.4142,1.4142) node[right]{$\vec{y_2}$}
    (1.4142,-1.4142) node[right]{$\vec{y_4}$}
    
    (0,-1.4142) circle (0.375);
  \end{wittendiagram}\]
  \caption{Pictorial representation of the disconnected "eye" diagram.}
\end{figure}

The integrals present in this contribution (which we will refer to it as the disconnected "eye" diagram) are nothing but the "eye" diagram that we just faced for the 2-point function, eq. (\ref{2pfphi3regquantum5}). Therefore, replacing its known result we find that the disconnected "eye" diagrams in the 4-point function can be written as:

\begin{align}
    I(\vec{y_1},\vec{y_2},\vec{y_3},\vec{y_4}) = &-\frac{2\nu c_\Delta}{\lvert\vec{y_1}-\vec{y_2}\rvert^{2\Delta}}\frac{2\nu c_\Delta}{\lvert\vec{y_3}-\vec{y_4}\rvert^{2\Delta}}\frac{\Pi(\kappa)}{\nu}\ln{\Bigl(\frac{\varepsilon}{\lvert\vec{y_3}-\vec{y_4}\rvert}\Bigl)}+(\vec{y_1}\leftrightarrow\vec{y_3})\nonumber\\
    &+(\vec{y_1}\leftrightarrow\vec{y_4})+(\vec{y_2}\leftrightarrow\vec{y_3})+(\vec{y_2}\leftrightarrow\vec{y_4})+(\vec{y_1}\leftrightarrow\vec{y_3},\vec{y_2}\leftrightarrow\vec{y_4})
\end{align}\

As in the previous cases, the presence of the logarithm in this result seems to break the conformal structure expected for contributions to the 4-point function of a CFT, however as we will see next when we consider the complete correlator, we will realize that the result just found corresponds to the expansion of not any conformal anomaly, but to the exact same anomaly dictated by the 2-point function, realization which will result in a natural and consistent renormalization scheme of both IR and UV divergences for the holographic 2-, 3- and 4-point functions (and thus expected to hold for higher point functions) equivalent to those schemes used in ordinary QFTs.

\subsubsection{Correlator Renormalization}

Replacing the result just found then for the disconnected "eye" diagram back into the holographic 4-point functions, we find that they can be written as:

\begin{align}
    \langle O_\Delta(\vec{y_1})O_\Delta(\vec{y_2})O_\Delta(\vec{y_3})O_\Delta(\vec{y_4})\rangle_{\text{CFT}} =&\frac{2\nu c_\Delta}{\lvert\vec{y_1}-\vec{y_2}\rvert^{2\Delta}}\frac{2\nu c_\Delta}{\lvert\vec{y_3}-\vec{y_4}\rvert^{2\Delta}}+(\vec{y_2}\leftrightarrow\vec{y_3})+(\vec{y_2}\leftrightarrow\vec{y_4})\nonumber\\
    &\hspace{-2cm}+\lambda^2\int\int K(x_1,\vec{y_1})K(x_1,\vec{y_2})G(x_1,x_2)K(x_2,\vec{y_3})K(x_2,\vec{y_4})\times3\nonumber\\
    &\hspace{-2cm}-\frac{2\nu c_\Delta}{\lvert\vec{y_1}-\vec{y_2}\rvert^{2\Delta}}\frac{2\nu c_\Delta}{\lvert\vec{y_3}-\vec{y_4}\rvert^{2\Delta}}\frac{\Pi(\kappa)}{\nu}\ln{\Bigl(\frac{\varepsilon}{\lvert\vec{y_3}-\vec{y_4}\rvert}\Bigl)}+(\vec{y_1}\leftrightarrow\vec{y_3})\nonumber\\
    &\hspace{-2cm}+(\vec{y_1}\leftrightarrow\vec{y_4})+(\vec{y_2}\leftrightarrow\vec{y_3})+(\vec{y_2}\leftrightarrow\vec{y_4})+(\vec{y_1}\leftrightarrow\vec{y_3},\vec{y_2}\leftrightarrow\vec{y_4})\nonumber\\
    &\hspace{-2cm}+ \mathcal{O}(\lambda^4)\nonumber\\
    \langle O_\Delta(\vec{y_1})O_\Delta(\vec{y_2})O_\Delta(\vec{y_3})O_\Delta(\vec{y_4})\rangle_{\text{CFT,con}} =&\nonumber\\
    &\hspace{-2.5cm}\lambda^2\int\int K(x_1,\vec{y_1})K(x_1,\vec{y_2})G(x_1,x_2)K(x_2,\vec{y_3})K(x_2,\vec{y_4})\times3+\mathcal{O}(\lambda^4)
\end{align}\

Now, up to this same order in the coupling constant, it is easy to see that the tree-level terms together with the disconnected "eye" diagrams can be factorized into the form:

\begin{align}\label{phi34pftreedisceye}
    &\frac{2\nu c_\Delta}{\lvert\vec{y_1}-\vec{y_2}\rvert^{2\Delta}}\frac{2\nu c_\Delta}{\lvert\vec{y_3}-\vec{y_4}\rvert^{2\Delta}}+(\vec{y_2}\leftrightarrow\vec{y_3})+(\vec{y_2}\leftrightarrow\vec{y_4})-\frac{2\nu c_\Delta}{\lvert\vec{y_1}-\vec{y_2}\rvert^{2\Delta}}\frac{2\nu c_\Delta}{\lvert\vec{y_3}-\vec{y_4}\rvert^{2\Delta}}\frac{\Pi(\kappa)}{\nu}\ln{\Bigl(\frac{\varepsilon}{\lvert\vec{y_3}-\vec{y_4}\rvert}\Bigl)}\nonumber\\
    &+(\vec{y_1}\leftrightarrow\vec{y_3})+(\vec{y_1}\leftrightarrow\vec{y_4})+(\vec{y_2}\leftrightarrow\vec{y_3})+(\vec{y_2}\leftrightarrow\vec{y_4})+(\vec{y_1}\leftrightarrow\vec{y_3},\vec{y_2}\leftrightarrow\vec{y_4})\nonumber\\
    &\hspace{2cm}=\varepsilon^{-\frac{2\Pi(\kappa)}{\nu}}\times\frac{2\nu c_\Delta}{\lvert\vec{y_1}-\vec{y_2}\rvert^{2\Delta-\frac{\Pi(\kappa)}{\nu}}}\frac{2\nu c_\Delta}{\lvert\vec{y_3}-\vec{y_4}\rvert^{2\Delta-\frac{\Pi(\kappa)}{\nu}}}+(\vec{y_2}\leftrightarrow\vec{y_3})+(\vec{y_2}\leftrightarrow\vec{y_4})+\mathcal{O}(\lambda^4)
\end{align}\

Notice that these contributions, apart from the IR rescaling, are nothing but the derivation of eq. (\ref{conformalform4pf}) where the scaling dimension $\Delta$ has been replaced by $\Delta-\frac{\Pi(\kappa)}{2\nu}$. Therefore, following the same steps shown there, we can rewrite eq. (\ref{phi34pftreedisceye}) in the form of:

\begin{align}
    &\frac{2\nu c_\Delta}{\lvert\vec{y_1}-\vec{y_2}\rvert^{2\Delta}}\frac{2\nu c_\Delta}{\lvert\vec{y_3}-\vec{y_4}\rvert^{2\Delta}}+(\vec{y_2}\leftrightarrow\vec{y_3})+(\vec{y_2}\leftrightarrow\vec{y_4})-\frac{2\nu c_\Delta}{\lvert\vec{y_1}-\vec{y_2}\rvert^{2\Delta}}\frac{2\nu c_\Delta}{\lvert\vec{y_3}-\vec{y_4}\rvert^{2\Delta}}\frac{\Pi(\kappa)}{\nu}\ln{\Bigl(\frac{\varepsilon}{\lvert\vec{y_3}-\vec{y_4}\rvert}\Bigl)}\nonumber\\
    &+(\vec{y_1}\leftrightarrow\vec{y_3})+(\vec{y_1}\leftrightarrow\vec{y_4})+(\vec{y_2}\leftrightarrow\vec{y_3})+(\vec{y_2}\leftrightarrow\vec{y_4})+(\vec{y_1}\leftrightarrow\vec{y_3},\vec{y_2}\leftrightarrow\vec{y_4})\nonumber\\
    &\hspace{2cm}=\varepsilon^{-\frac{2\Pi(\kappa)}{\nu}}\times\frac{(2\nu c_\Delta)^2}{\prod_{i<j}\lvert\vec{y_{ij}}\rvert^{\frac{2}{3}\bigl(\Delta-\frac{\Pi(\kappa)}{2\nu}\bigl)}}u^{-\frac{2}{3}\bigl(\Delta-\frac{\Pi(\kappa)}{2\nu}\bigl)}v^{\frac{1}{3}\bigl(\Delta-\frac{\Pi(\kappa)}{2\nu}\bigl)}+\Bigl(u,v\rightarrow\frac{1}{u},\frac{v}{u}\Bigl)\nonumber\\
    &\hspace{2.5cm}+(u,v\rightarrow v,u)+\mathcal{O}(\lambda^4)
\end{align}\

With these contributions to the 4-point functions written in this form, along with the known result for the connected scalar exchange diagram eq. (\ref{i4phi3}), we can express the correlators as:

\begin{align}\label{regiruv4pf}
    &\langle O_\Delta(\vec{y_1})O_\Delta(\vec{y_2})O_\Delta(\vec{y_3})O_\Delta(\vec{y_4})\rangle_{\text{CFT}}\nonumber\\
    &\hspace{1cm}=\varepsilon^{-\frac{2\Pi(\kappa)}{\nu}}\times\frac{(2\nu c_\Delta)^2}{\prod_{i<j}\lvert\vec{y_{ij}}\rvert^{\frac{2}{3}\bigl(\Delta-\frac{\Pi(\kappa)}{2\nu}\bigl)}}u^{-\frac{2}{3}\bigl(\Delta-\frac{\Pi(\kappa)}{2\nu}\bigl)}v^{\frac{1}{3}\bigl(\Delta-\frac{\Pi(\kappa)}{2\nu}\bigl)}+\Bigl(u,v\rightarrow\frac{1}{u},\frac{v}{u}\Bigl)+(u,v\rightarrow v,u)\nonumber\\
    &\hspace{1.5cm}+\lambda^2c^4_\Delta\frac{\pi^\frac{d}{2}}{8}\frac{\Gamma(\frac{3\Delta-d}{2})\Gamma(\frac{\Delta}{2})}{\Gamma(\Delta)^4}\frac{u^\frac{\Delta}{3}v^\frac{\Delta}{3}}{\prod_{i<j}\lvert\vec{y_{ij}}\rvert^{\frac{2\Delta}{3}}}\Bigl[\sum_{k=0}^\infty\frac{\Gamma(\frac{3\Delta-d}{2}+k)}{\Gamma(\nu+1+k)\ k!}H\Bigl(\Delta,\Delta,\frac{\Delta}{2}+1-k,2\Delta;u,v\Bigl)\nonumber\\
   &\hspace{3.5cm}-\sum_{k=0}^\infty\frac{\Gamma(2\Delta-\frac{d}{2}+k)}{\Gamma(\frac{\Delta}{2}+1+k)\Gamma(\frac{3\Delta-d}{2}+1+k)}H(\Delta,\Delta,1-k,2\Delta;u,v)\Bigl]\nonumber\\
   &\hspace{1.5cm}+\Bigl(u,v\rightarrow\frac{1}{u},\frac{v}{u}\Bigl) + (u,v\rightarrow v,u)+\mathcal{O}(\lambda^4)\nonumber\\
    &\langle O_\Delta(\vec{y_1})O_\Delta(\vec{y_2})O_\Delta(\vec{y_3})O_\Delta(\vec{y_4})\rangle_{\text{CFT,con}}\nonumber\\ &\hspace{1cm}=\lambda^2c^4_\Delta\frac{\pi^\frac{d}{2}}{8}\frac{\Gamma(\frac{3\Delta-d}{2})\Gamma(\frac{\Delta}{2})}{\Gamma(\Delta)^4}\frac{u^\frac{\Delta}{3}v^\frac{\Delta}{3}}{\prod_{i<j}\lvert\vec{y_{ij}}\rvert^{\frac{2\Delta}{3}}}\Bigl[\sum_{k=0}^\infty\frac{\Gamma(\frac{3\Delta-d}{2}+k)}{\Gamma(\nu+1+k)\ k!}H\Bigl(\Delta,\Delta,\frac{\Delta}{2}+1-k,2\Delta;u,v\Bigl)\nonumber\\
   &\hspace{3cm}-\sum_{k=0}^\infty\frac{\Gamma(2\Delta-\frac{d}{2}+k)}{\Gamma(\frac{\Delta}{2}+1+k)\Gamma(\frac{3\Delta-d}{2}+1+k)}H(\Delta,\Delta,1-k,2\Delta;u,v)\Bigl]\nonumber\\
   &\hspace{1.5cm}+\Bigl(u,v\rightarrow\frac{1}{u},\frac{v}{u}\Bigl) + (u,v\rightarrow v,u)+\mathcal{O}(\lambda^4)
\end{align}\

With the correlators written in this form, it is direct to see what are the effects of the quantum corrections coming from the off-shell part of the AdS path integral to the 4-point function found previously under the classical approximation of the AdS/CFT correspondence. Indeed, up to order $\lambda^4$ in the coupling constant, they contribute with an overall rescaling to the correlator along with a shift in its scaling dimension, just like for the 2-point functions. What is remarkable however about eq. (\ref{regiruv4pf}) is that the resulting rescaling and anomalous dimension of the 4-point functions are exactly the same as those dictated by the 2-point function! This fact implies that the very same redefinition of the bulk's parameters done for the 2-point function not only have the effect of renormalizing the divergences present there, but also for the divergences present in the 4-point function. Take for example the divergent anomalous dimension. The redefinition of the bulk's mass parameter $m^2$ as $m^2+\delta m^2$ in the AdS bulk action (where $\delta m^2=\mathcal{O}(\lambda^2)$) adds, up to order $\lambda^3$, new counterterms interactions to the holographic 4-point functions of the form:

\begin{align}
    &\langle O_\Delta(\vec{y_1})O_\Delta(\vec{y_2})O_\Delta(\vec{y_3})O_\Delta(\vec{y_4})\rangle_{\text{CFT}}\rightarrow\langle O_\Delta(\vec{y_1})O_\Delta(\vec{y_2})O_\Delta(\vec{y_3})O_\Delta(\vec{y_4})\rangle_{\text{CFT}}\nonumber\\
    &\hspace{6.8cm}-\frac{2\nu c_\Delta}{\lvert\vec{y_1}-\vec{y_2}\rvert^{2\Delta}}\delta m^2\int_{x_0=\varepsilon} d^{d+1}x\sqrt{g}\ K(x,\vec{y_3})K(x,\vec{y_4})\nonumber\\
    &\hspace{6.8cm}+(\vec{y_1}\leftrightarrow\vec{y_3})+(\vec{y_1}\leftrightarrow\vec{y_4})+(\vec{y_2}\leftrightarrow\vec{y_3})+(\vec{y_2}\leftrightarrow\vec{y_4})\nonumber\\
    &\hspace{6.8cm}+(\vec{y_1}\leftrightarrow\vec{y_3},\vec{y_2}\leftrightarrow\vec{y_4})+\mathcal{O}(\lambda^4)\nonumber\\
    &\langle O_\Delta(\vec{y_1})O_\Delta(\vec{y_2})O_\Delta(\vec{y_3})O_\Delta(\vec{y_4})\rangle_{\text{CFT,con}}\rightarrow\langle O_\Delta(\vec{y_1})O_\Delta(\vec{y_2})O_\Delta(\vec{y_3})O_\Delta(\vec{y_4})\rangle_{\text{CFT,con}}+\mathcal{O}(\lambda^4)
\end{align}\

But notice that these new contributions to the correlators have exactly the same form as the disconnected "eye" diagrams, where the coefficient $\Pi(\kappa)$ has been replaced by $-\delta m^2$. This implies that considering such counterterm interactions coming from the redefinition of $m^2$ in a earlier step in the computation of the 4-point function, both contributions can be exactly factorized resulting for the current expressions in the replacement of $\Pi(\kappa)\rightarrow\Pi(\kappa)-\delta m^2$. Therefore, the exact same choice for the counterterm $\delta m^2$ as $\delta m^2=\Pi_\infty(\kappa)$ made in the renormalization of the 2-point function also renormalizes the UV-divergences of the anomalous dimension present in the 4-point function, resulting in the UV-renormalized correlators:

\begin{align}
    &\langle O_\Delta(\vec{y_1})O_\Delta(\vec{y_2})O_\Delta(\vec{y_3})O_\Delta(\vec{y_4})\rangle_{\text{CFT}}\nonumber\\
    &\hspace{1cm}=\varepsilon^{-\frac{2\Pi_0(0)}{\nu}}\times\frac{(2\nu c_\Delta)^2}{\prod_{i<j}\lvert\vec{y_{ij}}\rvert^{\frac{2}{3}\bigl(\Delta-\frac{\Pi_0(0)}{2\nu}\bigl)}}u^{-\frac{2}{3}\bigl(\Delta-\frac{\Pi_0(0)}{2\nu}\bigl)}v^{\frac{1}{3}\bigl(\Delta-\frac{\Pi_0(0)}{2\nu}\bigl)}+\Bigl(u,v\rightarrow\frac{1}{u},\frac{v}{u}\Bigl)+(u,v\rightarrow v,u)\nonumber\\
    &\hspace{1.5cm}+\lambda^2c^4_\Delta\frac{\pi^\frac{d}{2}}{8}\frac{\Gamma(\frac{3\Delta-d}{2})\Gamma(\frac{\Delta}{2})}{\Gamma(\Delta)^4}\frac{u^\frac{\Delta}{3}v^\frac{\Delta}{3}}{\prod_{i<j}\lvert\vec{y_{ij}}\rvert^{\frac{2\Delta}{3}}}\Bigl[\sum_{k=0}^\infty\frac{\Gamma(\frac{3\Delta-d}{2}+k)}{\Gamma(\nu+1+k)\ k!}H\Bigl(\Delta,\Delta,\frac{\Delta}{2}+1-k,2\Delta;u,v\Bigl)\nonumber\\
   &\hspace{3.5cm}-\sum_{k=0}^\infty\frac{\Gamma(2\Delta-\frac{d}{2}+k)}{\Gamma(\frac{\Delta}{2}+1+k)\Gamma(\frac{3\Delta-d}{2}+1+k)}H(\Delta,\Delta,1-k,2\Delta;u,v)\Bigl]\nonumber\\
   &\hspace{1.5cm}+\Bigl(u,v\rightarrow\frac{1}{u},\frac{v}{u}\Bigl) + (u,v\rightarrow v,u)+\mathcal{O}(\lambda^4)\nonumber\\
    &\langle O_\Delta(\vec{y_1})O_\Delta(\vec{y_2})O_\Delta(\vec{y_3})O_\Delta(\vec{y_4})\rangle_{\text{CFT,con}}\nonumber\\ &\hspace{1cm}=\lambda^2c^4_\Delta\frac{\pi^\frac{d}{2}}{8}\frac{\Gamma(\frac{3\Delta-d}{2})\Gamma(\frac{\Delta}{2})}{\Gamma(\Delta)^4}\frac{u^\frac{\Delta}{3}v^\frac{\Delta}{3}}{\prod_{i<j}\lvert\vec{y_{ij}}\rvert^{\frac{2\Delta}{3}}}\Bigl[\sum_{k=0}^\infty\frac{\Gamma(\frac{3\Delta-d}{2}+k)}{\Gamma(\nu+1+k)\ k!}H\Bigl(\Delta,\Delta,\frac{\Delta}{2}+1-k,2\Delta;u,v\Bigl)\nonumber\\
   &\hspace{3cm}-\sum_{k=0}^\infty\frac{\Gamma(2\Delta-\frac{d}{2}+k)}{\Gamma(\frac{\Delta}{2}+1+k)\Gamma(\frac{3\Delta-d}{2}+1+k)}H(\Delta,\Delta,1-k,2\Delta;u,v)\Bigl]\nonumber\\
   &\hspace{1.5cm}+\Bigl(u,v\rightarrow\frac{1}{u},\frac{v}{u}\Bigl) + (u,v\rightarrow v,u)+\mathcal{O}(\lambda^4)
\end{align}\

where we safely took $\kappa=0$. Finally, the redefinition of the bulk field $\Phi(x)$ as $\sqrt{Z(\lambda)}\Phi(x)$ in the AdS bulk action adds a new factor to the holographic 4-point functions of the form:

\begin{align}
    &\langle O_\Delta(\vec{y_1})O_\Delta(\vec{y_2})O_\Delta(\vec{y_3})O_\Delta(\vec{y_4})\rangle_{\text{CFT}}\nonumber\\
    &\hspace{1cm}=\frac{\varepsilon^{-\frac{2\Pi_0(0)}{\nu}}}{Z(\lambda)^2}\times\frac{(2\nu c_\Delta)^2}{\prod_{i<j}\lvert\vec{y_{ij}}\rvert^{\frac{2}{3}\bigl(\Delta-\frac{\Pi_0(0)}{2\nu}\bigl)}}u^{-\frac{2}{3}\bigl(\Delta-\frac{\Pi_0(0)}{2\nu}\bigl)}v^{\frac{1}{3}\bigl(\Delta-\frac{\Pi_0(0)}{2\nu}\bigl)}+\Bigl(u,v\rightarrow\frac{1}{u},\frac{v}{u}\Bigl)+(u,v\rightarrow v,u)\nonumber\\
    &\hspace{1.5cm}+\lambda^2c^4_\Delta\frac{\pi^\frac{d}{2}}{8}\frac{\Gamma(\frac{3\Delta-d}{2})\Gamma(\frac{\Delta}{2})}{\Gamma(\Delta)^4}\frac{u^\frac{\Delta}{3}v^\frac{\Delta}{3}}{\prod_{i<j}\lvert\vec{y_{ij}}\rvert^{\frac{2\Delta}{3}}}\Bigl[\sum_{k=0}^\infty\frac{\Gamma(\frac{3\Delta-d}{2}+k)}{\Gamma(\nu+1+k)\ k!}H\Bigl(\Delta,\Delta,\frac{\Delta}{2}+1-k,2\Delta;u,v\Bigl)\nonumber\\
   &\hspace{3.5cm}-\sum_{k=0}^\infty\frac{\Gamma(2\Delta-\frac{d}{2}+k)}{\Gamma(\frac{\Delta}{2}+1+k)\Gamma(\frac{3\Delta-d}{2}+1+k)}H(\Delta,\Delta,1-k,2\Delta;u,v)\Bigl]\nonumber\\
   &\hspace{1.5cm}+\Bigl(u,v\rightarrow\frac{1}{u},\frac{v}{u}\Bigl) + (u,v\rightarrow v,u)+\mathcal{O}(\lambda^4)\nonumber\\
    &\langle O_\Delta(\vec{y_1})O_\Delta(\vec{y_2})O_\Delta(\vec{y_3})O_\Delta(\vec{y_4})\rangle_{\text{CFT,con}}\nonumber\\ &\hspace{1cm}=\lambda^2c^4_\Delta\frac{\pi^\frac{d}{2}}{8}\frac{\Gamma(\frac{3\Delta-d}{2})\Gamma(\frac{\Delta}{2})}{\Gamma(\Delta)^4}\frac{u^\frac{\Delta}{3}v^\frac{\Delta}{3}}{\prod_{i<j}\lvert\vec{y_{ij}}\rvert^{\frac{2\Delta}{3}}}\Bigl[\sum_{k=0}^\infty\frac{\Gamma(\frac{3\Delta-d}{2}+k)}{\Gamma(\nu+1+k)\ k!}H\Bigl(\Delta,\Delta,\frac{\Delta}{2}+1-k,2\Delta;u,v\Bigl)\nonumber\\
   &\hspace{3cm}-\sum_{k=0}^\infty\frac{\Gamma(2\Delta-\frac{d}{2}+k)}{\Gamma(\frac{\Delta}{2}+1+k)\Gamma(\frac{3\Delta-d}{2}+1+k)}H(\Delta,\Delta,1-k,2\Delta;u,v)\Bigl]\nonumber\\
   &\hspace{1.5cm}+\Bigl(u,v\rightarrow\frac{1}{u},\frac{v}{u}\Bigl) + (u,v\rightarrow v,u)+\mathcal{O}(\lambda^4)
\end{align}\

where in the last terms of both correlators we used that $Z(\lambda)=1+\mathcal{O}(\lambda^2)$, ignoring contributions of the order $\lambda^4$. With the correlators written in this form, it is direct to see that the exact same choice for the counterterm $Z(\lambda)$ as $Z(\lambda)=\varepsilon^{-\frac{\Pi_0(0)}{\nu}}$ made in the renormalization of the 2-point function also renormalizes the IR-divergence of the overall rescaling of the 4-point function, resulting in both IR and UV renormalized holographic 4-point functions:

\begin{align}
    &\langle O_\Delta(\vec{y_1})O_\Delta(\vec{y_2})O_\Delta(\vec{y_3})O_\Delta(\vec{y_4})\rangle_{\text{CFT}}\nonumber\\
    &\hspace{1cm}=\frac{(2\nu c_\Delta)^2}{\prod_{i<j}\lvert\vec{y_{ij}}\rvert^{\frac{2}{3}\bigl(\Delta-\frac{\Pi_0(0)}{2\nu}\bigl)}}u^{-\frac{2}{3}\bigl(\Delta-\frac{\Pi_0(0)}{2\nu}\bigl)}v^{\frac{1}{3}\bigl(\Delta-\frac{\Pi_0(0)}{2\nu}\bigl)}+\Bigl(u,v\rightarrow\frac{1}{u},\frac{v}{u}\Bigl)+(u,v\rightarrow v,u)\nonumber\\
    &\hspace{1.5cm}+\lambda^2c^4_\Delta\frac{\pi^\frac{d}{2}}{8}\frac{\Gamma(\frac{3\Delta-d}{2})\Gamma(\frac{\Delta}{2})}{\Gamma(\Delta)^4}\frac{u^\frac{\Delta}{3}v^\frac{\Delta}{3}}{\prod_{i<j}\lvert\vec{y_{ij}}\rvert^{\frac{2\Delta}{3}}}\Bigl[\sum_{k=0}^\infty\frac{\Gamma(\frac{3\Delta-d}{2}+k)}{\Gamma(\nu+1+k)\ k!}H\Bigl(\Delta,\Delta,\frac{\Delta}{2}+1-k,2\Delta;u,v\Bigl)\nonumber\\
   &\hspace{3.5cm}-\sum_{k=0}^\infty\frac{\Gamma(2\Delta-\frac{d}{2}+k)}{\Gamma(\frac{\Delta}{2}+1+k)\Gamma(\frac{3\Delta-d}{2}+1+k)}H(\Delta,\Delta,1-k,2\Delta;u,v)\Bigl]\nonumber\\
   &\hspace{1.5cm}+\Bigl(u,v\rightarrow\frac{1}{u},\frac{v}{u}\Bigl) + (u,v\rightarrow v,u)+\mathcal{O}(\lambda^4)\nonumber\\
    &\langle O_\Delta(\vec{y_1})O_\Delta(\vec{y_2})O_\Delta(\vec{y_3})O_\Delta(\vec{y_4})\rangle_{\text{CFT,con}}\nonumber\\ &\hspace{1cm}=\lambda^2c^4_\Delta\frac{\pi^\frac{d}{2}}{8}\frac{\Gamma(\frac{3\Delta-d}{2})\Gamma(\frac{\Delta}{2})}{\Gamma(\Delta)^4}\frac{u^\frac{\Delta}{3}v^\frac{\Delta}{3}}{\prod_{i<j}\lvert\vec{y_{ij}}\rvert^{\frac{2\Delta}{3}}}\Bigl[\sum_{k=0}^\infty\frac{\Gamma(\frac{3\Delta-d}{2}+k)}{\Gamma(\nu+1+k)\ k!}H\Bigl(\Delta,\Delta,\frac{\Delta}{2}+1-k,2\Delta;u,v\Bigl)\nonumber\\
   &\hspace{3cm}-\sum_{k=0}^\infty\frac{\Gamma(2\Delta-\frac{d}{2}+k)}{\Gamma(\frac{\Delta}{2}+1+k)\Gamma(\frac{3\Delta-d}{2}+1+k)}H(\Delta,\Delta,1-k,2\Delta;u,v)\Bigl]\nonumber\\
   &\hspace{1.5cm}+\Bigl(u,v\rightarrow\frac{1}{u},\frac{v}{u}\Bigl) + (u,v\rightarrow v,u)+\mathcal{O}(\lambda^4)
\end{align}\

where we safely took $\varepsilon=\kappa=0$. This result concludes the renormalization of the holographic correlators coming from a $\Phi^3$ theory on AdS.

\subsection{Renormalized Correlators}

The objective of this section is to summarize the key points of the recent renormalization study of the quantum corrected holographic correlators resulting from the consideration of a self-interacting scalar $\Phi^3$ theory on a fixed AdS background through the use of the AdS/CFT correspondence. As we saw, these correlators were infrared divergent as their different contributions approached the conformal boundary of the AdS space, and also ultraviolet divergent as their loops integrals involving the bulk-bulk propagator got integrated at coincident points. In order to compute finite and predictive correlators, these divergences demanded not only a delicate regularization scheme but also a delicate renormalization scheme, in order to absorb in a sensitive way the corresponding infinities. The infrared divergences present at the on-shell level of the AdS path integral were both regulated and renormalized through the holographic renormalization procedure with the addition of a covariant boundary term in the AdS action. This procedure for the infrared divergences present at the off-shell level of the AdS path integral naturally translated into their regularization by simply solving the loops contributions to the correlators up to the same radial regulator introduced in the holographic renormalization. While for the ultraviolet divergences present at the off-shell level of the AdS path integral, a point-splitting approach was taken, resulting in regularized bulk-bulk propagators which conserved their symmetry under AdS transformations. By explicitly computing these regularized correlators, their nice form allowed us to renormalize them in exactly the same spirit as it is done for ordinary QFTs, that is, by understating the parameters of the theory, $\Phi(x)$, $m^2$ and $\lambda,$ not as physical constants but bare quantities, opening the possibility of a renormalization scheme through their definition. This turned out to be indeed the case, where the redefinition of these parameters in the AdS bulk action as:

\begin{equation}
    \Phi(x)\rightarrow\sqrt{Z(\lambda)}\Phi(x),\hspace{0.5cm}m^2\rightarrow m^2+\delta m^2,\hspace{0.5cm}\lambda\rightarrow\lambda+\delta\lambda
\end{equation}\

exactly renormalized every single divergence present in the holographic n-point functions through the convenient choice of the counterterms $Z(\lambda)$, $\delta m^2$ and $\delta\lambda$ as:

\begin{equation}
    Z(\lambda)=\varepsilon^{-\frac{\Pi_0(0)}{\nu}},\hspace{0.5cm}\delta m^2=\Pi_\infty(\kappa),\hspace{0.5cm}\delta\lambda=-\lambda^3C^\infty_{GGG}(\kappa)
\end{equation}\

where $\varepsilon$ and $\kappa$ are the IR and UV regulators introduced in the regularization scheme, $\Pi_\infty(\kappa)$ and $\Pi_0(0)$ are the UV-divergent and UV-convergent parts of the 1PI contributions $\Pi(\kappa)$:

\begin{align}
    \Pi(\kappa)=\frac{\lambda^2}{2}\pi^\frac{d+1}{2}\Bigl(\frac{2^{-\Delta}c_\Delta}{2\nu}\Bigl)^2\sum_{k=0}^\infty &a^{(2)}_k\frac{\Gamma(\frac{3\Delta-d}{2}+k)\Gamma(\frac{\Delta}{2}+k)}{\Gamma(\Delta+k)\Gamma(\Delta+\frac{1}{2}+k)}\Bigl(\frac{1}{1+\kappa}\Bigl)^{2\Delta+2k}\nonumber\\
    &\hspace{-1cm}\times\ _3F_2\Bigl(d-1,\frac{3\Delta-d}{2}+k,\frac{\Delta}{2}+k;\Delta+k,\Delta+\frac{1}{2}+k;\Bigl(\frac{1}{1+\kappa}\Bigl)^2\Bigl)
\end{align}\

where the coefficient $a_k^{(2)}$ was defined in eq. (\ref{a2kcoef}), and where $C^\infty_{GGG}(\kappa)$ is the UV-divergent part of $C_{GGG}(\kappa)$ coming from the "O" diagram in the 3-point function:

\begin{equation}\label{cggg}
    C_{GGG}(\kappa)=\frac{2}{\pi^\frac{d}{2}}\frac{\Gamma(\Delta)^3}{\Gamma(\frac{3\Delta-d}{2})\Gamma(\frac{\Delta}{2})^3}\int\int\int \tilde{K}^\Delta(x_1,\hat{n})\tilde{K}^\Delta(x_2,\vec{0})x^\Delta_{3,0}G_\kappa(x_1,x_2)G_\kappa(x_2,x_3)G_\kappa(x_3,x_1)
\end{equation}\

where $\hat{n}$ is a unit vector and the notation $\int\int\int\equiv\int d^{d+1}x_1\sqrt{g}\int d^{d+1}x_2\sqrt{g}\int d^{d+1}x_3\sqrt{g}$ is understood. The renormalized correlators up to order $\lambda^3$ in the coupling constant resulting from the redefinition of the bulk's theory parameters, along with their convenient choice for the counterterms, can be summarized into the holographic 1-, 2-, 3- and 4-point functions:

\begin{align}
    \text{1-pt fn:}\hspace{0.25cm}&\langle O_\Delta(\vec{y_1})\rangle_{\text{CFT}} = \langle O_\Delta(\vec{y_1})\rangle_{\text{CFT,con}} = 0\nonumber\\
    \text{2-pt fn:}\hspace{0.25cm}&\langle O_\Delta(\vec{y_1})O_\Delta(\vec{y_2})\rangle_{\text{CFT}} = \langle O_\Delta(\vec{y_1})O_\Delta(\vec{y_2})\rangle_{\text{CFT,con}} = \frac{2\nu c_\Delta}{\lvert\vec{y_1}-\vec{y_2}\rvert^{2\Delta-\frac{\Pi_0(0)}{\nu}}}+\mathcal{O}(\lambda^4)\nonumber\\
    \text{3-pt fn:}\hspace{0.25cm}&\langle O_\Delta(\vec{y_1})O_\Delta(\vec{y_2})O_\Delta(\vec{y_3})\rangle_{\text{CFT}} = \langle O_\Delta(\vec{y_1})O_\Delta(\vec{y_2})O_\Delta(\vec{y_3})\rangle_{\text{CFT,con}}=\nonumber\\
    &\hspace{1.5cm}-\lambda c_\Delta^3\frac{\pi^\frac{d}{2}}{2}\frac{\Gamma(\frac{3\Delta-d}{2}-\frac{3\Pi_0(0)}{4\nu})}{\Gamma(\Delta-\frac{\Pi_0(0)}{2\nu})^3}\frac{\Gamma(\frac{\Delta}{2}-\frac{\Pi_0(0)}{4\nu})^3}{\lvert\vec{y_1}-\vec{y_2}\rvert^{\Delta-\frac{\Pi_0(0)}{2\nu}} \lvert\vec{y_2}-\vec{y_3}\rvert^{\Delta-\frac{\Pi_0(0)}{2\nu}} \lvert\vec{y_3}-\vec{y_1}\rvert^{\Delta-\frac{\Pi_0(0)}{2\nu}}}\nonumber\\
    &\hspace{1.5cm}-\lambda^3c^3_\Delta C^0_{GGG}(0)\frac{\pi^\frac{d}{2}}{2}\frac{\Gamma(\frac{3\Delta-d}{2})}{\Gamma(\Delta)^3}\frac{\Gamma(\frac{\Delta}{2})^3}{\lvert\vec{y_1}-\vec{y_2}\rvert^\Delta \lvert\vec{y_2}-\vec{y_3}\rvert^\Delta \lvert\vec{y_3}-\vec{y_1}\rvert^\Delta}+\mathcal{O}(\lambda^4)\nonumber\\
    \text{4-pt fn:}\hspace{0.25cm}&\langle O_\Delta(\vec{y_1})O_\Delta(\vec{y_2})O_\Delta(\vec{y_3})O_\Delta(\vec{y_4})\rangle_{\text{CFT}}=\nonumber\\
    &\hspace{1cm}\frac{(2\nu c_\Delta)^2}{\prod_{i<j}\lvert\vec{y_{ij}}\rvert^{\frac{2}{3}\bigl(\Delta-\frac{\Pi_0(0)}{2\nu}\bigl)}}u^{-\frac{2}{3}\bigl(\Delta-\frac{\Pi_0(0)}{2\nu}\bigl)}v^{\frac{1}{3}\bigl(\Delta-\frac{\Pi_0(0)}{2\nu}\bigl)}+\Bigl(u,v\rightarrow\frac{1}{u},\frac{v}{u}\Bigl)+(u,v\rightarrow v,u)\nonumber\\
    &\hspace{1cm}+\lambda^2c^4_\Delta\frac{\pi^\frac{d}{2}}{8}\frac{\Gamma(\frac{3\Delta-d}{2})\Gamma(\frac{\Delta}{2})}{\Gamma(\Delta)^4}\frac{u^\frac{\Delta}{3}v^\frac{\Delta}{3}}{\prod_{i<j}\lvert\vec{y_{ij}}\rvert^{\frac{2\Delta}{3}}}\nonumber\\
    &\hspace{2cm}\times\Bigl[\sum_{k=0}^\infty\frac{\Gamma(\frac{3\Delta-d}{2}+k)}{\Gamma(\nu+1+k)\ k!}H\Bigl(\Delta,\Delta,\frac{\Delta}{2}+1-k,2\Delta;u,v\Bigl)\nonumber\\
   &\hspace{3.5cm}-\sum_{k=0}^\infty\frac{\Gamma(2\Delta-\frac{d}{2}+k)}{\Gamma(\frac{\Delta}{2}+1+k)\Gamma(\frac{3\Delta-d}{2}+1+k)}H(\Delta,\Delta,1-k,2\Delta;u,v)\Bigl]\nonumber\\
   &\hspace{1cm}+\Bigl(u,v\rightarrow\frac{1}{u},\frac{v}{u}\Bigl) + (u,v\rightarrow v,u)+\mathcal{O}(\lambda^4)\nonumber\\
    &\langle O_\Delta(\vec{y_1})O_\Delta(\vec{y_2})O_\Delta(\vec{y_3})O_\Delta(\vec{y_4})\rangle_{\text{CFT,con}} =\lambda^2c^4_\Delta\frac{\pi^\frac{d}{2}}{8}\frac{\Gamma(\frac{3\Delta-d}{2})\Gamma(\frac{\Delta}{2})}{\Gamma(\Delta)^4}\frac{u^\frac{\Delta}{3}v^\frac{\Delta}{3}}{\prod_{i<j}\lvert\vec{y_{ij}}\rvert^{\frac{2\Delta}{3}}}\nonumber\\
    &\hspace{1.5cm}\times\Bigl[\sum_{k=0}^\infty\frac{\Gamma(\frac{3\Delta-d}{2}+k)}{\Gamma(\nu+1+k)\ k!}H\Bigl(\Delta,\Delta,\frac{\Delta}{2}+1-k,2\Delta;u,v\Bigl)\nonumber\\
   &\hspace{4cm}-\sum_{k=0}^\infty\frac{\Gamma(2\Delta-\frac{d}{2}+k)}{\Gamma(\frac{\Delta}{2}+1+k)\Gamma(\frac{3\Delta-d}{2}+1+k)}H(\Delta,\Delta,1-k,2\Delta;u,v)\Bigl]\nonumber\\
   &\hspace{1.5cm}+\Bigl(u,v\rightarrow\frac{1}{u},\frac{v}{u}\Bigl) + (u,v\rightarrow v,u)+\mathcal{O}(\lambda^4)
\end{align}\

where $C^0_{GGG}(0)$ is the UV-convergent part of eq. (\ref{cggg}). The form of these correlators are exactly the expected for a conformal theory as it is dictated by eq. (\ref{summcorr2}) up to conformal anomalies, where their overall factors, scaling dimension and effective coupling constant receive small corrections coming from the 1PI loop diagrams resulting from a perturbatively approach in the parameter $\lambda$. These results, while showing the clear role of the quantum corrections to the holographic correlators, also greatly motivate and contribute to the belief of the validity of the AdS/CFT conjecture.

\section{\texorpdfstring{$\Phi^4$}{} Theory}

\subsection{Semiclassical Approximation}

We have reviewed in detail how starting from a $\Phi^3$ self-interacting theory on a fixed AdS background we can obtain the corresponding renormalized dual CFT correlators, through the use of the AdS/CFT correspondence at its full quantum nature. Continuing in the line of complicating the current picture with the intention to not only construct more interesting theories but also to further test the validity of the AdS/CFT correspondence, we will study another self-interacting scalar theory on the AdS side. Consider now the $\Phi^4$ theory defined in eq. (\ref{phi4theory}), together with its holographic renormalization eq. (\ref{actionren}):

\begin{equation}\
    Z_{\text{AdS}}[\varphi_0] = \int D\Phi\ e^{-S_{\text{AdS}}[\Phi]-\int d^dx\sqrt{\gamma}\ B(\Phi(x))\rvert_{x_0=0}}
\end{equation}\

where $S_{\text{AdS}}[\Phi]$ is the $\Phi^4$ action:

\begin{equation}
    S_{\text{AdS}}[\Phi] = \int d^{d+1}x\sqrt{g}\ \Bigl[\frac{1}{2}g^{\mu\nu}\partial_\mu\Phi(x)\partial_\nu\Phi(x)+\frac{1}{2}m^2\Phi^2(x)+\frac{\lambda}{4!}\Phi^4(x)\Bigl]
\end{equation}\

and where $B(\Phi(x))$ is the boundary term eq. (\ref{boundaryterm}), counterterm responsible for the renormalization of the infrared divergences coming from the on-shell part of the path integral:

\begin{align}
    B\bigl(\Phi(x)\bigl) = &\frac{1}{2}(d-\Delta)\Phi^2(x)+\nu\ln{(x_0)}C_\nu \Phi(x)\Box_\gamma^\nu\Phi(x)+\frac{1}{2}C_\nu \Phi(x)\Box_\gamma^\nu\Phi(x)\nonumber\\
    &+\frac{1}{4(\nu-1)}\Phi(x)\Box_\gamma\Phi(x)+\dotsb
\end{align}\

where the triple dots represent higher order derivative terms. We will proceed then in the same way as we did before for the saddle point approximation of the correspondence, looking at quantum fluctuations $h(x)$ around the classical solution $\phi(x)$ of the AdS action through the change of variable $\Phi(x)=\phi(x)+h(x)$, but now keeping track of every quantity resulting from this separation. Under this change of variable the AdS path integral transforms as:

\begin{equation}\label{phi4pisep1}
    Z_{\text{AdS}}[\varphi_0] = e^{-\int d^dx\sqrt{\gamma}\ B(\phi(x))\rvert_{x_0=0}}\int Dh\ e^{-S_{\text{AdS}}[\phi+h]}
\end{equation}\

where we used that the on-shell field is functionally fixed, i.e., $D\phi=0$, and postulated that the quantum fluctuations of the bulk field are only contained in the interior of the AdS space, vanishing sufficiently fast at its boundaries. In other words, all the non-normalizable behavior of $\Phi(x)$ is contained in $\phi(x)$. This assumption for the quantum fluctuations further allows us to write the resulting AdS action as:

\begin{equation}\label{phi4pisep2}
    S_{\text{AdS}}[\phi+h] = S_{\text{AdS}}[\phi] + S_{\text{AdS}}[h] +\lambda\int d^{d+1}x\sqrt{g}\ \Bigl[\frac{1}{6}\phi(x)h^3(x)+\frac{1}{4}\phi^2(x)h^2(x)\Bigl]
\end{equation}\

where we integrated by parts dropping the quantum fluctuations $h(x)$ being evaluated at the boundaries of the AdS space, used the classical equation satisfied by $\phi(x)$ and finally identified the original form of the AdS action now for the different fields. We see that the $\Phi^4$ AdS action doesn't act as a linear functional under the field's change of variable ($S_{\text{AdS}}[\phi+h]\neq S_{\text{AdS}}[\phi] + S_{\text{AdS}}[h]$) due to the presence of the last term in eq. (\ref{phi4pisep2}) coming from the self-interaction. This quantity, where the on-shell part of the bulk field is directly coupled to its quantum fluctuations, can be seen as a deformation to the linearity of the action and it will be precisely the responsible for the quantum corrections to the classical correlators found in the previous chapter. Indeed, replacing eq. (\ref{phi4pisep2}) back into eq. (\ref{phi4pisep1}), the resulting $\Phi^4$ AdS path integral under the change of variable $\Phi(x)=\phi(x)+h(x)$ is:

\begin{equation}
    Z_{\text{AdS}}[\varphi_0] = e^{-S_{\text{AdS}}[\phi]-\int d^dx\sqrt{\gamma}\ B(\phi(x))\bigl\rvert_{x_0=0}}\int Dh\ e^{-S_\text{AdS}[h]-\lambda\int d^{d+1}x\sqrt{g}\ \bigl[\frac{1}{6}\phi(x)h^3(x)+\frac{1}{4}\phi^2(x)h^2(x)\bigl]}
\end{equation}\

and if we think of the parameter $\lambda$, which mediates the strength of the self-interaction, as being in some sense "small", we can resort to perturbation theory solving for the form of the AdS path integral as an expansion in this parameter:

\begin{align}\label{phi4pisep3}
    Z_{\text{AdS}}[\varphi_0] = &e^{-S_{\text{AdS}}[\phi]-\int d^dx\sqrt{\gamma}\ B(\phi(x))\bigl\rvert_{x_0=0}}\Bigl[\int Dh\ e^{-S_\text{AdS}[h]}-\frac{\lambda}{6}\int\phi(x_1)\int Dh\ h^3(x_1)e^{-S_\text{AdS}[h]}\nonumber\\
    &-\frac{\lambda}{4}\int\phi^2(x_1)\int Dh\ h^2(x_1)e^{-S_\text{AdS}[h]}\nonumber\\
    &+\frac{\lambda^2}{72}\int\int\phi(x_1)\phi(x_2)\int Dh\ h^3(x_1)h^3(x_2)e^{-S_\text{AdS}[h]}\nonumber\\
    &+\frac{\lambda^2}{24}\int\int\phi(x_1)\phi^2(x_2)\int Dh\ h^3(x_1)h^2(x_2)e^{-S_\text{AdS}[h]}\nonumber\\
    &+\frac{\lambda^2}{32}\int\int\phi^2(x_1)\phi^2(x_2)\int Dh\ h^2(x_1)h^2(x_2)e^{-S_\text{AdS}[h]}\Bigl]+\mathcal{O}(\lambda^3)
\end{align}\

where we expanded the deformation term up to order $\lambda^2$ and separated the integral in $h$ into the 6 resulting terms, defining in the process $\int\int\dotsm\equiv\int d^{d+1}x_1\sqrt{g}\int d^{d+1}x_2\sqrt{g}\dotsm$ to keep the notation short. Note that in this expansion we come across path integrals in the field $h(x)$ of the form $\int Dh\ h(x_1)\dotsm h(x_n)e^{-S_{\text{AdS}}[h]}$. Since this field $h(x)$, unlike the complete bulk field $\Phi(x)$, is thought to be perfectly regular at the boundaries of the current space under consideration, these resulting path integrals can be solved in exactly the same way as the ones encountered in ordinary quantum field theories. This realization motivated us to define the ordinary n-point functions in the bulk $G_n(x_1,\dotsc,x_n)$ (eq. (\ref{npphi3})) where these quantities are expected to be solved, again, as an expansion in $\lambda$ with each resulting term involving only the bulk-bulk propagator. In terms of these functions each path integral in the field $h(x)$ present in eq. (\ref{phi4pisep3}) can be solved directly obtaining, up to order $\lambda^2$ in the coupling constant, the normalized AdS path integral:

\begin{align}\label{phi4pisep4}
    \frac{Z_{\text{AdS}}[\varphi_0]}{Z_{\text{AdS}}[\varphi_0=0]} = &e^{-S_{\text{AdS}}[\phi]-\int d^dx\sqrt{\gamma}\ B(\phi(x))\bigl\rvert_{x_0=0}}\Bigl[1-\frac{\lambda}{6}\int\phi(x_1)G_3(x_1,x_1,x_1)\nonumber\\
    &-\frac{\lambda}{4}\int\phi^2(x_1)G_2(x_1,x_1)+\frac{\lambda^2}{72}\int\int\phi(x_1)\phi(x_2)G_6(x_1,x_1,x_1,x_2,x_2,x_2)\nonumber\\
    &+\frac{\lambda^2}{24}\int\int\phi(x_1)\phi^2(x_2)G_5(x_1,x_1,x_1,x_2,x_2)\nonumber\\
    &+\frac{\lambda^2}{32}\int\int\phi^2(x_1)\phi^2(x_2)G_4(x_1,x_1,x_2,x_2)\Bigl]+\mathcal{O}(\lambda^3)
\end{align}\

where, since $\phi(x)\bigl\rvert_{\varphi_0=0}=0$ (eq. (\ref{solphi4})), we used that $Z_{\text{AdS}}[\varphi_0=0]=\int Dh\ e^{-S_{\text{AdS}}[h]}$. Now, as we just said, each one of these n-point functions in the bulk $G_n(x_1,\dotsc,x_n)$ can be solved in exactly the same way as in ordinary QFT, this is, using the same regular methods of adding an external source coupled to the field in the generating functional, then performing the resulting integral with the use again of perturbation theory and finally computing the desired n-point function through the corresponding derivatives of the source, which at the end of the calculation are set to 0. This process for the particular n-point functions present in eq. (\ref{phi4pisep4}) results in, first for $G_3(x_1,x_1,x_1)$ and $G_5(x_1,x_1,x_1,x_2,x_2)$ at all orders in $\lambda$:

\begin{equation}
    G_3(x_1,x_1,x_1) = G_5(x_1,x_1,x_1,x_2,x_2) = 0
\end{equation}\

then for $G_2(x_1,x_1)$ up to order $\lambda^1$:

\begin{equation}
    G_2(x_1,x_1) =G(x_1,x_1)-\frac{\lambda}{2}\int G^2(x_1,x_2)G(x_2,x_2) +\mathcal{O}(\lambda^2)
\end{equation}\

where $\int\equiv\int d^{d+1}x_2\sqrt{g}$, then for $G_6(x_1,x_1,x_1,x_2,x_2,x_2)$ up to order $\lambda^0$:

\begin{equation}
    G_6(x_1,x_1,x_1,x_2,x_2,x_2) = 9G(x_1,x_1)G(x_1,x_2)G(x_2,x_2)+6G^3(x_1,x_2) +\mathcal{O}(\lambda)
\end{equation}\

and finally for $G_4(x_1,x_1,x_2,x_2)$ up to order $\lambda^0$:

\begin{equation}
    G_4(x_1,x_1,x_2,x_2) = G(x_1,x_1)G(x_2,x_2)+2G^2(x_1,x_2)+\mathcal{O}(\lambda)
\end{equation}\

where the quantities $G(x,z)$ are the usual bulk-bulk propagators. These results allow us to finally express the complete expansion of the normalized AdS path integral up to order $\lambda^2$ in the self-interacting coupling constant as:

\begin{align}\label{phi4pisep5}
    \frac{Z_{\text{AdS}}[\varphi_0]}{Z_{\text{AdS}}[\varphi_0=0]} = &e^{-S_{\text{AdS}}[\phi]-\int d^dx\sqrt{\gamma}\ B(\phi(x))\bigl\rvert_{x_0=0}}\Bigl[1-\frac{\lambda}{4}\int\phi^2(x_1)G(x_1,x_1)\nonumber\\
    &+\frac{\lambda^2}{8}\int\int\phi^2(x_1)G^2(x_1,x_2)G(x_2,x_2)\nonumber\\
    &+\frac{\lambda^2}{8}\int\int\phi(x_1)\phi(x_2)G(x_1,x_1)G(x_1,x_2)G(x_2,x_2)\nonumber\\
    &+\frac{\lambda^2}{12}\int\int\phi(x_1)\phi(x_2)G^3(x_1,x_2)\nonumber\\
    &+\frac{\lambda^2}{32}\int\int\phi^2(x_1)\phi^2(x_2)G(x_1,x_1)G(x_2,x_2)\nonumber\\
    &+\frac{\lambda^2}{16}\int\int\phi^2(x_1)\phi^2(x_2)G^2(x_1,x_2)\Bigl]+\mathcal{O}(\lambda^3)
\end{align}\

remembering that the on-shell field $\phi(x)$ in this same $\lambda$ expansion is given by eq. (\ref{solphi4}):

\begin{equation}\label{solphi4ch4}
    \phi(x) = \int d^dy\ K(x,\vec{y})\varphi_0(\vec{y})-\frac{\lambda}{6}\int d^{d+1}x_1\sqrt{g}\ G(x,x_1)\Bigl[\int d^dy\ K(x_1,\vec{y})\varphi_0(\vec{y})\Bigl]^3+\mathcal{O}(\lambda^2)
\end{equation}\

Eq. (\ref{phi4pisep5}) together with eq. (\ref{solphi4ch4}) give us the concrete and explicit dependence of the $\Phi^4$ AdS path integral as a functional of the dual source $\varphi_0(\vec{y})$, ready to be differentiated with the intention to compute the quantum corrected CFT holographic correlators.

\subsection{Correlation Functions}

Plugging the explicit form of the on-shell field $\phi(x)$ (eq. (\ref{solphi4ch4})) into the normalized AdS path integral $\frac{Z_{\text{AdS}}[\varphi_0]}{Z_{\text{AdS}}[\varphi_0=0]}$ (eq. (\ref{phi4pisep5})) and keeping terms of order $\lambda^2$, the obtention of the holographic correlators eqs. (\ref{euccorr1}) and (\ref{euccorr2}) up to this order in the coupling constant is reduced to a simple exercise of taking derivatives, where since the on-shell part of the path integral is holographic renormalized the variation of this part is understood to be given by eq. (\ref{varzphi4explicit}). The resulting quantum corrected holographic correlators from this process for some primary scalar operator $O_\Delta(\vec{x})$ of scaling dimension $\Delta$ dual to a $\Phi^4$ self-interacting scalar field in AdS$_{d+1}$ are given by:

\begin{align}\label{quantumphi4corr1}
    \text{1-pt fn:}\hspace{0.25cm}&\langle O_\Delta(\vec{y_1})\rangle_{\text{CFT}} = \langle O_\Delta(\vec{y_1})\rangle_{\text{CFT,con}} = 0\nonumber\\
    \text{2-pt fn:}\hspace{0.25cm}&\langle O_\Delta(\vec{y_1})O_\Delta(\vec{y_2})\rangle_{\text{CFT}} = \langle O_\Delta(\vec{y_1})O_\Delta(\vec{y_2})\rangle_{\text{CFT,con}} = \frac{2\nu c_\Delta}{\lvert\vec{y_1}-\vec{y_2}\rvert^{2\Delta}}\nonumber\\
    &\hspace{3cm}-\frac{\lambda}{2}\int K(x_1,\vec{y_1})K(x_1,\vec{y_2})G(x_1,x_1)\nonumber\\
    &\hspace{3cm}+\frac{\lambda^2}{4}\int\int K(x_1,\vec{y_1})K(x_1,\vec{y_2})G^2(x_1,x_2)G(x_2,x_2)\nonumber\\
    &\hspace{3cm}+\frac{\lambda^2}{6}\int\int K(x_1,\vec{y_1})K(x_2,\vec{y_2})G^3(x_1,x_2)\nonumber\\
    &\hspace{3cm}+\frac{\lambda^2}{4}\int\int K(x_1,\vec{y_1})K(x_2,\vec{y_2})G(x_1,x_1)G(x_1,x_2)G(x_2,x_2) + \mathcal{O}(\lambda^3)\nonumber\\
    \text{3-pt fn:}\hspace{0.25cm}&\langle O_\Delta(\vec{y_1})O_\Delta(\vec{y_2})O_\Delta(\vec{y_3})\rangle_{\text{CFT}} = \langle O_\Delta(\vec{y_1})O_\Delta(\vec{y_2})O_\Delta(\vec{y_3})\rangle_{\text{CFT,con}} = 0\nonumber\\
    \text{4-pt fn:}\hspace{0.25cm}&\langle O_\Delta(\vec{y_1})O_\Delta(\vec{y_2})O_\Delta(\vec{y_3})O_\Delta(\vec{y_4})\rangle_{\text{CFT}} =\frac{2\nu c_\Delta}{\lvert\vec{y_1}-\vec{y_2}\rvert^{2\Delta}}\frac{2\nu c_\Delta}{\lvert\vec{y_3}-\vec{y_4}\rvert^{2\Delta}}+(\vec{y_2}\leftrightarrow\vec{y_3})+(\vec{y_2}\leftrightarrow\vec{y_4})\nonumber\\
    &\hspace{1cm}-\lambda\int K(x_1,\vec{y_1})K(x_1,\vec{y_2})K(x_1,\vec{y_3})K(x_1,\vec{y_4})\nonumber\\
    &\hspace{1cm}+\frac{\lambda^2}{2}\int\int K(x_1,\vec{y_1})K(x_2,\vec{y_2})K(x_2,\vec{y_3})K(x_2,\vec{y_4})G(x_1,x_1)G(x_1,x_2)\times4\nonumber\\
    &\hspace{1cm}+\frac{\lambda^2}{2}\int\int K(x_1,\vec{y_1})K(x_1,\vec{y_2})K(x_2,\vec{y_3})K(x_2,\vec{y_4})G^2(x_1,x_2)\times3\nonumber\\
    &\hspace{1cm}+\frac{\lambda^2}{4}\int\int K(x_1,\vec{y_1})K(x_1,\vec{y_2})K(x_2,\vec{y_3})K(x_2,\vec{y_4})G(x_1,x_1)G(x_2,x_2)\times3\nonumber\\
    &\hspace{1cm}+\frac{2\nu c_\Delta}{\lvert\vec{y_1}-\vec{y_2}\rvert^{2\Delta}}\Bigl[-\frac{\lambda}{2}\int K(x_1,\vec{y_3})K(x_1,\vec{y_4})G(x_1,x_1)\nonumber\\
    &\hspace{3cm}+\frac{\lambda^2}{4}\int\int K(x_1,\vec{y_3})K(x_1,\vec{y_4})G^2(x_1,x_2)G(x_2,x_2)\nonumber\\
    &\hspace{3cm}+\frac{\lambda^2}{6}\int\int K(x_1,\vec{y_3})K(x_2,\vec{y_4})G^3(x_1,x_2)\nonumber\\
    &\hspace{3cm}+\frac{\lambda^2}{4}\int\int K(x_1,\vec{y_3})K(x_2,\vec{y_4})G(x_1,x_1)G(x_1,x_2)G(x_2,x_2)\Bigl]\times6 + \mathcal{O}(\lambda^3)\nonumber\\
    &\langle O_\Delta(\vec{y_1})O_\Delta(\vec{y_2})O_\Delta(\vec{y_3})O_\Delta(\vec{y_4})\rangle_{\text{CFT,con}} = -\lambda\int K(x_1,\vec{y_1})K(x_1,\vec{y_2})K(x_1,\vec{y_3})K(x_1,\vec{y_4})\nonumber\\
    &\hspace{2cm}+\frac{\lambda^2}{2}\int\int K(x_1,\vec{y_1})K(x_2,\vec{y_2})K(x_2,\vec{y_3})K(x_2,\vec{y_4})G(x_1,x_1)G(x_1,x_2)\times4\nonumber\\
    &\hspace{2cm}+\frac{\lambda^2}{2}\int\int K(x_1,\vec{y_1})K(x_1,\vec{y_2})K(x_2,\vec{y_3})K(x_2,\vec{y_4})G^2(x_1,x_2)\times3 + \mathcal{O}(\lambda^3)
\end{align}\

where, with the intention to keep the notation short, we defined\\ $\int\int\dotsm\equiv\int d^{d+1}x_1\sqrt{g}\int d^{d+1}x_2\sqrt{g}\dotsm$ and represented the different permutations of the integrals as a multiplicative factor at the end of each. Notice the new contributions to the correlators in comparison with those obtained under the classical approximation of the AdS/CFT correspondence, eq. (\ref{phi4corr}). Pictorially, these new terms correspond exactly to the loops diagrams expected from a regular QFT $\Phi^4$ self-interacting theory resulting from a perturbative expansion in the self-interacting coupling constant, even agreeing with the same coefficients! Since these new integrals are contributing to specific correlators which are conjectured to be of the form dictated by eq. (\ref{summcorr2}), the functional form of their results is strongly conditioned purely from conformal symmetry arguments. We will proceed then to study these quantities in detail through their explicit computation.

\subsection{Holographic Dictionary}

Before jumping straight into the calculations of the new contributions to the holographic correlators, as a consistency check we want to see if the holographic dictionary defined in eqs. (\ref{dict1}) and (\ref{dict2}), which relates the desired boundary correlators for the operator $O_\Delta(\vec{x})$ with the simple and known bulk correlators for the field $\Phi(x)$, is still valid for the recently obtained quantum corrected 1-, 2-, 3- and 4-point functions.\par
As we did for the $\Phi^3$ theory case, to keep the discussion clean and short the validity of the holographic dictionary for the present case can be argued to hold without doing any computation through the understanding of why it holds for its classical counterpart in the first place. In that case, under the appropriate limits the values of the bulk tree-level n-point functions obtained from the usual Feynman rules of scalar $\Phi^4$ theories (eq. (\ref{hdphi4class})) matched exactly the boundary correlators obtained through the classical approximation of the AdS/CFT correspondence (eq. (\ref{phi4corr})) mainly because the latter essentially contained the same type of diagrams expected from Feynman rules as the former, even with the same coefficients, with the exception that the external legs of said diagrams had been replaced with bulk-boundary propagators. But this replacement precisely matched the effective dictionary coming from the nice expansions of the bulk-bulk propagators eqs. (\ref{expbubup1}) and (\ref{expbubup2}) which for the bulk correlators implied the simple recipe "replace external $G$'s with $K$'s", inevitably resulting in exactly the same correlators as those found through the classical approximation of the AdS/CFT correspondence. These facts unequivocally led us to relate the boundary n-point functions with the bulk n-point functions simply as the former being the extension of the internal points of the latter to the conformal boundary of the AdS space.\par
Having said this, it is straightforward to see that exactly the same is happening to the now quantum corrected boundary correlators for the operator $O_\Delta(\vec{x})$ just found eq. (\ref{quantumphi4corr1}). Indeed, these correlators correspond exactly to the ones obtained from Feynman rules where the external bulk-bulk propagators have been replaced with bulk-boundary propagators, but this is precisely the effective action of the holographic dictionary, implying that we can always formulate these boundary correlators as the appropriate limit of some bulk correlators which follow the Feynman rules of a $\Phi^4$ theory. Unsurprisingly, these bulk correlators are nothing but the correlators for the bulk field $\Phi$, confirming in this way for the present case the boundary/bulk n-point functions equivalence.

\subsection{Regularization Schemes}

We are one step away from fully diving into the computation of the quantum corrections to the holographic correlators dual to a $\Phi^4$ theory on AdS. The last matter that we will see before this, with the intention to keep these computations organized and clean, is the introduction of the appropriate regularization schemes that we will use in this study, necessary for the handling of the different divergent quantities (as we will see in detail) present in the n-point functions, eq. (\ref{quantumphi4corr1}).\par
As we discussed in section 4.1.5 where we introduced the different regulators to be used in this work, there are 2 type of divergences present in the quantum corrections to the holographic correlators: of the type IR (infrared) coming from the different loops integrals as their internal points $x_i$ being integrated approach the conformal boundary of the AdS space at $x_{i,0}=0$, and also of the type UV (ultraviolet) coming from the same loops integrals as the bulk-bulk propagator contained in them get integrated at more closer and closer points. We regularized the former type of divergence extending the scheme used for the classical contributions to the correlators, which for their quantum corrections naturally translated into simply integrating the loops up to the same radial regulator. For the latter type of divergence we regularized it through a point-splitting approach, where we kept the internal points of the bulk-bulk propagators from being integrated at coincident points by a small proper distance. For the $\Phi^3$ theory these regularization schemes not only handled the infinities satisfactorily, but also resulted in a natural and consistent renormalization scheme identical to those used in ordinary QFTs. These facts motivate us to treat the divergences present in the current loops integrals in exactly the same way. Therefore, adopting the same regularization schemes as before, the resulting holographic correlators from these procedures can be simply stated as:

\begin{align}
    \text{1-pt fn:}\hspace{0.25cm}&\langle O_\Delta(\vec{y_1})\rangle_{\text{CFT}} = \langle O_\Delta(\vec{y_1})\rangle_{\text{CFT,con}} = 0\nonumber\\
    \text{2-pt fn:}\hspace{0.25cm}&\langle O_\Delta(\vec{y_1})O_\Delta(\vec{y_2})\rangle_{\text{CFT}} = \langle O_\Delta(\vec{y_1})O_\Delta(\vec{y_2})\rangle_{\text{CFT,con}} = \frac{2\nu c_\Delta}{\lvert\vec{y_1}-\vec{y_2}\rvert^{2\Delta}}\nonumber\\
    &\hspace{3cm}-\frac{\lambda}{2}\int_\varepsilon K(x_1,\vec{y_1})K(x_1,\vec{y_2})G_\kappa(x_1,x_1)\nonumber\\
    &\hspace{3cm}+\frac{\lambda^2}{4}\int_\varepsilon\int_\varepsilon K(x_1,\vec{y_1})K(x_1,\vec{y_2})G^2_\kappa(x_1,x_2)G_\kappa(x_2,x_2)\nonumber\\
    &\hspace{3cm}+\frac{\lambda^2}{6}\int_\varepsilon\int_\varepsilon K(x_1,\vec{y_1})K(x_2,\vec{y_2})G^3_\kappa(x_1,x_2)\nonumber\\
    &\hspace{3cm}+\frac{\lambda^2}{4}\int_\varepsilon\int_\varepsilon K(x_1,\vec{y_1})K(x_2,\vec{y_2})G_\kappa(x_1,x_1)G_\kappa(x_1,x_2)G_\kappa(x_2,x_2) + \mathcal{O}(\lambda^3)\nonumber\\
    \text{3-pt fn:}\hspace{0.25cm}&\langle O_\Delta(\vec{y_1})O_\Delta(\vec{y_2})O_\Delta(\vec{y_3})\rangle_{\text{CFT}} = \langle O_\Delta(\vec{y_1})O_\Delta(\vec{y_2})O_\Delta(\vec{y_3})\rangle_{\text{CFT,con}} = 0\nonumber\\
    \text{4-pt fn:}\hspace{0.25cm}&\langle O_\Delta(\vec{y_1})O_\Delta(\vec{y_2})O_\Delta(\vec{y_3})O_\Delta(\vec{y_4})\rangle_{\text{CFT}} =\frac{2\nu c_\Delta}{\lvert\vec{y_1}-\vec{y_2}\rvert^{2\Delta}}\frac{2\nu c_\Delta}{\lvert\vec{y_3}-\vec{y_4}\rvert^{2\Delta}}+(\vec{y_2}\leftrightarrow\vec{y_3})+(\vec{y_2}\leftrightarrow\vec{y_4})\nonumber\\
    &\hspace{1cm}-\lambda\int K(x_1,\vec{y_1})K(x_1,\vec{y_2})K(x_1,\vec{y_3})K(x_1,\vec{y_4})\nonumber\\
    &\hspace{1cm}+\frac{\lambda^2}{2}\int_\varepsilon\int_\varepsilon K(x_1,\vec{y_1})K(x_2,\vec{y_2})K(x_2,\vec{y_3})K(x_2,\vec{y_4})G_\kappa(x_1,x_1)G_\kappa(x_1,x_2)\times4\nonumber\\
    &\hspace{1cm}+\frac{\lambda^2}{2}\int_\varepsilon\int_\varepsilon K(x_1,\vec{y_1})K(x_1,\vec{y_2})K(x_2,\vec{y_3})K(x_2,\vec{y_4})G^2_\kappa(x_1,x_2)\times3\nonumber\\
    &\hspace{1cm}+\frac{\lambda^2}{4}\int_\varepsilon\int_\varepsilon K(x_1,\vec{y_1})K(x_1,\vec{y_2})K(x_2,\vec{y_3})K(x_2,\vec{y_4})G_\kappa(x_1,x_1)G_\kappa(x_2,x_2)\times3\nonumber\\
    &\hspace{1cm}+\frac{2\nu c_\Delta}{\lvert\vec{y_1}-\vec{y_2}\rvert^{2\Delta}}\Bigl[-\frac{\lambda}{2}\int_\varepsilon K(x_1,\vec{y_3})K(x_1,\vec{y_4})G_\kappa(x_1,x_1)\nonumber\\
    &\hspace{2.5cm}+\frac{\lambda^2}{4}\int_\varepsilon\int_\varepsilon K(x_1,\vec{y_3})K(x_1,\vec{y_4})G^2_\kappa(x_1,x_2)G_\kappa(x_2,x_2)\nonumber\\
    &\hspace{2.5cm}+\frac{\lambda^2}{6}\int_\varepsilon\int_\varepsilon K(x_1,\vec{y_3})K(x_2,\vec{y_4})G^3_\kappa(x_1,x_2)\nonumber\\
    &\hspace{2.5cm}+\frac{\lambda^2}{4}\int_\varepsilon\int_\varepsilon K(x_1,\vec{y_3})K(x_2,\vec{y_4})G_\kappa(x_1,x_1)G_\kappa(x_1,x_2)G_\kappa(x_2,x_2)\Bigl]\times6 + \mathcal{O}(\lambda^3)\nonumber\\
    &\langle O_\Delta(\vec{y_1})O_\Delta(\vec{y_2})O_\Delta(\vec{y_3})O_\Delta(\vec{y_4})\rangle_{\text{CFT,con}} = -\lambda\int K(x_1,\vec{y_1})K(x_1,\vec{y_2})K(x_1,\vec{y_3})K(x_1,\vec{y_4})\nonumber\\
    &\hspace{2cm}+\frac{\lambda^2}{2}\int_\varepsilon\int_\varepsilon K(x_1,\vec{y_1})K(x_2,\vec{y_2})K(x_2,\vec{y_3})K(x_2,\vec{y_4})G_\kappa(x_1,x_1)G_\kappa(x_1,x_2)\times4\nonumber\\
    &\hspace{2cm}+\frac{\lambda^2}{2}\int_\varepsilon\int_\varepsilon K(x_1,\vec{y_1})K(x_1,\vec{y_2})K(x_2,\vec{y_3})K(x_2,\vec{y_4})G^2_\kappa(x_1,x_2)\times3 + \mathcal{O}(\lambda^3)
\end{align}\

where we IR- and UV-regularized the quantum contributions in eq. (\ref{quantumphi4corr1}). These are the integrals that we will finally compute, process which we will proceed to do next.

\subsection{2-Point Function}

The regularized holographic 2-point functions dual to a $\Phi^4$ self-interacting theory on AdS are given by:

\begin{align}
    &\langle O_\Delta(\vec{y_1})O_\Delta(\vec{y_2})\rangle_{\text{CFT}} = \langle O_\Delta(\vec{y_1})O_\Delta(\vec{y_2})\rangle_{\text{CFT,con}} = \frac{2\nu c_\Delta}{\lvert\vec{y_1}-\vec{y_2}\rvert^{2\Delta}}\nonumber\\
    &\hspace{4cm}-\frac{\lambda}{2}\int_\varepsilon K(x_1,\vec{y_1})K(x_1,\vec{y_2})G_\kappa(x_1,x_1)\nonumber\\
    &\hspace{4cm}+\frac{\lambda^2}{4}\int_\varepsilon\int_\varepsilon K(x_1,\vec{y_1})K(x_1,\vec{y_2})G^2_\kappa(x_1,x_2)G_\kappa(x_2,x_2)\nonumber\\
    &\hspace{4cm}+\frac{\lambda^2}{6}\int_\varepsilon\int_\varepsilon K(x_1,\vec{y_1})K(x_2,\vec{y_2})G^3_\kappa(x_1,x_2)\nonumber\\
    &\hspace{4cm}+\frac{\lambda^2}{4}\int_\varepsilon\int_\varepsilon K(x_1,\vec{y_1})K(x_2,\vec{y_2})G_\kappa(x_1,x_1)G_\kappa(x_1,x_2)G_\kappa(x_2,x_2) + \mathcal{O}(\lambda^3)
\end{align}\

from where we see the quantum corrections they receive coming from the loop integrals:

\begin{equation}
    I_1(\vec{y_1},\vec{y_2}) = -\frac{\lambda}{2}\int_\varepsilon K(x_1,\vec{y_1})K(x_1,\vec{y_2})G_\kappa(x_1,x_1)
\end{equation}

\begin{equation}
    I_2(\vec{y_1},\vec{y_2}) = \frac{\lambda^2}{4}\int_\varepsilon\int_\varepsilon K(x_1,\vec{y_1})K(x_1,\vec{y_2})G^2_\kappa(x_1,x_2)G_\kappa(x_2,x_2)
\end{equation}

\begin{equation}
    I_3(\vec{y_1},\vec{y_2}) = \frac{\lambda^2}{6}\int_\varepsilon\int_\varepsilon K(x_1,\vec{y_1})K(x_2,\vec{y_2})G^3_\kappa(x_1,x_2)
\end{equation}

\begin{equation}
    I_4(\vec{y_1},\vec{y_2}) = \frac{\lambda^2}{4}\int_\varepsilon\int_\varepsilon K(x_1,\vec{y_1})K(x_2,\vec{y_2})G_\kappa(x_1,x_1)G_\kappa(x_1,x_2)G_\kappa(x_2,x_2)
\end{equation}\

In order to compute the complete 2-point functions up to this order in the expansion of $\lambda$, we will proceed then to compute these quantities. We will start by studying $I_1$, then $I_2$, then $I_3$ and lastly, $I_4$.

\subsubsection{The "Head" Diagram}

\begin{figure}[h]
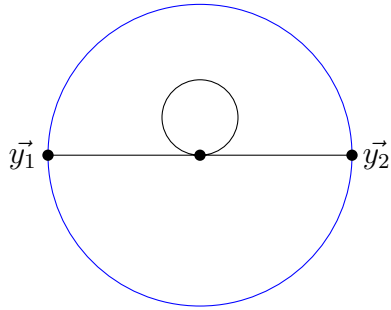

    \[\begin{wittendiagram}
    \draw (-2,0) node[vertex] -- (0,0)
    (0,0) node[vertex]
    (2,0) node[vertex] -- (0,0)
    
     (-2,0) node[left]{$\vec{y_1}$}
    (2,0) node[right]{$\vec{y_2}$}
    
    (0,0.5) circle (0.5) ;
  \end{wittendiagram}\]
  \caption{Pictorial representation of the "head" diagram.}
\end{figure}

In terms of the unnormalized bulk-boundary propagator we can rewrite the loop integral $I_1$ (which we will refer to it as the "head" diagram) as:

\begin{equation}
    I_1(\vec{y_1},\vec{y_2}) = -\frac{\lambda c_\Delta^2G_\kappa(1)}{2}\int_{x_{1,0}=\varepsilon}d^{d+1}x_1\sqrt{g}\ \tilde{K}^\Delta(x_1,\vec{y_1})\tilde{K}^\Delta(x_1,\vec{y_2})
\end{equation}\

where we used that the regularized bulk-bulk propagator being evaluated at coincident points is just a constant: $G_\kappa(x_1,x_1)=G_\kappa(\xi=1)\equiv G_\kappa(1)$. As we have seen repeatedly, the resulting integral is nothing but the D-function defined in eq. (\ref{regdfunc}), whose value is found to be given by eq. (\ref{d2funcreg2}). Therefore, replacing its known result we obtain that the "head" diagram can be written as:

\begin{equation}\label{headdiagres}
    I_1(\vec{y_1},\vec{y_2}) = \frac{2\nu c_\Delta}{\lvert\vec{y_1}-\vec{y_2}\rvert^{2\Delta}}\frac{\lambda G_\kappa(1)}{2\nu}\ln{\Bigl(\frac{\varepsilon}{\lvert\vec{y_1}-\vec{y_2}\rvert}\Bigl)}
\end{equation}\

where we have written it conveniently for the upcoming study. Similarly to the loops contributions for the 2-point functions dual to a $\Phi^3$ theory, the presence of the logarithm in this result seems to break its expected conformal structure as it was derived in eq. (\ref{summcorr2}), however as we will see when we consider the complete correlator, we will realize that the result just found has exactly the same interpretation as before, that is, the expansion of a conformal anomaly up to this same order in the self-interacting coupling constant.

\subsubsection{The "Eight" Diagram}

\begin{figure}[h]
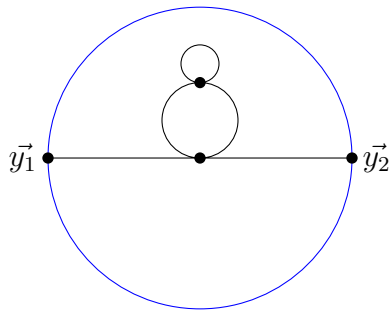

    \[\begin{wittendiagram}
    \draw (-2,0) node[vertex] -- (0,0)
    (0,0) node[vertex]
    (2,0) node[vertex] -- (0,0)
    (0,1) node[vertex]
    
     (-2,0) node[left]{$\vec{y_1}$}
    (2,0) node[right]{$\vec{y_2}$}
    
    (0,0.5) circle (0.5)
    (0,1.25) circle (0.25);
  \end{wittendiagram}\]
  \caption{Pictorial representation of the "eight" diagram.}
\end{figure}

In terms of the unnormalized bulk-boundary propagator we can rewrite the loop integral $I_2$ (which we will refer to it as the "eight" diagram) as:

\begin{equation}\label{unneightdig}
    I_2(\vec{y_1},\vec{y_2}) = \frac{\lambda^2c_\Delta^2G_\kappa(1)}{4}\int_{x_{1,0}=\varepsilon}d^{d+1}x_1\sqrt{g}\ \tilde{K}^\Delta(x_1,\vec{y_1})\tilde{K}^\Delta(x_1,\vec{y_2})\int_{x_{2,0}=\varepsilon}d^{d+1}x_2\sqrt{g}\ G^2_\kappa(x_1,x_2)
\end{equation}\

where we used that the regularized bulk-bulk propagator being evaluated at coincident points is just a constant. As we have done for past integrals, the solving strategy for this quantity will be brute force. We will start by first computing the $x_2$ integral hoping that the remaining integral in $x_1$ will be familiar to us, which as we will see will indeed be the case. The first question that we will be interested in answering is when in this $x_2$ integral we can safely take the IR-regulator $\varepsilon$ equal to 0. The infrared convergence region of this integral can be seen directly by studying how its integrand behaves as it approaches the boundary of AdS. In this case, using the explicit form of the metric and the known expansion of the bulk-bulk propagator, we obtain that:

\begin{equation}\label{cisg2int}
    \sqrt{g}\ G^2_\kappa(x_1,x_2) \underset{x_{2,0}\rightarrow0}{\sim} x_{2,0}^{-d-1}x_{2,0}^{2\Delta}=x_{2,0}^{-1+2\nu}
\end{equation}\

From here we conclude that the $x_2$ integral will be IR-convergent as long as $\nu\equiv\Delta-\frac{d}{2}$ is a positive number, but as it is discussed in section 3.1.4 in this work this will always be the case, implying that we can always take in this integral $\varepsilon=0$.\par
As we will see for every integral contributing to the holographic correlators, a remarkable feature of them is that their general structure can be derived by simply using AdS isometries transformations as change of variables. For the case of this $x_2$ integral performing the sequence of translation and rescaling, allows us to extract all the external dependence from the integral:

\begin{align}
    \int d^{d+1}x_2\sqrt{g}\ G^2_\kappa(x_1,x_2) &= \int d^{d+1}x_2\sqrt{g}\ G^2_\kappa\bigl((x_{1,0},\vec{0}),x_2\bigl)\nonumber\\
    &= \int d^{d+1}x_2\sqrt{g}\ G^2_\kappa\bigl((1,\vec{0}),x_2\bigl)\nonumber\\
    &= C_{G^2}(\kappa)
\end{align}\

where, using the invariance of the AdS measure and the bulk-bulk propagator, in the first equality we performed the translation $x_2\rightarrow x_2+\vec{x_1}$, in the second equality we performed the rescaling $x_2\rightarrow x_{1,0}x_2$, and in the final equality we noticed that the remaining integral of this sequence of change of variables is just a function of the UV-regulator $\kappa$ which we simply called $C_{G^2}(\kappa)$. This result is noteworthy, it is telling us that the integral $\int G^2$ is just a constant, where all the possible ultraviolet divergence coming from the bulk-bulk propagator being evaluated at coincident points is contained in it. In fact, notice that the power of $G$ did not play any role in this demonstration, which implies that this statement is true whenever the integral is IR-convergent. Doing the same infrared convergence study as in eq. (\ref{cisg2int}) now for a general power $n$ of the bulk-bulk propagator it is easy to verify that the integral will be IR-convergent whenever $n\Delta-d>0$, that is, for any $n>1$ or in the particular case $n=1$, whenever $\Delta>d$. This implies then the more general result:

\begin{equation}\label{intgnpropc}
    \int d^{d+1}x_2\sqrt{g}\ G^n_\kappa(x_1,x_2) = C_{G^n}(\kappa),\hspace{1cm}\text{for } n>1,\ \text{or } n=1,\Delta>d
\end{equation}\

where $C_{G^n}(\kappa)$ contains all the possible UV-divergence coming from the integral. The explicit form of this quantity is obtainable through the brute force calculation of the integral, process which will also show its ultraviolet convergence region. Let us then proceed to do this study for the particular integral that we are interested in computing, $\int G^2$. For this calculation, it will turn out to be useful to use the representation eq. (\ref{Grepdiv}) of the bulk-bulk propagator, where all its UV-divergence has been extracted out from the hypergeometric function, which as we saw in section 4.1.6, its square can be further expressed as in eq. (\ref{reggsquare}). Therefore, the integral of $\int G^2$ with the bulk-bulk propagator written in this form can be expressed as:

\begin{align}
    \int d^{d+1}x_2\sqrt{g}\ G^2_\kappa(x_1,x_2) = \Bigl(\frac{2^{-\Delta}c_\Delta}{2\nu}\Bigl)^2\sum_{k=0}^\infty a^{(2)}_k\sum_{l=0}^\infty&\frac{(d-1)_l}{l!}\Bigl(\frac{1}{1+\kappa}\Bigl)^{2\Delta+2k+2l}\nonumber\\
    &\times\int d^{d+1}x_2\sqrt{g}\ \xi^{2\Delta+2k+2l}(x_1,x_2)
\end{align}\

The complete study of this type of integrals can be found in section B.2 of Appendix B, concluding in its value in eq. (\ref{xiform}). Using this formula then for the particular value $\Delta_1=2\Delta+2k+2l$, we find that the result of the $x_2$ integral is given by:

\begin{align}
    \int d^{d+1}x_2\sqrt{g}\ G^2_\kappa(x_1,x_2) = &\pi^{\frac{d+1}{2}}\Bigl(\frac{2^{-\Delta}c_\Delta}{2\nu}\Bigl)^2\sum_{k=0}^\infty a^{(2)}_k\frac{\Gamma(\nu+k)}{\Gamma(\Delta+\frac{1}{2}+k)}\Bigl(\frac{1}{1+\kappa}\Bigl)^{2\Delta+2k}\nonumber\\
    &\hspace{3.5cm}\times\ _2F_1\Bigl(d-1,\nu+k;\Delta+\frac{1}{2}+k;\Bigl(\frac{1}{1+\kappa}\Bigl)^2\Bigl)\nonumber\\
    \equiv&C_{G^2}(\kappa)
\end{align}\

where we identified the representation of the hypergeometric function $_2F_1$ and denoted the constant result of the integral simply by $C_{G^2}(\kappa)$. This result for the $\int G^2$ integral obtained from its explicit computation has precisely the structure expected from AdS isometry arguments eq. (\ref{intgnpropc}), where in this case the value of the constant $C_{G^2}(\kappa)$ is found to be:

\begin{align}
    C_{G^2}(\kappa) = \pi^{\frac{d+1}{2}}\Bigl(\frac{2^{-\Delta}c_\Delta}{2\nu}\Bigl)^2\sum_{k=0}^\infty &a^{(2)}_k\frac{\Gamma(\nu+k)}{\Gamma(\Delta+\frac{1}{2}+k)}\Bigl(\frac{1}{1+\kappa}\Bigl)^{2\Delta+2k}\nonumber\\
    &\times\ _2F_1\Bigl(d-1,\nu+k;\Delta+\frac{1}{2}+k;\Bigl(\frac{1}{1+\kappa}\Bigl)^2\Bigl)
\end{align}\

where in turn the coefficient $a_k^{(2)}$ was defined in eq. (\ref{a2kcoef}). Notice how the ultraviolet convergence region of this quantity can be read directly from its hypergeometric function. Indeed, as $\kappa\rightarrow0$ the argument of this function goes to 1, which applying the corresponding convergence criteria introduced in section 4.1.5 implies the convergence condition $\Delta+\frac{1}{2}+k-(d-1)-(\nu+k)=\frac{3-d}{2}>0$, that is, $d<3$. In other words, the $\int G^2$ integral is UV-divergent for values of the dimension $d$ equal or greater than $3$, and UV-convergent otherwise, being able to safely take $\kappa=0$ in this case.\par
Remember that we are trying to compute the "eight" diagram eq. (\ref{unneightdig}). Replacing then the nice result just found for the $x_2$ integral back into the quantity we are trying to compute it reduces to:

\begin{equation}
    I_2(\vec{y_1},\vec{y_2}) = \frac{\lambda^2c_\Delta^2G_\kappa(1)C_{G^2}(\kappa)}{4}\int_{x_{1,0}=\varepsilon}d^{d+1}x_1\sqrt{g}\ \tilde{K}^\Delta(x_1,\vec{y_1})\tilde{K}^\Delta(x_1,\vec{y_2})
\end{equation}\

This resulting integral in $x_1$ is the known D-function eq. (\ref{regdfunc}), whose value is found to be given by eq. (\ref{d2funcreg2}). Therefore, replacing its known result we obtain that the "eight" diagram can be written as:

\begin{equation}\label{eightdiagres}
    I_2(\vec{y_1},\vec{y_2}) = -\frac{2\nu c_\Delta}{\lvert\vec{y_1}-\vec{y_2}\rvert^{2\Delta}}\frac{\lambda^2G_\kappa(1)C_{G^2}(\kappa)}{4\nu}\ln{\Bigl(\frac{\varepsilon}{\lvert\vec{y_1}-\vec{y_2}\rvert}\Bigl)}
\end{equation}\

where we have written it conveniently for the upcoming study. Again, as we will see when we consider the complete correlator, we will realize that the result just found has exactly the same interpretation as before, that is, the expansion of a conformal anomaly up to this same order in the self-interacting coupling constant.

\subsubsection{The "Sunset" Diagram}

\begin{figure}[h]
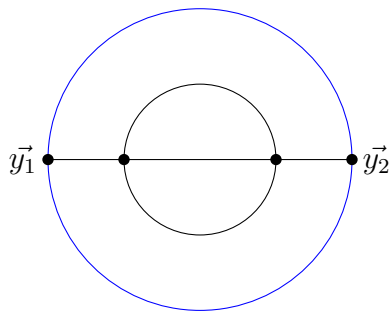

    \[\begin{wittendiagram}
    \draw (-2,0) node[vertex] -- (-1,0)
    (-1,0) node[vertex] -- (1,0)
    (1,0) node[vertex]
    (2,0) node[vertex] -- (1,0)
    
     (-2,0) node[left]{$\vec{y_1}$}
    (2,0) node[right]{$\vec{y_2}$}
    
    (0,0) circle (1) ;
  \end{wittendiagram}\]
  \caption{Pictorial representation of the "sunset" diagram.}
\end{figure}

In terms of the unnormalized bulk-boundary propagator we can rewrite the loop integral $I_3$ (which we will refer to it as the "sunset" diagram) as:

\begin{equation}\label{unnsunsetdig}
    I_3(\vec{y_1},\vec{y_2}) = \frac{\lambda^2c_\Delta^2}{6}\int_{x_{1,0}=\varepsilon}d^{d+1}x_1\sqrt{g}\ \tilde{K}^\Delta(x_1,\vec{y_1})\int_{x_{2,0}=\varepsilon}d^{d+1}x_2\sqrt{g}\ G^3_\kappa(x_1,x_2)\tilde{K}^\Delta(x_2,\vec{y_2})
\end{equation}\

Continuing in the same line we have followed so far for the integrals that we have stumble upon, we will try to solve these that are present in $I_3$ by brute force, first computing the $x_2$ integral and hoping that the remaining integral in $x_1$ will be familiar. As before, the first question that we will be interested in answering is when in this $x_2$ integral we can safely take the IR-regulator $\varepsilon$ equal to 0. But as we discussed in section 4.1.6 when we studied the "eye" diagram present in the holographic 2-point functions dual to a $\Phi^3$ theory, integrals of the form $\int G^nK$ with $n>1$ are always IR-convergent. Moreover, by AdS isometry arguments one can show that their value must be proportional to $K$ (eq. (\ref{intgnkpropk})), where all their UV-divergences are contained in the proportionality constant. The explicit form of these constants are obtainable through the brute force calculation of the integrals, process which will also show their ultraviolet convergence region. Let us then proceed to do this study for the particular integral that we are interested in computing, that is, $\int G^3K$. For this calculation, it will turn out to be useful to use the representation eq. (\ref{Grepdiv}) of the bulk-bulk propagator, where all its UV-divergence has been extracted out from the hypergeometric function. In terms of this representation then, we can write its regularized version cubed simply as:

\begin{equation}
    G^3_\kappa(x_1,x_2) = \Bigl(\frac{2^{-\Delta}c_\Delta}{2\nu}\Bigl)^3\frac{\xi_\kappa^{3\Delta}}{(1-\xi_\kappa^2)^\frac{3(d-1)}{2}}\Bigl[\ _2F_1\Bigl(\frac{\Delta-d}{2}+1,\frac{\Delta-d+1}{2};\nu+1;\xi_\kappa^2\Bigl)\Bigl]^3
\end{equation}\

where $\xi_\kappa=\frac{\xi}{1+\kappa}$. Using the triple sum property\\ $\sum_{k=0}^\infty\sum_{l=0}^\infty\sum_{n=0}^\infty a_{k,l,n}=\sum_{k=0}^\infty\sum_{l=0}^{k}\sum_{n=0}^{k-l} a_{k-l-n,l,n}$, the cube of the hypergeometric function can be written as a single sum in the regularized parameter $\xi_\kappa$:

\begin{equation}\label{2f1cube}
    \Bigl[\ _2F_1\Bigl(\frac{\Delta-d}{2}+1,\frac{\Delta-d+1}{2};\nu+1;\xi_\kappa^2\Bigl)\Bigl]^3 = \sum_{k=0}^\infty a_k^{(3)}\xi_\kappa^{2k}
\end{equation}\

where we defined the coefficient:

\begin{equation}\label{a3kcoef}
    a_k^{(3)}\equiv\sum_{l=0}^{k}\sum_{n=0}^{k-l}\frac{(\frac{\Delta-d}{2}+1)_{k-l-n}(\frac{\Delta-d+1}{2})_{k-l-n}}{(\nu+1)_{k-l-n}\ (k-l-n)!}\frac{(\frac{\Delta-d}{2}+1)_l(\frac{\Delta-d+1}{2})_l(\frac{\Delta-d}{2}+1)_n(\frac{\Delta-d+1}{2})_n}{(\nu+1)_l(\nu+1)_n\ l!\ n!}
\end{equation}\

This nice form for the cube of the hypergeometric function further allows us to express the cube of the regularized propagator in the form of:

\begin{equation}\label{reggcube}
    G^3_\kappa(x_1,x_2) = \Bigl(\frac{2^{-\Delta}c_\Delta}{2\nu}\Bigl)^3\sum_{k=0}^\infty a^{(3)}_k\sum_{l=0}^\infty\frac{(\frac{3(d-1)}{2})_l}{l!}\Bigl(\frac{\xi}{1+\kappa}\Bigl)^{3\Delta+2k+2l}
\end{equation}\

where we used that $\frac{1}{(1-\xi_\kappa^2)^\frac{3(d-1)}{2}}=\ _1F_0(\frac{3(d-1)}{2};\xi_\kappa^2)=\sum_{l=0}^\infty\frac{(\frac{3(d-1)}{2})_l}{l!}\xi_\kappa^{2l}$ and the explicit form of $\xi_\kappa$. Therefore, the integral of $\int G^3K$ with the bulk-bulk propagator written in this form can be expressed as:

\begin{align}
    \int d^{d+1}x_2\sqrt{g}\ G^3_\kappa(x_1,x_2)\tilde{K}^\Delta(x_2,\vec{y_2}) = &\Bigl(\frac{2^{-\Delta}c_\Delta}{2\nu}\Bigl)^3\sum_{k=0}^\infty a^{(3)}_k\sum_{l=0}^\infty\frac{(\frac{3(d-1)}{2})_l}{l!}\Bigl(\frac{1}{1+\kappa}\Bigl)^{3\Delta+2k+2l}\nonumber\\
    &\hspace{2cm}\times\int d^{d+1}x_2\sqrt{g}\ \xi^{3\Delta+2k+2l}\tilde{K}^\Delta(x_2,\vec{y_2})
\end{align}\

The complete study of this type of integrals can be found in section B.3 of Appendix B, concluding in its value in eq. (\ref{xikform}). Using this formula then for the particular values $\Delta_1=3\Delta+2k+2k$ and $\Delta_2=\Delta$, we find that the result of the $x_2$ integral is given by:

\begin{align}
    \int d^{d+1}x_2\sqrt{g}\ G^3_\kappa(x_1,x_2)\tilde{K}^\Delta(x_2,\vec{y_2}) = &\tilde{K}^\Delta(x_1,\vec{y_2})\pi^\frac{d+1}{2}\Bigl(\frac{2^{-\Delta}c_\Delta}{2\nu}\Bigl)^3\sum_{k=0}^\infty a^{(3)}_k\frac{\Gamma(2\Delta-\frac{d}{2}+k)\Gamma(\Delta+k)}{\Gamma(\frac{3\Delta}{2}+k)\Gamma(\frac{3\Delta+1}{2}+k)}\nonumber\\
    &\hspace{-5cm}\times\Bigl(\frac{1}{1+\kappa}\Bigl)^{3\Delta+2k}\ _3F_2\Bigl(\frac{3(d-1)}{2},2\Delta-\frac{d}{2}+k,\Delta+k;\frac{3\Delta}{2}+k,\frac{3\Delta+1}{2}+k;\Bigl(\frac{1}{1+\kappa}\Bigl)^2\Bigl)\nonumber\\
    \equiv &C_{G^3K}(\kappa)\tilde{K}^\Delta(x_1,\vec{y_2})
\end{align}\

where we identified the representation of the generalized hypergeometric function $_3F_2$ and denoted all the terms not dependent on the external points simply by $C_{G^3K}(\kappa)$. This result for the $\int G^3K$ integral obtained from its explicit computation has precisely the structure expected from AdS isometry arguments eq. (\ref{intgnkpropk}), where in this case the value of the constant $C_{G^3K}(\kappa)$ is found to be:

\begin{align}
    C_{G^3K}(\kappa) = \pi^\frac{d+1}{2}\Bigl(\frac{2^{-\Delta}c_\Delta}{2\nu}\Bigl)^3\sum_{k=0}^\infty &a^{(3)}_k\frac{\Gamma(2\Delta-\frac{d}{2}+k)\Gamma(\Delta+k)}{\Gamma(\frac{3\Delta}{2}+k)\Gamma(\frac{3\Delta+1}{2}+k)}\Bigl(\frac{1}{1+\kappa}\Bigl)^{3\Delta+2k}\nonumber\\
    &\hspace{-2cm}\times\ _3F_2\Bigl(\frac{3(d-1)}{2},2\Delta-\frac{d}{2}+k,\Delta+k;\frac{3\Delta}{2}+k,\frac{3\Delta+1}{2}+k;\Bigl(\frac{1}{1+\kappa}\Bigl)^2\Bigl)
\end{align}\

where in turn the coefficient $a^{(3)}_k$ was defined in eq. (\ref{a3kcoef}). Notice how the ultraviolet convergence region of this quantity can be read directly from its generalized hypergeometric function. Indeed, as $\kappa\rightarrow0$ the argument of this function goes to 1, which applying the corresponding convergence criteria introduced in section 4.1.5 implies the convergence condition $\frac{3\Delta}{2}+k+\frac{3\Delta+1}{2}+k-\frac{3(d-1)}{2}-(2\Delta-\frac{d}{2}+k)-(\Delta+k)=2-d>0$, that is, $d<2$. In other words, the $\int G^3K$ integral is UV-divergent for values of the dimension $d$ equal or greater than 2, and UV-convergent otherwise, being able to safely take $\kappa=0$ in this case.\par

Remember that we are trying to compute the "sunset" diagram eq. (\ref{unnsunsetdig}). Replacing then the nice result just found for the $x_2$ integral back into the quantity we are trying to compute it reduces to:

\begin{equation}
    I_3(\vec{y_1},\vec{y_2}) = \frac{\lambda^2c_\Delta^2C_{G^3K}(\kappa)}{6}\int_{x_{1,0}=\varepsilon}d^{d+1}x_1\sqrt{g}\ \tilde{K}^\Delta(x_1,\vec{y_1})\tilde{K}^\Delta(x_1,\vec{y_2})
\end{equation}\

This resulting integral in $x_1$ is the known D-function eq. (\ref{regdfunc}), whose value is found to be given by eq. (\ref{d2funcreg2}). Therefore, replacing its known result we obtain that the "sunset" diagram can be written as:

\begin{equation}\label{sunsetdiagres}
    I_3(\vec{y_1},\vec{y_2}) = -\frac{2\nu c_\Delta}{\lvert\vec{y_1}-\vec{y_2}\rvert^{2\Delta}}\frac{\lambda^2C_{G^3K}(\kappa)}{6\nu}\ln{\Bigl(\frac{\varepsilon}{\lvert\vec{y_1}-\vec{y_2}\rvert}\Bigl)}
\end{equation}\

where we have written it conveniently for the upcoming study. Again, as we will see when we consider the complete correlator, we will realize that the result just found has exactly the same interpretation as before, that is, the expansion of a conformal anomaly up to this same order in the self-interacting coupling constant.

\subsubsection{The "Double Head" Diagram}

\begin{figure}[h]
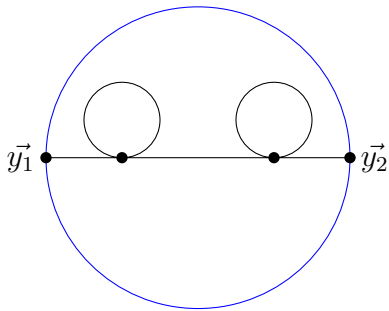

    \[\begin{wittendiagram}
    \draw (-2,0) node[vertex] -- (-1,0)
    (-1,0) node[vertex] -- (1,0)
    (1,0) node[vertex]
    (2,0) node[vertex] -- (1,0)
    
     (-2,0) node[left]{$\vec{y_1}$}
    (2,0) node[right]{$\vec{y_2}$}
    
    (-1,0.5) circle (0.5)
    (1,0.5) circle (0.5);
  \end{wittendiagram}\]
  \caption{Pictorial representation of the "double head" diagram.}
\end{figure}

In terms of the unnormalized bulk-boundary propagator we can rewrite the loop integral $I_4$ (which we will refer to it as the "double head" diagram) as:

\begin{equation}
    I_4(\vec{y_1},\vec{y_2}) = \frac{\lambda^2c_\Delta^2G^2_\kappa(1)}{4}\int_{x_{1,0}=\varepsilon}d^{d+1}x_1\sqrt{g}\ \tilde{K}^\Delta(x_1,\vec{y_1})\int_{x_{2,0}=\varepsilon}d^{d+1}x_2\sqrt{g}\ G_\kappa(x_1,x_2)\tilde{K}^\Delta(x_2,\vec{y_2})
\end{equation}\

where we used that the regularized bulk-bulk propagator being evaluated at coincident points is just a constant. These resulting integrals in $x_1$ and $x_2$, as it is discussed in detail in section B.6 of Appendix B, a simple power counting suggests that, if it were not for the $\varepsilon$-regulator, they would be divergent in the lower limits of integration of their radial coordinates $x_{1,0}$ and $x_{2,0}$. Precisely the role of this regulator is not only to tame these divergences but also to capture the correct behavior of these integrals hidden in them. The complete study of this quantity can be found in the section of the appendix mentioned above, concluding in its value in eq. (\ref{scintkgk}). Using this value then in our present case we find that the result of $I_4$ is given by:

\begin{equation}\label{doubheaddiagres}
    I_4(\vec{y_1},\vec{y_2}) = \frac{2\nu c_\Delta}{\lvert\vec{y_1}-\vec{y_2}\rvert^{2\Delta}}\frac{\lambda^2G^2_\kappa(1)}{8\nu^2}\ln^2{\Bigl(\frac{\varepsilon}{\lvert\vec{y_1}-\vec{y_2}\rvert}\Bigl)}
\end{equation}\

where we have written it conveniently for the upcoming study. As we will see next when we consider the complete correlator, we will realize that the result just found contributes to the exact same interpretation as before, that is, the expansion of a conformal anomaly up to this same order in $\lambda$.

\subsubsection{Correlator Renormalization}

Replacing the results just found then for the "head" diagram, the "eight" diagram, the "sunset" diagram and the "double head" diagram back into the holographic 2-point function, we find that they can be factorized into the form:

\begin{align}\label{2pfcorrrenst1}
    \langle O_\Delta(\vec{y_1})O_\Delta(\vec{y_2})\rangle_{\text{CFT}} =& \langle O_\Delta(\vec{y_1})O_\Delta(\vec{y_2})\rangle_{\text{CFT,con}}\nonumber\\
    &\hspace{-2cm}= \frac{2\nu c_\Delta}{\lvert\vec{y_1}-\vec{y_2}\rvert^{2\Delta}}\Bigl\{1-\frac{1}{\nu}\Bigl[-\frac{\lambda G_\kappa(1)}{2}+\frac{\lambda^2G_\kappa(1)C_{G^2}(\kappa)}{4}+\frac{\lambda^2C_{G^3K}(\kappa)}{6}\Bigl]\ln{\Bigl(\frac{\varepsilon}{\lvert\vec{y_1}-\vec{y_2}\rvert}\Bigl)}\nonumber\\
    &\hspace{0.8cm}+\frac{1}{2\nu^2}\Bigl[\frac{\lambda^2 G^2_\kappa(1)}{4}\Bigl]\ln^2{\Bigl(\frac{\varepsilon}{\lvert\vec{y_1}-\vec{y_2}\rvert}\Bigl)}\Bigl\} + \mathcal{O}(\lambda^3)
\end{align}\

Notice how the one-particle irreducible (1PI) diagrams are proportional to a single logarithm, while the one-time reducible (1TR) is proportional to a logarithm squared. Tracing the steps back this is due to the nice results of the integrals eqs. (\ref{intgnkpropk}) and (\ref{intgnpropc}), which for all the 1PI integrals allow to reduce them to the form $\int KK$ and for all the 1TR integrals to the form $\int K\int GK$. The terms inside the first square bracket in eq. (\ref{2pfcorrrenst1}) correspond exactly to the overall factor of this resulting $\int KK$ integral. As we did for the $\Phi^3$ theory then, this fact motivates us to define these quantities as the 1PI contributions to the correlators $\Pi(\kappa)$:

\begin{equation}\label{1piphi4}
    \Pi(\kappa) = -\frac{\lambda G_\kappa(1)}{2}+\frac{\lambda^2G_\kappa(1)C_{G^2}(\kappa)}{4}+\frac{\lambda^2C_{G^3K}(\kappa)}{6}
\end{equation}\

Moreover, the term inside the second square bracket in eq. (\ref{2pfcorrrenst1}) correspond exactly to the overall factor of the resulting $\int K\int GK$ integral. But notice that, up to this order in $\lambda$, it is nothing but the square of the 1PI contributions $\Pi(\kappa)$! This gives us a nice picture of the quantum corrections to the correlators as an expansion in $\Pi(\kappa)$ similar to those found in regular QFT theories, with the difference that in our current case the external legs of this diagrammatic expansion have been replaced by bulk-boundary propagators. In terms of $\Pi(\kappa)$ then, we can rewrite the 2-point functions as:

\begin{align}
    \langle O_\Delta(\vec{y_1})O_\Delta(\vec{y_2})\rangle_{\text{CFT}} =& \langle O_\Delta(\vec{y_1})O_\Delta(\vec{y_2})\rangle_{\text{CFT,con}}\nonumber\\
    =& \frac{2\nu c_\Delta}{\lvert\vec{y_1}-\vec{y_2}\rvert^{2\Delta}}\Bigl[1-\frac{\Pi(\kappa)}{\nu}\ln{\Bigl(\frac{\varepsilon}{\lvert\vec{y_1}-\vec{y_2}\rvert}\Bigl)}+\frac{\Pi^2(\kappa)}{2\nu^2}\ln^2{\Bigl(\frac{\varepsilon}{\lvert\vec{y_1}-\vec{y_2}\rvert}\Bigl)}\Bigl] + \mathcal{O}(\lambda^3)
\end{align}\

The terms inside the square bracket, up to this same order in $\lambda$, correspond to the known Taylor series of an exponent:

\begin{equation}
    1-\frac{\Pi(\kappa)}{\nu}\ln{\Bigl(\frac{\varepsilon}{\lvert\vec{y_1}-\vec{y_2}\rvert}\Bigl)}+\frac{\Pi^2(\kappa)}{2\nu^2}\ln^2{\Bigl(\frac{\varepsilon}{\lvert\vec{y_1}-\vec{y_2}\rvert}\Bigl)} = \Bigl(\frac{\varepsilon}{\lvert\vec{y_1}-\vec{y_2}\rvert}\Bigl)^{-\frac{\Pi(\kappa)}{\nu}}+\mathcal{O}(\lambda^3)
\end{equation}\

This fact allows us to express the regularized holographic 2-point functions in the nice compact form:

\begin{equation}\label{2pf2pfphi41}
    \langle O_\Delta(\vec{y_1})O_\Delta(\vec{y_2})\rangle_{\text{CFT}} =\langle O_\Delta(\vec{y_1})O_\Delta(\vec{y_2})\rangle_{\text{CFT,con}}= \varepsilon^{-\frac{\Pi(\kappa)}{\nu}}\frac{2\nu c_\Delta}{\lvert\vec{y_1}-\vec{y_2}\rvert^{2\Delta-\frac{\Pi(\kappa)}{\nu}}} + \mathcal{O}(\lambda^3)
\end{equation}\

With the correlators written in this form, it is direct to see that the effects of the quantum corrections coming from the off-shell part of the AdS path integral to the 2-point function found previously under the classical approximation of the AdS/CFT correspondence are the same as those obtained for a $\Phi^3$ theory. Indeed, they contribute with an overall rescaling to the correlator along with a shift in its scaling dimension. Now, of course as we take the understood limits $\varepsilon\rightarrow0$ and $\kappa\rightarrow0$ this correlator becomes divergent so it is necessary the introduction of a delicate renormalization scheme in order to absorb the respective infinities. For a $\Phi^3$ theory on AdS we saw that the nice form of the regularized correlator allowed us to renormalize it in exactly the same spirit as it is done for ordinary QFTs, through the redefinition of the bulk's theory parameters. It is satisfactory to find then that the exact same scheme also works for the present case. Take for example the anomalous dimension of the correlators, which is expected to be renormalized through a redefinition of the bulk's mass parameter $m^2$. Indeed, redefining this parameter in the AdS bulk action simply as $m^2\rightarrow m^2+\delta m^2$ (where the counterterm is expected to be of order $\delta m^2=\mathcal{O}(\lambda)$) adds, up to order $\lambda^2$, new counterterm interactions to the holographic 2-point function eq. (\ref{2pf2pfphi41}) of the form:

\begin{align}
    \langle O_\Delta(\vec{y_1})O_\Delta(\vec{y_2})\rangle_{\text{CFT}} &=\langle O_\Delta(\vec{y_1})O_\Delta(\vec{y_2})\rangle_{\text{CFT,con}}\nonumber\\
    &= \varepsilon^{-\frac{\Pi(\kappa)}{\nu}}\frac{2\nu c_\Delta}{\lvert\vec{y_1}-\vec{y_2}\rvert^{2\Delta-\frac{\Pi(\kappa)}{\nu}}} -\delta m^2\int_\varepsilon K(x,\vec{y_1})K(x,\vec{y_2})\nonumber\\
    &\hspace{0.5cm}+\delta m^2\frac{\lambda}{2}\int_\varepsilon K(x_1,\vec{y_1})G_\kappa(x_1,x_1)\int_\varepsilon G_\kappa(x_1,x_2)K(x_2,\vec{y_2})+(\vec{y_1}\leftrightarrow\vec{y_2})\nonumber\\
    &\hspace{0.5cm}+(\delta m^2)^2\int_\varepsilon K(x_1,\vec{y_1})\int_\varepsilon G_\kappa(x_1,x_2)K(x_2,\vec{y_2})+\mathcal{O}(\lambda^3)
\end{align}\

These new contributions to the correlators can be solved in terms of eqs. (\ref{d2funcreg2}) and (\ref{scintkgk}):

\begin{align}
    -\delta m^2\int_\varepsilon K(x,\vec{y_1})K(x,\vec{y_2}) &= \frac{2\nu c_\Delta}{\lvert\vec{y_1}-\vec{y_2}\rvert^{2\Delta}}\frac{\delta m^2}{\nu}\ln{\Bigl(\frac{\varepsilon}{\lvert\vec{y_1}-\vec{y_2}\rvert}\Bigl)}\nonumber\\
    \delta m^2\frac{\lambda}{2}\int_\varepsilon K(x_1,\vec{y_1})G_\kappa(x_1,x_1)\int_\varepsilon G_\kappa(x_1,x_2)K(x_2,\vec{y_2}) &= \frac{2\nu c_\Delta}{\lvert\vec{y_1}-\vec{y_2}\rvert^{2\Delta}}\frac{\delta m^2\lambda G_\kappa(1)}{4\nu^2}\ln^2{\Bigl(\frac{\varepsilon}{\lvert\vec{y_1}-\vec{y_2}\rvert}\Bigl)}\nonumber\\
    (\delta m^2)^2\int_\varepsilon K(x_1,\vec{y_1})\int_\varepsilon G_\kappa(x_1,x_2)K(x_2,\vec{y_2}) &= \frac{2\nu c_\Delta}{\lvert\vec{y_1}-\vec{y_2}\rvert^{2\Delta}}\frac{(\delta m^2)^2}{2\nu^2}\ln^2{\Bigl(\frac{\varepsilon}{\lvert\vec{y_1}-\vec{y_2}\rvert}\Bigl)}
\end{align}

which ultimately results in:

\begin{align}
    \langle O_\Delta(\vec{y_1})O_\Delta(\vec{y_2})\rangle_{\text{CFT}} =&\langle O_\Delta(\vec{y_1})O_\Delta(\vec{y_2})\rangle_{\text{CFT,con}}\nonumber\\
    =&\varepsilon^{\frac{\delta m^2-\Pi(\kappa)}{\nu}}\frac{2\nu c_\Delta}{\lvert\vec{y_1}-\vec{y_2}\rvert^{2\Delta+\frac{\delta m^2-\Pi(\kappa)}{\nu}}} + \mathcal{O}(\lambda^3)
\end{align}\

Therefore, denoting the 1PI contributions $\Pi(\kappa)$ as $\Pi(\kappa) = \Pi_\infty(\kappa)+\Pi_0(\kappa)$, where all its UV-divergent terms are contained in $\Pi_\infty(\kappa)$, the infinities present in the correlators coming from the ultraviolet divergences of the loops integrals can be renormalized away through the convenient choice of the counterterm $\delta m^2$ as:

\begin{equation}
    \delta m^2 = \Pi_\infty(\kappa)
\end{equation}\

resulting in the UV-renormalized holographic 2-point function:

\begin{equation}
    \langle O_\Delta(\vec{y_1})O_\Delta(\vec{y_2})\rangle_{\text{CFT}} =\langle O_\Delta(\vec{y_1})O_\Delta(\vec{y_2})\rangle_{\text{CFT,con}}=\varepsilon^{-\frac{\Pi_0(0)}{\nu}}\frac{2\nu c_\Delta}{\lvert\vec{y_1}-\vec{y_2}\rvert^{2\Delta-\frac{\Pi_0(0)}{\nu}}} + \mathcal{O}(\lambda^3)
\end{equation}\

where we safely took the limit $\kappa=0$. We are still half way in the renormalization process as we still have to deal with the infrared divergence of the correlators. However, as we saw for the $\Phi^3$ theory, this divergence is expected to be renormalized through a redefinition of the bulk's field $\Phi(x)$. Indeed, redefining it in the AdS bulk action simply as $\Phi(x)\rightarrow\sqrt{Z(\lambda)}\Phi(x)$ adds a new factor to the holographic 2-point function of the form:

\begin{equation}
    \langle O_\Delta(\vec{y_1})O_\Delta(\vec{y_2})\rangle_{\text{CFT}} =\langle O_\Delta(\vec{y_1})O_\Delta(\vec{y_2})\rangle_{\text{CFT,con}}=\frac{\varepsilon^{-\frac{\Pi_0(0)}{\nu}}}{Z(\lambda)}\frac{2\nu c_\Delta}{\lvert\vec{y_1}-\vec{y_2}\rvert^{2\Delta-\frac{\Pi_0(0)}{\nu}}} + \mathcal{O}(\lambda^3)
\end{equation}\

Therefore, the infinities present in the correlators coming from the infrared divergences of the loops integrals can be renormalized away through the convenient choice of the counterterm $Z(\lambda)$ as:

\begin{equation}
    Z(\lambda)=\varepsilon^{-\frac{\Pi_0(0)}{\nu}}=1-\frac{\Pi_0(0)}{\nu}\ln{(\varepsilon)}+\frac{\Pi_0(0)^2}{2\nu^2}\ln^2{(\varepsilon)}+\mathcal{O}(\lambda^3)
\end{equation}\

resulting in both IR and UV renormalized holographic 2-point functions:

\begin{equation}
    \langle O_\Delta(\vec{y_1})O_\Delta(\vec{y_2})\rangle_{\text{CFT}} =\langle O_\Delta(\vec{y_1})O_\Delta(\vec{y_2})\rangle_{\text{CFT,con}}=\frac{2\nu c_\Delta}{\lvert\vec{y_1}-\vec{y_2}\rvert^{2\Delta-\frac{\Pi_0(0)}{\nu}}} + \mathcal{O}(\lambda^3)
\end{equation}\

where the limits $\varepsilon=\kappa=0$ have been taken and where $\Pi_0(0)$ denote the UV-finite part of the 1P1 contributions $\Pi(\kappa)$.

\subsection{4-Point Function}

The regularized holographic 4-point functions dual to a $\Phi^4$ self-interacting theory on AdS are given by:

\begin{align}
    &\langle O_\Delta(\vec{y_1})O_\Delta(\vec{y_2})O_\Delta(\vec{y_3})O_\Delta(\vec{y_4})\rangle_{\text{CFT}} =\frac{2\nu c_\Delta}{\lvert\vec{y_1}-\vec{y_2}\rvert^{2\Delta}}\frac{2\nu c_\Delta}{\lvert\vec{y_3}-\vec{y_4}\rvert^{2\Delta}}+(\vec{y_2}\leftrightarrow\vec{y_3})+(\vec{y_2}\leftrightarrow\vec{y_4})\nonumber\\
    &\hspace{2cm}-\lambda\int K(x_1,\vec{y_1})K(x_1,\vec{y_2})K(x_1,\vec{y_3})K(x_1,\vec{y_4})\nonumber\\
    &\hspace{2cm}+\frac{\lambda^2}{2}\int_\varepsilon\int_\varepsilon K(x_1,\vec{y_1})K(x_2,\vec{y_2})K(x_2,\vec{y_3})K(x_2,\vec{y_4})G_\kappa(x_1,x_1)G_\kappa(x_1,x_2)\times4\nonumber\\
    &\hspace{2cm}+\frac{\lambda^2}{2}\int_\varepsilon\int_\varepsilon K(x_1,\vec{y_1})K(x_1,\vec{y_2})K(x_2,\vec{y_3})K(x_2,\vec{y_4})G^2_\kappa(x_1,x_2)\times3\nonumber\\
    &\hspace{2cm}+\frac{\lambda^2}{4}\int_\varepsilon\int_\varepsilon K(x_1,\vec{y_1})K(x_1,\vec{y_2})K(x_2,\vec{y_3})K(x_2,\vec{y_4})G_\kappa(x_1,x_1)G_\kappa(x_2,x_2)\times3\nonumber\\
    &\hspace{2cm}+\frac{2\nu c_\Delta}{\lvert\vec{y_1}-\vec{y_2}\rvert^{2\Delta}}\Bigl[-\frac{\lambda}{2}\int_\varepsilon K(x_1,\vec{y_3})K(x_1,\vec{y_4})G_\kappa(x_1,x_1)\nonumber\\
    &\hspace{4cm}+\frac{\lambda^2}{4}\int_\varepsilon\int_\varepsilon K(x_1,\vec{y_3})K(x_1,\vec{y_4})G^2_\kappa(x_1,x_2)G_\kappa(x_2,x_2)\nonumber\\
    &\hspace{4cm}+\frac{\lambda^2}{6}\int_\varepsilon\int_\varepsilon K(x_1,\vec{y_3})K(x_2,\vec{y_4})G^3_\kappa(x_1,x_2)\nonumber\\
    &\hspace{4cm}+\frac{\lambda^2}{4}\int_\varepsilon\int_\varepsilon K(x_1,\vec{y_3})K(x_2,\vec{y_4})G_\kappa(x_1,x_1)G_\kappa(x_1,x_2)G_\kappa(x_2,x_2)\Bigl]\times6 + \mathcal{O}(\lambda^3)\nonumber\\
    &\langle O_\Delta(\vec{y_1})O_\Delta(\vec{y_2})O_\Delta(\vec{y_3})O_\Delta(\vec{y_4})\rangle_{\text{CFT,con}} = -\lambda\int K(x_1,\vec{y_1})K(x_1,\vec{y_2})K(x_1,\vec{y_3})K(x_1,\vec{y_4})\nonumber\\
    &\hspace{3cm}+\frac{\lambda^2}{2}\int_\varepsilon\int_\varepsilon K(x_1,\vec{y_1})K(x_2,\vec{y_2})K(x_2,\vec{y_3})K(x_2,\vec{y_4})G_\kappa(x_1,x_1)G_\kappa(x_1,x_2)\times4\nonumber\\
    &\hspace{3cm}+\frac{\lambda^2}{2}\int_\varepsilon\int_\varepsilon K(x_1,\vec{y_1})K(x_1,\vec{y_2})K(x_2,\vec{y_3})K(x_2,\vec{y_4})G^2_\kappa(x_1,x_2)\times3 + \mathcal{O}(\lambda^3)
\end{align}\

from where we see the quantum corrections they receive coming from the loop integrals:

\begin{equation}
    I_1(\vec{y_1},\vec{y_2},\vec{y_3},\vec{y_4}) =\frac{\lambda^2}{2}\int_\varepsilon\int_\varepsilon K(x_1,\vec{y_1})K(x_2,\vec{y_2})K(x_2,\vec{y_3})K(x_2,\vec{y_4})G_\kappa(x_1,x_1)G_\kappa(x_1,x_2)\times4
\end{equation}

\begin{equation}
    I_2(\vec{y_1},\vec{y_2},\vec{y_3},\vec{y_4}) =\frac{\lambda^2}{2}\int_\varepsilon\int_\varepsilon K(x_1,\vec{y_1})K(x_1,\vec{y_2})K(x_2,\vec{y_3})K(x_2,\vec{y_4})G^2_\kappa(x_1,x_2)\times3
\end{equation}

\begin{equation}
    I_3(\vec{y_1},\vec{y_2},\vec{y_3},\vec{y_4}) =\frac{\lambda^2}{4}\int_\varepsilon\int_\varepsilon K(x_1,\vec{y_1})K(x_1,\vec{y_2})K(x_2,\vec{y_3})K(x_2,\vec{y_4})G_\kappa(x_1,x_1)G_\kappa(x_2,x_2)\times3
\end{equation}

\begin{align}
    I_4(\vec{y_1},\vec{y_2},\vec{y_3},\vec{y_4}) =&\frac{2\nu c_\Delta}{\lvert\vec{y_1}-\vec{y_2}\rvert^{2\Delta}}\Bigl[-\frac{\lambda}{2}\int_\varepsilon K(x_1,\vec{y_3})K(x_1,\vec{y_4})G_\kappa(x_1,x_1)\nonumber\\
    &\hspace{2.1cm}+\frac{\lambda^2}{4}\int_\varepsilon\int_\varepsilon K(x_1,\vec{y_3})K(x_1,\vec{y_4})G^2_\kappa(x_1,x_2)G_\kappa(x_2,x_2)\nonumber\\
    &\hspace{2.1cm}+\frac{\lambda^2}{6}\int_\varepsilon\int_\varepsilon K(x_1,\vec{y_3})K(x_2,\vec{y_4})G^3_\kappa(x_1,x_2)\nonumber\\
    &\hspace{2.1cm}+\frac{\lambda^2}{4}\int_\varepsilon\int_\varepsilon K(x_1,\vec{y_3})K(x_2,\vec{y_4})G_\kappa(x_1,x_1)G_\kappa(x_1,x_2)G_\kappa(x_2,x_2)\Bigl]\times6
\end{align}\

In order to compute the complete 4-point functions up to this order in the expansion of $\lambda$, we will proceed then to compute these quantities. We will start by studying $I_1$, then $I_2$, then $I_3$ and lastly, $I_4$.

\subsubsection{The Reducible "Head" Diagram}

\begin{figure}[h]
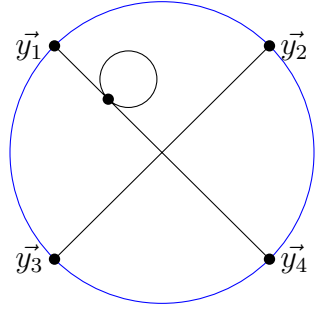

    \[\begin{wittendiagram}
    \draw (-1.4142,1.4142) node[vertex] -- (1.4142,-1.4142)
    (-1.4142,-1.4142) node[vertex] -- (1.4142,1.4142)
    (1.4142,1.4142) node[vertex]
    (1.4142,-1.4142) node[vertex]
    (-0.7071,0.7071) node[vertex]
    
    (-1.4142,1.4142) node[left]{$\vec{y_1}$}
    (-1.4142,-1.4142) node[left]{$\vec{y_3}$}
    (1.4142,1.4142) node[right]{$\vec{y_2}$}
    (1.4142,-1.4142) node[right]{$\vec{y_4}$}
    
    (-0.4419,0.9722) circle (0.375);
  \end{wittendiagram}\]
  \caption{Pictorial representation of the reducible "head" diagram.}
\end{figure}

In terms of the unnormalized bulk-boundary propagator we can rewrite the loop integral $I_1$ (which we will refer to it as the reducible "head" diagram) as:

\begin{align}
    I_1(\vec{y_1},\vec{y_2},\vec{y_3},\vec{y_4}) =&\frac{\lambda^2c_\Delta^4G_\kappa(1)}{2}\int_{x_{2,0}=\varepsilon}d^{d+1}x_2\sqrt{g}\ \tilde{K}^\Delta(x_2,\vec{y_2})\tilde{K}^\Delta(x_2,\vec{y_3})\tilde{K}^\Delta(x_2,\vec{y_4})\nonumber\\
    &\hspace{-1cm}\times\int_{x_{1,0}=\varepsilon}d^{d+1}x_1\sqrt{g}\ G_\kappa(x_2,x_1)\tilde{K}^\Delta(x_1,\vec{y_1})+(\vec{y_1}\leftrightarrow\vec{y_2})+(\vec{y_1}\leftrightarrow\vec{y_3})+(\vec{y_1}\leftrightarrow\vec{y_4})
\end{align}

where we used that the regularized bulk-bulk propagator being evaluated at coincident points is just a constant. As we have seen repeatedly, the value of the $x_1$ integral is given by eq. (\ref{scintgk}). Therefore, replacing its known result:

\begin{align}
    I_1(\vec{y_1},\vec{y_2},\vec{y_3},\vec{y_4}) =&\frac{\lambda c_\Delta^4\Pi(\kappa)}{2\nu}\int d^{d+1}x_1\sqrt{g}\ \tilde{K}^\Delta(x_1,\vec{y_1})\tilde{K}^\Delta(x_1,\vec{y_2})\tilde{K}^\Delta(x_1,\vec{y_3})\tilde{K}^\Delta(x_1,\vec{y_4})\nonumber\\
    &\hspace{1.5cm}\times\ln{\bigl(\varepsilon\tilde{K}(x_1,\vec{y_1})\bigl)}+(\vec{y_1}\leftrightarrow\vec{y_2})+(\vec{y_1}\leftrightarrow\vec{y_3})+(\vec{y_1}\leftrightarrow\vec{y_4})+\mathcal{O}(\lambda^3)
\end{align}\

where we wrote, up to order $\lambda^2$, the constant factors in terms of $\Pi(\kappa)$ (eq. (\ref{1piphi4})), called the integrated variable $x_2\rightarrow x_1$, and performed a simple power counting in the radial coordinate $x_{1,0}$ as it approaches the boundaries, realizing that the integral is IR-convergent and therefore allowing us to simply take the regulator $\varepsilon$ equal to 0. As we will see shortly, instead of trying to compute this integral it will turn out to be much simpler to keep it in this form for now, since later when grouping it with the other contributions to the 4-point function it will result in a natural factorization into a known integral. Let us proceed then to study the second loop contribution, $I_2$.

\subsubsection{The "Scalar Exchange" Diagram}

\begin{figure}[h]
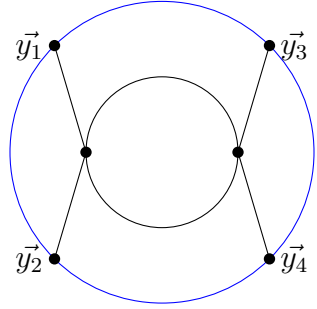

    \[\begin{wittendiagram}
    \draw (-1.4142,1.4142) node[vertex] -- (-1,0)
    (-1.4142,-1.4142) node[vertex] -- (-1,0)
    (1.4142,1.4142) node[vertex] -- (1,0)
    (1.4142,-1.4142) node[vertex] -- (1,0)
    (-1,0) node[vertex]
    (1,0) node[vertex]
    
    (-1.4142,1.4142) node[left]{$\vec{y_1}$}
    (-1.4142,-1.4142) node[left]{$\vec{y_2}$}
    (1.4142,1.4142) node[right]{$\vec{y_3}$}
    (1.4142,-1.4142) node[right]{$\vec{y_4}$}
    
    (0,0) circle (1);
  \end{wittendiagram}\]
  \caption{Pictorial representation of the "scalar exchange" diagram.}
\end{figure}

In terms of the unnormalized bulk-boundary propagator we can rewrite the loop integral $I_2$ (which we will refer to it as the "scalar exchange" diagram) as:

\begin{align}\label{qsediag1}
    I_2(\vec{y_1},\vec{y_2},\vec{y_3},\vec{y_4}) =&\frac{\lambda^2c_\Delta^4}{2}\int_{x_{1,0}=\varepsilon}d^{d+1}x_1\sqrt{g}\ \tilde{K}^\Delta(x_1,\vec{y_1})\tilde{K}^\Delta(x_1,\vec{y_2})\nonumber\\
    &\hspace{-1cm}\times\int_{x_{2,0}=\varepsilon}d^{d+1}x_2\sqrt{g}\ G^2_\kappa(x_1,x_2)\tilde{K}^\Delta(x_2,\vec{y_3})\tilde{K}^\Delta(x_2,\vec{y_4})+(\vec{y_2}\leftrightarrow\vec{y_3})+(\vec{y_2}\leftrightarrow\vec{y_4})
\end{align}\

Continuing under the same approach as before, we will solve the $x_2$ integral first hoping that the remaining integral in $x_1$ will be familiar. As we will see, this will be indeed the case in a similar way as how it happens for the tree-level scalar exchange diagram present in the $\Phi^3$ theory, eq. (\ref{phi34pfclass1}). The first question that we will be interested in answering however is when in this $x_2$ integral we can safely take the IR-regulator $\varepsilon$ equal to 0. The infrared convergence region of this integral can be seen directly by studying how its integrand behaves as it approaches the boundary of AdS. In this case, using the explicit form of the metric and the known expansions of the propagators, we obtain that:

\begin{equation}
    \sqrt{g}\ G^2_\kappa(x_1,x_2)\tilde{K}^\Delta(x_2,\vec{y_3})\tilde{K}^\Delta(x_2,\vec{y_4}) \underset{x_{2,0}\rightarrow0}{\sim} x_{2,0}^{-d-1}x_{2,0}^{2\Delta}x_{2,0}^{d-\Delta}x_{2,0}^\Delta= x_{2,0}^{-1+2\Delta}
\end{equation}\

From here we conclude that the $x_2$ integral will be IR-convergent as long as $\Delta$ is a positive number, but in this work this will always be the case, implying that we can always take in this integral $\varepsilon=0$.\par
For the computation of the $x_2$ integral it will turn out to be useful to use the representation eq. (\ref{Grepdiv}) of the bulk-bulk propagator, where all its UV-divergence has been extracted out from the hypergeometric function, which as we saw in section 4.1.6, its square can be further expressed as in eq. (\ref{reggsquare}). Therefore, the integral of $\int G^2KK$ with the bulk-bulk propagator written in this form can be expressed as:

\begin{align}
    \int d^{d+1}x_2\sqrt{g}\ G^2_\kappa(x_1,x_2)\tilde{K}^\Delta(x_2,\vec{y_3})\tilde{K}^\Delta(x_2,\vec{y_4}) = &\Bigl(\frac{2^{-\Delta}c_\Delta}{2\nu}\Bigl)^2\sum_{k=0}^\infty a^{(2)}_k\sum_{l=0}^\infty\frac{(d-1)_l}{l!}\Bigl(\frac{1}{1+\kappa}\Bigl)^{2\Delta+2k+2l}\nonumber\\
    &\hspace{-1cm}\times\int d^{d+1}x_2\sqrt{g}\ \xi^{2\Delta+2k+2l}\tilde{K}^\Delta(x_2,\vec{y_3})\tilde{K}^\Delta(x_2,\vec{y_4})
\end{align}\

The complete study of this type of integrals can be found in section B.4 of Appendix B, concluding in its value in eq. (\ref{xikkform}). Using this formula then for the particular values $\Delta_1=2\Delta+2k+2l$, $\Delta_2=\Delta$ and $\Delta_3=\Delta$, we find that the result of the $x_2$ integral is given by:

\begin{align}\label{qsediag2}
    \int d^{d+1}x_2\sqrt{g}\ G^2_\kappa(x_1,x_2)\tilde{K}^\Delta(x_2,\vec{y_3})\tilde{K}^\Delta(x_2,\vec{y_4})&\nonumber\\
    &\hspace{-6cm}=\tilde{K}^{\Delta}(x_1,\vec{y_3})\tilde{K}^{\Delta}(x_1,\vec{y_4})\pi^{\frac{d+1}{2}}\Bigl(\frac{2^{-\Delta}c_\Delta}{2\nu}\Bigl)^2\sum_{l=0}^\infty\frac{\Gamma(2\Delta-\frac{d}{2}+l)\Gamma(\Delta+l)}{\Gamma(\Delta+\frac{1}{2}+l)\Gamma(2\Delta+l)}\Bigl(\frac{1}{1+\kappa}\Bigl)^{2\Delta+2l}\nonumber\\
    &\hspace{-5.5cm}\times\ _2F_1\Bigl(\Delta,\Delta;2\Delta+l;1-\tilde{K}(x_1,\vec{y_3})\tilde{K}(x_1,\vec{y_4})\lvert\vec{y_{34}}\rvert^2\Bigl)\sum_{k=0}^{l} a^{(2)}_k\frac{(d-1)_{l-k}}{(l-k)!}
\end{align}\

where we used the double sum property $\sum_{l=0}^\infty\sum_{k=0}^\infty a_{l,k}=\sum_{l=0}^\infty\sum_{k=0}^l a_{l-k,k}$. A nice clue on how to proceed with the calculations is to note that if we are able to write this hypergeometric function being summed as a power series in $\tilde{K}(x_1,\vec{y_3})\tilde{K}(x_1,\vec{y_4})\lvert\vec{y_{34}}\rvert^2$ the value of the $x_2$ integral would consist of a sum of 2 bulk-boundary propagators of different scaling dimensions, value which when replaced back into the original integral eq. (\ref{qsediag1}) would result in a sum of integrals of 4 bulk-boundary propagators in the $x_1$ variable. But we have already discussed these integrals involving only bulk-boundary propagators, these are precisely the D-functions reviewed in detail in Appendix A. Therefore, if we are able to write the hypergeometric function in eq. (\ref{qsediag2}) as a series in the bulk-boundary propagators, we can solve for eq. (\ref{qsediag1}) in terms of D-functions similar to how we did for the scalar exchange diagram present in the $\Phi^3$ theory. But for our present case this can be easily achieved using the known linear transformation of the hypergeometric function \cite{HypFunc3}:

\begin{align}
    _2F_1(a,b;a+b+n;z)=&\frac{\Gamma(n)\Gamma(a+b+n)}{\Gamma(a+n)\Gamma(b+n)}\sum_{i=0}^{n-1}\frac{(a)_i(b)_i}{(1-n)_i\ i!}(1-z)^i\nonumber\\
    &\hspace{-3cm}+\frac{\Gamma(a+b+n)}{\Gamma(a)\Gamma(b)}(z-1)^n\sum_{i=0}^\infty\frac{(a+n)_i(b+n)_i}{(i+n)!\ i!}\nonumber\\
    &\hspace{-2cm}\times\bigl[-\ln{(1-z)}+\psi(i+1)+\psi(i+n+1)-\psi(i+n+a)-\psi(i+n+b)\bigl](1-z)^i
\end{align}\

for some integer value $n$, where $\psi(x)$ is the digamma function. This transformation for the particular values $a=\Delta$, $b=\Delta$, $n=l$ and $z=1-\tilde{K}(x_1,\vec{y_3})\tilde{K}(x_1,\vec{y_4})\lvert\vec{y_{34}}\rvert^2$, let us rewrite eq. (\ref{qsediag2}) as:

\begin{align}\label{qsediag3}
    \int d^{d+1}x_2\sqrt{g}\ G^2_\kappa(x_1,x_2)\tilde{K}^\Delta(x_2,\vec{y_3})\tilde{K}^\Delta(x_2,\vec{y_4})&=\tilde{K}^{\Delta}(x_1,\vec{y_3})\tilde{K}^{\Delta}(x_1,\vec{y_4})\pi^{\frac{d+1}{2}}\Bigl(\frac{2^{-\Delta}c_\Delta}{2\nu}\Bigl)^2\nonumber\\
    &\hspace{-7cm}\times\Bigl\{\sum_{l=0}^\infty\frac{\Gamma(2\Delta-\frac{d}{2}+l)\Gamma(l)}{\Gamma(\Delta+\frac{1}{2}+l)\Gamma(\Delta+l)}\Bigl(\frac{1}{1+\kappa}\Bigl)^{2\Delta+2l}\sum_{i=0}^{l-1}\frac{(\Delta)_i(\Delta)_i}{(1-l)_i\ i!}\bigl[\tilde{K}(x_1,\vec{y_3})\tilde{K}(x_1,\vec{y_4})\lvert\vec{y_{34}}\rvert^2\bigl]^i\nonumber\\
    &\hspace{-6.2cm}+\frac{1}{\Gamma(\Delta)^2}\sum_{l=0}^\infty\frac{\Gamma(2\Delta-\frac{d}{2}+l)\Gamma(\Delta+l)}{\Gamma(\Delta+\frac{1}{2}+l)}\Bigl(\frac{1}{1+\kappa}\Bigl)^{2\Delta+2l}\bigl[-\tilde{K}(x_1,\vec{y_3})\tilde{K}(x_1,\vec{y_4})\lvert\vec{y_{34}}\rvert^2\bigl]^l\nonumber\\
    &\hspace{-5.5cm}\times\sum_{i=0}^\infty\frac{(\Delta+l)_i(\Delta+l)_i}{(i+l)!\ i!}\bigl[-\ln{\bigl(\tilde{K}(x_1,\vec{y_3})\tilde{K}(x_1,\vec{y_4})\lvert\vec{y_{34}}\rvert^2\bigl)}+\psi(i+1)+\psi(i+l+1)\nonumber\\
    &\hspace{-3cm}-2\psi(i+l+\Delta)\bigl]\bigl[\tilde{K}(x_1,\vec{y_3})\tilde{K}(x_1,\vec{y_4})\lvert\vec{y_{34}}\rvert^2\bigl]^i\Bigl\}\sum_{k=0}^{l} a^{(2)}_k\frac{(d-1)_{l-k}}{(l-k)!}
\end{align}\

Using known properties of sums and Pochhammer symbols, the first sum can be rewritten as:

\begin{align}\label{qphi44pfsum1}
    &\sum_{l=0}^\infty\frac{\Gamma(2\Delta-\frac{d}{2}+l)\Gamma(l)}{\Gamma(\Delta+\frac{1}{2}+l)\Gamma(\Delta+l)}\Bigl(\frac{1}{1+\kappa}\Bigl)^{2\Delta+2l}\sum_{i=0}^{l-1}\frac{(\Delta)_i(\Delta)_i}{(1-l)_i\ i!}\bigl[\tilde{K}(x_1,\vec{y_3})\tilde{K}(x_1,\vec{y_4})\lvert\vec{y_{34}}\rvert^2\bigl]^i\nonumber\\
    &\times\sum_{k=0}^{l} a^{(2)}_k\frac{(d-1)_{l-k}}{(l-k)!}=\sum_{i=0}^\infty a_i(\kappa)\bigl[\tilde{K}(x_1,\vec{y_3})\tilde{K}(x_1,\vec{y_4})\lvert\vec{y_{34}}\rvert^2\bigl]^i
\end{align}\

where we used that the $l=0$ term is 0 and redefined $\sum_{l=1}^\infty a_l\rightarrow\sum_{l=0}^\infty a_{l+1}$, used that $\sum_{l=0}^\infty\sum_{i=0}^l a_{l,i}=\sum_{l=0}^\infty\sum_{i=0}^\infty a_{l+i,i}$, rewrote and simplified terms accordingly and finally defined the coefficient $a_i(\kappa)$:

\begin{align}
    a_i(\kappa) = &\frac{(\Delta)_i(\Delta)_i}{i!}\frac{\Gamma(2\Delta-\frac{d}{2}+1+i)}{\Gamma(\Delta+\frac{3}{2}+i)\Gamma(\Delta+1+i)}\Bigl(\frac{1}{1+\kappa}\Bigl)^{2\Delta+2i+2}(-1)^i\nonumber\\
    &\times\sum_{l=0}^\infty\frac{(2\Delta-\frac{d}{2}+1+i)_l(1)_l(1)_l}{(\Delta+\frac{3}{2}+i)_l(\Delta+1+i)_l\ l!}\Bigl(\frac{1}{1+\kappa}\Bigl)^{2l}\sum_{k=0}^{l+i+1} a^{(2)}_k\frac{(d-1)_{l+i+1-k}}{(l+i+1-k)!}
\end{align}\

The UV-finiteness of this coefficient can be determined by noticing that for every value of $i$ it consists in a sum of terms of the form:

\begin{align}\label{aicoefk}
    a_i(\kappa) = &(d-1)\frac{(\Delta)_i(\Delta)_i(d)_i}{(2)_i\ i!}\frac{\Gamma(2\Delta-\frac{d}{2}+1+i)}{\Gamma(\Delta+\frac{3}{2}+i)\Gamma(\Delta+1+i)}\Bigl(\frac{1}{1+\kappa}\Bigl)^{2\Delta+2i+2}(-1)^i\nonumber\\
    &\times\ _4F_3\Bigl(2\Delta-\frac{d}{2}+1+i,1,1,d+i;\Delta+\frac{3}{2}+i,\Delta+1+i,2+i;\Bigl(\frac{1}{1+\kappa}\Bigl)^2\Bigl)+\dotsb
\end{align}\

where we identified the representation of the generalized hypergeometric function $_4F_3$. The triple dots represent other contributions to the coefficient, whose UV-convergence conditions turn out to be exactly the same as the term in eq. (\ref{aicoefk}). Therefore, the ultraviolet convergence region of the $a_i(\kappa)$ coefficient can be read directly from this generalized hypergeometric function. Indeed, as $\kappa\rightarrow0$ the argument of this function goes to 1, which applying the corresponding convergence criteria introduced in section 4.1.5 implies the convergence condition $\Delta+\frac{3}{2}+i+\Delta+1+i+2+i-(2\Delta-\frac{d}{2}+1+i)-1-1-(d+i)=\frac{3-d}{2}+i>0$, that is $d<3+2i$. In other words, for $d<3$ all the coefficients $a_i(\kappa)$ are UV-finite, for $d=3,4$ only the coefficient $a_0(\kappa)$ is UV-divergent, for $d=5,6$ the coefficients $a_0(\kappa)$ and $a_1(\kappa)$ are UV-divergent, for $d=7,8$ the coefficients $a_0(\kappa)$, $a_1(\kappa)$ and $a_2(\kappa)$ are UV-divergent, and so on.\par
Similarly for the second sum in eq. (\ref{qsediag3}) it can be rewritten as:

\begin{align}\label{qphi44pfsum2}
    &\frac{1}{\Gamma(\Delta)^2}\sum_{l=0}^\infty\frac{\Gamma(2\Delta-\frac{d}{2}+l)\Gamma(\Delta+l)}{\Gamma(\Delta+\frac{1}{2}+l)}\Bigl(\frac{1}{1+\kappa}\Bigl)^{2\Delta+2l}\bigl[-\tilde{K}(x_1,\vec{y_3})\tilde{K}(x_1,\vec{y_4})\lvert\vec{y_{34}}\rvert^2\bigl]^l\nonumber\\
    &\hspace{1.9cm}\times\sum_{i=0}^\infty\frac{(\Delta+l)_i(\Delta+l)_i}{(i+l)!\ i!}\bigl[-\ln{\bigl(\tilde{K}(x_1,\vec{y_3})\tilde{K}(x_1,\vec{y_4})\lvert\vec{y_{34}}\rvert^2\bigl)}+\psi(i+1)+\psi(i+l+1)\nonumber\\
    &\hspace{5cm}-2\psi(i+l+\Delta)\bigl]\bigl[\tilde{K}(x_1,\vec{y_3})\tilde{K}(x_1,\vec{y_4})\lvert\vec{y_{34}}\rvert^2\bigl]^i\sum_{k=0}^{l} a^{(2)}_k\frac{(d-1)_{l-k}}{(l-k)!}\nonumber\\
    &\hspace{1cm}=\sum_{i=0}^\infty\Bigl[b_i(\kappa)+c_i(\kappa)\ln{\bigl(\tilde{K}(x_1,\vec{y_3})\tilde{K}(x_1,\vec{y_4})\lvert\vec{y_{34}}\rvert^2\bigl)}\Bigl]\bigl[\tilde{K}(x_1,\vec{y_3})\tilde{K}(x_1,\vec{y_4})\lvert\vec{y_{34}}\rvert^2\bigl]^i
\end{align}\

where we used that $\sum_{i=0}^\infty\sum_{l=0}^\infty a_{i,l}=\sum_{i=0}^\infty\sum_{l=0}^i a_{i-l,l}$, rewrote and simplified terms accordingly and finally defined the coefficients $b_i(\kappa)$ and $c_i(\kappa)$:

\begin{align}
    b_i(\kappa) = \frac{\Gamma(2\Delta-\frac{d}{2})}{\Gamma(\Delta)\Gamma(\Delta+\frac{1}{2})}\frac{(\Delta)_i(\Delta)_i}{i!^2}\sum_{l=0}^i&\frac{(2\Delta-\frac{d}{2})_l(-i)_l}{(\Delta+\frac{1}{2})_l(\Delta)_l}\Bigl(\frac{1}{1+\kappa}\Bigl)^{2\Delta+2l}\nonumber\\
    &\hspace{-1cm}\times\bigl[\psi(i-l+1)+\psi(i+1)-2\psi(i+\Delta)\bigl]\sum_{k=0}^{l} a^{(2)}_k\frac{(d-1)_{l-k}}{(l-k)!}
\end{align}
\begin{equation}
    c_i(\kappa) = -\frac{\Gamma(2\Delta-\frac{d}{2})}{\Gamma(\Delta)\Gamma(\Delta+\frac{1}{2})}\frac{(\Delta)_i(\Delta)_i}{i!^2}\sum_{l=0}^i\frac{(2\Delta-\frac{d}{2})_l(-i)_l}{(\Delta+\frac{1}{2})_l(\Delta)_l}\Bigl(\frac{1}{1+\kappa}\Bigl)^{2\Delta+2l}\sum_{k=0}^{l} a^{(2)}_k\frac{(d-1)_{l-k}}{(l-k)!}
\end{equation}\

The UV-finiteness of these coefficients can be determined by noticing that for every value of $i$ they consist in a terminating sum of finite coefficients. Therefore, they are always UV-finite and we can always take in these coefficients $\kappa=0$. All the UV-divergence of the integral $\int G^2KK$ is contained then in the coefficient $a_i(\kappa)$.\par
Eq. (\ref{qphi44pfsum1}) together with eq. (\ref{qphi44pfsum2}) allow us to express the result for the $\int G^2KK$ integral eq. (\ref{qsediag3}) in the nice form:

\begin{align}
    \int d^{d+1}x_2\sqrt{g}\ G^2_\kappa(x_1,x_2)\tilde{K}^\Delta(x_2,\vec{y_3})\tilde{K}^\Delta(x_2,\vec{y_4})=&\tilde{K}^{\Delta}(x_1,\vec{y_3})\tilde{K}^{\Delta}(x_1,\vec{y_4})\pi^{\frac{d+1}{2}}\Bigl(\frac{2^{-\Delta}c_\Delta}{2\nu}\Bigl)^2\nonumber\\
    &\hspace{-2cm}\times\sum_{i=0}^\infty\Bigl[ a_i(\kappa)+b_i(0)+c_i(0)\ln{\bigl(\tilde{K}(x_1,\vec{y_3})\tilde{K}(x_1,\vec{y_4})\lvert\vec{y_{34}}\rvert^2\bigl)}\Bigl]\nonumber\\
    &\hspace{0.2cm}\times\bigl[\tilde{K}(x_1,\vec{y_3})\tilde{K}(x_1,\vec{y_4})\lvert\vec{y_{34}}\rvert^2\bigl]^i
\end{align}

where we safely took $\kappa=0$ in the coefficients $b_i(\kappa)$ and $c_i(\kappa)$. Remember that we are trying to compute $I_2$, eq. (\ref{qsediag1}). Replacing then the result for the $x_2$ integral just found back into this quantity that we are trying to compute, we find that it reduces to:

\begin{align}
    I_2(\vec{y_1},\vec{y_2},\vec{y_3},\vec{y_4}) =&\frac{\lambda^2c_\Delta^4}{2}\pi^\frac{d+1}{2}\Bigl(\frac{2^{-\Delta}c_\Delta}{2\nu}\Bigl)^2\nonumber\\
    &\hspace{-1cm}\times\sum_{i=0}^\infty\int d^{d+1}x_1\sqrt{g}\ \tilde{K}^\Delta(x_1,\vec{y_1})\tilde{K}^\Delta(x_1,\vec{y_2})\tilde{K}^{\Delta+i}(x_1,\vec{y_3})\tilde{K}^{\Delta+i}(x_1,\vec{y_4})\lvert\vec{y_{34}}\rvert^{2i}\nonumber\\
    &\hspace{-1cm}\times\Bigl[ a_i(\kappa)+b_i(0)+c_i(0)\ln{\bigl(\tilde{K}(x_1,\vec{y_3})\tilde{K}(x_1,\vec{y_4})\lvert\vec{y_{34}}\rvert^2\bigl)}\Bigl]+(\vec{y_2}\leftrightarrow\vec{y_3})+(\vec{y_2}\leftrightarrow\vec{y_4})
\end{align}\

As we mentioned before, the solving strategy for this resulting integral in $x_1$ is identical in spirit as the one used for the tree-level scalar exchange diagram present in the $\Phi^3$ theory, that is, writing it as a sum of D-functions. For our current case, this can be achieved by noticing that the logarithmic term can be written as:

\begin{equation}
    \ln{\bigl(\tilde{K}(x_1,\vec{y_3})\tilde{K}(x_1,\vec{y_4})\lvert\vec{y_{34}}\rvert^2\bigl)} = \frac{d}{d\alpha}\bigl[\tilde{K}^\alpha(x_1,\vec{y_3})\tilde{K}^\alpha(x_1,\vec{y_4})\lvert\vec{y_{34}}\rvert^{2\alpha}\bigl]\Bigl\rvert_{\alpha=0}
\end{equation}\

representation that let us express $I_2$ in the form of:

\begin{align}
    &I_2(\vec{y_1},\vec{y_2},\vec{y_3},\vec{y_4}) =\frac{\lambda^2c_\Delta^4}{2}\pi^\frac{d+1}{2}\Bigl(\frac{2^{-\Delta}c_\Delta}{2\nu}\Bigl)^2\nonumber\\
    &\hspace{0.5cm}\times\sum_{i=0}^\infty\Bigl\{\bigl[a_i(\kappa)+b_i(0)\bigl]\int d^{d+1}x_1\sqrt{g}\ \tilde{K}^\Delta(x_1,\vec{y_1})\tilde{K}^\Delta(x_1,\vec{y_2})\tilde{K}^{\Delta+i}(x_1,\vec{y_3})\tilde{K}^{\Delta+i}(x_1,\vec{y_4})\lvert\vec{y_{34}}\rvert^{2i}\nonumber\\
    &\hspace{0.5cm}+c_i(0)\frac{d}{d\alpha}\Bigl[\int d^{d+1}x_1\sqrt{g}\ \tilde{K}^\Delta(x_1,\vec{y_1})\tilde{K}^\Delta(x_1,\vec{y_2})\tilde{K}^{\Delta+i+\alpha}(x_1,\vec{y_3})\tilde{K}^{\Delta+i+\alpha}(x_1,\vec{y_4})\lvert\vec{y_{34}}\rvert^{2i+2\alpha}\Bigl]\Bigl\rvert_{\alpha=0}\Bigl\}\nonumber\\
    &\hspace{0.5cm}+(\vec{y_2}\leftrightarrow\vec{y_3})+(\vec{y_2}\leftrightarrow\vec{y_4})
\end{align}\

where we separated the integral in two. Finally then, the definition of the D-functions allows us to write these integrals in the desired form:

\begin{align}
    I_2(\vec{y_1},\vec{y_2},\vec{y_3},\vec{y_4}) =&\frac{\lambda^2c_\Delta^4}{2}\pi^\frac{d+1}{2}\Bigl(\frac{2^{-\Delta}c_\Delta}{2\nu}\Bigl)^2\sum_{i=0}^\infty\Bigl\{\bigl[a_i(\kappa)+b_i(0)\bigl]D_{\Delta\Delta\Delta+i\Delta+i}(\vec{y_1},\vec{y_2},\vec{y_3},\vec{y_4})\lvert\vec{y_{34}}\rvert^{2i}\nonumber\\
    &\hspace{3.5cm}+c_i(0)\frac{d}{d\alpha}\bigl[D_{\Delta\Delta\Delta+i+\alpha\Delta+i+\alpha}(\vec{y_1},\vec{y_2},\vec{y_3},\vec{y_4})\lvert\vec{y_{34}}\rvert^{2i+2\alpha}\bigl]\Bigl\rvert_{\alpha=0}\Bigl\}\nonumber\\
    &+(\vec{y_2}\leftrightarrow\vec{y_3})+(\vec{y_2}\leftrightarrow\vec{y_4})
\end{align}\

All the UV-divergence of this quantity are contained in the coefficients $a_i(\kappa)$. Since each one of these coefficients is proportional to a certain D-function, as we will see when we consider the complete correlator this fact will lead to the renormalization conditions of the theory.

\subsubsection{The Disconnected "Heads" Diagram}

\begin{figure}[h]
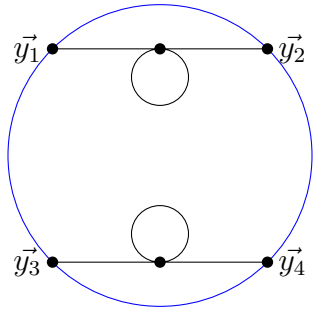

    \[\begin{wittendiagram}
    \draw (-1.4142,1.4142) node[vertex] -- (1.4142,1.4142)
    (-1.4142,-1.4142) node[vertex] -- (1.4142,-1.4142)
    (1.4142,1.4142) node[vertex]
    (1.4142,-1.4142) node[vertex]
    (0,1.4142) node[vertex]
    (0,-1.4142) node[vertex]
    
    (-1.4142,1.4142) node[left]{$\vec{y_1}$}
    (-1.4142,-1.4142) node[left]{$\vec{y_3}$}
    (1.4142,1.4142) node[right]{$\vec{y_2}$}
    (1.4142,-1.4142) node[right]{$\vec{y_4}$}
    
    (0,1.0392) circle (0.375)
    (0,-1.0392) circle (0.375);
  \end{wittendiagram}\]
  \caption{Pictorial representation of the disconnected "heads" diagram.}
\end{figure}

In terms of the unnormalized bulk-boundary propagator we can rewrite the loop integral $I_3$ (which we will refer to it as the disconnected "heads" diagram) as:

\begin{align}
    I_3(\vec{y_1},\vec{y_2},\vec{y_3},\vec{y_4}) =&\frac{\lambda^2c_\Delta^4G_\kappa^2(1)}{4}\int_{x_{1,0}=\varepsilon}d^{d+1}x_1\sqrt{g}\ \tilde{K}^\Delta(x_1,\vec{y_1})\tilde{K}^\Delta(x_1,\vec{y_2})\nonumber\\
    &\times\int_{x_{2,0}=\varepsilon}d^{d+1}x_2\sqrt{g}\ \tilde{K}^\Delta(x_2,\vec{y_3})\tilde{K}^\Delta(x_2,\vec{y_4})+(\vec{y_2}\leftrightarrow\vec{y_3})+(\vec{y_2}\leftrightarrow\vec{y_4})
\end{align}\

where we used that the regularized bulk-bulk propagator being evaluated at coincident points is just a constant. As we have seen repeatedly, the values of the $x_1$ and $x_2$ integrals are given by eq. (\ref{d2funcreg2}). Therefore, replacing their known results:

\begin{align}
    I_3(\vec{y_1},\vec{y_2},\vec{y_3},\vec{y_4}) =&\frac{\Pi^2(\kappa)}{\nu^2}\frac{2\nu c_\Delta}{\lvert\vec{y_1}-\vec{y_2}\rvert^{2\Delta}}\ln{\Bigl(\frac{\varepsilon}{\lvert\vec{y_1}-\vec{y_2}\rvert}\Bigl)}\frac{2\nu c_\Delta}{\lvert\vec{y_3}-\vec{y_4}\rvert^{2\Delta}}\ln{\Bigl(\frac{\varepsilon}{\lvert\vec{y_3}-\vec{y_4}\rvert}\Bigl)}\nonumber\\
    &+(\vec{y_2}\leftrightarrow\vec{y_3})+(\vec{y_2}\leftrightarrow\vec{y_4})+\mathcal{O}(\lambda^3)
\end{align}\

where we wrote, up to order $\lambda^2$, the constant factors in terms of $\Pi(\kappa)$ (eq. (\ref{1piphi4})). As we will see when we consider the complete correlator, we will realize that the result just found has exactly the expected interpretation, that is up to this order in $\lambda$, the expansion of the same conformal anomaly dictated by the 2-point function.

\subsubsection{The Disconnected Diagrams}

\begin{figure}[h]
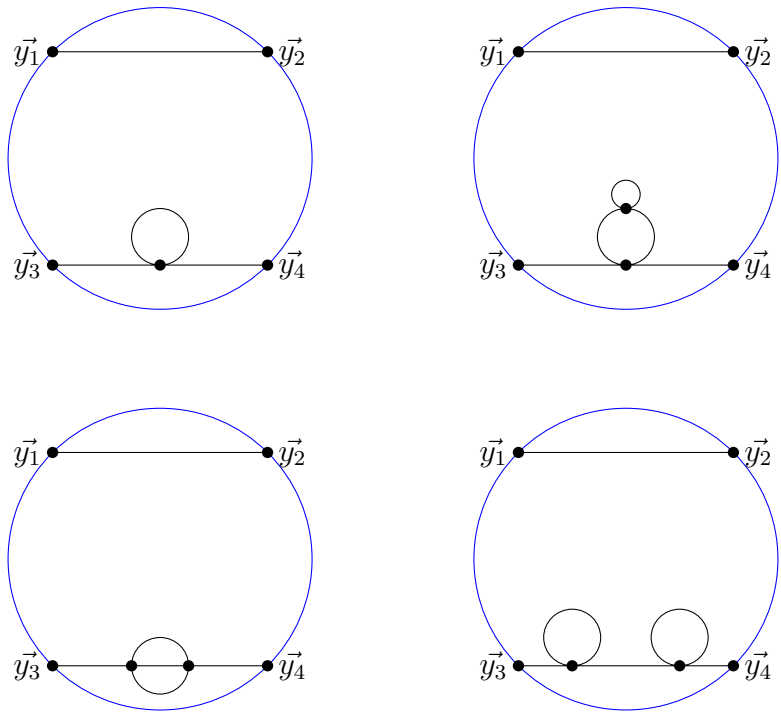

    \[\begin{wittendiagram}
    \draw (-1.4142,1.4142) node[vertex] -- (1.4142,1.4142)
    (-1.4142,-1.4142) node[vertex] -- (1.4142,-1.4142)
    (1.4142,1.4142) node[vertex]
    (1.4142,-1.4142) node[vertex]
    (0,-1.4142) node[vertex]
    
    (-1.4142,1.4142) node[left]{$\vec{y_1}$}
    (-1.4142,-1.4142) node[left]{$\vec{y_3}$}
    (1.4142,1.4142) node[right]{$\vec{y_2}$}
    (1.4142,-1.4142) node[right]{$\vec{y_4}$}
    
    (0,-1.0392) circle (0.375);
  \end{wittendiagram}\hspace{2cm}
  \begin{wittendiagram}
    \draw (-1.4142,1.4142) node[vertex] -- (1.4142,1.4142)
    (-1.4142,-1.4142) node[vertex] -- (1.4142,-1.4142)
    (1.4142,1.4142) node[vertex]
    (1.4142,-1.4142) node[vertex]
    (0,-1.4142) node[vertex]
    (0,-0.6642) node[vertex]
    
    (-1.4142,1.4142) node[left]{$\vec{y_1}$}
    (-1.4142,-1.4142) node[left]{$\vec{y_3}$}
    (1.4142,1.4142) node[right]{$\vec{y_2}$}
    (1.4142,-1.4142) node[right]{$\vec{y_4}$}
    
    (0,-1.0392) circle (0.375)
    (0,-0.4767) circle (0.1875);
  \end{wittendiagram}\]\vspace{1cm}
  \[\begin{wittendiagram}
    \draw (-1.4142,1.4142) node[vertex] -- (1.4142,1.4142)
    (-1.4142,-1.4142) node[vertex] -- (1.4142,-1.4142)
    (1.4142,1.4142) node[vertex]
    (1.4142,-1.4142) node[vertex]
    (-0.375,-1.4142) node[vertex]
    (0.375,-1.4142) node[vertex]
    
    (-1.4142,1.4142) node[left]{$\vec{y_1}$}
    (-1.4142,-1.4142) node[left]{$\vec{y_3}$}
    (1.4142,1.4142) node[right]{$\vec{y_2}$}
    (1.4142,-1.4142) node[right]{$\vec{y_4}$}
    
    (0,-1.4142) circle (0.375);
  \end{wittendiagram}\hspace{2cm}
  \begin{wittendiagram}
    \draw (-1.4142,1.4142) node[vertex] -- (1.4142,1.4142)
    (-1.4142,-1.4142) node[vertex] -- (1.4142,-1.4142)
    (1.4142,1.4142) node[vertex]
    (1.4142,-1.4142) node[vertex]
    (-0.7071,-1.4142) node[vertex]
    (0.7071,-1.4142) node[vertex]
    
    (-1.4142,1.4142) node[left]{$\vec{y_1}$}
    (-1.4142,-1.4142) node[left]{$\vec{y_3}$}
    (1.4142,1.4142) node[right]{$\vec{y_2}$}
    (1.4142,-1.4142) node[right]{$\vec{y_4}$}
    
    (-0.7071,-1.0392) circle (0.375)
    (0.7071,-1.0392) circle (0.375);
  \end{wittendiagram}\]
  \caption{Pictorial representation of the disconnected diagrams.}
\end{figure}

Finally for the loop integrals $I_4$ (which we will refer to them simply as the disconnected diagrams), notice that the terms inside the square bracket are nothing but the loop integrals present in the 2-point functions (the "head", "eight", "sunset" and "double head" diagrams) whose value were found to be given by eqs. (\ref{headdiagres}), (\ref{eightdiagres}), (\ref{sunsetdiagres}) and (\ref{doubheaddiagres}) respectively. Therefore, replacing their known results we find that $I_4$ can be expressed as:

\begin{align}
    I_4(\vec{y_1},\vec{y_2},\vec{y_3},\vec{y_4}) =&-\frac{2\nu c_\Delta}{\lvert\vec{y_1}-\vec{y_2}\rvert^{2\Delta}}\frac{2\nu c_\Delta}{\lvert\vec{y_3}-\vec{y_4}\rvert^{2\Delta}}\frac{\Pi(\kappa)}{\nu}\ln{\Bigl(\frac{\varepsilon}{\lvert\vec{y_3}-\vec{y_4}\rvert}\Bigl)}\times6\nonumber\\
    &+\frac{2\nu c_\Delta}{\lvert\vec{y_1}-\vec{y_2}\rvert^{2\Delta}}\frac{2\nu c_\Delta}{\lvert\vec{y_3}-\vec{y_4}\rvert^{2\Delta}}\frac{\Pi^2(\kappa)}{2\nu^2}\ln^2{\Bigl(\frac{\varepsilon}{\lvert\vec{y_3}-\vec{y_4}\rvert}\Bigl)}\times6+\mathcal{O}(\lambda^3)
\end{align}\

where we wrote, up to order $\lambda^2$, the constant factors in terms of $\Pi(\kappa)$. Next we will see how all the contributions to the 4-point functions nicely factorize to give rise to the same conformal anomaly dictated by the 2-point function, together with a correction to its coupling constant. This realization, similarly to the $\Phi^3$ study, will provide us with a natural and consistent renormalization scheme of both IR and UV divergences for all the holographic n-point functions, equivalent to those schemes used in ordinary QFTs.

\subsubsection{Correlator Renormalization}

Replacing the results just found then for $I_1$, $I_2$, $I_3$ and $I_4$ back into the holographic 4-point functions, we find that they can be written as:

\begin{align}
    &\langle O_\Delta(\vec{y_1})O_\Delta(\vec{y_2})O_\Delta(\vec{y_3})O_\Delta(\vec{y_4})\rangle_{\text{CFT}} =\frac{2\nu c_\Delta}{\lvert\vec{y_1}-\vec{y_2}\rvert^{2\Delta}}\frac{2\nu c_\Delta}{\lvert\vec{y_3}-\vec{y_4}\rvert^{2\Delta}}+(\vec{y_2}\leftrightarrow\vec{y_3})+(\vec{y_2}\leftrightarrow\vec{y_4})\nonumber\\
    &\hspace{2cm}-\lambda c_\Delta^4\int d^{d+1}x_1\sqrt{g}\ \tilde{K}^\Delta(x_1,\vec{y_1})\tilde{K}^\Delta(x_1,\vec{y_2})\tilde{K}^\Delta(x_1,\vec{y_3})\tilde{K}^\Delta(x_1,\vec{y_4})\nonumber\\
    &\hspace{2cm}+\frac{\lambda c_\Delta^4\Pi(\kappa)}{2\nu}\int d^{d+1}x_1\sqrt{g}\ \tilde{K}^\Delta(x_1,\vec{y_1})\ln{\bigl(\varepsilon\tilde{K}(x_1,\vec{y_1})\bigl)}K(x_1,\vec{y_2})K(x_1,\vec{y_3})K(x_1,\vec{y_4})\nonumber\\
    &\hspace{2cm}+(\vec{y_1}\leftrightarrow\vec{y_2})+(\vec{y_1}\leftrightarrow\vec{y_3})+(\vec{y_1}\leftrightarrow\vec{y_4})\nonumber\\
    &\hspace{3cm}+\frac{\lambda^2c_\Delta^4}{2}\pi^\frac{d+1}{2}\Bigl(\frac{2^{-\Delta}c_\Delta}{2\nu}\Bigl)^2\sum_{i=0}^\infty\Bigl\{\bigl[a_i(\kappa)+b_i(0)\bigl]D_{\Delta\Delta\Delta+i\Delta+i}(\vec{y_1},\vec{y_2},\vec{y_3},\vec{y_4})\lvert\vec{y_{34}}\rvert^{2i}\nonumber\\
    &\hspace{6.5cm}+c_i(0)\frac{d}{d\alpha}\bigl[D_{\Delta\Delta\Delta+i+\alpha\Delta+i+\alpha}(\vec{y_1},\vec{y_2},\vec{y_3},\vec{y_4})\lvert\vec{y_{34}}\rvert^{2i+2\alpha}\bigl]\Bigl\rvert_{\alpha=0}\Bigl\}\nonumber\\
    &\hspace{2cm}+(\vec{y_2}\leftrightarrow\vec{y_3})+(\vec{y_2}\leftrightarrow\vec{y_4})\nonumber\\
    &\hspace{2cm}+\frac{\Pi^2(\kappa)}{\nu^2}\frac{2\nu c_\Delta}{\lvert\vec{y_1}-\vec{y_2}\rvert^{2\Delta}}\ln{\Bigl(\frac{\varepsilon}{\lvert\vec{y_1}-\vec{y_2}\rvert}\Bigl)}\frac{2\nu c_\Delta}{\lvert\vec{y_3}-\vec{y_4}\rvert^{2\Delta}}\ln{\Bigl(\frac{\varepsilon}{\lvert\vec{y_3}-\vec{y_4}\rvert}\Bigl)}\nonumber\\
    &\hspace{2cm}+(\vec{y_2}\leftrightarrow\vec{y_3})+(\vec{y_2}\leftrightarrow\vec{y_4})\nonumber\\
    &\hspace{2cm}-\frac{2\nu c_\Delta}{\lvert\vec{y_1}-\vec{y_2}\rvert^{2\Delta}}\frac{2\nu c_\Delta}{\lvert\vec{y_3}-\vec{y_4}\rvert^{2\Delta}}\frac{\Pi(\kappa)}{\nu}\ln{\Bigl(\frac{\varepsilon}{\lvert\vec{y_3}-\vec{y_4}\rvert}\Bigl)}+(\vec{y_1}\leftrightarrow\vec{y_3})+(\vec{y_1}\leftrightarrow\vec{y_4})\nonumber\\
    &\hspace{2cm}+(\vec{y_2}\leftrightarrow\vec{y_3})+(\vec{y_2}\leftrightarrow\vec{y_4})+(\vec{y_1}\leftrightarrow\vec{y_3},\vec{y_2}\leftrightarrow\vec{y_4})\nonumber\\
    &\hspace{2cm}+\frac{2\nu c_\Delta}{\lvert\vec{y_1}-\vec{y_2}\rvert^{2\Delta}}\frac{2\nu c_\Delta}{\lvert\vec{y_3}-\vec{y_4}\rvert^{2\Delta}}\frac{\Pi^2(\kappa)}{2\nu^2}\ln^2{\Bigl(\frac{\varepsilon}{\lvert\vec{y_3}-\vec{y_4}\rvert}\Bigl)}+(\vec{y_1}\leftrightarrow\vec{y_3})+(\vec{y_1}\leftrightarrow\vec{y_4})\nonumber\\
    &\hspace{2cm}+(\vec{y_2}\leftrightarrow\vec{y_3})+(\vec{y_2}\leftrightarrow\vec{y_4})+(\vec{y_1}\leftrightarrow\vec{y_3},\vec{y_2}\leftrightarrow\vec{y_4})+\mathcal{O}(\lambda^3)\nonumber\\
    &\langle O_\Delta(\vec{y_1})O_\Delta(\vec{y_2})O_\Delta(\vec{y_3})O_\Delta(\vec{y_4})\rangle_{\text{CFT,con}} =\nonumber\\
    &\hspace{2cm}-\lambda c_\Delta^4\int d^{d+1}x_1\sqrt{g}\ \tilde{K}^\Delta(x_1,\vec{y_1})\tilde{K}^\Delta(x_1,\vec{y_2})\tilde{K}^\Delta(x_1,\vec{y_3})\tilde{K}^\Delta(x_1,\vec{y_4})\nonumber\\
    &\hspace{2cm}+\frac{\lambda c_\Delta^4\Pi(\kappa)}{2\nu}\int d^{d+1}x_1\sqrt{g}\ \tilde{K}^\Delta(x_1,\vec{y_1})\ln{\bigl(\varepsilon\tilde{K}(x_1,\vec{y_1})\bigl)}K(x_1,\vec{y_2})K(x_1,\vec{y_3})K(x_1,\vec{y_4})\nonumber\\
    &\hspace{2cm}+(\vec{y_1}\leftrightarrow\vec{y_2})+(\vec{y_1}\leftrightarrow\vec{y_3})+(\vec{y_1}\leftrightarrow\vec{y_4})\nonumber\\
    &\hspace{2cm}+\frac{\lambda^2c_\Delta^4}{2}\pi^\frac{d+1}{2}\Bigl(\frac{2^{-\Delta}c_\Delta}{2\nu}\Bigl)^2\sum_{i=0}^\infty\Bigl\{\bigl[a_i(\kappa)+b_i(0)\bigl]D_{\Delta\Delta\Delta+i\Delta+i}(\vec{y_1},\vec{y_2},\vec{y_3},\vec{y_4})\lvert\vec{y_{34}}\rvert^{2i}\nonumber\\
    &\hspace{6.5cm}+c_i(0)\frac{d}{d\alpha}\bigl[D_{\Delta\Delta\Delta+i+\alpha\Delta+i+\alpha}(\vec{y_1},\vec{y_2},\vec{y_3},\vec{y_4})\lvert\vec{y_{34}}\rvert^{2i+2\alpha}\bigl]\Bigl\rvert_{\alpha=0}\Bigl\}\nonumber\\
    &\hspace{2cm}+(\vec{y_2}\leftrightarrow\vec{y_3})+(\vec{y_2}\leftrightarrow\vec{y_4})+\mathcal{O}(\lambda^3)
\end{align}\

Now, up to this same order in the coupling constant, it is easy to see that the disconnected contributions to the correlators can be factorized into the form:

\begin{align}\label{phi44pfdisc}
    &\frac{2\nu c_\Delta}{\lvert\vec{y_1}-\vec{y_2}\rvert^{2\Delta}}\frac{2\nu c_\Delta}{\lvert\vec{y_3}-\vec{y_4}\rvert^{2\Delta}}\times3
    +\frac{\Pi^2(\kappa)}{\nu^2}\frac{2\nu c_\Delta}{\lvert\vec{y_1}-\vec{y_2}\rvert^{2\Delta}}\ln{\Bigl(\frac{\varepsilon}{\lvert\vec{y_1}-\vec{y_2}\rvert}\Bigl)}\frac{2\nu c_\Delta}{\lvert\vec{y_3}-\vec{y_4}\rvert^{2\Delta}}\ln{\Bigl(\frac{\varepsilon}{\lvert\vec{y_3}-\vec{y_4}\rvert}\Bigl)}\times3\nonumber\\
    &-\frac{2\nu c_\Delta}{\lvert\vec{y_1}-\vec{y_2}\rvert^{2\Delta}}\frac{2\nu c_\Delta}{\lvert\vec{y_3}-\vec{y_4}\rvert^{2\Delta}}\frac{\Pi(\kappa)}{\nu}\ln{\Bigl(\frac{\varepsilon}{\lvert\vec{y_3}-\vec{y_4}\rvert}\Bigl)}\times6\nonumber\\
    &+\frac{2\nu c_\Delta}{\lvert\vec{y_1}-\vec{y_2}\rvert^{2\Delta}}\frac{2\nu c_\Delta}{\lvert\vec{y_3}-\vec{y_4}\rvert^{2\Delta}}\frac{\Pi^2(\kappa)}{2\nu^2}\ln^2{\Bigl(\frac{\varepsilon}{\lvert\vec{y_3}-\vec{y_4}\rvert}\Bigl)}\times6\nonumber\\
    &\hspace{1.4cm}=\varepsilon^{-\frac{2\Pi(\kappa)}{\nu}}\frac{2\nu c_\Delta}{\lvert\vec{y_1}-\vec{y_2}\rvert^{2\Delta-\frac{\Pi(\kappa)}{\nu}}}\frac{2\nu c_\Delta}{\lvert\vec{y_3}-\vec{y_4}\rvert^{2\Delta-\frac{\Pi(\kappa)}{\nu}}}+(\vec{y_2}\leftrightarrow\vec{y_3})+(\vec{y_2}\leftrightarrow\vec{y_4})+\mathcal{O}(\lambda^3)
\end{align}\

Notice that these contributions, apart from the IR rescaling, are nothing but the derivation of eq. (\ref{conformalform4pf}) where the scaling dimension $\Delta$ has been replaced by $\Delta-\frac{\Pi(\kappa)}{2\nu}$. Therefore, following the same steps shown there, we can rewrite eq. (\ref{phi44pfdisc}) in the form of: 

\begin{align}\label{phi44pfdisccon5}
    &\frac{2\nu c_\Delta}{\lvert\vec{y_1}-\vec{y_2}\rvert^{2\Delta}}\frac{2\nu c_\Delta}{\lvert\vec{y_3}-\vec{y_4}\rvert^{2\Delta}}\times3
    +\frac{\Pi^2(\kappa)}{\nu^2}\frac{2\nu c_\Delta}{\lvert\vec{y_1}-\vec{y_2}\rvert^{2\Delta}}\ln{\Bigl(\frac{\varepsilon}{\lvert\vec{y_1}-\vec{y_2}\rvert}\Bigl)}\frac{2\nu c_\Delta}{\lvert\vec{y_3}-\vec{y_4}\rvert^{2\Delta}}\ln{\Bigl(\frac{\varepsilon}{\lvert\vec{y_3}-\vec{y_4}\rvert}\Bigl)}\times3\nonumber\\
    &-\frac{2\nu c_\Delta}{\lvert\vec{y_1}-\vec{y_2}\rvert^{2\Delta}}\frac{2\nu c_\Delta}{\lvert\vec{y_3}-\vec{y_4}\rvert^{2\Delta}}\frac{\Pi(\kappa)}{\nu}\ln{\Bigl(\frac{\varepsilon}{\lvert\vec{y_3}-\vec{y_4}\rvert}\Bigl)}\times6\nonumber\\
    &+\frac{2\nu c_\Delta}{\lvert\vec{y_1}-\vec{y_2}\rvert^{2\Delta}}\frac{2\nu c_\Delta}{\lvert\vec{y_3}-\vec{y_4}\rvert^{2\Delta}}\frac{\Pi^2(\kappa)}{2\nu^2}\ln^2{\Bigl(\frac{\varepsilon}{\lvert\vec{y_3}-\vec{y_4}\rvert}\Bigl)}\times6\nonumber\\
    &\hspace{2cm}=\varepsilon^{-\frac{2\Pi(\kappa)}{\nu}}\times\frac{(2\nu c_\Delta)^2}{\prod_{i<j}\lvert\vec{y_{ij}}\rvert^{\frac{2}{3}\bigl(\Delta-\frac{\Pi(\kappa)}{2\nu}\bigl)}}u^{-\frac{2}{3}\bigl(\Delta-\frac{\Pi(\kappa)}{2\nu}\bigl)}v^{\frac{1}{3}\bigl(\Delta-\frac{\Pi(\kappa)}{2\nu}\bigl)}+\Bigl(u,v\rightarrow\frac{1}{u},\frac{v}{u}\Bigl)\nonumber\\
    &\hspace{2.5cm}+(u,v\rightarrow v,u)+\mathcal{O}(\lambda^3)
\end{align}\

Similarly for the connected contributions to the correlators (ignoring the "scalar exchange" diagram), it can be directly seen that they can be factorized into the form:

\begin{align}\label{phi44pfconcon5}
    &-\lambda c_\Delta^4\int d^{d+1}x_1\sqrt{g}\ \tilde{K}^\Delta(x_1,\vec{y_1})\tilde{K}^\Delta(x_1,\vec{y_2})\tilde{K}^\Delta(x_1,\vec{y_3})\tilde{K}^\Delta(x_1,\vec{y_4})\nonumber\\
    &+\frac{\lambda c_\Delta^4\Pi(\kappa)}{2\nu}\int d^{d+1}x_1\sqrt{g}\ \tilde{K}^\Delta(x_1,\vec{y_1})\ln{\bigl(\varepsilon\tilde{K}(x_1,\vec{y_1})\bigl)}K(x_1,\vec{y_2})K(x_1,\vec{y_3})K(x_1,\vec{y_4})\times4\nonumber\\
    &\hspace{1cm}=\varepsilon^{-\frac{2\Pi(\kappa)}{\nu}}\times-\lambda c_\Delta^4D_{\Delta-\frac{\Pi(\kappa)}{2\nu}\Delta-\frac{\Pi(\kappa)}{2\nu}\Delta-\frac{\Pi(\kappa)}{2\nu}\Delta-\frac{\Pi(\kappa)}{2\nu}}(\vec{y_1},\vec{y_2},\vec{y_3},\vec{y_4})+\mathcal{O}(\lambda^3)
\end{align}\

where we wrote the resulting integral in terms of the D-function. The results eqs. (\ref{phi44pfdisccon5}) and (\ref{phi44pfconcon5}) allow us to express the 4-point functions in the nice compact forms:

\begin{align}\label{4pfcorrrenmidstep1}
    &\langle O_\Delta(\vec{y_1})O_\Delta(\vec{y_2})O_\Delta(\vec{y_3})O_\Delta(\vec{y_4})\rangle_{\text{CFT}} =\varepsilon^{-\frac{2\Pi(\kappa)}{\nu}}\times\frac{(2\nu c_\Delta)^2}{\prod_{i<j}\lvert\vec{y_{ij}}\rvert^{\frac{2}{3}\bigl(\Delta-\frac{\Pi(\kappa)}{2\nu}\bigl)}}u^{-\frac{2}{3}\bigl(\Delta-\frac{\Pi(\kappa)}{2\nu}\bigl)}v^{\frac{1}{3}\bigl(\Delta-\frac{\Pi(\kappa)}{2\nu}\bigl)}\nonumber\\
    &\hspace{3cm}+\Bigl(u,v\rightarrow\frac{1}{u},\frac{v}{u}\Bigl)+(u,v\rightarrow v,u)\nonumber\\
    &\hspace{3cm}+\varepsilon^{-\frac{2\Pi(\kappa)}{\nu}}\times-\lambda c_\Delta^4D_{\Delta-\frac{\Pi(\kappa)}{2\nu}\Delta-\frac{\Pi(\kappa)}{2\nu}\Delta-\frac{\Pi(\kappa)}{2\nu}\Delta-\frac{\Pi(\kappa)}{2\nu}}(\vec{y_1},\vec{y_2},\vec{y_3},\vec{y_4})\nonumber\\
    &\hspace{3cm}+\frac{\lambda^2c_\Delta^4}{2}\pi^\frac{d+1}{2}\Bigl(\frac{2^{-\Delta}c_\Delta}{2\nu}\Bigl)^2\sum_{i=0}^\infty\Bigl\{\bigl[a_i(\kappa)+b_i(0)\bigl]D_{\Delta\Delta\Delta+i\Delta+i}(\vec{y_1},\vec{y_2},\vec{y_3},\vec{y_4})\lvert\vec{y_{34}}\rvert^{2i}\nonumber\\
    &\hspace{6.5cm}+c_i(0)\frac{d}{d\alpha}\bigl[D_{\Delta\Delta\Delta+i+\alpha\Delta+i+\alpha}(\vec{y_1},\vec{y_2},\vec{y_3},\vec{y_4})\lvert\vec{y_{34}}\rvert^{2i+2\alpha}\bigl]\Bigl\rvert_{\alpha=0}\Bigl\}\nonumber\\
    &\hspace{3cm}+(\vec{y_2}\leftrightarrow\vec{y_3})+(\vec{y_2}\leftrightarrow\vec{y_4})+\mathcal{O}(\lambda^3)\nonumber\\
    &\langle O_\Delta(\vec{y_1})O_\Delta(\vec{y_2})O_\Delta(\vec{y_3})O_\Delta(\vec{y_4})\rangle_{\text{CFT,con}} = \varepsilon^{-\frac{2\Pi(\kappa)}{\nu}}\times-\lambda c_\Delta^4D_{\Delta-\frac{\Pi(\kappa)}{2\nu}\Delta-\frac{\Pi(\kappa)}{2\nu}\Delta-\frac{\Pi(\kappa)}{2\nu}\Delta-\frac{\Pi(\kappa)}{2\nu}}(\vec{y_1},\vec{y_2},\vec{y_3},\vec{y_4})\nonumber\\
    &\hspace{3cm}+\frac{\lambda^2c_\Delta^4}{2}\pi^\frac{d+1}{2}\Bigl(\frac{2^{-\Delta}c_\Delta}{2\nu}\Bigl)^2\sum_{i=0}^\infty\Bigl\{\bigl[a_i(\kappa)+b_i(0)\bigl]D_{\Delta\Delta\Delta+i\Delta+i}(\vec{y_1},\vec{y_2},\vec{y_3},\vec{y_4})\lvert\vec{y_{34}}\rvert^{2i}\nonumber\\
    &\hspace{6.5cm}+c_i(0)\frac{d}{d\alpha}\bigl[D_{\Delta\Delta\Delta+i+\alpha\Delta+i+\alpha}(\vec{y_1},\vec{y_2},\vec{y_3},\vec{y_4})\lvert\vec{y_{34}}\rvert^{2i+2\alpha}\bigl]\Bigl\rvert_{\alpha=0}\Bigl\}\nonumber\\
    &\hspace{3cm}+(\vec{y_2}\leftrightarrow\vec{y_3})+(\vec{y_2}\leftrightarrow\vec{y_4})+\mathcal{O}(\lambda^3)
\end{align}\

With the correlators written in this form, it is direct to see what are the effects of the quantum corrections coming from the off-shell part of the AdS path integral to the 4-point functions found previously under the classical approximation of the AdS/CFT correspondence. Indeed, up to order $\lambda^2$ in the coupling constant, they contribute with an overall rescaling to the correlator along with a shift in its scaling dimension, just like for the 2-point function, with the difference as we will see briefly that in the current case the effective self-interacting coupling constant between the bulk fields also receives a correction coming from the "scalar exchange" diagram. What is remarkable however about eq. (\ref{4pfcorrrenmidstep1}) is that the resulting rescaling and anomalous dimension of the 4-point functions are exactly the same as those dictated by the 2-point function! This fact implies that the very same redefinitions of the bulk's parameters done for the 2-point function not only have the effect of renormalizing the divergences present there, but also for the divergences present in the 4-point function, where now a redefinition of the coupling constant $\lambda$ is also needed. To see this, consider a redefinition of the bulk's self-interacting coupling constant $\lambda$ in the AdS bulk action of the form:

\begin{equation}
    \lambda\rightarrow\lambda+\delta\lambda
\end{equation}\

where the counterterm is expected to be of order $\delta\lambda=\mathcal{O}(\lambda^2)$. This redefinition of $\lambda$ adds, up to order $\lambda^2$, a new counterterm interaction to the holographic 4-point functions eq. (\ref{4pfcorrrenmidstep1}) of the form:

\begin{align}
    &\langle O_\Delta(\vec{y_1})O_\Delta(\vec{y_2})O_\Delta(\vec{y_3})O_\Delta(\vec{y_4})\rangle_{\text{CFT}},\ \langle O_\Delta(\vec{y_1})O_\Delta(\vec{y_2})O_\Delta(\vec{y_3})O_\Delta(\vec{y_4})\rangle_{\text{CFT,con}}\nonumber\\
    &\hspace{1.5cm}\rightarrow\langle O_\Delta(\vec{y_1})O_\Delta(\vec{y_2})O_\Delta(\vec{y_3})O_\Delta(\vec{y_4})\rangle_{\text{CFT}},\ \langle O_\Delta(\vec{y_1})O_\Delta(\vec{y_2})O_\Delta(\vec{y_3})O_\Delta(\vec{y_4})\rangle_{\text{CFT,con}}\nonumber\\
    &\hspace{2.2cm}-\delta\lambda\int d^{d+1}x_1\sqrt{g}\ K(x_1,\vec{y_1})K(x_1,\vec{y_2})K(x_1,\vec{y_3})K(x_1,\vec{y_4})+\mathcal{O}(\lambda^3)\nonumber\\
    &\hspace{1.6cm}=\langle O_\Delta(\vec{y_1})O_\Delta(\vec{y_2})O_\Delta(\vec{y_3})O_\Delta(\vec{y_4})\rangle_{\text{CFT}},\ \langle O_\Delta(\vec{y_1})O_\Delta(\vec{y_2})O_\Delta(\vec{y_3})O_\Delta(\vec{y_4})\rangle_{\text{CFT,con}}\nonumber\\
    &\hspace{2.3cm}-\delta\lambda c_\Delta^4D_{\Delta\Delta\Delta\Delta}(\vec{y_1},\vec{y_2},\vec{y_3},\vec{y_4})+\mathcal{O}(\lambda^3)
\end{align}\

Given the form of this counterterm interaction we conclude that the UV-divergences coming from each $a_i(\kappa)$ coefficient present in the "scalar exchange" diagram will be renormalizable through a redefinition of $\lambda$ for only those coefficients that are writable as terms proportional to $D_{\Delta\Delta\Delta\Delta}$. Take for example the coefficient $i=0$. It is direct to see that this is indeed possible in this case:

\begin{equation}\label{iequal0coef}
    a_0(\kappa)D_{\Delta\Delta\Delta\Delta}+(\vec{y_2}\leftrightarrow\vec{y_3})+(\vec{y_2}\leftrightarrow\vec{y_4}) = 3a_0(\kappa)D_{\Delta\Delta\Delta\Delta}
\end{equation}\

resulting in a contribution proportional to $D_{\Delta\Delta\Delta\Delta}$ and thus renormalizable. For the coefficient $i=1$ it turns out that this is also possible thanks to the nice property eq. (\ref{4dfuncrecrel}) satisfied by the D-functions:

\begin{align}
    &a_1(\kappa)D_{\Delta\Delta\Delta+1\Delta+1}\lvert\vec{y_{34}}\rvert^2+(\vec{y_2}\leftrightarrow\vec{y_3})+(\vec{y_2}\leftrightarrow\vec{y_4}) =\nonumber\\
    &\hspace{4cm}-a_1(\kappa)\frac{(2\Delta-\frac{d}{2})}{\Delta^2}\Bigl(y_{34}^2\frac{\partial}{\partial y_{34}^2}+y_{24}^2\frac{\partial}{\partial y_{24}^2}+y_{23}^2\frac{\partial}{\partial y_{23}^2}\Bigl)D_{\Delta\Delta\Delta\Delta}
\end{align}\

But notice from eq. (\ref{4dfunc}) that the D-function can be written as $D_{\Delta\Delta\Delta\Delta}=f(u,v)\prod_{i<j}y_{ij}^{-\frac{2\Delta}{3}}$, therefore using the product rule for derivatives:

\begin{equation}\label{iequal1coef}
    a_1(\kappa)D_{\Delta\Delta\Delta+1\Delta+1}\lvert\vec{y_{34}}\rvert^2+(\vec{y_2}\leftrightarrow\vec{y_3})+(\vec{y_2}\leftrightarrow\vec{y_4})=a_1(\kappa)\frac{(2\Delta-\frac{d}{2})}{\Delta}D_{\Delta\Delta\Delta\Delta}
\end{equation}\

where when differentiating the quantity $f(u,v)$ we used the chain rules $y_{34}^2\frac{\partial}{\partial y_{34}^2}=u\frac{\partial}{\partial u}$, $y_{24}^2\frac{\partial}{\partial y_{24}^2}=-u\frac{\partial}{\partial u}-v\frac{\partial}{\partial v}$ and $y_{23}^2\frac{\partial}{\partial y_{23}^2}=v\frac{\partial}{\partial v}$, resulting in contributions that cancel each other out. Just like the coefficient $i=0$, it results in a contribution proportional to $D_{\Delta\Delta\Delta\Delta}$ and thus renormalizable. However, for the terms coming from the $i=2$ coefficients and higher this feature is no longer true and the UV-divergences present in these coefficients cannot be renormalized under the current renormalization scheme through a redefinition of $\lambda$. Remembering that $a_0(\kappa)$ is UV-divergent for $d>2$, $a_1(\kappa)$ is UV-divergent for $d>4$, $a_2(\kappa)$ is UV-divergent for $d>6$, and so on, since the only coefficients that are renormalizable are $a_0(\kappa)$ and $a_1(\kappa)$, this fact gives us directly the region of renormalization of the theory: CFT theories of dimensions greater than 6 dual to $\Phi^4$ theories on AdS are non-renormalizable.\par
From now on then we will be only interested in $\Phi^4$ theories with $d<7$, which together with the results eqs. (\ref{iequal0coef}) and (\ref{iequal1coef}) allow us to express the explicit forms of the 4-point functions as:

\begin{align}
    &\langle O_\Delta(\vec{y_1})O_\Delta(\vec{y_2})O_\Delta(\vec{y_3})O_\Delta(\vec{y_4})\rangle_{\text{CFT}} =\varepsilon^{-\frac{2\Pi(\kappa)}{\nu}}\times\frac{(2\nu c_\Delta)^2}{\prod_{i<j}\lvert\vec{y_{ij}}\rvert^{\frac{2}{3}\bigl(\Delta-\frac{\Pi(\kappa)}{2\nu}\bigl)}}u^{-\frac{2}{3}\bigl(\Delta-\frac{\Pi(\kappa)}{2\nu}\bigl)}v^{\frac{1}{3}\bigl(\Delta-\frac{\Pi(\kappa)}{2\nu}\bigl)}\nonumber\\
    &\hspace{1.5cm}+\Bigl(u,v\rightarrow\frac{1}{u},\frac{v}{u}\Bigl)+(u,v\rightarrow v,u)\nonumber\\
    &\hspace{1.5cm}+\varepsilon^{-\frac{2\Pi(\kappa)}{\nu}}\times-\lambda c_\Delta^4D_{\Delta-\frac{\Pi(\kappa)}{2\nu}\Delta-\frac{\Pi(\kappa)}{2\nu}\Delta-\frac{\Pi(\kappa)}{2\nu}\Delta-\frac{\Pi(\kappa)}{2\nu}}(\vec{y_1},\vec{y_2},\vec{y_3},\vec{y_4})\nonumber\\
    &\hspace{1.5cm}+\Bigl\{\frac{\lambda^2}{2}\pi^\frac{d+1}{2}\Bigl(\frac{2^{-\Delta}c_\Delta}{2\nu}\Bigl)^2\Bigl[3a_0(\kappa)+a_1(\kappa)\frac{(2\Delta-\frac{d}{2})}{\Delta}\Bigl]-\delta\lambda\Bigl\}c_\Delta^4D_{\Delta\Delta\Delta\Delta}(\vec{y_1},\vec{y_2},\vec{y_3},\vec{y_4})\nonumber\\
    &\hspace{1.5cm}+\frac{\lambda^2c_\Delta^4}{2}\pi^\frac{d+1}{2}\Bigl(\frac{2^{-\Delta}c_\Delta}{2\nu}\Bigl)^2\sum_{i=2}^\infty a_i(0)D_{\Delta\Delta\Delta+i\Delta+i}(\vec{y_1},\vec{y_2},\vec{y_3},\vec{y_4})\lvert\vec{y_{34}}\rvert^{2i}+(\vec{y_2}\leftrightarrow\vec{y_3})+(\vec{y_2}\leftrightarrow\vec{y_4})\nonumber\\
    &\hspace{1.5cm}+\frac{\lambda^2c_\Delta^4}{2}\pi^\frac{d+1}{2}\Bigl(\frac{2^{-\Delta}c_\Delta}{2\nu}\Bigl)^2\sum_{i=0}^\infty\Bigl\{b_i(0)D_{\Delta\Delta\Delta+i\Delta+i}(\vec{y_1},\vec{y_2},\vec{y_3},\vec{y_4})\lvert\vec{y_{34}}\rvert^{2i}\nonumber\\
    &\hspace{6.5cm}+c_i(0)\frac{d}{d\alpha}\bigl[D_{\Delta\Delta\Delta+i+\alpha\Delta+i+\alpha}(\vec{y_1},\vec{y_2},\vec{y_3},\vec{y_4})\lvert\vec{y_{34}}\rvert^{2i+2\alpha}\bigl]\Bigl\rvert_{\alpha=0}\Bigl\}\nonumber\\
    &\hspace{1.5cm}+(\vec{y_2}\leftrightarrow\vec{y_3})+(\vec{y_2}\leftrightarrow\vec{y_4})+\mathcal{O}(\lambda^3)\nonumber\\
    &\langle O_\Delta(\vec{y_1})O_\Delta(\vec{y_2})O_\Delta(\vec{y_3})O_\Delta(\vec{y_4})\rangle_{\text{CFT,con}} = \varepsilon^{-\frac{2\Pi(\kappa)}{\nu}}\times-\lambda c_\Delta^4D_{\Delta-\frac{\Pi(\kappa)}{2\nu}\Delta-\frac{\Pi(\kappa)}{2\nu}\Delta-\frac{\Pi(\kappa)}{2\nu}\Delta-\frac{\Pi(\kappa)}{2\nu}}(\vec{y_1},\vec{y_2},\vec{y_3},\vec{y_4})\nonumber\\
    &\hspace{1.5cm}+\Bigl\{\frac{\lambda^2}{2}\pi^\frac{d+1}{2}\Bigl(\frac{2^{-\Delta}c_\Delta}{2\nu}\Bigl)^2\Bigl[3a_0(\kappa)+a_1(\kappa)\frac{(2\Delta-\frac{d}{2})}{\Delta}\Bigl]-\delta\lambda\Bigl\}c_\Delta^4D_{\Delta\Delta\Delta\Delta}(\vec{y_1},\vec{y_2},\vec{y_3},\vec{y_4})\nonumber\\
    &\hspace{1.5cm}+\frac{\lambda^2c_\Delta^4}{2}\pi^\frac{d+1}{2}\Bigl(\frac{2^{-\Delta}c_\Delta}{2\nu}\Bigl)^2\sum_{i=2}^\infty a_i(0)D_{\Delta\Delta\Delta+i\Delta+i}(\vec{y_1},\vec{y_2},\vec{y_3},\vec{y_4})\lvert\vec{y_{34}}\rvert^{2i}+(\vec{y_2}\leftrightarrow\vec{y_3})+(\vec{y_2}\leftrightarrow\vec{y_4})\nonumber\\
    &\hspace{1.5cm}+\frac{\lambda^2c_\Delta^4}{2}\pi^\frac{d+1}{2}\Bigl(\frac{2^{-\Delta}c_\Delta}{2\nu}\Bigl)^2\sum_{i=0}^\infty\Bigl\{b_i(0)D_{\Delta\Delta\Delta+i\Delta+i}(\vec{y_1},\vec{y_2},\vec{y_3},\vec{y_4})\lvert\vec{y_{34}}\rvert^{2i}\nonumber\\
    &\hspace{6.5cm}+c_i(0)\frac{d}{d\alpha}\bigl[D_{\Delta\Delta\Delta+i+\alpha\Delta+i+\alpha}(\vec{y_1},\vec{y_2},\vec{y_3},\vec{y_4})\lvert\vec{y_{34}}\rvert^{2i+2\alpha}\bigl]\Bigl\rvert_{\alpha=0}\Bigl\}\nonumber\\
    &\hspace{1.5cm}+(\vec{y_2}\leftrightarrow\vec{y_3})+(\vec{y_2}\leftrightarrow\vec{y_4})+\mathcal{O}(\lambda^3)
\end{align}\

where since $d<7$, we safely took $\kappa=0$ for the coefficients $a_{i>1}(\kappa)$. Therefore, denoting the coefficients $a_0(\kappa)$ and $a_1(\kappa)$ as $a_0(\kappa)=a^{(\infty)}_0(\kappa)+a^{(0)}_0(\kappa)$ and $a_1(\kappa)=a^{(\infty)}_1(\kappa)+a^{(0)}_1(\kappa)$ respectively, where all their UV-divergent terms are contained in $a^{(\infty)}_0(\kappa)$ and $a^{(\infty)}_1(\kappa)$, the infinities present in the correlators coming from the ultraviolet divergence of the "scalar exchange" diagram can be renormalized away through the convenient choice of the counterterm $\delta\lambda$ as:

\begin{equation}
    \delta\lambda = \frac{\lambda^2}{2}\pi^\frac{d+1}{2}\Bigl(\frac{2^{-\Delta}c_\Delta}{2\nu}\Bigl)^2\Bigl[3a^{(\infty)}_0(\kappa)+a^{(\infty)}_1(\kappa)\frac{(2\Delta-\frac{d}{2})}{\Delta}\Bigl]
\end{equation}\

resulting in the partially renormalized holographic 4-point functions:

\begin{align}
    &\langle O_\Delta(\vec{y_1})O_\Delta(\vec{y_2})O_\Delta(\vec{y_3})O_\Delta(\vec{y_4})\rangle_{\text{CFT}} =\varepsilon^{-\frac{2\Pi(\kappa)}{\nu}}\times\frac{(2\nu c_\Delta)^2}{\prod_{i<j}\lvert\vec{y_{ij}}\rvert^{\frac{2}{3}\bigl(\Delta-\frac{\Pi(\kappa)}{2\nu}\bigl)}}u^{-\frac{2}{3}\bigl(\Delta-\frac{\Pi(\kappa)}{2\nu}\bigl)}v^{\frac{1}{3}\bigl(\Delta-\frac{\Pi(\kappa)}{2\nu}\bigl)}\nonumber\\
    &\hspace{1.5cm}+\Bigl(u,v\rightarrow\frac{1}{u},\frac{v}{u}\Bigl)+(u,v\rightarrow v,u)\nonumber\\
    &\hspace{1.5cm}+\varepsilon^{-\frac{2\Pi(\kappa)}{\nu}}\times-\lambda c_\Delta^4D_{\Delta-\frac{\Pi(\kappa)}{2\nu}\Delta-\frac{\Pi(\kappa)}{2\nu}\Delta-\frac{\Pi(\kappa)}{2\nu}\Delta-\frac{\Pi(\kappa)}{2\nu}}(\vec{y_1},\vec{y_2},\vec{y_3},\vec{y_4})\nonumber\\
    &\hspace{1.5cm}+\frac{\lambda^2c_\Delta^4}{2}\pi^\frac{d+1}{2}\Bigl(\frac{2^{-\Delta}c_\Delta}{2\nu}\Bigl)^2\Bigl[3a^{(0)}_0(0)+a^{(0)}_1(0)\frac{(2\Delta-\frac{d}{2})}{\Delta}\Bigl]D_{\Delta\Delta\Delta\Delta}(\vec{y_1},\vec{y_2},\vec{y_3},\vec{y_4})\nonumber\\
    &\hspace{1.5cm}+\frac{\lambda^2c_\Delta^4}{2}\pi^\frac{d+1}{2}\Bigl(\frac{2^{-\Delta}c_\Delta}{2\nu}\Bigl)^2\sum_{i=2}^\infty a_i(0)D_{\Delta\Delta\Delta+i\Delta+i}(\vec{y_1},\vec{y_2},\vec{y_3},\vec{y_4})\lvert\vec{y_{34}}\rvert^{2i}+(\vec{y_2}\leftrightarrow\vec{y_3})+(\vec{y_2}\leftrightarrow\vec{y_4})\nonumber\\
    &\hspace{1.5cm}+\frac{\lambda^2c_\Delta^4}{2}\pi^\frac{d+1}{2}\Bigl(\frac{2^{-\Delta}c_\Delta}{2\nu}\Bigl)^2\sum_{i=0}^\infty\Bigl\{b_i(0)D_{\Delta\Delta\Delta+i\Delta+i}(\vec{y_1},\vec{y_2},\vec{y_3},\vec{y_4})\lvert\vec{y_{34}}\rvert^{2i}\nonumber\\
    &\hspace{6.5cm}+c_i(0)\frac{d}{d\alpha}\bigl[D_{\Delta\Delta\Delta+i+\alpha\Delta+i+\alpha}(\vec{y_1},\vec{y_2},\vec{y_3},\vec{y_4})\lvert\vec{y_{34}}\rvert^{2i+2\alpha}\bigl]\Bigl\rvert_{\alpha=0}\Bigl\}\nonumber\\
    &\hspace{1.5cm}+(\vec{y_2}\leftrightarrow\vec{y_3})+(\vec{y_2}\leftrightarrow\vec{y_4})+\mathcal{O}(\lambda^3)\nonumber\\
    &\langle O_\Delta(\vec{y_1})O_\Delta(\vec{y_2})O_\Delta(\vec{y_3})O_\Delta(\vec{y_4})\rangle_{\text{CFT,con}} = \varepsilon^{-\frac{2\Pi(\kappa)}{\nu}}\times-\lambda c_\Delta^4D_{\Delta-\frac{\Pi(\kappa)}{2\nu}\Delta-\frac{\Pi(\kappa)}{2\nu}\Delta-\frac{\Pi(\kappa)}{2\nu}\Delta-\frac{\Pi(\kappa)}{2\nu}}(\vec{y_1},\vec{y_2},\vec{y_3},\vec{y_4})\nonumber\\
    &\hspace{1.5cm}+\frac{\lambda^2c_\Delta^4}{2}\pi^\frac{d+1}{2}\Bigl(\frac{2^{-\Delta}c_\Delta}{2\nu}\Bigl)^2\Bigl[3a^{(0)}_0(0)+a^{(0)}_1(0)\frac{(2\Delta-\frac{d}{2})}{\Delta}\Bigl]D_{\Delta\Delta\Delta\Delta}(\vec{y_1},\vec{y_2},\vec{y_3},\vec{y_4})\nonumber\\
    &\hspace{1.5cm}+\frac{\lambda^2c_\Delta^4}{2}\pi^\frac{d+1}{2}\Bigl(\frac{2^{-\Delta}c_\Delta}{2\nu}\Bigl)^2\sum_{i=2}^\infty a_i(0)D_{\Delta\Delta\Delta+i\Delta+i}(\vec{y_1},\vec{y_2},\vec{y_3},\vec{y_4})\lvert\vec{y_{34}}\rvert^{2i}+(\vec{y_2}\leftrightarrow\vec{y_3})+(\vec{y_2}\leftrightarrow\vec{y_4})\nonumber\\
    &\hspace{1.5cm}+\frac{\lambda^2c_\Delta^4}{2}\pi^\frac{d+1}{2}\Bigl(\frac{2^{-\Delta}c_\Delta}{2\nu}\Bigl)^2\sum_{i=0}^\infty\Bigl\{b_i(0)D_{\Delta\Delta\Delta+i\Delta+i}(\vec{y_1},\vec{y_2},\vec{y_3},\vec{y_4})\lvert\vec{y_{34}}\rvert^{2i}\nonumber\\
    &\hspace{6.5cm}+c_i(0)\frac{d}{d\alpha}\bigl[D_{\Delta\Delta\Delta+i+\alpha\Delta+i+\alpha}(\vec{y_1},\vec{y_2},\vec{y_3},\vec{y_4})\lvert\vec{y_{34}}\rvert^{2i+2\alpha}\bigl]\Bigl\rvert_{\alpha=0}\Bigl\}\nonumber\\
    &\hspace{1.5cm}+(\vec{y_2}\leftrightarrow\vec{y_3})+(\vec{y_2}\leftrightarrow\vec{y_4})+\mathcal{O}(\lambda^3)
\end{align}\

We are not done with the renormalization process as we still have to deal with the other divergences of the correlators. However, as we already anticipated, the very same redefinitions for the bulk's parameters introduced in the study of the 2-point function exactly renormalize the divergences present in the current case for the 4-point function. Take for example the divergent anomalous dimension. The redefinition of the bulk's mass parameter $m^2$ as $m^2+\delta m^2$ in the AdS bulk action (where $\delta m^2=\mathcal{O}(\lambda)$) adds, up to order $\lambda^2$, new counterterms interactions to the holographic 4-point functions of the form:

\begin{align}
    &\langle O_\Delta(\vec{y_1})O_\Delta(\vec{y_2})O_\Delta(\vec{y_3})O_\Delta(\vec{y_4})\rangle_{\text{CFT}}\rightarrow\nonumber\langle O_\Delta(\vec{y_1})O_\Delta(\vec{y_2})O_\Delta(\vec{y_3})O_\Delta(\vec{y_4})\rangle_{\text{CFT}}\\
    &\hspace{3cm}+\lambda\delta m^2\int_\varepsilon\int_\varepsilon K(x_1,\vec{y_1})G(x_1,x_2)K(x_2,\vec{y_2})K(x_2,\vec{y_3})K(x_2,\vec{y_4})\times4\nonumber\\
    &\hspace{3cm}+\delta m^2\frac{\lambda}{2}\int_\varepsilon\int_\varepsilon K(x_1,\vec{y_1})G_\kappa(x_1,x_1)K(x_1,\vec{y_2})K(x_2,\vec{y_3})K(x_2,\vec{y_4})\times6\nonumber\\
    &\hspace{3cm}+(\delta m^2)^2\int_\varepsilon\int_\varepsilon K(x_1,\vec{y_1})K(x_1,\vec{y_2})K(x_2,\vec{y_3})K(x_2,\vec{y_4})\times3\nonumber\\
    &\hspace{3cm}+\frac{2\nu c_\Delta}{\lvert\vec{y_1}-\vec{y_2}\rvert^{2\Delta}}\Bigl[-\delta m^2\int_\varepsilon K(x_1,\vec{y_3})K(x_1,\vec{y_4})\nonumber\\
    &\hspace{5.6cm}+\delta m^2\frac{\lambda}{2}\int_\varepsilon\int_\varepsilon K(x_1,\vec{y_3})G_\kappa(x_1,x_1)G_\kappa(x_1,x_2)K(x_2,\vec{y_4})\times2\nonumber\\
    &\hspace{5.6cm}+(\delta m^2)^2\int_\varepsilon\int_\varepsilon K(x_1,\vec{y_3})G_\kappa(x_1,x_2)K(x_2,\vec{y_4})\Bigl]\times6+\mathcal{O}(\lambda^3)\nonumber\\
    &\langle O_\Delta(\vec{y_1})O_\Delta(\vec{y_2})O_\Delta(\vec{y_3})O_\Delta(\vec{y_4})\rangle_{\text{CFT,con}}\rightarrow\langle O_\Delta(\vec{y_1})O_\Delta(\vec{y_2})O_\Delta(\vec{y_3})O_\Delta(\vec{y_4})\rangle_{\text{CFT,con}}\nonumber\\
    &\hspace{3cm}+\lambda\delta m^2\int_\varepsilon\int_\varepsilon K(x_1,\vec{y_1})G(x_1,x_2)K(x_2,\vec{y_2})K(x_2,\vec{y_3})K(x_2,\vec{y_4})\times4+\mathcal{O}(\lambda^3)
\end{align}\

It turns out that considering these new contributions to the correlators coming from the redefinition of $m^2$ in a earlier step in the computation of the 4-point function exactly factorize with those terms dependent on $\Pi(\kappa)$, resulting for the current expressions in the replacement of $\Pi(\kappa)\rightarrow\Pi(\kappa)-\delta m^2$. Therefore, the exact same choice for the counterterm $\delta m^2$ as $\delta m^2=\Pi_\infty(\kappa)$ made in the renormalization of the 2-point function also renormalizes the UV-divergences of the anomalous dimension present in the 4-point functions. Moreover, the redefinition of the bulk field $\Phi(x)$ as $\Phi(x)\rightarrow\sqrt{Z(\lambda)}\Phi(x)$ in the AdS bulk action adds a factor $\frac{1}{Z(\lambda)^2}$ to the holographic 4-point functions. It is also direct to see then that the exact same choice for the counterterm $Z(\lambda)$ as $Z(\lambda)=\varepsilon^{-\frac{\Pi_0(0)}{\nu}}$ made in the renormalization of the 2-point function also renormalizes the IR-divergence of the overall rescaling of the 4-point functions, resulting in both IR and UV renormalized correlators:

\begin{align}\label{renorm4pfphi4quantum}
    &\langle O_\Delta(\vec{y_1})O_\Delta(\vec{y_2})O_\Delta(\vec{y_3})O_\Delta(\vec{y_4})\rangle_{\text{CFT}} =\frac{(2\nu c_\Delta)^2}{\prod_{i<j}\lvert\vec{y_{ij}}\rvert^{\frac{2}{3}\bigl(\Delta-\frac{\Pi_0(0)}{2\nu}\bigl)}}u^{-\frac{2}{3}\bigl(\Delta-\frac{\Pi_0(0)}{2\nu}\bigl)}v^{\frac{1}{3}\bigl(\Delta-\frac{\Pi_0(0)}{2\nu}\bigl)}\nonumber\\
    &\hspace{1.5cm}+\Bigl(u,v\rightarrow\frac{1}{u},\frac{v}{u}\Bigl)+(u,v\rightarrow v,u)-\lambda c_\Delta^4D_{\Delta-\frac{\Pi_0(0)}{2\nu}\Delta-\frac{\Pi_0(0)}{2\nu}\Delta-\frac{\Pi_0(0)}{2\nu}\Delta-\frac{\Pi_0(0)}{2\nu}}(\vec{y_1},\vec{y_2},\vec{y_3},\vec{y_4})\nonumber\\
    &\hspace{1.5cm}+\frac{\lambda^2c_\Delta^4}{2}\pi^\frac{d+1}{2}\Bigl(\frac{2^{-\Delta}c_\Delta}{2\nu}\Bigl)^2\Bigl[3a^{(0)}_0(0)+a^{(0)}_1(0)\frac{(2\Delta-\frac{d}{2})}{\Delta}\Bigl]D_{\Delta\Delta\Delta\Delta}(\vec{y_1},\vec{y_2},\vec{y_3},\vec{y_4})\nonumber\\
    &\hspace{1.5cm}+\frac{\lambda^2c_\Delta^4}{2}\pi^\frac{d+1}{2}\Bigl(\frac{2^{-\Delta}c_\Delta}{2\nu}\Bigl)^2\sum_{i=2}^\infty a_i(0)D_{\Delta\Delta\Delta+i\Delta+i}(\vec{y_1},\vec{y_2},\vec{y_3},\vec{y_4})\lvert\vec{y_{34}}\rvert^{2i}+(\vec{y_2}\leftrightarrow\vec{y_3})+(\vec{y_2}\leftrightarrow\vec{y_4})\nonumber\\
    &\hspace{1.5cm}+\frac{\lambda^2c_\Delta^4}{2}\pi^\frac{d+1}{2}\Bigl(\frac{2^{-\Delta}c_\Delta}{2\nu}\Bigl)^2\sum_{i=0}^\infty\Bigl\{b_i(0)D_{\Delta\Delta\Delta+i\Delta+i}(\vec{y_1},\vec{y_2},\vec{y_3},\vec{y_4})\lvert\vec{y_{34}}\rvert^{2i}\nonumber\\
    &\hspace{6.5cm}+c_i(0)\frac{d}{d\alpha}\bigl[D_{\Delta\Delta\Delta+i+\alpha\Delta+i+\alpha}(\vec{y_1},\vec{y_2},\vec{y_3},\vec{y_4})\lvert\vec{y_{34}}\rvert^{2i+2\alpha}\bigl]\Bigl\rvert_{\alpha=0}\Bigl\}\nonumber\\
    &\hspace{1.5cm}+(\vec{y_2}\leftrightarrow\vec{y_3})+(\vec{y_2}\leftrightarrow\vec{y_4})+\mathcal{O}(\lambda^3)\nonumber\\
    &\langle O_\Delta(\vec{y_1})O_\Delta(\vec{y_2})O_\Delta(\vec{y_3})O_\Delta(\vec{y_4})\rangle_{\text{CFT,con}} = -\lambda c_\Delta^4D_{\Delta-\frac{\Pi_0(0)}{2\nu}\Delta-\frac{\Pi_0(0)}{2\nu}\Delta-\frac{\Pi_0(0)}{2\nu}\Delta-\frac{\Pi_0(0)}{2\nu}}(\vec{y_1},\vec{y_2},\vec{y_3},\vec{y_4})\nonumber\\
    &\hspace{1.5cm}+\frac{\lambda^2c_\Delta^4}{2}\pi^\frac{d+1}{2}\Bigl(\frac{2^{-\Delta}c_\Delta}{2\nu}\Bigl)^2\Bigl[3a^{(0)}_0(0)+a^{(0)}_1(0)\frac{(2\Delta-\frac{d}{2})}{\Delta}\Bigl]D_{\Delta\Delta\Delta\Delta}(\vec{y_1},\vec{y_2},\vec{y_3},\vec{y_4})\nonumber\\
    &\hspace{1.5cm}+\frac{\lambda^2c_\Delta^4}{2}\pi^\frac{d+1}{2}\Bigl(\frac{2^{-\Delta}c_\Delta}{2\nu}\Bigl)^2\sum_{i=2}^\infty a_i(0)D_{\Delta\Delta\Delta+i\Delta+i}(\vec{y_1},\vec{y_2},\vec{y_3},\vec{y_4})\lvert\vec{y_{34}}\rvert^{2i}+(\vec{y_2}\leftrightarrow\vec{y_3})+(\vec{y_2}\leftrightarrow\vec{y_4})\nonumber\\
    &\hspace{1.5cm}+\frac{\lambda^2c_\Delta^4}{2}\pi^\frac{d+1}{2}\Bigl(\frac{2^{-\Delta}c_\Delta}{2\nu}\Bigl)^2\sum_{i=0}^\infty\Bigl\{b_i(0)D_{\Delta\Delta\Delta+i\Delta+i}(\vec{y_1},\vec{y_2},\vec{y_3},\vec{y_4})\lvert\vec{y_{34}}\rvert^{2i}\nonumber\\
    &\hspace{6.5cm}+c_i(0)\frac{d}{d\alpha}\bigl[D_{\Delta\Delta\Delta+i+\alpha\Delta+i+\alpha}(\vec{y_1},\vec{y_2},\vec{y_3},\vec{y_4})\lvert\vec{y_{34}}\rvert^{2i+2\alpha}\bigl]\Bigl\rvert_{\alpha=0}\Bigl\}\nonumber\\
    &\hspace{1.5cm}+(\vec{y_2}\leftrightarrow\vec{y_3})+(\vec{y_2}\leftrightarrow\vec{y_4})+\mathcal{O}(\lambda^3)
\end{align}\

where the limits $\varepsilon=\kappa=0$ have been taken. The complete study of these type of D-functions can be found in section A.3 of Appendix A, concluding in its value in eq. (\ref{4dfuncgeneral}). Using this formula then in our present case we find that the different D-functions present in eq. (\ref{renorm4pfphi4quantum}) can be written as:

\begin{equation}
    D_{\Delta\Delta\Delta\Delta} = \frac{\pi^{\frac{d}{2}}}{2}\frac{\Gamma(2\Delta-\frac{d}{2})}{\Gamma(\Delta)^4}\frac{u^{\frac{\Delta}{3}}v^{\frac{\Delta}{3}}}{\prod_{i<j}\lvert\vec{y_i}-\vec{y_j}\rvert^{\frac{2\Delta}{3}}}H(\Delta,\Delta,1,2\Delta;u,v)
\end{equation}

\begin{align}
    D_{\Delta-\frac{\Pi_0(0)}{2\nu}\Delta-\frac{\Pi_0(0)}{2\nu}\Delta-\frac{\Pi_0(0)}{2\nu}\Delta-\frac{\Pi_0(0)}{2\nu}} = &\frac{\pi^{\frac{d}{2}}}{2}\frac{\Gamma(2\Delta-\frac{d}{2}-\frac{\Pi_0(0)}{\nu})}{\Gamma(\Delta-\frac{\Pi_0(0)}{2\nu})^4}\frac{u^{\frac{1}{3}\bigl(\Delta-\frac{\Pi_0(0)}{2\nu}\bigl)}v^{\frac{1}{3}\bigl(\Delta-\frac{\Pi_0(0)}{2\nu}\bigl)}}{\prod_{i<j}\lvert\vec{y_i}-\vec{y_j}\rvert^{\frac{2}{3}\bigl(\Delta-\frac{\Pi_0(0)}{2\nu}\bigl)}}\nonumber\\
    &\times H\Bigl(\Delta-\frac{\Pi_0(0)}{2\nu},\Delta-\frac{\Pi_0(0)}{2\nu},1,2\Delta-\frac{\Pi_0(0)}{\nu};u,v\Bigl)
\end{align}

\begin{equation}
    D_{\Delta\Delta\Delta+i\Delta+i}\lvert\vec{y_{34}}\rvert^{2i} =\frac{\pi^\frac{d}{2}}{2}\frac{\Gamma(2\Delta-\frac{d}{2}+i)}{\Gamma(\Delta)^2\Gamma(\Delta+i)^2}\frac{u^{\frac{\Delta}{3}}v^{\frac{\Delta}{3}}}{\prod_{i<j}\lvert\vec{y_{ij}}\rvert^{\frac{2\Delta}{3}}}H(\Delta,\Delta,1-i,2\Delta;u,v)
\end{equation}

\begin{equation}
    D_{\Delta\Delta\Delta+i+\alpha\Delta+i+\alpha}\lvert\vec{y_{34}}\rvert^{2i+2\alpha} =\frac{\pi^\frac{d}{2}}{2}\frac{\Gamma(2\Delta-\frac{d}{2}+i+\alpha)}{\Gamma(\Delta)^2\Gamma(\Delta+i+\alpha)^2}\frac{u^{\frac{\Delta}{3}}v^{\frac{\Delta}{3}}}{\prod_{i<j}\lvert\vec{y_{ij}}\rvert^{\frac{2\Delta}{3}}}H(\Delta,\Delta,1-i-\alpha,2\Delta;u,v)
\end{equation}\

which further allow us to express the renormalized 4-point functions in their explicit forms:

\begin{align}
    &\langle O_\Delta(\vec{y_1})O_\Delta(\vec{y_2})O_\Delta(\vec{y_3})O_\Delta(\vec{y_4})\rangle_{\text{CFT}} =\frac{(2\nu c_\Delta)^2}{\prod_{i<j}\lvert\vec{y_{ij}}\rvert^{\frac{2}{3}\bigl(\Delta-\frac{\Pi_0(0)}{2\nu}\bigl)}}u^{-\frac{2}{3}\bigl(\Delta-\frac{\Pi_0(0)}{2\nu}\bigl)}v^{\frac{1}{3}\bigl(\Delta-\frac{\Pi_0(0)}{2\nu}\bigl)}\nonumber\\
    &\hspace{1cm}+\Bigl(u,v\rightarrow\frac{1}{u},\frac{v}{u}\Bigl)+(u,v\rightarrow v,u)-\lambda c_\Delta^4\frac{\pi^{\frac{d}{2}}}{2}\frac{\Gamma(2\Delta-\frac{d}{2}-\frac{\Pi_0(0)}{\nu})}{\Gamma(\Delta-\frac{\Pi_0(0)}{2\nu})^4}\frac{u^{\frac{1}{3}\bigl(\Delta-\frac{\Pi_0(0)}{2\nu}\bigl)}v^{\frac{1}{3}\bigl(\Delta-\frac{\Pi_0(0)}{2\nu}\bigl)}}{\prod_{i<j}\lvert\vec{y_i}-\vec{y_j}\rvert^{\frac{2}{3}\bigl(\Delta-\frac{\Pi_0(0)}{2\nu}\bigl)}}\nonumber\\
    &\hspace{1.5cm}\times H\Bigl(\Delta-\frac{\Pi_0(0)}{2\nu},\Delta-\frac{\Pi_0(0)}{2\nu},1,2\Delta-\frac{\Pi_0(0)}{\nu};u,v\Bigl)\nonumber\\
    &\hspace{1cm}+\frac{\lambda^2c_\Delta^4}{4}\pi^{d+\frac{1}{2}}\Bigl(\frac{2^{-\Delta}c_\Delta}{2\nu}\Bigl)^2\Bigl[3a^{(0)}_0(0)+a^{(0)}_1(0)\frac{(2\Delta-\frac{d}{2})}{\Delta}\Bigl]\frac{\Gamma(2\Delta-\frac{d}{2})}{\Gamma(\Delta)^4}\frac{u^{\frac{\Delta}{3}}v^{\frac{\Delta}{3}}}{\prod_{i<j}\lvert\vec{y_i}-\vec{y_j}\rvert^{\frac{2\Delta}{3}}}\nonumber\\
    &\hspace{1.5cm}\times H(\Delta,\Delta,1,2\Delta;u,v)\nonumber\\
    &\hspace{1cm}+\frac{\lambda^2c_\Delta^4}{4}\pi^{d+\frac{1}{2}}\Bigl(\frac{2^{-\Delta}c_\Delta}{2\nu}\Bigl)^2\frac{1}{\Gamma(\Delta)^2}\frac{u^{\frac{\Delta}{3}}v^{\frac{\Delta}{3}}}{\prod_{i<j}\lvert\vec{y_{ij}}\rvert^{\frac{2\Delta}{3}}}\sum_{i=2}^\infty a_i(0)\frac{\Gamma(2\Delta-\frac{d}{2}+i)}{\Gamma(\Delta+i)^2}H(\Delta,\Delta,1-i,2\Delta;u,v)\nonumber\\
    &\hspace{1cm}+\Bigl(u,v\rightarrow\frac{1}{u},\frac{v}{u}\Bigl)+(u,v\rightarrow v,u)\nonumber\\
    &\hspace{1cm}+\frac{\lambda^2c_\Delta^4}{4}\pi^{d+\frac{1}{2}}\Bigl(\frac{2^{-\Delta}c_\Delta}{2\nu}\Bigl)^2\frac{1}{\Gamma(\Delta)^2}\frac{u^{\frac{\Delta}{3}}v^{\frac{\Delta}{3}}}{\prod_{i<j}\lvert\vec{y_{ij}}\rvert^{\frac{2\Delta}{3}}}\sum_{i=0}^\infty\Bigl\{b_i(0)\frac{\Gamma(2\Delta-\frac{d}{2}+i)}{\Gamma(\Delta+i)^2}H(\Delta,\Delta,1-i,2\Delta;u,v)\nonumber\\
    &\hspace{3cm}+c_i(0)\frac{d}{d\alpha}\Bigl[\frac{\Gamma(2\Delta-\frac{d}{2}+i+\alpha)}{\Gamma(\Delta+i+\alpha)^2}H(\Delta,\Delta,1-i-\alpha,2\Delta;u,v)\Bigl]\Bigl\rvert_{\alpha=0}\Bigl\}\nonumber\\
    &\hspace{1cm}+\Bigl(u,v\rightarrow\frac{1}{u},\frac{v}{u}\Bigl)+(u,v\rightarrow v,u)+\mathcal{O}(\lambda^3)
\end{align}
\begin{align}
    &\langle O_\Delta(\vec{y_1})O_\Delta(\vec{y_2})O_\Delta(\vec{y_3})O_\Delta(\vec{y_4})\rangle_{\text{CFT,con}} = -\lambda c_\Delta^4\frac{\pi^{\frac{d}{2}}}{2}\frac{\Gamma(2\Delta-\frac{d}{2}-\frac{\Pi_0(0)}{\nu})}{\Gamma(\Delta-\frac{\Pi_0(0)}{2\nu})^4}\frac{u^{\frac{1}{3}\bigl(\Delta-\frac{\Pi_0(0)}{2\nu}\bigl)}v^{\frac{1}{3}\bigl(\Delta-\frac{\Pi_0(0)}{2\nu}\bigl)}}{\prod_{i<j}\lvert\vec{y_i}-\vec{y_j}\rvert^{\frac{2}{3}\bigl(\Delta-\frac{\Pi_0(0)}{2\nu}\bigl)}}\nonumber\\
    &\hspace{1.5cm}\times H\Bigl(\Delta-\frac{\Pi_0(0)}{2\nu},\Delta-\frac{\Pi_0(0)}{2\nu},1,2\Delta-\frac{\Pi_0(0)}{\nu};u,v\Bigl)\nonumber\\
    &\hspace{1cm}+\frac{\lambda^2c_\Delta^4}{4}\pi^{d+\frac{1}{2}}\Bigl(\frac{2^{-\Delta}c_\Delta}{2\nu}\Bigl)^2\Bigl[3a^{(0)}_0(0)+a^{(0)}_1(0)\frac{(2\Delta-\frac{d}{2})}{\Delta}\Bigl]\frac{\Gamma(2\Delta-\frac{d}{2})}{\Gamma(\Delta)^4}\frac{u^{\frac{\Delta}{3}}v^{\frac{\Delta}{3}}}{\prod_{i<j}\lvert\vec{y_i}-\vec{y_j}\rvert^{\frac{2\Delta}{3}}}\nonumber\\
    &\hspace{1.5cm}\times H(\Delta,\Delta,1,2\Delta;u,v)\nonumber\\
    &\hspace{1cm}+\frac{\lambda^2c_\Delta^4}{4}\pi^{d+\frac{1}{2}}\Bigl(\frac{2^{-\Delta}c_\Delta}{2\nu}\Bigl)^2\frac{1}{\Gamma(\Delta)^2}\frac{u^{\frac{\Delta}{3}}v^{\frac{\Delta}{3}}}{\prod_{i<j}\lvert\vec{y_{ij}}\rvert^{\frac{2\Delta}{3}}}\sum_{i=2}^\infty a_i(0)\frac{\Gamma(2\Delta-\frac{d}{2}+i)}{\Gamma(\Delta+i)^2}H(\Delta,\Delta,1-i,2\Delta;u,v)\nonumber\\
    &\hspace{1cm}+\Bigl(u,v\rightarrow\frac{1}{u},\frac{v}{u}\Bigl)+(u,v\rightarrow v,u)\nonumber\\
    &\hspace{1cm}+\frac{\lambda^2c_\Delta^4}{4}\pi^{d+\frac{1}{2}}\Bigl(\frac{2^{-\Delta}c_\Delta}{2\nu}\Bigl)^2\frac{1}{\Gamma(\Delta)^2}\frac{u^{\frac{\Delta}{3}}v^{\frac{\Delta}{3}}}{\prod_{i<j}\lvert\vec{y_{ij}}\rvert^{\frac{2\Delta}{3}}}\sum_{i=0}^\infty\Bigl\{b_i(0)\frac{\Gamma(2\Delta-\frac{d}{2}+i)}{\Gamma(\Delta+i)^2}H(\Delta,\Delta,1-i,2\Delta;u,v)\nonumber\\
    &\hspace{3cm}+c_i(0)\frac{d}{d\alpha}\Bigl[\frac{\Gamma(2\Delta-\frac{d}{2}+i+\alpha)}{\Gamma(\Delta+i+\alpha)^2}H(\Delta,\Delta,1-i-\alpha,2\Delta;u,v)\Bigl]\Bigl\rvert_{\alpha=0}\Bigl\}\nonumber\\
    &\hspace{1cm}+\Bigl(u,v\rightarrow\frac{1}{u},\frac{v}{u}\Bigl)+(u,v\rightarrow v,u)+\mathcal{O}(\lambda^3)
\end{align}\

where we wrote the permutations in terms of the cross ratios $u$ and $v$. This result concludes the renormalization of the holographic correlators coming from a $\Phi^4$ theory on AdS.

\subsection{Renormalized Correlators}

The objective of this section is to summarize the key points of the recent renormalization study of the quantum corrected holographic correlators resulting from the consideration of a self-interacting scalar $\Phi^4$ theory on a fixed AdS background through the use of the AdS/CFT correspondence. As we saw, these correlators were infrared divergent as their different contributions approached the conformal boundary of the AdS space, and also ultraviolet divergent as their loops integrals involving the bulk-bulk propagator got integrated at coincident points. In order to compute finite and predictive correlators, these divergences demanded not only a delicate regularization scheme but also a delicate renormalization scheme, in order to absorb in a sensitive way the corresponding infinities. The infrared divergences present at the on-shell level of the AdS path integral were both regulated and renormalized through the holographic renormalization procedure with the addition of a covariant boundary term in the AdS action. This procedure for the infrared divergences present at the off-shell level of the AdS path integral naturally translated into their regularization by simply solving the loops contributions to the correlators up to the same radial regulator introduced in the holographic renormalization. While for the ultraviolet divergences present at the off-shell level of the AdS path integral, a point-splitting approach was taken, resulting in regularized bulk-bulk propagators which conserved their symmetry under AdS transformations. By explicitly computing these regularized correlators, their nice form for $d<7$ allowed us to renormalize them in exactly the same spirit as it is done for ordinary QFTs, that is, by understating the parameters of the theory, $\Phi(x)$, $m^2$ and $\lambda,$ not as physical constants but bare quantities, opening the possibility of a renormalization scheme through their definition. This turned out to be indeed the case, where the redefinition of these parameters in the AdS bulk action as:

\begin{equation}
    \Phi(x)\rightarrow\sqrt{Z(\lambda)}\Phi(x),\hspace{0.5cm}m^2\rightarrow m^2+\delta m^2,\hspace{0.5cm}\lambda\rightarrow\lambda+\delta\lambda
\end{equation}\

exactly renormalized every single divergence present in the holographic n-point functions through the convenient choice of the counterterms $Z(\lambda)$, $\delta m^2$ and $\delta\lambda$ as:

\begin{equation}
    Z(\lambda)=\varepsilon^{-\frac{\Pi_0(0)}{\nu}},\hspace{0.5cm}\delta m^2=\Pi_\infty(\kappa),\hspace{0.5cm}\delta\lambda = \frac{\lambda^2}{2}\pi^\frac{d+1}{2}\Bigl(\frac{2^{-\Delta}c_\Delta}{2\nu}\Bigl)^2\Bigl[3a^{(\infty)}_0(\kappa)+a^{(\infty)}_1(\kappa)\frac{(2\Delta-\frac{d}{2})}{\Delta}\Bigl]
\end{equation}\

where $\varepsilon$ and $\kappa$ are the IR and UV regulators introduced in the regularization scheme, $\Pi_\infty(\kappa)$ and $\Pi_0(0)$ are the UV-divergent and UV-convergent parts of the 1PI contributions $\Pi(\kappa)$:

\begin{align}\label{1picontribphi4final}
    \Pi(\kappa) = &-\frac{\lambda G_\kappa(1)}{2}+\frac{\lambda^2G_\kappa(1)}{4}\pi^{\frac{d+1}{2}}\Bigl(\frac{2^{-\Delta}c_\Delta}{2\nu}\Bigl)^2\sum_{k=0}^\infty a^{(2)}_k\frac{\Gamma(\nu+k)}{\Gamma(\Delta+\frac{1}{2}+k)}\Bigl(\frac{1}{1+\kappa}\Bigl)^{2\Delta+2k}\nonumber\\
    &\hspace{6cm}\times\ _2F_1\Bigl(d-1,\nu+k;\Delta+\frac{1}{2}+k;\Bigl(\frac{1}{1+\kappa}\Bigl)^2\Bigl)\nonumber\\
    &+\frac{\lambda^2}{6}\pi^\frac{d+1}{2}\Bigl(\frac{2^{-\Delta}c_\Delta}{2\nu}\Bigl)^3\sum_{k=0}^\infty a^{(3)}_k\frac{\Gamma(2\Delta-\frac{d}{2}+k)\Gamma(\Delta+k)}{\Gamma(\frac{3\Delta}{2}+k)\Gamma(\frac{3\Delta+1}{2}+k)}\Bigl(\frac{1}{1+\kappa}\Bigl)^{3\Delta+2k}\nonumber\\
    &\hspace{2cm}\times\ _3F_2\Bigl(\frac{3(d-1)}{2},2\Delta-\frac{d}{2}+k,\Delta+k;\frac{3\Delta}{2}+k,\frac{3\Delta+1}{2}+k;\Bigl(\frac{1}{1+\kappa}\Bigl)^2\Bigl)
\end{align}\

where the coefficients $a^{(2)}_k$ and $a^{(3)}_k$ present in $\Pi(\kappa)$ were defined in eqs. (\ref{a2kcoef}) and (\ref{a3kcoef}) respectively, and where $a_0^{(\infty)}(\kappa)$ and $a_1^{(\infty)}(\kappa)$ present in $\delta\lambda$ are the UV-divergent parts of the $i=0$ and $i=1$ coefficients $a_i(\kappa)$:

\begin{align}\label{ai0coeffin}
    a_i(\kappa) = &\frac{(\Delta)_i(\Delta)_i}{i!}\frac{\Gamma(2\Delta-\frac{d}{2}+1+i)}{\Gamma(\Delta+\frac{3}{2}+i)\Gamma(\Delta+1+i)}\Bigl(\frac{1}{1+\kappa}\Bigl)^{2\Delta+2i+2}(-1)^i\nonumber\\
    &\times\sum_{l=0}^\infty\frac{(2\Delta-\frac{d}{2}+1+i)_l(1)_l(1)_l}{(\Delta+\frac{3}{2}+i)_l(\Delta+1+i)_l\ l!}\Bigl(\frac{1}{1+\kappa}\Bigl)^{2l}\sum_{k=0}^{l+i+1} a^{(2)}_k\frac{(d-1)_{l+i+1-k}}{(l+i+1-k)!}
\end{align}\

The renormalized correlators for $d<7$ and up to order $\lambda^2$ in the coupling constant resulting from these redefinition of the bulk's theory parameters, along with their convenient choice for the counterterms, can be summarized into the holographic 1-, 2-, 3- and 4-point functions:

\begin{align}\label{phi4corrparte1}
    \text{1-pt fn:}\hspace{0.25cm}&\langle O_\Delta(\vec{y_1})\rangle_{\text{CFT}} = \langle O_\Delta(\vec{y_1})\rangle_{\text{CFT,con}} = 0\nonumber\\
    \text{2-pt fn:}\hspace{0.25cm}&\langle O_\Delta(\vec{y_1})O_\Delta(\vec{y_2})\rangle_{\text{CFT}} = \langle O_\Delta(\vec{y_1})O_\Delta(\vec{y_2})\rangle_{\text{CFT,con}} = \frac{2\nu c_\Delta}{\lvert\vec{y_1}-\vec{y_2}\rvert^{2\Delta-\frac{\Pi_0(0)}{\nu}}}+\mathcal{O}(\lambda^3)\nonumber\\
    \text{3-pt fn:}\hspace{0.25cm}&\langle O_\Delta(\vec{y_1})O_\Delta(\vec{y_2})O_\Delta(\vec{y_3})\rangle_{\text{CFT}} = \langle O_\Delta(\vec{y_1})O_\Delta(\vec{y_2})O_\Delta(\vec{y_3})\rangle_{\text{CFT,con}}=0\nonumber\\
    \text{4-pt fn:}\hspace{0.25cm}&\langle O_\Delta(\vec{y_1})O_\Delta(\vec{y_2})O_\Delta(\vec{y_3})O_\Delta(\vec{y_4})\rangle_{\text{CFT}}=\frac{(2\nu c_\Delta)^2}{\prod_{i<j}\lvert\vec{y_{ij}}\rvert^{\frac{2}{3}\bigl(\Delta-\frac{\Pi_0(0)}{2\nu}\bigl)}}u^{-\frac{2}{3}\bigl(\Delta-\frac{\Pi_0(0)}{2\nu}\bigl)}v^{\frac{1}{3}\bigl(\Delta-\frac{\Pi_0(0)}{2\nu}\bigl)}\nonumber\\
    &+\Bigl(u,v\rightarrow\frac{1}{u},\frac{v}{u}\Bigl)+(u,v\rightarrow v,u)-\lambda c_\Delta^4\frac{\pi^{\frac{d}{2}}}{2}\frac{\Gamma(2\Delta-\frac{d}{2}-\frac{\Pi_0(0)}{\nu})}{\Gamma(\Delta-\frac{\Pi_0(0)}{2\nu})^4}\frac{u^{\frac{1}{3}\bigl(\Delta-\frac{\Pi_0(0)}{2\nu}\bigl)}v^{\frac{1}{3}\bigl(\Delta-\frac{\Pi_0(0)}{2\nu}\bigl)}}{\prod_{i<j}\lvert\vec{y_i}-\vec{y_j}\rvert^{\frac{2}{3}\bigl(\Delta-\frac{\Pi_0(0)}{2\nu}\bigl)}}\nonumber\\
    &\hspace{1.5cm}\times H\Bigl(\Delta-\frac{\Pi_0(0)}{2\nu},\Delta-\frac{\Pi_0(0)}{2\nu},1,2\Delta-\frac{\Pi_0(0)}{\nu};u,v\Bigl)\nonumber\\
    &+\frac{\lambda^2c_\Delta^4}{4}\pi^{d+\frac{1}{2}}\Bigl(\frac{2^{-\Delta}c_\Delta}{2\nu}\Bigl)^2\Bigl[3a^{(0)}_0(0)+a^{(0)}_1(0)\frac{(2\Delta-\frac{d}{2})}{\Delta}\Bigl]\frac{\Gamma(2\Delta-\frac{d}{2})}{\Gamma(\Delta)^4}\frac{u^{\frac{\Delta}{3}}v^{\frac{\Delta}{3}}}{\prod_{i<j}\lvert\vec{y_i}-\vec{y_j}\rvert^{\frac{2\Delta}{3}}}\nonumber\\
    &\hspace{1.5cm}\times H(\Delta,\Delta,1,2\Delta;u,v)\nonumber\\
    &+\frac{\lambda^2c_\Delta^4}{4}\pi^{d+\frac{1}{2}}\Bigl(\frac{2^{-\Delta}c_\Delta}{2\nu}\Bigl)^2\frac{1}{\Gamma(\Delta)^2}\frac{u^{\frac{\Delta}{3}}v^{\frac{\Delta}{3}}}{\prod_{i<j}\lvert\vec{y_{ij}}\rvert^{\frac{2\Delta}{3}}}\sum_{i=2}^\infty a_i(0)\frac{\Gamma(2\Delta-\frac{d}{2}+i)}{\Gamma(\Delta+i)^2}H(\Delta,\Delta,1-i,2\Delta;u,v)\nonumber\\
    &+\Bigl(u,v\rightarrow\frac{1}{u},\frac{v}{u}\Bigl)+(u,v\rightarrow v,u)\nonumber\\
    &+\frac{\lambda^2c_\Delta^4}{4}\pi^{d+\frac{1}{2}}\Bigl(\frac{2^{-\Delta}c_\Delta}{2\nu}\Bigl)^2\frac{1}{\Gamma(\Delta)^2}\frac{u^{\frac{\Delta}{3}}v^{\frac{\Delta}{3}}}{\prod_{i<j}\lvert\vec{y_{ij}}\rvert^{\frac{2\Delta}{3}}}\sum_{i=0}^\infty\Bigl\{b_i(0)\frac{\Gamma(2\Delta-\frac{d}{2}+i)}{\Gamma(\Delta+i)^2}H(\Delta,\Delta,1-i,2\Delta;u,v)\nonumber\\
    &\hspace{3cm}+c_i(0)\frac{d}{d\alpha}\Bigl[\frac{\Gamma(2\Delta-\frac{d}{2}+i+\alpha)}{\Gamma(\Delta+i+\alpha)^2}H(\Delta,\Delta,1-i-\alpha,2\Delta;u,v)\Bigl]\Bigl\rvert_{\alpha=0}\Bigl\}\nonumber\\
    &+\Bigl(u,v\rightarrow\frac{1}{u},\frac{v}{u}\Bigl)+(u,v\rightarrow v,u)+\mathcal{O}(\lambda^3)
\end{align}
\begin{align}\label{phi4corrparte2}
    &\langle O_\Delta(\vec{y_1})O_\Delta(\vec{y_2})O_\Delta(\vec{y_3})O_\Delta(\vec{y_4})\rangle_{\text{CFT,con}} =-\lambda c_\Delta^4\frac{\pi^{\frac{d}{2}}}{2}\frac{\Gamma(2\Delta-\frac{d}{2}-\frac{\Pi_0(0)}{\nu})}{\Gamma(\Delta-\frac{\Pi_0(0)}{2\nu})^4}\frac{u^{\frac{1}{3}\bigl(\Delta-\frac{\Pi_0(0)}{2\nu}\bigl)}v^{\frac{1}{3}\bigl(\Delta-\frac{\Pi_0(0)}{2\nu}\bigl)}}{\prod_{i<j}\lvert\vec{y_i}-\vec{y_j}\rvert^{\frac{2}{3}\bigl(\Delta-\frac{\Pi_0(0)}{2\nu}\bigl)}}\nonumber\\
    &\hspace{1.5cm}\times H\Bigl(\Delta-\frac{\Pi_0(0)}{2\nu},\Delta-\frac{\Pi_0(0)}{2\nu},1,2\Delta-\frac{\Pi_0(0)}{\nu};u,v\Bigl)\nonumber\\
    &+\frac{\lambda^2c_\Delta^4}{4}\pi^{d+\frac{1}{2}}\Bigl(\frac{2^{-\Delta}c_\Delta}{2\nu}\Bigl)^2\Bigl[3a^{(0)}_0(0)+a^{(0)}_1(0)\frac{(2\Delta-\frac{d}{2})}{\Delta}\Bigl]\frac{\Gamma(2\Delta-\frac{d}{2})}{\Gamma(\Delta)^4}\frac{u^{\frac{\Delta}{3}}v^{\frac{\Delta}{3}}}{\prod_{i<j}\lvert\vec{y_i}-\vec{y_j}\rvert^{\frac{2\Delta}{3}}}\nonumber\\
    &\hspace{1.5cm}\times H(\Delta,\Delta,1,2\Delta;u,v)\nonumber\\
    &+\frac{\lambda^2c_\Delta^4}{4}\pi^{d+\frac{1}{2}}\Bigl(\frac{2^{-\Delta}c_\Delta}{2\nu}\Bigl)^2\frac{1}{\Gamma(\Delta)^2}\frac{u^{\frac{\Delta}{3}}v^{\frac{\Delta}{3}}}{\prod_{i<j}\lvert\vec{y_{ij}}\rvert^{\frac{2\Delta}{3}}}\sum_{i=2}^\infty a_i(0)\frac{\Gamma(2\Delta-\frac{d}{2}+i)}{\Gamma(\Delta+i)^2}H(\Delta,\Delta,1-i,2\Delta;u,v)\nonumber\\
    &+\Bigl(u,v\rightarrow\frac{1}{u},\frac{v}{u}\Bigl)+(u,v\rightarrow v,u)\nonumber\\
    &+\frac{\lambda^2c_\Delta^4}{4}\pi^{d+\frac{1}{2}}\Bigl(\frac{2^{-\Delta}c_\Delta}{2\nu}\Bigl)^2\frac{1}{\Gamma(\Delta)^2}\frac{u^{\frac{\Delta}{3}}v^{\frac{\Delta}{3}}}{\prod_{i<j}\lvert\vec{y_{ij}}\rvert^{\frac{2\Delta}{3}}}\sum_{i=0}^\infty\Bigl\{b_i(0)\frac{\Gamma(2\Delta-\frac{d}{2}+i)}{\Gamma(\Delta+i)^2}H(\Delta,\Delta,1-i,2\Delta;u,v)\nonumber\\
    &\hspace{3cm}+c_i(0)\frac{d}{d\alpha}\Bigl[\frac{\Gamma(2\Delta-\frac{d}{2}+i+\alpha)}{\Gamma(\Delta+i+\alpha)^2}H(\Delta,\Delta,1-i-\alpha,2\Delta;u,v)\Bigl]\Bigl\rvert_{\alpha=0}\Bigl\}\nonumber\\
    &+\Bigl(u,v\rightarrow\frac{1}{u},\frac{v}{u}\Bigl)+(u,v\rightarrow v,u)+\mathcal{O}(\lambda^3)
\end{align}\

where the coefficients $b_i(0)$ and $c_i(0)$ are given by:

\begin{align}\label{bi0coeffin}
    b_i(0) = \frac{\Gamma(2\Delta-\frac{d}{2})}{\Gamma(\Delta)\Gamma(\Delta+\frac{1}{2})}\frac{(\Delta)_i(\Delta)_i}{i!^2}\sum_{l=0}^i&\frac{(2\Delta-\frac{d}{2})_l(-i)_l}{(\Delta+\frac{1}{2})_l(\Delta)_l}\nonumber\\
    &\hspace{-1cm}\times\bigl[\psi(i-l+1)+\psi(i+1)-2\psi(i+\Delta)\bigl]\sum_{k=0}^{l} a^{(2)}_k\frac{(d-1)_{l-k}}{(l-k)!}
\end{align}
\begin{equation}\label{ci0coeffin}
    c_i(0) = -\frac{\Gamma(2\Delta-\frac{d}{2})}{\Gamma(\Delta)\Gamma(\Delta+\frac{1}{2})}\frac{(\Delta)_i(\Delta)_i}{i!^2}\sum_{l=0}^i\frac{(2\Delta-\frac{d}{2})_l(-i)_l}{(\Delta+\frac{1}{2})_l(\Delta)_l}\sum_{k=0}^{l} a^{(2)}_k\frac{(d-1)_{l-k}}{(l-k)!}
\end{equation}\

The form of these correlators are exactly the expected for a conformal theory as it is dictated by eq. (\ref{summcorr2}) up to conformal anomalies, where their overall factors, scaling dimension and effective coupling constant receive small corrections coming from the 1PI loop diagrams resulting from a perturbatively approach in the parameter $\lambda$. These results, while showing the clear role of the quantum corrections to the holographic correlators, also greatly motivate and contribute to the belief of the validity of the AdS/CFT conjecture.

\subsection{Tachyonic Fields on AdS: A Concrete Example}

So far we have studied a general $\Phi^4$ theory on a fixed AdS$_{d+1}$ background, obtaining its dual renormalized CFT$_d$ correlators through the use of the AdS/CFT correspondence, where the only real restriction we have imposed so far is that the parameter $\Delta$ must be greater than $\frac{d}{2}$, obtaining that the resulting holographic theory is renormalizable for dimensions lower than 7. With the intention to study this duality in a more concrete way, as an example we will focus on what are perhaps the most interesting cases. These are, given a certain boundary theory of dimension $d$, the corresponding correlators of lowest integer scaling dimension $\Delta$. In particular, we will focus on those cases where $\Delta<d$, i.e., $\nu<\frac{d}{2}$. Since $\nu=\sqrt{(\frac{d}{2})^2+m^2}$, these cases are composed of bulk theories of negative mass parameter $m^2$, also known as tachyonic fields. Having QFT theories on flat space as a background intuition one may be worried about stability issues of the bulk fields, however the nice boundary conditions of AdS spaces are such that the resulting field theories of negative mass parameter are perfectly stable \cite{Witten}.\par
As a concrete example of CFT theories dual to $\Phi^4$ theories on AdS then, we will focus on those that have an integer value of $\Delta$ such that $\frac{d}{2}<\Delta<d<7$, i.e., for $d=3$ in the case $\Delta=2$, for $d=4$ in the case $\Delta=3$, for $d=5$ in the cases $\Delta=3$ and $\Delta=4$ and finally for $d=6$ in the cases $\Delta=4$ and $\Delta=5$. Notice how all these cases can be summarized simply as either $\Delta=d-1$ or $\Delta=d-2$. What is remarkable of these particular values of $\Delta$ is the simple form that the bulk-bulk propagator takes. Indeed, from its representation eq. (\ref{Grepdiv}) and using that $_2F_1(0,b;c;z)=1$, it is direct to see that it reduces to:

\begin{equation}
    G(x_1,x_2) = \frac{2^{-\Delta}c_\Delta}{2\nu}\frac{\xi^\Delta}{(1-\xi^2)^{\frac{d-1}{2}}},\hspace{1cm}\Delta=d-1\text{ or }\Delta=d-2
\end{equation}\

Furthermore, in a similar way it is direct to see that the coefficients $a_k^{(2)}$ and $a_k^{(3)}$ defined in eqs. (\ref{a2kcoef}) and (\ref{a3kcoef}), which are present in the different quantities defined in the correlators eqs. (\ref{phi4corrparte1}) and (\ref{phi4corrparte2}), take the simple values $a_k^{(2)}=\delta_{k,0}$ and $a_k^{(3)}=\delta_{k,0}$ allowing us to express the 1PI contributions $\Pi(\kappa)$ in the nice closed form:

\begin{align}
    \Pi(\kappa) = &-\frac{\lambda G_\kappa(1)}{2}\nonumber\\
    &+\frac{\lambda^2G_\kappa(1)}{4}\pi^{\frac{d+1}{2}}\Bigl(\frac{2^{-\Delta}c_\Delta}{2\nu}\Bigl)^2\frac{\Gamma(\nu)}{\Gamma(\Delta+\frac{1}{2})}\Bigl(\frac{1}{1+\kappa}\Bigl)^{2\Delta}\ _2F_1\Bigl(d-1,\nu;\Delta+\frac{1}{2};\Bigl(\frac{1}{1+\kappa}\Bigl)^2\Bigl)\nonumber\\
    &+\frac{\lambda^2}{6}\pi^\frac{d+1}{2}\Bigl(\frac{2^{-\Delta}c_\Delta}{2\nu}\Bigl)^3\frac{\Gamma(2\Delta-\frac{d}{2})\Gamma(\Delta)}{\Gamma(\frac{3\Delta}{2})\Gamma(\frac{3\Delta+1}{2})}\Bigl(\frac{1}{1+\kappa}\Bigl)^{3\Delta}\nonumber\\
    &\hspace{5cm}\times\ _3F_2\Bigl(\frac{3(d-1)}{2},2\Delta-\frac{d}{2},\Delta;\frac{3\Delta}{2},\frac{3\Delta+1}{2};\Bigl(\frac{1}{1+\kappa}\Bigl)^2\Bigl)
\end{align}\

along with the coefficients $a_i(\kappa)$, $b_i(0)$ and $c_i(0)$ present in the 4-point functions, defined in eqs. (\ref{ai0coeffin}), (\ref{bi0coeffin}) and (\ref{ci0coeffin}):

\begin{align}
    a_i(\kappa) = &(d-1)\frac{(\Delta)_i(\Delta)_i(d)_i}{(2)_i\ i!}\frac{\Gamma(2\Delta-\frac{d}{2}+1+i)}{\Gamma(\Delta+\frac{3}{2}+i)\Gamma(\Delta+1+i)}\Bigl(\frac{1}{1+\kappa}\Bigl)^{2\Delta+2i+2}(-1)^i\nonumber\\
    &\times\ _4F_3\Bigl(2\Delta-\frac{d}{2}+1+i,1,1,d+i;\Delta+\frac{3}{2}+i,\Delta+1+i,2+i;\Bigl(\frac{1}{1+\kappa}\Bigl)^2\Bigl)
\end{align}
\begin{align}
    b_i(0) = &\frac{\Gamma(2\Delta-\frac{d}{2})}{\Gamma(\Delta)\Gamma(\Delta+\frac{1}{2})}\frac{(\Delta)_i(\Delta)_i}{i!^2}\bigl[\psi(i+1)-2\psi(i+\Delta)\bigl]\ _3F_2\Bigl(2\Delta-\frac{d}{2},-i,d-1;\Delta+\frac{1}{2},\Delta;1\Bigl)\nonumber\\
    &+\frac{\Gamma(2\Delta-\frac{d}{2})}{\Gamma(\Delta)\Gamma(\Delta+\frac{1}{2})}\frac{(\Delta)_i(\Delta)_i}{i!^2}\sum_{l=0}^i\frac{(2\Delta-\frac{d}{2})_l(-i)_l(d-1)_l}{(\Delta+\frac{1}{2})_l(\Delta)_l\ l!}\psi(i-l+1)
\end{align}
\begin{equation}
    c_i(0) = -\frac{\Gamma(2\Delta-\frac{d}{2})}{\Gamma(\Delta)\Gamma(\Delta+\frac{1}{2})}\frac{(\Delta)_i(\Delta)_i}{i!^2}\ _3F_2\Bigl(2\Delta-\frac{d}{2},-i,d-1;\Delta+\frac{1}{2},\Delta;1\Bigl)
\end{equation}\

where we wrote them in terms of generalized hypergeometric functions. The previous expressions, including $\Pi(\kappa)$, are valid whenever $\Delta=d-1$ or $\Delta=d-2$ which, as we argued, contain all the cases where $\Delta$ is an integer such that $\frac{d}{2}<\Delta<d<7$. To illustrate how starting from these expressions one can obtain the exact holographic renormalized correlators together with the corresponding convenient counterterms, we will proceed to study one particular case.

\subsubsection{Case $\Delta=2$ and $d=3$}

In this case $\nu=\frac{1}{2}$ and $c_\Delta=\frac{1}{\pi^2}$, therefore it is straightforward to see that $\Pi(\kappa)$ reduces to:

\begin{align}
    \Pi(\kappa) = &-\frac{\lambda G_\kappa(1)}{2}+\frac{\lambda^2}{48\pi^2}G_\kappa(1)\Bigl(\frac{1}{1+\kappa}\Bigl)^4\ _2F_1\Bigl(2,\frac{1}{2};\frac{5}{2};\Bigl(\frac{1}{1+\kappa}\Bigl)^2\Bigl)\nonumber\\
    &+\frac{\lambda^2}{1920\pi^4}\Bigl(\frac{1}{1+\kappa}\Bigl)^{6}\ _2F_1\Bigl(\frac{5}{2},2;\frac{7}{2};\Bigl(\frac{1}{1+\kappa}\Bigl)^2\Bigl)
\end{align}\

As we did for the study of a general $\Phi^4$ theory, we need to identify the UV-divergent and UV-convergent parts of this quantity, $\Pi_\infty(\kappa)$ and $\Pi_0(0)$ respectively. Using the software Mathematica we obtain that every quantity in $\Pi(\kappa)$ dependent on the UV-regulator $\kappa$ can be expanded as:

\begin{equation}
    G_\kappa(1) = \frac{1}{8\pi^2\kappa}-\frac{1}{16\pi^2}+\frac{\kappa}{32\pi^2}+\mathcal{O}(\kappa^2)
\end{equation}
\begin{equation}
    \Bigl(\frac{1}{1+\kappa}\Bigl)^4 = 1 -4\kappa + \mathcal{O}(\kappa^2)
\end{equation}
\begin{equation}
    \Bigl(\frac{1}{1+\kappa}\Bigl)^6 = 1 -6\kappa + \mathcal{O}(\kappa^2)
\end{equation}
\begin{align}
    _2F_1\Bigl(2,\frac{1}{2};\frac{5}{2};\Bigl(\frac{1}{1+\kappa}\Bigl)^2\Bigl) = &\frac{3}{4}(-1+\ln{2}-\ln{\kappa})+\frac{3}{8}\kappa(-3+4\ln{2}-4\ln{\kappa})\nonumber\\
    &+\mathcal{O}(\kappa^2,\kappa^2\ln{\kappa})
\end{align}
\begin{align}
    _2F_1\Bigl(\frac{5}{2},2;\frac{7}{2};\Bigl(\frac{1}{1+\kappa}\Bigl)^2\Bigl) = &\frac{5}{4\kappa}-\frac{5}{8}(-19+6\ln{2}-6\ln{\kappa})-\frac{5}{16}\kappa(-107+60\ln{2}-60\ln{\kappa})\nonumber\\
    &+\mathcal{O}(\kappa^2,\kappa^2\ln{\kappa})
\end{align}\

which results in the corresponding expansion for $\Pi(\kappa)$:

\begin{align}
    \Pi(\kappa) = &-\frac{\lambda}{16\pi^2\kappa}+\frac{\lambda^2}{1536\pi^4}\Bigl(-\frac{2}{\kappa}+\frac{3}{\kappa}\ln{2}-\frac{3}{\kappa}\ln{\kappa}+\frac{21}{2}\ln{\kappa}\Bigl)+\frac{\lambda}{32\pi^2}+\frac{\lambda^2}{3072\pi^4}(25-21\ln{2})\nonumber\\
    &+\mathcal{O}(\kappa,\kappa\ln{\kappa})
\end{align}\

With the 1PI contributions written in this form, its UV-divergent and UV-convergent parts are identified as:

\begin{equation}
    \Pi_\infty(\kappa) = -\frac{\lambda}{16\pi^2\kappa}+\frac{\lambda^2}{1536\pi^4}\Bigl(-\frac{2}{\kappa}+\frac{3}{\kappa}\ln{2}-\frac{3}{\kappa}\ln{\kappa}+\frac{21}{2}\ln{\kappa}\Bigl)
\end{equation}
\begin{equation}
    \Pi_0(0) = \frac{\lambda}{32\pi^2}+\frac{\lambda^2}{3072\pi^4}(25-21\ln{2})
\end{equation}\

For the case of the $a_i(\kappa)$ coefficient present in the 4-point function, when $\Delta=2$ and $d=3$ it is straightforward to see that it reduces to:

\begin{equation}
    a_i(\kappa) = \frac{(2)_i(-1)^i}{i!}\Bigl(\frac{1}{1+\kappa}\Bigl)^{6+2i}\ _2F_1\Bigl(1,1;2+i;\Bigl(\frac{1}{1+\kappa}\Bigl)^2\Bigl)
\end{equation}\

We need to identify the UV-divergent and UV-convergent parts of every coefficient, $a^{(\infty)}_i(\kappa)$ and $a^{(0)}_i(0)$ respectively. Since $d=3$ only the $i=0$ coefficient is expected to be divergent. Indeed, using the software Mathematica we obtain that the $i=0$ coefficient can be expanded as:

\begin{equation}
    a_0(\kappa) = \Bigl(\frac{1}{1+\kappa}\Bigl)^6\ _2F_1\Bigl(1,1;2;\Bigl(\frac{1}{1+\kappa}\Bigl)^2\Bigl)=-\ln{\kappa}-\ln{2}+\mathcal{O}(\kappa,\kappa\ln{\kappa})
\end{equation}\

from where we see its UV-divergent and UV-convergent parts:

\begin{equation}
    a^{(\infty)}_0(\kappa) = -\ln{\kappa},\hspace{0.5cm}a^{(0)}_0(0) = -\ln{2}
\end{equation}\

While for the $i>0$ coefficients, safely taking $\kappa=0$ they can be written as:

\begin{equation}
    a_{i>0}(0) = \frac{(2)_i(-1)^i}{i!}\ _2F_1(1,1;2+i;1) = \frac{(2)_i(2)_i\Gamma(i)}{i!^3}(-1)^i
\end{equation}\

where we used that $_2F_1(a,b;c;1)=\frac{\Gamma(c)\Gamma(c-a-b)}{\Gamma(c-a)\Gamma(c-b)}$. Similarly for the coefficients $b_i(0)$ and $c_i(0)$ present in the 4-point function, when $\Delta=2$ and $d=3$ it is straightforward to see that they reduce to:

\begin{equation}
    b_i(0) =-2\frac{(2)_i(2)_i}{i!^2}\frac{1}{i+1}\delta_{i,0}+\frac{(2)_i(2)_i}{i!^2}\frac{(-1)^{i+1}}{i}(1-\delta_{i,0})
\end{equation}
\begin{equation}
    c_i(0) = -\frac{(2)_i(2)_i}{i!^2}\delta_{i,0}
\end{equation}\

where we solved $\sum_{l=0}^i\frac{(-i)_l}{l!}\psi(i-l+1)=\psi(i+1)\delta_{i,0}+\frac{(-1)^{i+1}}{i}(1-\delta_{i,0})$, used that $\psi(i+1)-\psi(i+2)=-\frac{1}{i+1}$ and also that $_1F_0(-i;1)=\delta_{i,0}$.\par
Finally then, using these particular results for the 1PI contributions $\Pi(\kappa)$ and for the coefficients $a_i(\kappa)$, $b_i(0)$ and $c_i(0)$ in the summarized study of a general $\Phi^4$ theory found in section 4.2.7, we conclude that for $\Delta=2$ and $d=3$, conveniently choosing the counterterms $Z(\lambda)$, $\delta m^2$ and $\delta\lambda$ as:

\begin{equation}
    Z(\lambda)=\varepsilon^{-\frac{\lambda}{16\pi^2}-\frac{\lambda^2}{1536\pi^4}(25-21\ln{2})}
\end{equation}
\begin{equation}
    \delta m^2 = -\frac{\lambda}{16\pi^2\kappa}+\frac{\lambda^2}{1536\pi^4}\Bigl(-\frac{2}{\kappa}+\frac{3}{\kappa}\ln{2}-\frac{3}{\kappa}\ln{\kappa}+\frac{21}{2}\ln{\kappa}\Bigl)
\end{equation}
\begin{equation}
    \delta\lambda = -\frac{3\lambda^2}{32\pi^2}\ln{\kappa}
\end{equation}\

result, up to order $\lambda^2$ in the coupling constant, in the holographic renormalized 1-, 2-, 3- and 4-point functions:

\begin{align}
    \text{1-pt fn:}\hspace{0.25cm}&\langle O_\Delta(\vec{y_1})\rangle_{\text{CFT}} = \langle O_\Delta(\vec{y_1})\rangle_{\text{CFT,con}} = 0\nonumber\\
    \text{2-pt fn:}\hspace{0.25cm}&\langle O_\Delta(\vec{y_1})O_\Delta(\vec{y_2})\rangle_{\text{CFT}} = \langle O_\Delta(\vec{y_1})O_\Delta(\vec{y_2})\rangle_{\text{CFT,con}} = \frac{1}{\pi^2}\frac{1}{\lvert\vec{y_1}-\vec{y_2}\rvert^{4-2\Pi_0(0)}}+\mathcal{O}(\lambda^3)\nonumber\\
    \text{3-pt fn:}\hspace{0.25cm}&\langle O_\Delta(\vec{y_1})O_\Delta(\vec{y_2})O_\Delta(\vec{y_3})\rangle_{\text{CFT}} = \langle O_\Delta(\vec{y_1})O_\Delta(\vec{y_2})O_\Delta(\vec{y_3})\rangle_{\text{CFT,con}}=0\nonumber\\
    \text{4-pt fn:}\hspace{0.25cm}&\langle O_\Delta(\vec{y_1})O_\Delta(\vec{y_2})O_\Delta(\vec{y_3})O_\Delta(\vec{y_4})\rangle_{\text{CFT}}=\frac{1}{\pi^4}\frac{1}{\prod_{i<j}\lvert\vec{y_{ij}}\rvert^{\frac{2}{3}\bigl(2-\Pi_0(0)\bigl)}}u^{-\frac{2}{3}\bigl(2-\Pi_0(0)\bigl)}v^{\frac{1}{3}\bigl(2-\Pi_0(0)\bigl)}\nonumber\\
    &+\Bigl(u,v\rightarrow\frac{1}{u},\frac{v}{u}\Bigl)+(u,v\rightarrow v,u)-\frac{\lambda}{2\pi^\frac{13}{2}} \frac{\Gamma(\frac{5}{2}-2\Pi_0(0))}{\Gamma(2-\Pi_0(0))^4}\frac{u^{\frac{1}{3}\bigl(2-\Pi_0(0)\bigl)}v^{\frac{1}{3}\bigl(2-\Pi_0(0)\bigl)}}{\prod_{i<j}\lvert\vec{y_i}-\vec{y_j}\rvert^{\frac{2}{3}\bigl(2-\Pi_0(0)\bigl)}}\nonumber\\
    &\hspace{0.5cm}\times H\bigl(2-\Pi_0(0),2-\Pi_0(0),1,4-2\Pi_0(0);u,v\bigl)\nonumber\\
    &-(2+\ln{2})\frac{9\lambda^2}{256\pi^8}\frac{u^{\frac{2}{3}}v^{\frac{2}{3}}}{\prod_{i<j}\lvert\vec{y_{ij}}\rvert^{\frac{4}{3}}}H(2,2,1,4;u,v)\nonumber\\
    &-\frac{\lambda^2}{64\pi^\frac{17}{2}}\frac{u^{\frac{2}{3}}v^{\frac{2}{3}}}{\prod_{i<j}\lvert\vec{y_{ij}}\rvert^{\frac{4}{3}}}\frac{d}{d\alpha}\Bigl[\frac{\Gamma(\frac{5}{2}+\alpha)}{\Gamma(2+\alpha)^2}H(2,2,1-\alpha,4;u,v)\Bigl]\Bigl\rvert_{\alpha=0}\nonumber\\
    &+\Bigl(u,v\rightarrow\frac{1}{u},\frac{v}{u}\Bigl)+(u,v\rightarrow v,u)+\mathcal{O}(\lambda^3)
\end{align}
\begin{align}
    &\langle O_\Delta(\vec{y_1})O_\Delta(\vec{y_2})O_\Delta(\vec{y_3})O_\Delta(\vec{y_4})\rangle_{\text{CFT,con}} =-\frac{\lambda}{2\pi^\frac{13}{2}} \frac{\Gamma(\frac{5}{2}-2\Pi_0(0))}{\Gamma(2-\Pi_0(0))^4}\frac{u^{\frac{1}{3}\bigl(2-\Pi_0(0)\bigl)}v^{\frac{1}{3}\bigl(2-\Pi_0(0)\bigl)}}{\prod_{i<j}\lvert\vec{y_i}-\vec{y_j}\rvert^{\frac{2}{3}\bigl(2-\Pi_0(0)\bigl)}}\nonumber\\
    &\times H\bigl(2-\Pi_0(0),2-\Pi_0(0),1,4-2\Pi_0(0);u,v\bigl)-(2+\ln{2})\frac{9\lambda^2}{256\pi^8}\frac{u^{\frac{2}{3}}v^{\frac{2}{3}}}{\prod_{i<j}\lvert\vec{y_{ij}}\rvert^{\frac{4}{3}}}H(2,2,1,4;u,v)\nonumber\\
    &-\frac{\lambda^2}{64\pi^\frac{17}{2}}\frac{u^{\frac{2}{3}}v^{\frac{2}{3}}}{\prod_{i<j}\lvert\vec{y_{ij}}\rvert^{\frac{4}{3}}}\frac{d}{d\alpha}\Bigl[\frac{\Gamma(\frac{5}{2}+\alpha)}{\Gamma(2+\alpha)^2}H(2,2,1-\alpha,4;u,v)\Bigl]\Bigl\rvert_{\alpha=0}+\Bigl(u,v\rightarrow\frac{1}{u},\frac{v}{u}\Bigl)+(u,v\rightarrow v,u)\nonumber\\
    &+\mathcal{O}(\lambda^3)
\end{align}\

where $\Pi_0(0) = \frac{\lambda}{32\pi^2}+\frac{\lambda^2}{3072\pi^4}(25-21\ln{2})$. Notice how in the present case
the 4-point functions consist in a finite number of terms, as compared to those obtained for general values of $\Delta$ and $d$ summarized in section 4.2.7. This is due to the nice form that the coefficients $a_i$, $b_i$ and $c_i$ take for the particular case $\Delta=2$ and $d=3$, where $c_i$ becomes a Kronecker delta, while every coefficient $b_{i>0}$ exactly cancel those contributions coming from the coefficients $a_{i>0}$, resulting in a finite number of contributions to the 4-point functions coming only from the coefficients $a_0$, $b_0$ and $c_0$.

\newpage


\chapter*{Conclusions}
\addcontentsline{toc}{chapter}{Conclusions}
The holographic nature of gravity is captured by the AdS/CFT correspondence in a remarkable way: through delicate manipulations we can obtain correlators for some CFT in $d$ dimensions using as a starting point a field theory on a AdS space in $d+1$ dimensions. Now, these manipulations are delicate because the CFT correlators are obtained from the boundary behavior of the of the field theory on AdS, where both the metric and the fields living in it diverge. The key point here is that the CFT theory is hidden in these infrared divergences of the bulk theory, requiring a sensitive renormalization treatment to its obtention. This fact is no surprise  due to the strong/weak duality between both theories, and it is precisely what has already been known for some time for the classical approximation of the bulk theory on AdS, motivating the entire holographic renormalization program.\par
In this thesis we went beyond the classical approximation, considering the quantum corrections to the holographic correlators. The first result here of our work was to find that for scalar $\Phi^3$ and $\Phi^4$ theories on AdS, these quantum corrections to the CFT correlators correspond exactly to the expected diagrams obtained from the known Feynman rules for these theories, where in our case the external legs of the diagrams have been extended to the conformal boundary of the AdS space. This allowed us to define the holographic dictionary between the CFT correlators in the boundary with the AdS correlators in the bulk, simply understanding the former as the limit of the latter.\par
Then, the second result of our work was to develop the necessary formulas for the computation of every single one of these quantum corrections present in the holographic correlators, facing not only the expected infrared divergences but also the ultraviolet divergences coming from the evaluation of the bulk-bulk propagators at coincident points in the AdS space. Since the physics of the holography is precisely hidden in these divergences this motivated us to introduce regularization schemes, extrapolating the used for the infrared divergences at the classical level now for the loop integrals, and taking a point-splitting approach for the new ultraviolet divergences not seen at tree-level calculations. What is remarkable of these schemes is that the resulting holographic 2-point functions correspond exactly to those of a CFT under anomalies, in particular with the presence of an anomalous scaling dimension receiving contributions from its one-particle irreducible loop diagrams. In practice this occurs because the formulas found for the vertices $\int G^n\sim1$ and $\int G^nK\sim K$ allow us to write the 2-point function in the 1PI expansion $\langle OO\rangle\sim\frac{1}{y^{2\Delta}}+\Pi\int KK+\Pi^2\int K\int GK+\dotsb$, where the constant $\Pi$ is obtained from the 1PI contributions. This expansion for the correlator along with the formulas found for the vertices $\int KK\sim\frac{\ln{(y)}}{y^{2\Delta}}$, $\int K\int GK\sim\frac{\ln^2{(y)}}{y^{2\Delta}}$, etc, result exactly in the Taylor series of an exponent, allowing us to factorize the 1PI expansion in the form of $\langle OO\rangle\sim\frac{1}{y^{2\Delta+\gamma}}$, where $\gamma=-\frac{\Pi}{\nu}$ is precisely what is known as the anomalous dimension. In this way, the role of the quantum corrections to the holographic 2-point function can be appreciated, being able to also notice both the role of its IR and UV divergences: the IR give rise to the logarithmic structure of the correlator, while the UV fix the value of the anomalous exponent.\par
For the holographic 3- and 4-point functions, the consideration of the formulas found for the vertices $\int GK\sim K\ln(\tilde{K})$ and $\int G^nKK\sim KKf(\tilde{K}\tilde{K}y^2)$ result in the exact same anomalous structure for the correlators. In summary, for scalar $\Phi^3$ and $\Phi^4$ theories on a fixed AdS$_{d+1}$ background it is obtained that the resulting quantum corrected CFT$_d$ correlators can be factorized into the forms $\langle O\rangle=0$, $\langle OO\rangle=\frac{C_2}{y^{2\Delta+\gamma}}$, $\langle OOO\rangle=\frac{C_3}{\prod_{i<j}y_{ij}^{\Delta+\frac{\gamma}{2}}}$ and $\langle OOOO\rangle=\frac{C_4(u,v)}{\prod_{i<j}y_{ij}^{\frac{2}{3}(\Delta+\frac{\gamma}{2})}}$, where for a $\Phi^4$ theory $C_3=0$. These correlators correspond precisely to those of a CFT under the anomaly $\Delta\rightarrow\Delta+\frac{\gamma}{2}$.\par
One of the consequences that we found of this nice factorization of the 1PI expansions in the correlators is that the different IR and UV divergences contained in the quantities $\gamma$, $C_2$, $C_3$ and $C_4(u,v)$ can be renormalized in exactly the same way as it is done for ordinary field theories, this is, through a redefinition of the parameters of the theory in the bulk, where for a $\Phi^4$ theory the only restriction is that $d<7$. This might be related with the fact that for $d>6$ there are not SCFT since $d=6$ is the highest dimension in which the superconformal algebra exist \cite{Minwalla}. In holographic terms, our results deliver a possible answer to this fact: there are not SCFT in $d>6$ since the resulting theories in the bulk are non-renormalizable.\par
Scalar field theories are of course not the only ones that we can consider on a AdS space. Furthermore, since in general the theory that one will be interested in studying in the bulk will come from the low energy limit of a certain string theory, the AdS space will not be fixed either but it will present quantum fluctuations which will contribute to the holographic correlators through diagrams containing gravitons. Non-scalar theories and quantum fluctuations of the AdS metric are not studied in this work, however many of the computations that appear in these cases effectively reduce to those addressed here for scalar theories, therefore we expect that the different schemes, formulas and tools developed in this thesis will be of great interest for these cases, not only at the classical level of the correspondence but at its full quantum nature.

\newpage


\chapter*{Appendices}
\addcontentsline{toc}{chapter}{Appendices}

\newpage


\appendix

\chapter{D-functions}
Throughout this work one encounters many integrals in which the difficulty of their resolution justify a separate treatment from the main text as to not lose the focus of discussion. Those that fall under this category that we will study in this appendix are the ones involving only bulk-boundary propagators, integrals which are also known as D-functions in the literature. Pictorially, these functions correspond to diagrams that only have external lines which coincide at some point in the interior of the AdS space and that extend up to some arbitrary point on its conformal boundary. By definition, they are represented by:

\begin{equation}\label{dfunc}
    D_{\Delta_1\dotsm\Delta_n}(\vec{y_1},\dotsc,\vec{y_n}) \equiv \int d^{d+1}x\sqrt{g}\ \tilde{K}^{\Delta_1}(x,\vec{y_1})\dotsm \tilde{K}^{\Delta_n}(x,\vec{y_n})
\end{equation}\

where the quantities $\tilde{K}^{\Delta_i}(x,\vec{y_i})$ are the unnormalized bulk-boundary propagators of scaling dimension $\Delta_i$:

\begin{equation}\label{unbubop}
    \tilde{K}^{\Delta_i}(x,\vec{y_i}) = \Bigl[\frac{x_0}{(x-\vec{y_i})^2}\Bigl]^{\Delta_i}
\end{equation}\

We will proceed, following the work by Muck and Viswanathan \cite{Muck}, to partially compute the general integral present in eq. (\ref{dfunc}) to write down a more useful representation for the D-functions, with the intention to later study the concrete D-functions encountered in this work more easily.

\section{General Case}

As we said, we will solve the general case  eq. (\ref{dfunc}) up to a point in which it makes it easier to study the concrete cases we are interested in, these are the cases with $n=3$, $n=4$ and for last the special case of $n=2$. We will start then by explicitly writing the form of the bulk-boundary propagators in the definition of the general D-function:

\begin{equation}
    D_{\Delta_1\dotsm\Delta_n}(\vec{y_1},\dotsc,\vec{y_n}) = \int d^{d+1}x\ \frac{x_0^{\sum\Delta_i-d-1}}{\prod \bigl[(x-\vec{y_i})^2\bigl]^{\Delta_i}}
\end{equation}\

We can make progress in the computation of this integral grouping the factors in the denominator using Feynman parametrization:

\begin{equation}\label{feynpam}
   \frac{1}{A_1^{\alpha_1}\dotsm A_n^{\alpha_n}} = \frac{\Gamma(\sum \alpha_i)}{\prod\Gamma(\alpha_i)}\int_0^1d^nu\ \frac{\delta(\sum u_i-1)u_1^{\alpha_1-1}\dotsm u_n^{\alpha_n-1}}{(u_1A_1+\dotsb+u_nA_n)^{\sum \alpha_i}}
\end{equation}\

which results in:

\begin{align}
   \frac{1}{\prod \bigl[(x-\vec{y_i})^2\bigl]^{\Delta_i}} &= \frac{\Gamma(\sum\Delta_i)}{\prod\Gamma(\Delta_i)}\int_0^1d^nu\ \frac{\delta(\sum u_i-1)u_1^{\Delta_1-1}\dotsm u_n^{\Delta_n-1}}{\bigl[u_1(x-\vec{y_1})^2+\dotsb+u_n(x-\vec{y_n})^2\bigl]^{\sum\Delta_i}}\nonumber\\
   &= \frac{\Gamma(\sum\Delta_i)}{\prod\Gamma(\Delta_i)}\int_0^1d^nu\ \frac{\delta(\sum u_i-1)u_1^{\Delta_1-1}\dotsm u_n^{\Delta_n-1}}{\bigl[x_0^2+(\vec{x}-\sum u_i\vec{y_i})^2+\sum u_i\vec{y_i}^2-(\sum u_i\vec{y_i})^2\bigl]^{\sum\Delta_i}}\nonumber\\
   &= \frac{\Gamma(\sum\Delta_i)}{\prod\Gamma(\Delta_i)}\int_0^1d^nu\ \frac{\delta(\sum u_i-1)u_1^{\Delta_1-1}\dotsm u_n^{\Delta_n-1}}{\bigl[x_0^2+(\vec{x}-\sum u_i\vec{y_i})^2+\sum_{i<j}u_iu_j\vec{y_{ij}}^2\bigl]^{\sum\Delta_i}}
\end{align}\

where we completed squares, used the fact that the delta function forces $\sum u_i=1$ and simplified the resulting sums, defining in the process the quantity $\vec{y_{ij}}^2\equiv \lvert\vec{y_i}-\vec{y_j}\rvert^2$. Replacing this representation for the denominator back into the original integral and performing the translation $\vec{x}\rightarrow\vec{x}+\sum u_i\vec{y_i}$:

\begin{align}
    D_{\Delta_1\dotsm\Delta_n}(\vec{y_1},\dotsc,\vec{y_n}) &= \frac{\Gamma(\sum\Delta_i)}{\prod\Gamma(\Delta_i)}\int_0^1d^nu\int d^{d+1}x\ \frac{\delta(\sum u_i-1)u_1^{\Delta_1-1}\dotsm u_n^{\Delta_n-1}x_0^{\sum\Delta_i-d-1}}{\bigl[x_0^2+(\vec{x}-\sum u_i\vec{y_i})^2+\sum_{i<j}u_iu_j\vec{y_{ij}}^2\bigl]^{\sum\Delta_i}}\nonumber\\
    &= \frac{\Gamma(\sum\Delta_i)}{\prod\Gamma(\Delta_i)}\int_0^1d^nu\int d^{d+1}x\ \frac{\delta(\sum u_i-1)u_1^{\Delta_1-1}\dotsm u_n^{\Delta_n-1}x_0^{\sum\Delta_i-d-1}}{(x_0^2+\vec{x}^2+\sum_{i<j}u_iu_j\vec{y_{ij}}^2)^{\sum\Delta_i}}
\end{align}\

The resulting integral in the $\vec{x}$ variable can be simply done using spherical coordinates, which can be seen more easily by doing first the rescaling $\vec{x}\rightarrow\vec{x}\sqrt{x_0^2+\sum_{i<j}u_iu_j\vec{y_{ij}}^2}$:

\begin{align}
    D_{\Delta_1\dotsm\Delta_n}(\vec{y_1},\dotsc,\vec{y_n}) = \frac{\Gamma(\sum\Delta_i)}{\prod\Gamma(\Delta_i)}\int_0^1d^nu&\int_0^\infty dx_0\ \frac{\delta(\sum u_i-1)u_1^{\Delta_1-1}\dotsm u_n^{\Delta_n-1}x_0^{\sum\Delta_i-d-1}}{(x_0^2+\sum_{i<j}u_iu_j\vec{y_{ij}}^2)^{\sum\Delta_i-\frac{d}{2}}}\nonumber\\
    &\times\int d^dx\ \frac{1}{(1+\vec{x}^2)^{\sum\Delta_i}}
\end{align}\

Computing the value of this last integral gives:

\begin{equation}
    \int d^dx\ \frac{1}{(1+\vec{x}^2)^{\sum\Delta_i}} = \pi^{\frac{d}{2}}\frac{\Gamma(\sum\Delta_i-\frac{d}{2})}{\Gamma(\sum\Delta_i)}
\end{equation}\

Then the D-function reduces to:

\begin{equation}
    D_{\Delta_1\dotsm\Delta_n}(\vec{y_1},\dotsc,\vec{y_n}) = \pi^{\frac{d}{2}}\frac{\Gamma(\sum\Delta_i-\frac{d}{2})}{\prod\Gamma(\Delta_i)}\int_0^1d^nu\int_0^\infty dx_0\ \frac{\delta(\sum u_i-1)u_1^{\Delta_1-1}\dotsm u_n^{\Delta_n-1}x_0^{\sum\Delta_i-d-1}}{(x_0^2+\sum_{i<j}u_iu_j\vec{y_{ij}}^2)^{\sum\Delta_i-\frac{d}{2}}}
\end{equation}\

In the same way, the resulting integral now in the $x_0$ variable can be simply done which can be seen more easily by doing the rescaling $x_0\rightarrow x_0\sqrt{\sum_{i<j}u_iu_j\vec{y_{ij}}^2}$:

\begin{equation}
    D_{\Delta_1\dotsm\Delta_n}(\vec{y_1},\dotsc,\vec{y_n}) = \pi^{\frac{d}{2}}\frac{\Gamma(\sum\Delta_i-\frac{d}{2})}{\prod\Gamma(\Delta_i)}\int_0^1d^nu\ \frac{\delta(\sum u_i-1)u_1^{\Delta_1-1}\dotsm u_n^{\Delta_n-1}}{(\sum_{i<j}u_iu_j\vec{y_{ij}}^2)^{\frac{\sum\Delta_i}{2}}}\int_0^\infty dx_0\ \frac{x_0^{\sum\Delta_i-d-1}}{(1+x_0^2)^{\sum\Delta_i-\frac{d}{2}}}
\end{equation}\

The value of this last integral is:

\begin{equation}
    \int_0^\infty dx_0\ \frac{x_0^{\sum\Delta_i-d-1}}{(1+x_0^2)^{\sum\Delta_i-\frac{d}{2}}} = \frac{1}{2}\frac{\Gamma(\frac{\sum\Delta_i-d}{2})\Gamma(\frac{\sum\Delta_i}{2})}{\Gamma(\sum\Delta_i-\frac{d}{2})}
\end{equation}\

Then the D-function further reduces to:

\begin{equation}\label{gendfuncmedrep}
    D_{\Delta_1\dotsm\Delta_n}(\vec{y_1},\dotsc,\vec{y_n}) = \frac{\pi^{\frac{d}{2}}}{2}\frac{\Gamma(\frac{\sum\Delta_i-d}{2})\Gamma(\frac{\sum\Delta_i}{2})}{\prod\Gamma(\Delta_i)}\int_0^1d^nu\ \frac{\delta(\sum u_i-1)u_1^{\Delta_1-1}\dotsm u_n^{\Delta_n-1}}{(\sum_{i<j}u_iu_j\vec{y_{ij}}^2)^{\frac{\sum\Delta_i}{2}}}
\end{equation}\

This representation for the general D-function is already friendly enough to be used for the different cases we are interested in, and in fact we will use this specific form to prove some nice properties of these functions, but there is still one more simplification we can do which consists of making the change of variable $u_1\rightarrow u_1$, $u_{i>1}\rightarrow u_1u_{i>1}$ and then trivially performing the $u_1$ integral through the Dirac delta function, process which finally brings the expression for the general D-function to the form it will be mainly used:

\begin{equation}\label{dfuncrep}
    D_{\Delta_1\dotsm\Delta_n}(\vec{y_1},\dotsc,\vec{y_n}) = \frac{\pi^{\frac{d}{2}}}{2}\frac{\Gamma(\frac{\sum\Delta_i-d}{2})\Gamma(\frac{\sum\Delta_i}{2})}{\prod\Gamma(\Delta_i)}\int_0^\infty d^{n-1}u\ \frac{u_2^{\Delta_2-1}\dotsm u_n^{\Delta_n-1}}{(\sum_{1<i}u_i\vec{y_{1i}}^2+\sum_{1<i<j}u_iu_j\vec{y_{ij}}^2)^{\frac{\sum\Delta_i}{2}}}
\end{equation}\

Next we will see how this expression allows us to directly compute some of the integrals encountered in our study of holographic correlators.

\section{Case \texorpdfstring{$n=3$}{}}

The particular case of eq. (\ref{dfunc}) with $n=3$, being the integral of 3 bulk-boundary propagators, is encountered in our work as contributions to the holographic CFT 3-point functions from considering $\Phi^3$ theories on AdS. Now, as we showed in eq. (\ref{summcorr}), the functional form of CFT 3-point functions is completely fixed purely from the highly restrictive conformal symmetries. It is very satisfactory then to check that these contributions do indeed have this expected form. Let us calculate this particular case using as a starting point the representation for the D-function just obtained eq. (\ref{dfuncrep}):

\begin{equation}
        D_{\Delta_1\Delta_2\Delta_3}(\vec{y_1},\vec{y_2},\vec{y_3}) = \frac{\pi^{\frac{d}{2}}}{2}\frac{\Gamma(\frac{\sum\Delta_i-d}{2})\Gamma(\frac{\sum\Delta_i}{2})}{\prod\Gamma(\Delta_i)}\int_0^\infty d^2u\ \frac{u_2^{\Delta_2-1}u_3^{\Delta_3-1}}{(u_2\vec{y_{12}}^2+u_3\vec{y_{13}}^2+u_2u_3\vec{y_{23}}^2)^{\frac{\sum\Delta_i}{2}}}
\end{equation}\

where $i=1,2,3$. The resulting integral in the $u_3$ variable can be simply done by doing first the rescaling $u_3\rightarrow u_3\Bigl(\frac{u_2\vec{y_{12}}^2}{\vec{y_{13}}^2+u_2\vec{y_{23}}^2}\Bigl)$:

\begin{align}
        D_{\Delta_1\Delta_2\Delta_3}(\vec{y_1},\vec{y_2},\vec{y_3}) = &\frac{\pi^{\frac{d}{2}}}{2}\frac{\Gamma(\frac{\sum\Delta_i-d}{2})\Gamma(\frac{\sum\Delta_i}{2})}{\prod\Gamma(\Delta_i)}\frac{1}{\lvert\vec{y_1}-\vec{y_2}\rvert^{\Delta_1+\Delta_2-\Delta_3}}\nonumber\\
        &\hspace{1cm}\times\int_0^\infty du_2\ \frac{u_2^{\frac{\Delta_2+\Delta_3-\Delta_1}{2}-1}}{(\vec{y_{13}}^2+u_2\vec{y_{23}}^2)^{\Delta_3}}\int_0^\infty du_3\ \frac{u_3^{\Delta_3-1}}{(1+u_3)^{\frac{\sum\Delta_i}{2}}}
\end{align}\

Computing the value of this last integral gives:

\begin{equation}
    \int_0^\infty du_3\ \frac{u_3^{\Delta_3-1}}{(1+u_3)^{\frac{\sum\Delta_i}{2}}} = \frac{\Gamma(\Delta_3)\Gamma(\frac{\Delta_1+\Delta_2-\Delta_3}{2})}{\Gamma(\frac{\sum\Delta_i}{2})}
\end{equation}\

Then the D-function reduces to:

\begin{equation}
        D_{\Delta_1\Delta_2\Delta_3}(\vec{y_1},\vec{y_2},\vec{y_3}) = \frac{\pi^{\frac{d}{2}}}{2}\frac{\Gamma(\frac{\sum\Delta_i-d}{2})}{\Gamma(\Delta_1)\Gamma(\Delta_2)}\frac{\Gamma(\frac{\Delta_1+\Delta_2-\Delta_3}{2})}{\lvert\vec{y_1}-\vec{y_2}\rvert^{\Delta_1+\Delta_2-\Delta_3}}\int_0^\infty du_2\ \frac{u_2^{\frac{\Delta_2+\Delta_3-\Delta_1}{2}-1}}{(\vec{y_{13}}^2+u_2\vec{y_{23}}^2)^{\Delta_3}}
\end{equation}\

In exactly the same way, the resulting integral now in the $u_2$ variable can be simply done by doing first the rescaling $u_2\rightarrow u_2\frac{\vec{y_{13}}^2}{\vec{y_{23}}^2}$:

\begin{align}
        D_{\Delta_1\Delta_2\Delta_3}(\vec{y_1},\vec{y_2},\vec{y_3}) = &\frac{\pi^{\frac{d}{2}}}{2}\frac{\Gamma(\frac{\sum\Delta_i-d}{2})}{\Gamma(\Delta_1)\Gamma(\Delta_2)}\frac{\Gamma(\frac{\Delta_1+\Delta_2-\Delta_3}{2})}{\lvert\vec{y_1}-\vec{y_2}\rvert^{\Delta_1+\Delta_2-\Delta_3}\lvert\vec{y_2}-\vec{y_3}\rvert^{\Delta_2+\Delta_3-\Delta_1}\lvert\vec{y_3}-\vec{y_1}\rvert^{\Delta_3+\Delta_1-\Delta_2}}\nonumber\\
        &\hspace{3cm}\times\int_0^\infty du_2\ \frac{u_2^{\frac{\Delta_2+\Delta_3-\Delta_1}{2}-1}}{(1+u_2)^{\Delta_3}}
\end{align}\

The value of this last integral is:

\begin{equation}
    \int_0^\infty du_2\ \frac{u_2^{\frac{\Delta_2+\Delta_3-\Delta_1}{2}-1}}{(1+u_2)^{\Delta_3}} = \frac{\Gamma(\frac{\Delta_2+\Delta_3-\Delta_1}{2})\Gamma(\frac{\Delta_3+\Delta_1-\Delta_2}{2})}{\Gamma(\Delta_3)}
\end{equation}\

Allowing us to compute the D-function completely:

\begin{equation}
        D_{\Delta_1\Delta_2\Delta_3}(\vec{y_1},\vec{y_2},\vec{y_3}) = \frac{\pi^\frac{d}{2}}{2}\frac{\Gamma(\frac{\sum\Delta_i-d}{2})}{\prod\Gamma(\Delta_i)}\frac{\Gamma(\frac{\Delta_1+\Delta_2-\Delta_3}{2})\Gamma(\frac{\Delta_2+\Delta_3-\Delta_1}{2})\Gamma(\frac{\Delta_3+\Delta_1-\Delta_2}{2})}{\lvert\vec{y_1}-\vec{y_2}\rvert^{\Delta_1+\Delta_2-\Delta_3}\lvert\vec{y_2}-\vec{y_3}\rvert^{\Delta_2+\Delta_3-\Delta_1}\lvert\vec{y_3}-\vec{y_1}\rvert^{\Delta_3+\Delta_1-\Delta_2}}
\end{equation}\

This result is consistent with was found by Muck and Viswanathan \cite{Muck} and also independently by Freedman et. al. \cite{Freedman1}. The form of this result, of course, satisfactorily agrees with the form expected for CFT 3-point functions derived in eq. (\ref{summcorr}). In the particular case where all the scaling dimensions $\Delta_i$ are equal $\Delta_i=\Delta$, the D-function further reduces to:

\begin{equation}\label{3dfunc}
        D_{\Delta\Delta\Delta}(\vec{y_1},\vec{y_2},\vec{y_3}) = \frac{\pi^\frac{d}{2}}{2}\frac{\Gamma(\frac{3\Delta-d}{2})}{\Gamma(\Delta)^3}\frac{\Gamma(\frac{\Delta}{2})^3}{\lvert\vec{y_1}-\vec{y_2}\rvert^\Delta \lvert\vec{y_2}-\vec{y_3}\rvert^\Delta \lvert\vec{y_3}-\vec{y_1}\rvert^\Delta}
\end{equation}\

agreeing again, unsurprisingly, with the expected form derived in eq. (\ref{summcorr2}).

\section{Case \texorpdfstring{$n=4$}{}}

Continuing our study of D-functions, the particular case of eq. (\ref{dfunc}) now with $n=4$, being the integral of 4 bulk-boundary propagators, is encountered in our work as contributions to the holographic CFT 4-point functions from both $\Phi^3$ and $\Phi^4$ theories on AdS. Now, the functional form of CFT 4-point functions is also strongly restricted purely from conformal symmetry arguments, as we showed in eq. (\ref{summcorr}). It is very satisfactory then to check that this D-function do indeed have this non-trivial form. Just like how we did before, let us calculate this particular case using as a starting point the representation for the D-function eq. (\ref{dfuncrep}):

\begin{align}
        D_{\Delta_1\Delta_2\Delta_3\Delta_4}(\vec{y_1},\vec{y_2},\vec{y_3},\vec{y_4}) = &\frac{\pi^{\frac{d}{2}}}{2}\frac{\Gamma(\frac{\sum\Delta_i-d}{2})\Gamma(\frac{\sum\Delta_i}{2})}{\prod \Gamma(\Delta_i)}\nonumber\\
        &\hspace{-2cm}\times\int_0^\infty d^3u\ \frac{u_2^{\Delta_2-1}u_3^{\Delta_3-1}u_4^{\Delta_4-1}}{(u_2\vec{y_{12}}^2+u_3\vec{y_{13}}^2+u_4\vec{y_{14}}^2+u_2u_3\vec{y_{23}}^2+u_2u_4\vec{y_{24}}^2+u_3u_4\vec{y_{34}}^2)^{\frac{\sum\Delta_i}{2}}}
\end{align}\

where $i=1,2,3,4$. The resulting integral in the $u_4$ variable can be simply done by doing first the rescaling $u_4\rightarrow u_4\Bigl(\frac{u_2\vec{y_{12}}^2+u_3\vec{y_{13}}^2+u_2u_3\vec{y_{23}}^2}{\vec{y_{14}}^2+u_2\vec{y_{24}}^2+u_3\vec{y_{34}}^2}\Bigl)$

\begin{align}
        D_{\Delta_1\Delta_2\Delta_3\Delta_4}(\vec{y_1},\vec{y_2},\vec{y_3},\vec{y_4}) = &\frac{\pi^{\frac{d}{2}}}{2}\frac{\Gamma(\frac{\sum\Delta_i-d}{2})\Gamma(\frac{\sum\Delta_i}{2})}{\prod \Gamma(\Delta_i)}\nonumber\\
        &\hspace{-2cm}\times\int_0^\infty d^2u\ \frac{u_2^{\Delta_2-1}u_3^{\Delta_3-1}}{(u_2\vec{y_{12}}^2+u_3\vec{y_{13}}^2+u_2u_3\vec{y_{23}}^2)^{\frac{\sum\Delta_i}{2}-\Delta_4}(\vec{y_{14}}^2+u_2\vec{y_{24}}^2+u_3\vec{y_{34}}^2)^{\Delta_4}}\nonumber\\
        &\hspace{-2cm}\times\int_0^\infty du_4\ \frac{u_4^{\Delta_4-1}}{(1+u_4)^{\frac{\sum\Delta_i}{2}}}
\end{align}\

Computing the value of this last integral gives:

\begin{equation}
    \int_0^\infty du_4\ \frac{u_4^{\Delta_4-1}}{(1+u_4)^{\frac{\sum\Delta_i}{2}}} = \frac{\Gamma(\Delta_4)\Gamma(\frac{\sum\Delta_i}{2}-\Delta_4)}{\Gamma(\frac{\sum\Delta_i}{2})}
\end{equation}\

Then the D-function reduces to:

\begin{align}
        D_{\Delta_1\Delta_2\Delta_3\Delta_4}(\vec{y_1},\vec{y_2},\vec{y_3},\vec{y_4}) = &\frac{\pi^{\frac{d}{2}}}{2}\frac{\Gamma(\frac{\sum\Delta_i-d}{2})\Gamma(\frac{\sum\Delta_i}{2}-\Delta_4)}{\Gamma(\Delta_1)\Gamma(\Delta_2)\Gamma(\Delta_3)}\nonumber\\
        &\hspace{-2cm}\times\int_0^\infty d^2u\ \frac{u_2^{\Delta_2-1}u_3^{\Delta_3-1}}{(u_2\vec{y_{12}}^2+u_3\vec{y_{13}}^2+u_2u_3\vec{y_{23}}^2)^{\frac{\sum\Delta_i}{2}-\Delta_4}(\vec{y_{14}}^2+u_2\vec{y_{24}}^2+u_3\vec{y_{34}}^2)^{\Delta_4}}
\end{align}\

The resulting integral now in the $u_3$ variable can be done more simply by doing first the rescaling $u_3\rightarrow u_3\Bigl(\frac{\vec{y_{14}}^2+u_2\vec{y_{24}}^2}{\vec{y_{34}}^2}\Bigl)$:

\begin{align}
        D_{\Delta_1\Delta_2\Delta_3\Delta_4}(\vec{y_1},\vec{y_2},\vec{y_3},\vec{y_4}) = &\frac{\pi^{\frac{d}{2}}}{2}\frac{\Gamma(\frac{\sum\Delta_i-d}{2})\Gamma(\frac{\sum\Delta_i}{2}-\Delta_4)}{\Gamma(\Delta_1)\Gamma(\Delta_2)\Gamma(\Delta_3)}\nonumber\\
        &\hspace{-2cm}\times\frac{1}{\lvert\vec{y_1}-\vec{y_2}\rvert^{\sum\Delta_i-2\Delta_4}\lvert\vec{y_3}-\vec{y_4}\rvert^{2\Delta_3}\lvert\vec{y_1}-\vec{y_4}\rvert^{2\Delta_4-2\Delta_3}}\int_0^\infty du_2\ \frac{u_2^{\Delta_2+\Delta_4-\frac{\sum\Delta_i}{2}-1}}{(1+u_2\frac{\vec{y_{24}}^2}{\vec{y_{14}}^2})^{\Delta_4-\Delta_3}}\nonumber\\
        &\hspace{-2cm}\times\int_0^\infty du_3\ \frac{u_3^{\Delta_3-1}}{(1+u_3)^{\Delta_4}\bigl[1+u_3(\frac{\vec{y_{13}}^2+u_2\vec{y_{23}}^2}{u_2\vec{y_{12}}^2})(\frac{\vec{y_{14}}^2+u_2\vec{y_{24}}^2}{\vec{y_{34}}^2})\bigl]^{\frac{\sum\Delta_i}{2}-\Delta_4}}
\end{align}\

This last integral is nothing more than the integral representation of Gauss' hypergeometric function $_2F_1$, up to constant factors:

\begin{equation}
    _2F_1(a,b;c;1-z) = \frac{\Gamma(c)}{\Gamma(b)\Gamma(c-b)}\int_0^\infty dx\ \frac{x^{b-1}}{(1+x)^{c-a}(1+xz)^a}
\end{equation}\

Writing it then in terms of this function allows us to further reduce the expression for the D-function to:

\begin{align}
        D_{\Delta_1\Delta_2\Delta_3\Delta_4}(\vec{y_1},\vec{y_2},\vec{y_3},\vec{y_4}) = &\frac{\pi^{\frac{d}{2}}}{2}\frac{\Gamma(\frac{\sum\Delta_i-d}{2})\Gamma(\frac{\sum\Delta_i}{2}-\Delta_3)\Gamma(\frac{\sum\Delta_i}{2}-\Delta_4)}{\Gamma(\Delta_1)\Gamma(\Delta_2)\Gamma(\frac{\sum\Delta_i}{2})}\nonumber\\
        &\hspace{-2cm}\times\frac{1}{\lvert\vec{y_1}-\vec{y_2}\rvert^{\sum\Delta_i-2\Delta_4}\lvert\vec{y_3}-\vec{y_4}\rvert^{2\Delta_3}\lvert\vec{y_1}-\vec{y_4}\rvert^{2\Delta_4-2\Delta_3}}\int_0^\infty du_2\ \frac{u_2^{\Delta_2+\Delta_4-\frac{\sum\Delta_i}{2}-1}}{(1+u_2\frac{\vec{y_{24}}^2}{\vec{y_{14}}^2})^{\Delta_4-\Delta_3}}\nonumber\\
        &\hspace{-2cm}\times\ _2F_1\Bigl(\frac{\sum\Delta_i}{2}-\Delta_4,\Delta_3;\frac{\sum\Delta_i}{2};1-\Bigl[\frac{\vec{y_{13}}^2+u_2\vec{y_{23}}^2}{u_2\vec{y_{12}}^2}\Bigl]\Bigl[\frac{\vec{y_{14}}^2+u_2\vec{y_{24}}^2}{\vec{y_{34}}^2}\Bigl]\Bigl)
\end{align}\

Let us make one last change of variable $u_2\frac{\vec{y_{24}}^2}{\vec{y_{14}}^2}=x$, and conveniently write the external points ratios in terms of the conformal invariant cross ratios $u$ and $v$ defined in eq. (\ref{crossr}). This process brings the expression for the D-function into the form:

\begin{align}\label{dfunc4medstep}
        D_{\Delta_1\Delta_2\Delta_3\Delta_4}(\vec{y_1},\vec{y_2},\vec{y_3},\vec{y_4}) = &\frac{\pi^{\frac{d}{2}}}{2}\frac{\Gamma(\frac{\sum\Delta_i-d}{2})\Gamma(\frac{\sum\Delta_i}{2}-\Delta_3)\Gamma(\frac{\sum\Delta_i}{2}-\Delta_4)}{\Gamma(\Delta_1)\Gamma(\Delta_2)\Gamma(\frac{\sum\Delta_i}{2})}\frac{u^{\frac{\Delta_4}{2}-\frac{\Delta_3}{2}-\frac{\sum\Delta_i}{6}}v^{\frac{\Delta_2}{2}+\frac{\Delta_3}{2}-\frac{\sum\Delta_i}{6}}}{\prod_{i<j}\lvert\vec{y_i}-\vec{y_j}\rvert^{\Delta_i+\Delta_j-\frac{\sum\Delta_n}{3}}}\nonumber\\
        &\hspace{-2cm}\times\int_0^\infty dx\ \frac{x^{\Delta_2+\Delta_4-\frac{\sum\Delta_i}{2}-1}}{(1+x)^{\Delta_4-\Delta_3}}\ _2F_1\Bigl(\frac{\sum\Delta_i}{2}-\Delta_4,\Delta_3;\frac{\sum\Delta_i}{2};1-\frac{1}{u}\Bigl[1+v+\frac{1}{x}+vx\Bigl]\Bigl)
\end{align}\

which is, of course, consistent with \cite{Muck}. Written this way, it is already clear enough that it satisfactorily agrees with the form expected for CFT 4-point functions derived in eq. (\ref{summcorr})! The remaining integral in the $x$ variable can be solved in terms of the $H(\dotsc;u,v)$ function defined by Dolan and Osborn in \cite{Osborn2}, which simply consists in a particular series expansion on the cross ratios $u$ and $v$:

\begin{align}
    &\int_0^\infty dx\ \frac{x^{\Delta_2+\Delta_4-\frac{\sum\Delta_i}{2}-1}}{(1+x)^{\Delta_4-\Delta_3}}\ _2F_1\Bigl(\frac{\sum\Delta_i}{2}-\Delta_4,\Delta_3;\frac{\sum\Delta_i}{2};1-\frac{1}{u}\Bigl[1+v+\frac{1}{x}+vx\Bigl]\Bigl)\nonumber\\
    &\hspace{2cm}=\frac{\Gamma(\frac{\sum\Delta_i}{2})}{\Gamma(\Delta_3)\Gamma(\Delta_4)\Gamma(\frac{\sum\Delta_i}{2}-\Delta_3)\Gamma(\frac{\sum\Delta_i}{2}-\Delta_4)}u^{\frac{\sum\Delta_i}{2}-\Delta_4}\nonumber\\
    &\hspace{4cm}\times H\Bigl(\Delta_2,\frac{\sum\Delta_i}{2}-\Delta_4,\Delta_1+\Delta_2-\frac{\sum\Delta_i}{2}+1,\Delta_1+\Delta_2;u,v\Bigl)
\end{align}\

Plugging this result into eq. (\ref{dfunc4medstep}) finally brings the D-function to the final form that we will be interested in this work:

\begin{align}\label{4dfuncgeneral}
        D_{\Delta_1\Delta_2\Delta_3\Delta_4}(\vec{y_1},\vec{y_2},\vec{y_3},\vec{y_4}) = &\frac{\pi^{\frac{d}{2}}}{2}\frac{\Gamma(\frac{\sum\Delta_i-d}{2})}{\prod\Gamma(\Delta_i)}\frac{u^{\frac{\Delta_1}{2}+\frac{\Delta_2}{2}-\frac{\sum\Delta_i}{6}}v^{\frac{\Delta_2}{2}+\frac{\Delta_3}{2}-\frac{\sum\Delta_i}{6}}}{\prod_{i<j}\lvert\vec{y_i}-\vec{y_j}\rvert^{\Delta_i+\Delta_j-\frac{\sum\Delta_n}{3}}}\nonumber\\
        &\hspace{-1cm}\times H\Bigl(\Delta_2,\frac{\sum\Delta_i}{2}-\Delta_4,\Delta_1+\Delta_2-\frac{\sum\Delta_i}{2}+1,\Delta_1+\Delta_2;u,v\Bigl)
\end{align}\

which is consistent with what was found by Dolan and Osborn in \cite{Osborn3}. In the particular case where all the scaling dimensions $\Delta_i$ are equal $\Delta_i=\Delta$, the D-function takes the relatively simple form:

\begin{equation}\label{4dfunc}
        D_{\Delta\Delta\Delta\Delta}(\vec{y_1},\vec{y_2},\vec{y_3},\vec{y_4}) = \frac{\pi^{\frac{d}{2}}}{2}\frac{\Gamma(2\Delta-\frac{d}{2})}{\Gamma(\Delta)^4}\frac{u^{\frac{\Delta}{3}}v^{\frac{\Delta}{3}}}{\prod_{i<j}\lvert\vec{y_i}-\vec{y_j}\rvert^{\frac{2\Delta}{3}}}H(\Delta,\Delta,1,2\Delta;u,v)
\end{equation}\

agreeing again, unsurprisingly, with the expected form derived in eq. (\ref{summcorr2}).\par
There is a nice derivative recurrence relation followed by these quantities which relate D-functions of different scaling dimensions that can be proved starting from their representation in eq. (\ref{gendfuncmedrep}). Indeed, differentiating both sides of this representation with respect to $y_{12}^2$ it is direct to see that it reduces to \cite{Freedman2}:

\begin{equation}\label{4dfuncrecrel}
    \frac{\partial}{\partial y_{12}^2}D_{\Delta_1\Delta_2\Delta_3\Delta_4}(\vec{y_1},\vec{y_2},\vec{y_3},\vec{y_4}) = -\frac{\Delta_1\Delta_2}{(\frac{\sum\Delta_i-d}{2})}D_{\Delta_1+1\Delta_2+1\Delta_3\Delta_4}(\vec{y_1},\vec{y_2},\vec{y_3},\vec{y_4})
\end{equation}\

and generalizing this equality iteratively:

\begin{equation}
    \frac{\partial^n}{\partial (y_{12}^2)^n}D_{\Delta_1\Delta_2\Delta_3\Delta_4}(\vec{y_1},\vec{y_2},\vec{y_3},\vec{y_4}) = \frac{(-1)^n(\Delta_1)_n(\Delta_2)_n}{(\frac{\sum\Delta_i-d}{2})_n}D_{\Delta_1+n\Delta_2+n\Delta_3\Delta_4}(\vec{y_1},\vec{y_2},\vec{y_3},\vec{y_4})
\end{equation}\

This nice property of the D-functions will be useful in the study of the quantum corrected holographic 4-point functions dual to a $\Phi^4$ theory on AdS.

\section{Special Case \texorpdfstring{$n=2$}{}}

The last D-function we will be interested in is a delicate one in which, as we will see shortly, its proper treatment will reveal the correct structure of the quantum corrections to the holographic correlators. It is the particular case of eq. (\ref{dfunc}) with $n=2$, representing the integral of 2 bulk-boundary propagators. This function is encountered in our work as loops one-particle irreducible contributions to the holographic CFT 2-point functions from both $\Phi^3$ and $\Phi^4$ theories on AdS. Since the functional form of these CFT 2-point functions is completely fixed purely from conformal symmetry arguments, as we showed in eq. (\ref{summcorr}), and also CFT loops diagrams can very well be UV-divergent, we expect these particular D-functions to have the functional form of CFT 2-point correlators up to eventual anomalies coming from the possible need to IR-regularize them (IR in this case due to the weak/strong duality). We can quickly convince ourselves that this is indeed the case if we try to naively compute this D-function in the same way we have done it for the previous ones, using as a starting point the representation for the D-function eq. (\ref{dfuncrep}):

\begin{align}
    D_{\Delta_1\Delta_2}(\vec{y_1},\vec{y_2}) &= \frac{\pi^{\frac{d}{2}}}{2}\frac{\Gamma(\frac{\Delta_1+ \Delta_2-d}{2})\Gamma(\frac{\Delta_1+\Delta_2}{2})}{\Gamma(\Delta_1)\Gamma(\Delta_2)}\int_0^\infty du_2\ \frac{u_2^{\Delta_2-1}}{(u_2\vec{y_{12}}^2)^{\frac{\Delta_1+\Delta_2}{2}}}\nonumber\\
    &= \frac{\pi^{\frac{d}{2}}}{2}\frac{\Gamma(\frac{\Delta_1+ \Delta_2-d}{2})\Gamma(\frac{\Delta_1+\Delta_2}{2})}{\Gamma(\Delta_1)\Gamma(\Delta_2)}\frac{1}{\lvert\vec{y_1}-\vec{y_2}\rvert^{\Delta_1+\Delta_2}}\int_0^\infty du_2\ u_2^{-1+\frac{\Delta_2-\Delta_1}{2}}
\end{align}\

The resulting integral in the $u_2$ variable is clearly divergent for any pair of values $\Delta_1$ and $\Delta_2$, polynomically divergent for $\Delta_1\neq\Delta_2$ and logarithmically divergent for $\Delta_1=\Delta_2$. In fact, this same behavior can already be seen from the original definition of the D-function as the integrated radial coordinate $x_0$ approaches the conformal boundary of the AdS space at $x_0=0$. Indeed, the integrand of eq. (\ref{dfunc}) with $n=2$ in this limit where $x_0\rightarrow0$ behaves like:

\begin{align}\label{integrandexp}
    \sqrt{g}\ \tilde{K}^{\Delta_1}(x,\vec{y_1})\tilde{K}^{\Delta_2}(x,\vec{y_2}) \underset{x_0\rightarrow0}{=} &x_0^{-d-1}\Bigl[c_{\Delta_1}^{-1}x_0^{d-\Delta_1}\delta^d(\vec{x}-\vec{y_1})+\dotsb+x_0^{\Delta_1}\frac{1}{\lvert\vec{x}-\vec{y_1}\rvert^{2\Delta_1}}+\dotsb\Bigl]\nonumber\\
    &\hspace{0.5cm}\times\Bigl[c_{\Delta_2}^{-1}x_0^{d-\Delta_2}\delta^d(\vec{x}-\vec{y_2})+\dotsb+x_0^{\Delta_2}\frac{1}{\lvert\vec{x}-\vec{y_2}\rvert^{2\Delta_2}}+\dotsb\Bigl]\nonumber\\
    =\ &(\text{contact terms}) + x_0^{-1+\Delta_2-\Delta_1}\frac{c_{\Delta_1}^{-1}\delta^d(\vec{x}-\vec{y_1})}{\lvert\vec{x}-\vec{y_2}\rvert^{2\Delta_2}}\nonumber\\
    &+x_0^{-1+\Delta_1-\Delta_2}\frac{c_{\Delta_2}^{-1}\delta^d(\vec{x}-\vec{y_2})}{\lvert\vec{x}-\vec{y_1}\rvert^{2\Delta_1}} + (\text{subleading terms})
\end{align}\

where we used the expansion of the bulk-boundary propagator eq. (\ref{expbubop}). Again, the resulting integral now in the radial coordinate $x_0$ is clearly divergent as it approaches $x_0=0$, polynomically divergent for $\Delta_1\neq\Delta_2$ and logarithmically divergent for $\Delta_1=\Delta_2$. The realization of this fact strongly suggests the regularization of the $x_0$ integral in its lower limit of integration since in this region is where the divergences are emerging. But we have already encountered the need for such a regulator in what would appears to be a completely different context, when regularizing the variation of the AdS path integral in the holographic renormalization procedure with the intention to compute the tree-level contributions to the holographic correlators, introducing the IR-regulator $\varepsilon$ for the first time in eq. (\ref{regvarz}). The regularization scheme in this case correctly captures the structure of these classical contributions, so it is extremely satisfying to find that it also does it for the quantum contributions to the correlators as well. In fact, keeping track of this regulator carefully throughout the manipulations of the AdS path integral, it naturally IR-regularize every single one of the loops integrals found in the study of the holographic n-point functions! All this discussion is to motivate the definition of the regularized version of the D-function which is the quantity that one actually encounters:

\begin{equation}\label{d2funcreg1}
    D^{(\varepsilon)}_{\Delta_1\Delta_2}(\vec{y_1},\vec{y_2}) \equiv \int_{x_0=\varepsilon} d^{d+1}x\sqrt{g}\ \tilde{K}^{\Delta_1}(x,\vec{y_1})\tilde{K}^{\Delta_2}(x,\vec{y_2})
\end{equation}\

We will proceed to show how one can easily compute the value of this integral using a very clever trick, in the case where both scaling dimensions are the same $\Delta_1=\Delta_2=\Delta$ which is, of course, the case in which one is most interested. For the case when $\Delta_1\neq\Delta_2$ the arguments presented here easily generalize. We will be interested then in the value of the integral:

\begin{equation}\label{regdfunc}
    D^{(\varepsilon)}_{\Delta\Delta}(\vec{y_1},\vec{y_2}) = \int_{x_0=\varepsilon} d^{d+1}x\sqrt{g}\ \tilde{K}^\Delta(x,\vec{y_1})\tilde{K}^\Delta(x,\vec{y_2})
\end{equation}\

The strategy will be to extract all the dependence on the external points of the integrand through AdS isometry transformations eq. (\ref{adsiso}), remaining dependence only in the limits of integration, for then solving the differential equation followed by the resulting integral instead of solving the integral directly, which will turn out to be much simpler to do. With these goals in mind then, we will start computing eq. (\ref{regdfunc}) by performing the translation $\vec{x}\rightarrow\vec{x}+\vec{y_2}$:

\begin{equation}
    D^{(\varepsilon)}_{\Delta\Delta}(\vec{y_1},\vec{y_2}) = \int_{x_0=\varepsilon} d^{d+1}x\sqrt{g}\ \tilde{K}^\Delta(x,\vec{y_{12}})\tilde{K}^\Delta(x,\vec{0})
\end{equation}\

where we used that under AdS isometry transformations the AdS measure is invariant and that the bulk-boundary propagator transforms according to eq. (\ref{bubotr}). The leftover dependency of the integrand on the external points can be removed by doing the rescaling $x\rightarrow \lvert\vec{y_1}-\vec{y_2}\rvert x$:

\begin{align}\label{kkintdfunc}
        D^{(\varepsilon)}_{\Delta\Delta}(\vec{y_1},\vec{y_2}) &= \int_{x_0=\frac{\varepsilon}{\lvert\vec{y_1}-\vec{y_2}\rvert}}d^{d+1}x\sqrt{g}\ \frac{1}{\lvert\vec{y_1}-\vec{y_2}\rvert^\Delta}\tilde{K}^\Delta\Bigl(x,\frac{\vec{y_{12}}}{\lvert\vec{y_1}-\vec{y_2}\rvert}\Bigl)\frac{1}{\lvert\vec{y_1}-\vec{y_2}\rvert^\Delta}\tilde{K}^\Delta(x,\vec{0})\nonumber\\
        &= \frac{1}{\lvert\vec{y_1}-\vec{y_2}\rvert^{2\Delta}}\int_{x_0=\sigma}d^{d+1}x\sqrt{g}\ \tilde{K}^\Delta(x,\hat{n})\tilde{K}^\Delta(x,\vec{0})
\end{align}\

where we used the transformation rules of the measure and bulk-boundary propagator, and defined the quantities $\sigma\equiv\frac{\varepsilon}{\lvert\vec{y_1}-\vec{y_2}\rvert}$ and $\hat{n}=\frac{\vec{y_{12}}}{\lvert\vec{y_1}-\vec{y_2}\rvert}$, $\hat{n}$ being a unit vector pointing in the direction of $\vec{y_{12}}$. Notice how under these transformations we managed to extract all the dependence of the external points of the integrand, remaining only in the lower limit of integration of the $x_0$ integral in the form of $\sigma$. It remains to compute the value of this last integral in the limit $\varepsilon\rightarrow0$, which in terms of $\sigma$ translates to $\sigma\rightarrow0$. The key realization here is that in this limit the differential equation in $\sigma$ satisfied by the integral is much easier to solve than the integral itself thanks to the convenient presence of Dirac deltas in the expansion of the integrand coming from the bulk-boundary propagators, as it can be seen from eq. (\ref{integrandexp}). Indeed, differentiating the integral in eq. (\ref{kkintdfunc}) with respect to $\sigma$ in the limit $\sigma\rightarrow0$:

\begin{align}
    \frac{d}{d\sigma}\Bigl[\int_{x_0=\sigma}d^{d+1}x\sqrt{g}\ \tilde{K}^\Delta(x,\hat{n})\tilde{K}^\Delta(x,\vec{0})\Bigl] = &-\int d^dx\sqrt{g}\ \tilde{K}^\Delta(x,\hat{n})\tilde{K}^\Delta(x,\vec{0})\Bigl\rvert_{x_0=\sigma}\nonumber\\
    \underset{\sigma\rightarrow0}{=} &(\text{contact terms})\nonumber\\
    &-\frac{1}{c_\Delta\sigma}\int d^dx\ \Bigl[\frac{\delta^d(\vec{x}-\hat{n})}{\lvert\vec{x}\rvert^{2\Delta}}+\frac{\delta^d(\vec{x})}{\lvert\vec{x}-\hat{n}\rvert^{2\Delta}}\Bigl] + \mathcal{O}(\sigma^{-1<})\nonumber\\
    = &(\text{contact terms})-\frac{2}{c_\Delta\sigma}+ \mathcal{O}(\sigma^{-1<})
\end{align}\

where in the first line we used the fundamental theorem of calculus, in the second line we expanded the integrand using eq. (\ref{integrandexp}) and in the last line we trivially computed the integrals with the Dirac deltas coming from the bulk-boundary propagators, using the fact that the vector $\hat{n}$ is unitary. Notice how the resulting equation for the integral is very easy to solve! Simply integrating both sides with respect to $\sigma$ we find, up to integration constants, that the value of the integral is given by:

\begin{equation}
    \int_{x_0=\sigma}d^{d+1}x\sqrt{g}\ \tilde{K}^\Delta(x,\hat{n})\tilde{K}^\Delta(x,\vec{0}) = (\text{contact terms})-\frac{2}{c_\Delta}\ln{(\sigma)}+\mathcal{O}(\sigma^{0<})
\end{equation}\

Notice how the subleading terms of order $\mathcal{O}(\sigma^{0<})$ simply go to 0 in the limit $\sigma\rightarrow0$. Therefore, plugging this result for the integral back into eq. (\ref{kkintdfunc}) and remembering that $\sigma\equiv\frac{\varepsilon}{\lvert\vec{y_1}-\vec{y_2}\rvert}$, the value we find for the regularized D-function eq. (\ref{regdfunc}) (up to contact terms which can always be renormalized with appropriate local counterterms) is given by:

\begin{equation}\label{d2funcreg2}
    D^{(\varepsilon)}_{\Delta\Delta}(\vec{y_1},\vec{y_2}) = -\frac{2c_\Delta^{-1}}{\lvert\vec{y_1}-\vec{y_2}\rvert^{2\Delta}}\ln{\Bigl(\frac{\varepsilon}{\lvert\vec{y_1}-\vec{y_2}\rvert}\Bigl)}
\end{equation}\

This result seems to break the conformal structure expected for contributions to the 2-point function of a CFT, however as it is discussed in the main text, its form corresponds exactly to the expansion of a conformal anomaly, realization which will lead not only to a clear picture of the role of the quantum corrections to the holographic correlators, but also to a clear picture in their renormalization scheme.

\newpage


\chapter{Integral Formulas}
In Appendix A we studied the set of integrals encountered in our work of holographic CFT correlators that only involve bulk-boundary propagators, integrals which are known in the AdS/CFT literature as D-functions. There we not only presented a possible approach on how to solve these quantities but also, since they represent contributions to specific CFT correlators, we checked that they have precisely the functional form expected from conformal symmetry arguments. In the present appendix we will continue the study of the integrals encountered in our work but now focusing on those that also contain the complicated bulk-bulk propagator. One come across these integrals mainly from the quantum corrections to the holographic correlators, which pictorially correspond to loops diagrams in the interior of the AdS space. The representation of the bulk-bulk propagator however as a hypergeometric function in the $\xi$ variable, eq. (\ref{G}), makes these integrals extremely difficult to evaluate. Fortunately, one can express this hypergeometric function in a much useful form thanks to the fact that in the entire region of integration its argument is constrained between the values 0 and 1, allowing us to write it in its convergent power series representation which greatly facilitates the resulting integrals to solve. The claim of the codomain of $\xi$ can be seen directly from the Euclidean distance between the two points of the bulk-bulk propagator $G(x,z)$:

\begin{align}
    0&\leq(x-z)^2\nonumber\\
    &= (x_0-z_0)^2+(\vec{x}-\vec{z})^2\nonumber\\
    &= x_0^2-2x_0z_0+z_0^2+(\vec{x}-\vec{z})^2
\end{align}\

Adding $2x_0z_0$ to both sides of the inequality and then dividing by $x_0^2+z_0^2+(\vec{x}-\vec{z})^2$:

\begin{align}
    \implies &0\leq2x_0z_0\leq x_0^2+z_0^2+(\vec{x}-\vec{z})^2\nonumber\\
    \implies &0\leq\frac{2x_0z_0}{x_0^2+z_0^2+(\vec{x}-\vec{z})^2}\leq 1
\end{align}\

where, since both radial coordinates $x_0$ and $z_0$ are always positive, we trivially added 0 as the lower limit of the inequality. The quantity between the inequality symbols is precisely $\xi$, confirming our claim. This fact let us express the bulk-bulk propagator as a power series in this variable:

\begin{equation}\label{Gseries}
    G(x,z) = \frac{2^{-\Delta}c_\Delta}{2\nu}\sum_{k=0}^\infty \frac{(\frac{\Delta}{2})_k(\frac{\Delta+1}{2})_k}{(\nu+1)_k\ k!}\xi^{\Delta+2k},\hspace{0.5cm} \xi = \frac{2x_0z_0}{x_0^2+z_0^2+(\vec{x}-\vec{z})^2}
\end{equation}\

where we introduced the Pochhammer symbol $(a)_k = \frac{\Gamma(a+k)}{\Gamma(a)}$. The strategy of resolution for these complicated integrals involving the bulk-bulk propagator will then be to write $G(x,z)$ in its power series representation, which will lead to simpler integrals involving powers of $\xi$ and bulk-boundary propagators. It will turn out that, this process for the particular integrals encountered in this work, when written explicitly most of them will have the form:

\begin{equation}\label{genintb2}
    I(z_0,\vec{y_1},\vec{y_2}) = \int d^{d+1}x\ \frac{x_0^a}{(x_0^2+z_0^2+\lvert\vec{x}-\vec{y_1}\rvert^2)^b(x_0^2+\lvert\vec{x}-\vec{y_2}\rvert^2)^c}
\end{equation}\

We will then proceed to compute the general formula for this integral, to later use it and study more easily the concrete cases encountered in our work of holographic correlators in which we are interested.

\section{General Formula}

As we said, we will solve for the general formula of eq. (\ref{genintb2}) with the intention to use it for the concrete integrals involving the bulk-bulk propagator that we are interested in, propagator which written as a power series in $\xi$ will result in simpler integrals involving powers of this parameters and bulk-boundary propagators. We will start then the computation of eq. (\ref{genintb2}) by grouping the factors in the denominator using Feynman parametrization eq. (\ref{feynpam}), resulting in:

\begin{align}
    I(z_0,\vec{y_1},\vec{y_2}) &= \frac{\Gamma(b+c)}{\Gamma(b)\Gamma(c)}\int_0^1 d^2u\int d^{d+1}x\ \frac{\delta(u_1+u_2-1)u_1^{b-1}u_2^{c-1}x_0^a}{\bigl[u_1(x_0^2+z_0^2+\lvert\vec{x}-\vec{y_1}\rvert^2)+u_2(x_0^2+\lvert\vec{x}-\vec{y_2}\rvert^2)\bigl]^{b+c}}\nonumber\\
    &\hspace{-1.5cm}= \frac{\Gamma(b+c)}{\Gamma(b)\Gamma(c)}\int_0^1 d^2u\int d^{d+1}x\ \frac{\delta(u_1+u_2-1)u_1^{b-1}u_2^{c-1}x_0^a}{\bigl[x_0^2+(\vec{x}-u_1\vec{y_1}-u_2\vec{y_2})^2+u_1\vec{y_1}^2+u_2\vec{y_2}^2-(u_1\vec{y_1}+u_2\vec{y_2})^2+u_1z_0^2\bigl]^{b+c}}\nonumber\\
    &= \frac{\Gamma(b+c)}{\Gamma(b)\Gamma(c)}\int_0^1 d^2u\int d^{d+1}x\ \frac{\delta(u_1+u_2-1)u_1^{b-1}u_2^{c-1}x_0^a}{\bigl[x_0^2+(\vec{x}-u_1\vec{y_1}-u_2\vec{y_2})^2+u_1u_2\vec{y_{12}}^2+u_1z_0^2\bigl]^{b+c}}
\end{align}\

where we completed squares, used the fact that the delta function forces $u_1+u_2=1$ and simplified the resulting terms, defining in the process the quantity $\vec{y_{12}}^2\equiv \lvert\vec{y_1}-\vec{y_2}\rvert^2$. The resulting integral in the $\vec{x}$ variable can be simply done by first doing the translation $\vec{x}\rightarrow\vec{x}+u_1\vec{y_1}+u_2\vec{y_2}$:

\begin{equation}
    I(z_0,\vec{y_1},\vec{y_2}) = \frac{\Gamma(b+c)}{\Gamma(b)\Gamma(c)}\int_0^1 d^2u\int d^{d+1}x\ \frac{\delta(u_1+u_2-1)u_1^{b-1}u_2^{c-1}x_0^a}{(x_0^2+\vec{x}^2+u_1u_2\vec{y_{12}}^2+u_1z_0^2)^{b+c}}
\end{equation}\

followed by the rescaling $\vec{x}\rightarrow\vec{x}\sqrt{x_0^2+u_1u_2\vec{y_{12}}^2+u_1z_0^2}$:

\begin{equation}
    I(z_0,\vec{y_1},\vec{y_2}) = \frac{\Gamma(b+c)}{\Gamma(b)\Gamma(c)}\int_0^1 d^2u\int_0^\infty dx_0\ \frac{\delta(u_1+u_2-1)u_1^{b-1}u_2^{c-1}x_0^a}{(x_0^2+u_1u_2\vec{y_{12}}^2+u_1z_0^2)^{b+c-\frac{d}{2}}}\int d^dx\ \frac{1}{(1+\vec{x}^2)^{b+c}}
\end{equation}\

Computing the value of this last integral gives:

\begin{equation}
    \int d^dx\ \frac{1}{(1+\vec{x}^2)^{b+c}} = \pi^{\frac{d}{2}}\frac{\Gamma(b+c-\frac{d}{2})}{\Gamma(b+c)}
\end{equation}\

Then $I(z_0,\vec{y_1},\vec{y_2})$ reduces to:

\begin{equation}
    I(z_0,\vec{y_1},\vec{y_2}) = \pi^{\frac{d}{2}}\frac{\Gamma(b+c-\frac{d}{2})}{\Gamma(b)\Gamma(c)}\int_0^1 d^2u\int_0^\infty dx_0\ \frac{\delta(u_1+u_2-1)u_1^{b-1}u_2^{c-1}x_0^a}{(x_0^2+u_1u_2\vec{y_{12}}^2+u_1z_0^2)^{b+c-\frac{d}{2}}}
\end{equation}\

In the same way, the resulting integral now in the $x_0$ variable can be done more simply by doing first the rescaling $x_0\rightarrow x_0\sqrt{u_1u_2\vec{y_{12}}^2+u_1z_0^2}$:

\begin{equation}
    I(z_0,\vec{y_1},\vec{y_2}) = \pi^{\frac{d}{2}}\frac{\Gamma(b+c-\frac{d}{2})}{\Gamma(b)\Gamma(c)}\int_0^1 d^2u\ \frac{\delta(u_1+u_2-1)u_1^{b-1}u_2^{c-1}}{(u_1u_2\vec{y_{12}}^2+u_1z_0^2)^{b+c-\frac{a+d+1}{2}}}\int_0^\infty dx_0\ \frac{x_0^a}{(1+x_0^2)^{b+c-\frac{d}{2}}}
\end{equation}\

The value of this last integral is:

\begin{equation}
    \int_0^\infty dx_0\ \frac{x_0^a}{(1+x_0^2)^{b+c-\frac{d}{2}}} = \frac{1}{2}\frac{\Gamma(\frac{a+1}{2})\Gamma(b+c-\frac{a+d+1}{2})}{\Gamma(b+c-\frac{d}{2})}
\end{equation}\

Then $I(z_0,\vec{y_1},\vec{y_2})$ further reduces to:

\begin{align}
    I(z_0,\vec{y_1},\vec{y_2}) &= \frac{\pi^{\frac{d}{2}}}{2}\frac{\Gamma(\frac{a+1}{2})\Gamma(b+c-\frac{a+d+1}{2})}{\Gamma(b)\Gamma(c)}\int_0^1 d^2u\ \frac{\delta(u_1+u_2-1)u_1^{\frac{a+d+1}{2}-c-1}u_2^{c-1}}{(z_0^2+u_2\vec{y_{12}}^2)^{b+c-\frac{a+d+1}{2}}}\nonumber\\
    &= \frac{\pi^{\frac{d}{2}}}{2}\frac{\Gamma(\frac{a+1}{2})\Gamma(b+c-\frac{a+d+1}{2})}{\Gamma(b)\Gamma(c)}\int_0^1 du_1\ \frac{u_1^{\frac{a+d+1}{2}-c-1}(1-u_1)^{c-1}}{(z_0^2+\vec{y_{12}}^2-u_1\vec{y_{12}}^2)^{b+c-\frac{a+d+1}{2}}}\nonumber\\
    &= \frac{\pi^{\frac{d}{2}}}{2}\frac{\Gamma(\frac{a+1}{2})\Gamma(b+c-\frac{a+d+1}{2})}{\Gamma(b)\Gamma(c)(z_0^2+\vec{y_{12}}^2)^{b+c-\frac{a+d+1}{2}}}\int_0^1 du_1\ \frac{u_1^{\frac{a+d+1}{2}-c-1}(1-u_1)^{c-1}}{\bigl(1-u_1\frac{\vec{y_{12}}^2}{z_0^2+\vec{y_{12}}^2}\bigl)^{b+c-\frac{a+d+1}{2}}}
\end{align}\

where we trivially computed the $u_2$ integral through the Dirac delta function and extracted the $z_0^2+\vec{y_{12}}^2$ factor from the denominator. This last integral is nothing more than the integral representation of Gauss' hypergeometric function $_2F_1$, up to constant factors \cite{HypFunc1}:

\begin{equation}
    _2F_1(a,b;c;z) = \frac{\Gamma(c)}{\Gamma(b)\Gamma(c-b)}\int_0^1dx\ \frac{x^{b-1}(1-x)^{c-b-1}}{(1-xz)^a}
\end{equation}\

Writing it then in terms of this function allows us to finally solve for the general formula of eq. (\ref{genintb2}):

\begin{align}\label{generalformula}
    \int d^{d+1}x\ \frac{x_0^a}{(x_0^2+z_0^2+\lvert\vec{x}-\vec{y_1}\rvert^2)^b(x_0^2+\lvert\vec{x}-\vec{y_2}\rvert^2)^c} = &\frac{\pi^{\frac{d}{2}}}{2}\frac{\Gamma(\frac{a+1}{2})\Gamma(b+c-\frac{a+d+1}{2})\Gamma(\frac{a+d+1}{2}-c)}{\Gamma(b)\Gamma(\frac{a+d+1}{2})(z_0^2+\lvert\vec{y_1}-\vec{y_2}\rvert^2)^{b+c-\frac{a+d+1}{2}}}\nonumber\\
    &\hspace{-5cm}\times\ _2F_1\Bigl(b+c-\frac{a+d+1}{2},\frac{a+d+1}{2}-c;\frac{a+d+1}{2};1-\frac{z_0^2}{z_0^2+\lvert\vec{y_1}-\vec{y_2}\rvert^2}\Bigl)
\end{align}\

Next we will see how this formula can be applied to the different resulting integrals encountered in our study of holographic correlators coming from writing the bulk-bulk propagator in its power series representation.

\section{\texorpdfstring{$\int\xi$}{}-type Integrals}

The first type of integrals we are going to consider are those resulting from $\int G^n$, being the integral of the $n$-th power of the bulk-bulk propagator. The particular case with $n=2$ is encountered in our work in the "eight" or "double-scoop" loop diagram coming from $\Phi^4$ theories on AdS. Using the power series representation of $G(x,z)$ as an expansion in the variable $\xi$, these type of integrals simply reduce to a sum of integrals of powers of $\xi$, each one having the general form:

\begin{equation}\label{intxi}
    \int d^{d+1}x\sqrt{g}\ \xi^{\Delta_1}(x,z)
\end{equation}\

for some number $\Delta_1$. Each one of these integrals can be solved directly from the formula eq. (\ref{generalformula}) under the particular case $c=\lvert\vec{y_1}-\vec{y_2}\rvert=0$, where the formula simply reduces to:

\begin{equation}\label{redgenform}
    \int d^{d+1}x\ \frac{x_0^a}{(x_0^2+z_0^2+\lvert\vec{x}-\vec{y_1}\rvert^2)^b} = \frac{\pi^{\frac{d}{2}}}{2}\frac{\Gamma(\frac{a+1}{2})\Gamma(b-\frac{a+d+1}{2})}{\Gamma(b)z_0^{2b-a-d-1}}
\end{equation}\

where we used that $_2F_1(\dotsc;0)=1$. Let us see how eq. (\ref{intxi}) can be directly solved from the reduced version of the general formula eq. (\ref{redgenform}):

\begin{align}
    \int d^{d+1}x\sqrt{g}\ \xi^{\Delta_1}(x,z) &= \int d^{d+1}x\ x_0^{-d-1}\Bigl(\frac{2x_0z_0}{x_0^2+z_0^2+\lvert\vec{x}-\vec{z}\rvert^2}\Bigl)^{\Delta_1}\nonumber\\
    &= 2^{\Delta_1}z_0^{\Delta_1}\int d^{d+1}x\ \frac{x_0^{\Delta_1-d-1}}{(x_0^2+z_0^2+\lvert\vec{x}-\vec{z}\rvert^2)^{\Delta_1}}
\end{align}\

This last integral is precisely the reduced formula eq. (\ref{redgenform}) under the particular case $a=\Delta_1-d-1$ and $b=\Delta_1$. Therefore, replacing its value:

\begin{align}
    \int d^{d+1}x\sqrt{g}\ \xi^{\Delta_1}(x,z) &= 2^{\Delta_1}z_0^{\Delta_1}\frac{\pi^{\frac{d}{2}}}{2}\frac{\Gamma(\frac{\Delta_1-d}{2})\Gamma(\frac{\Delta_1}{2})}{\Gamma(\Delta_1)z_0^{\Delta_1}}\nonumber\\
    &= 2^{\Delta_1}\frac{\pi^{\frac{d}{2}}}{2}\frac{\Gamma(\frac{\Delta_1-d}{2})\Gamma(\frac{\Delta_1}{2})}{\Gamma(\Delta_1)}
\end{align}\

In principle, the form of this result is already nice enough to be used, but for the specific calculations that we want to carry out it will be useful to simplify it using what is known as Legendre duplication formula \cite{LegDupFor}:

\begin{equation}\label{LDF}
    \Gamma(z)\Gamma\Bigl(z+\frac{1}{2}\Bigl) = 2^{1-2z}\sqrt{\pi}\ \Gamma(2z)
\end{equation}\

which, for the particular value $z=\frac{\Delta_1}{2}$, allows us to write the final result of the integral as:

\begin{equation}\label{xiform}
    \int d^{d+1}x\sqrt{g}\ \xi^{\Delta_1}(x,z) = \pi^{\frac{d+1}{2}}\frac{\Gamma(\frac{\Delta_1-d}{2})}{\Gamma(\frac{\Delta_1+1}{2})}
\end{equation}\

Notice how the value of this integral is just a number, not dependent of the external point. Since the integral $\int G^n$ can be written as a sum of integrals of powers of $\xi$ this result implies that, whenever $\int G^n$ is convergent, its value is just a constant! As it is discussed in the main text, this fact is completely consistent with the expected purely from AdS isometry arguments.

\section{\texorpdfstring{$\int\xi K$}{}-type Integrals}

Another type of integrals that we are going to consider are those resulting from $\int G^nK$, being the integral of the $n$-th power of the bulk-bulk propagator times the bulk-boundary propagator. Many particular cases of these integrals are found throughout our study of holographic correlators, e.g., the case with $n=2$ in the "eye" loop diagram coming from $\Phi^3$ theories, the case with $n=3$ in the "sunset" loop diagram coming from $\Phi^4$ theories, and of course the special case with $n=1$ in the one-time reducible diagrams coming from both theories. Using the power series representation of $G(x,z)$ as an expansion in the variable $\xi$, these type of integrals simply reduce to a sum of integrals of powers of $\xi$ times $K$, each one having the general form:

\begin{equation}
    \int d^{d+1}x\sqrt{g}\ \xi^{\Delta_1}(x,z)\tilde{K}^{\Delta_2}(x,\vec{y_2})
\end{equation}\

for some numbers $\Delta_1$ and $\Delta_2$, where $\tilde{K}^{\Delta_2}(x,\vec{y_2})$ is the unnormalized bulk-boundary propagator of scaling dimension $\Delta_2$, eq. (\ref{unbubop}). Each one of these integrals can be solved directly from the general formula eq. (\ref{generalformula}). Let us see how this is done:

\begin{align}
    \int d^{d+1}x\sqrt{g}\ \xi^{\Delta_1}(x,z)\tilde{K}^{\Delta_2}(x,\vec{y_2}) &= \int d^{d+1}x\ x_0^{-d-1} \Bigl(\frac{2x_0z_0}{x_0^2+z_0^2+\lvert\vec{x}-\vec{z}\rvert^2}\Bigl)^{\Delta_1}\Bigl(\frac{x_0}{x_0^2+\lvert\vec{x}-\vec{y_2}\vert^2}\Bigl)^{\Delta_2}\nonumber\\
    &= 2^{\Delta_1}z_0^{\Delta_1}\int d^{d+1}x\ \frac{x_0^{\Delta_1+\Delta_2-d-1}}{(x_0^2+z_0^2+\lvert\vec{x}-\vec{z}\rvert^2)^{\Delta_1}(x_0^2+\lvert\vec{x}-\vec{y_2}\rvert^2)^{\Delta_2}}
\end{align}\

This last integral is precisely the general formula eq. (\ref{generalformula}) under the particular case $a=\Delta_1+\Delta_2-d-1$, $b=\Delta_1$, $c=\Delta_2$. Therefore, replacing its value:

\begin{align}
    \int d^{d+1}x\sqrt{g}\ \xi^{\Delta_1}(x,z)\tilde{K}^{\Delta_2}(x,\vec{y_2}) = &2^{\Delta_1}z_0^{\Delta_1}\frac{\pi^{\frac{d}{2}}}{2}\frac{\Gamma(\frac{\Delta_1+\Delta_2-d}{2})\Gamma(\frac{\Delta_1+\Delta_2}{2})\Gamma(\frac{\Delta_1-\Delta_2}{2})}{\Gamma(\Delta_1)\Gamma(\frac{\Delta_1+\Delta_2}{2})(z_0^2+\lvert\vec{z}-\vec{y_2}\rvert^2)^{\frac{\Delta_1+\Delta_2}{2}}}\nonumber\\
    &\times\ _2F_1\Bigl(\frac{\Delta_1+\Delta_2}{2},\frac{\Delta_1-\Delta_2}{2};\frac{\Delta_1+\Delta_2}{2};1-\frac{z_0^2}{z_0^2+\lvert\vec{z}-\vec{y_2}\rvert^2}\Bigl)\nonumber\\
    = &2^{\Delta_1}z_0^{\Delta_1}\frac{\pi^{\frac{d}{2}}}{2}\frac{\Gamma(\frac{\Delta_1+\Delta_2-d}{2})\Gamma(\frac{\Delta_1-\Delta_2}{2})}{\Gamma(\Delta_1)(z_0^2+\lvert\vec{z}-\vec{y_2}\rvert^2)^{\frac{\Delta_1+\Delta_2}{2}}}\nonumber\\
    &\times\ _1F_0\Bigl(\frac{\Delta_1-\Delta_2}{2};1-\frac{z_0^2}{z_0^2+\lvert\vec{z}-\vec{y_2}\rvert^2}\Bigl)\nonumber\\
    = &2^{\Delta_1}z_0^{\Delta_1}\frac{\pi^{\frac{d}{2}}}{2}\frac{\Gamma(\frac{\Delta_1+\Delta_2-d}{2})\Gamma(\frac{\Delta_1-\Delta_2}{2})}{\Gamma(\Delta_1)(z_0^2+\lvert\vec{z}-\vec{y_2}\rvert^2)^{\frac{\Delta_1+\Delta_2}{2}}}\Bigl(\frac{z_0^2}{z_0^2+\lvert\vec{z}-\vec{y_2}\rvert^2}\Bigl)^{\frac{\Delta_2-\Delta_1}{2}}\nonumber\\
    = &2^{\Delta_1}\frac{\pi^{\frac{d}{2}}}{2}\frac{\Gamma(\frac{\Delta_1+\Delta_2-d}{2})\Gamma(\frac{\Delta_1-\Delta_2}{2})}{\Gamma(\Delta_1)}\tilde{K}^{\Delta_2}(z,\vec{y_2})
\end{align}\

where we used that $_2F_1(a,b;a;z)=\ _1F_0(b;z)=(1-z)^{-b}$. In principle, the form of this result is already nice enough to be used, but for the specific calculations that we want to carry out it will be useful to simplify it using Legendre duplication formula eq. (\ref{LDF}) which, for the particular value $z=\frac{\Delta_1}{2}$, allows us to write the final result of the integral as:

\begin{equation}\label{xikform}
    \int d^{d+1}x\sqrt{g}\ \xi^{\Delta_1}(x,z)\tilde{K}^{\Delta_2}(x,\vec{y_2}) = \pi^{\frac{d+1}{2}}\frac{\Gamma(\frac{\Delta_1+\Delta_2-d}{2})\Gamma(\frac{\Delta_1-\Delta_2}{2})}{\Gamma(\frac{\Delta_1}{2})\Gamma(\frac{\Delta_1+1}{2})}\tilde{K}^{\Delta_2}(z,\vec{y_2})
\end{equation}\

Notice how the value of this integral is proportional to a bulk-boundary propagator of the same scaling dimension as the one being integrated. Since the integral $\int G^nK$ can be written as a sum of integrals of powers of $\xi$ times $K$ this result implies that, whenever $\int G^nK$ is convergent, its value will be proportional to $K$!. As it is discussed in the main text, this fact is completely consistent with the expected purely from AdS isometry arguments.

\section{\texorpdfstring{$\int \xi KK$}{}-type Integrals}

The most challenging type of integrals that we are going to consider in this work are those resulting from $\int G^nKK$, being the integral of the $n$-power of the bulk-bulk propagator times two bulk-boundary propagators. The particular cases with $n=1$ and $n=2$ are encountered in the scalar exchange diagrams coming from $\Phi^3$ and $\Phi^4$ theories on AdS, respectively. Using the power series representation of $G(x,z)$ as an expansion in the variable $\xi$, these type of integrals simply reduce to a sum of integrals of powers of $\xi$ times $KK$, each one having the general form:

\begin{equation}
    \int d^{d+1}x\sqrt{g}\ \xi^{\Delta_1}(x,z)\tilde{K}^{\Delta_2}(x,\vec{y_2})\tilde{K}^{\Delta_3}(x,\vec{y_3})
\end{equation}\

for some numbers $\Delta_1$, $\Delta_2$ and $\Delta_3$, where $\tilde{K}^{\Delta_i}(x,\vec{y_i})$ is the unnormalized bulk-boundary propagator of scaling dimension $\Delta_i$, eq. (\ref{unbubop}). Each one of these integrals can be solved directly from the general formula eq. (\ref{generalformula}) after a couple of AdS isometry transformations, where one tries to simplify the expression for one of the bulk-boundary propagators by translating its external boundary dependence to $\vec{0}$ and then referring its interior point to infinity through an inversion transformation, resulting for the bulk-boundary propagator in a simple power of the radial coordinate $x_0$. Let us see how this is done. Under this spirit then of simplifying the expression for one of the bulk-boundary propagators we will start by performing the translation $\vec{x}\rightarrow \vec{x}+\vec{y_3}$, also conveniently defining in the process $z=z'+\vec{y_3}$:

\begin{equation}
    \int d^{d+1}x\sqrt{g}\ \xi^{\Delta_1}(x,z)\tilde{K}^{\Delta_2}(x,\vec{y_2})\tilde{K}^{\Delta_3}(x,\vec{y_3}) = \int d^{d+1}x\sqrt{g}\ \xi^{\Delta_1}(x,z')\tilde{K}^{\Delta_2}(x,\vec{y_{23}})\tilde{K}^{\Delta_3}(x,\vec{0})
\end{equation}\

where we used that under AdS isometry transformations both the AdS measure and variable $\xi$ are invariant and that the bulk-boundary propagator transforms according to eq. (\ref{bubotr}). As we anticipated, the form of the second bulk-boundary propagator which has its external boundary dependence translated to $\vec{0}$, can be greatly simplified to a simple power of $x_0$ through an inversion of its interior point. Therefore, performing the inversion $x^\mu\rightarrow\frac{x^\mu}{x^2}$, also conveniently defining in the process $z'^\mu=\frac{z''^\mu}{z''^2}$ and $\vec{y_{23}}=\frac{\vec{y_{23}}'}{\vec{y_{23}}'^2}$:

\begin{align}
        &\int d^{d+1}x\sqrt{g}\ \xi^{\Delta_1}(x,z)\tilde{K}^{\Delta_2}(x,\vec{y_2})\tilde{K}^{\Delta_3}(x,\vec{y_3})\nonumber\\
        &\hspace{3cm}= \int d^{d+1}x\sqrt{g}\ \xi^{\Delta_1}(x,z'')\lvert\vec{y_{23}}'\rvert^{2\Delta_2}\tilde{K}^{\Delta_2}(x,\vec{y_{23}}')x_0^{\Delta_3}\nonumber\\
        &\hspace{3cm}= \lvert\vec{y_{23}}'\rvert^{2\Delta_2}\int d^{d+1}x\ x_0^{-d-1} \Bigl(\frac{2x_0z''_0}{x_0^2+z''^2_0+\lvert\vec{x}-\vec{z''}\rvert^2}\Bigl)^{\Delta_1}\Bigl(\frac{x_0}{x_0^2+\lvert\vec{x}-\vec{y_{23}}'\rvert^2}\Bigl)^{\Delta_2}x_0^{\Delta_3}\nonumber\\
        &\hspace{3cm}= 2^{\Delta_1}z''^{\Delta_1}_0\lvert\vec{y_{23}}'\rvert^{2\Delta_2}\int d^{d+1}x\ \frac{x_0^{\sum\Delta_i-d-1}}{(x_0^2+z''^2_0+\lvert\vec{x}-\vec{z''}\rvert^2)^{\Delta_1}(x_0^2+\lvert\vec{x}-\vec{y_{23}}'\rvert^2)^{\Delta_2}}
\end{align}\

where, again, we used the transformation rules of the measure, the variable $\xi$ and the bulk-boundary propagator. Notice how this last integral is precisely the general formula eq. (\ref{generalformula}) under the particular case $a=\sum\Delta_i-d-1$, $b=\Delta_1$ and $c=\Delta_2$. Therefore, replacing its value:

\begin{align}\label{intgkkmedstp}
        &\int d^{d+1}x\sqrt{g}\ \xi^{\Delta_1}(x,z)\tilde{K}^{\Delta_2}(x,\vec{y_2})\tilde{K}^{\Delta_3}(x,\vec{y_3})\nonumber\\
        &\hspace{3cm}= 2^{\Delta_1}z''^{\Delta_1}_0\lvert\vec{y_{23}}'\rvert^{2\Delta_2}\frac{\pi^{\frac{d}{2}}}{2}\frac{\Gamma(\frac{\sum\Delta_i-d}{2})\Gamma(\frac{\Delta_1+\Delta_2-\Delta_3}{2})\Gamma(\frac{\Delta_3+\Delta_1-\Delta_2}{2})}{\Gamma(\Delta_1)\Gamma(\frac{\sum\Delta_i}{2})(z''^2_0+\lvert\vec{z''}-\vec{y_{23}}'\rvert^2)^{\frac{\Delta_1+\Delta_2-\Delta_3}{2}}}\nonumber\\
        &\hspace{3.4cm}\times\ _2F_1\Bigl(\frac{\Delta_1+\Delta_2-\Delta_3}{2},\frac{\Delta_3+\Delta_1-\Delta_2}{2};\frac{\sum\Delta_i}{2};1-\frac{z''^2_0}{z''^2_0+\lvert\vec{z''}-\vec{y_{23}}'\rvert^2}\Bigl)
\end{align}\

There are many transformation formulas satisfied by the hypergeometric functions that allows one to relate hypergeometric functions of different parameters and arguments.
Among them is a linear transformation of the argument known as Euler's transformation \cite{HypFunc2}:

\begin{equation}\label{eultrans}
    _2F_1(a,b;c;z) = (1-z)^{c-a-b}\ _2F_1(c-a,c-b;c;z)
\end{equation}\

which, for the hypergeometric function present in eq. (\ref{intgkkmedstp}), this transformation will turn out to be extremely revealing since it will allows us to write the final result of the integral in terms of the original bulk-boundary propagators. Indeed, performing this transformation:

\begin{align}
        &\int d^{d+1}x\sqrt{g}\ \xi^{\Delta_1}(x,z)\tilde{K}^{\Delta_2}(x,\vec{y_2})\tilde{K}^{\Delta_3}(x,\vec{y_3})\nonumber\\
        &\hspace{3cm}= 2^{\Delta_1}z''^{\Delta_1}_0\lvert\vec{y_{23}}'\rvert^{2\Delta_2}\frac{\pi^{\frac{d}{2}}}{2}\frac{\Gamma(\frac{\sum\Delta_i-d}{2})\Gamma(\frac{\Delta_1+\Delta_2-\Delta_3}{2})\Gamma(\frac{\Delta_3+\Delta_1-\Delta_2}{2})}{\Gamma(\Delta_1)\Gamma(\frac{\sum\Delta_i}{2})(z''^2_0+\lvert\vec{z''}-\vec{y_{23}}'\rvert^2)^{\frac{\Delta_1+\Delta_2-\Delta_3}{2}}}\nonumber\\
        &\hspace{3.4cm}\times\Bigl(\frac{z''^2_0}{z''^2_0+\lvert\vec{z''}-\vec{y_{23}}'\rvert^2}\Bigl)^{\frac{\Delta_2+\Delta_3-\Delta_1}{2}}\ _2F_1\Bigl(\Delta_2,\Delta_3;\frac{\sum\Delta_i}{2};1-\frac{z''^2_0}{z''^2_0+\lvert\vec{z''}-\vec{y_{23}}'\rvert^2}\Bigl)\nonumber\\
        &\hspace{3cm}= 2^{\Delta_1}\frac{\pi^{\frac{d}{2}}}{2}\frac{\Gamma(\frac{\sum\Delta_i-d}{2})\Gamma(\frac{\Delta_1+\Delta_2-\Delta_3}{2})\Gamma(\frac{\Delta_3+\Delta_1-\Delta_2}{2})}{\Gamma(\Delta_1)\Gamma(\frac{\sum\Delta_i}{2})}\nonumber\\
        &\hspace{3.4cm}\times\lvert\vec{y_{23}}'\rvert^{2\Delta_2}\frac{z''^{\Delta_2+\Delta_3}_0}{(z''^2_0+\lvert\vec{z''}-\vec{y_{23}}'\rvert^2)^{\Delta_2}}\ _2F_1\Bigl(\Delta_2,\Delta_3;\frac{\sum\Delta_i}{2};1-\frac{z''^2_0}{z''^2_0+\lvert\vec{z''}-\vec{y_{23}}'\rvert^2}\Bigl)\nonumber\\
        &\hspace{3cm}= 2^{\Delta_1}\frac{\pi^{\frac{d}{2}}}{2}\frac{\Gamma(\frac{\sum\Delta_i-d}{2})\Gamma(\frac{\Delta_1+\Delta_2-\Delta_3}{2})\Gamma(\frac{\Delta_3+\Delta_1-\Delta_2}{2})}{\Gamma(\Delta_1)\Gamma(\frac{\sum\Delta_i}{2})}\nonumber\\
        &\hspace{3.4cm}\times\lvert\vec{y_{23}}'\rvert^{2\Delta_2}\tilde{K}^{\Delta_2}(z'',\vec{y_{23}}')z''^{\Delta_3}_0\ _2F_1\Bigl(\Delta_2,\Delta_3;\frac{\sum\Delta_i}{2};1-\tilde{K}(z'',\vec{y_{23}}')z''_0\Bigl)
\end{align}\

and remembering that $\vec{y_{23}}'=\frac{\vec{y_{23}}}{\vec{y_{23}}^2}$ and $z''^\mu=\frac{z'^\mu}{z'^2}$, where $z'=z-\vec{y_3}$:

\begin{align}
        &\int d^{d+1}x\sqrt{g}\ \xi^{\Delta_1}(x,z)\tilde{K}^{\Delta_2}(x,\vec{y_2})\tilde{K}^{\Delta_3}(x,\vec{y_3})\nonumber\\
        &\hspace{3cm}= 2^{\Delta_1}\frac{\pi^{\frac{d}{2}}}{2}\frac{\Gamma(\frac{\sum\Delta_i-d}{2})\Gamma(\frac{\Delta_1+\Delta_2-\Delta_3}{2})\Gamma(\frac{\Delta_3+\Delta_1-\Delta_2}{2})}{\Gamma(\Delta_1)\Gamma(\frac{\sum\Delta_i}{2})}\nonumber\\
        &\hspace{3.4cm}\times\tilde{K}^{\Delta_2}(z',\vec{y_{23}})\tilde{K}^{\Delta_3}(z',\vec{0})\ _2F_1\Bigl(\Delta_2,\Delta_3;\frac{\sum\Delta_i}{2};1-\tilde{K}(z',\vec{y_{23}})\tilde{K}(z',\vec{0})\lvert\vec{y_{23}}\rvert^2\Bigl)\nonumber\\
        &\hspace{3cm}= 2^{\Delta_1}\frac{\pi^{\frac{d}{2}}}{2}\frac{\Gamma(\frac{\sum\Delta_i-d}{2})\Gamma(\frac{\Delta_1+\Delta_2-\Delta_3}{2})\Gamma(\frac{\Delta_3+\Delta_1-\Delta_2}{2})}{\Gamma(\Delta_1)\Gamma(\frac{\sum\Delta_i}{2})}\nonumber\\
        &\hspace{3.4cm}\times\tilde{K}^{\Delta_2}(z,\vec{y_2})\tilde{K}^{\Delta_3}(z,\vec{y_3})\ _2F_1\Bigl(\Delta_2,\Delta_3;\frac{\sum\Delta_i}{2};1-\tilde{K}(z,\vec{y_2})\tilde{K}(z,\vec{y_3})\lvert\vec{y_{23}}\rvert^2\Bigl)
\end{align}\

where we used the transformation rules of the bulk-boundary propagator. The form of this result is already nice enough to be used, but for the specific calculations that we want to carry out it will be useful to simplify it using Legendre duplication formula eq. (\ref{LDF}) which, for the particular value $z=\frac{\Delta_1}{2}$, allows us to write the final result of the integral as:

\begin{align}\label{xikkform}
        &\int d^{d+1}x\sqrt{g}\ \xi^{\Delta_1}(x,z)\tilde{K}^{\Delta_2}(x,\vec{y_2})\tilde{K}^{\Delta_3}(x,\vec{y_3})\nonumber\\
        &\hspace{3cm}= \pi^{\frac{d+1}{2}}\frac{\Gamma(\frac{\sum\Delta_i-d}{2})\Gamma(\frac{\Delta_1+\Delta_2-\Delta_3}{2})\Gamma(\frac{\Delta_3+\Delta_1-\Delta_2}{2})}{\Gamma(\frac{\Delta_1}{2})\Gamma(\frac{\Delta_1+1}{2})\Gamma(\frac{\sum\Delta_i}{2})}\nonumber\\
        &\hspace{3.4cm}\times\tilde{K}^{\Delta_2}(z,\vec{y_2})\tilde{K}^{\Delta_3}(z,\vec{y_3})\ _2F_1\Bigl(\Delta_2,\Delta_3;\frac{\sum\Delta_i}{2};1-\tilde{K}(z,\vec{y_2})\tilde{K}(z,\vec{y_3})\lvert\vec{y_{23}}\rvert^2\Bigl)
\end{align}\

Notice how the value of this integral is proportional to bulk-boundary propagators of the same scaling dimension as the ones being integrated times some function of them. Since the integral $\int G^nKK$ can be written as a sum of integrals of powers of $\xi$ times $KK$ this result implies that, whenever $\int G^nKK$ is convergent, its value will be proportional to $KKf(\tilde{K}\tilde{K}y^2)$, where the function $f(\tilde{K}\tilde{K}y^2)$ can in general be written as a series in its argument.

\section{Special Case \texorpdfstring{$\int GK$}{}}

There are many integrals involving the bulk-bulk propagator that are encountered in the study of holographic correlators, however the formulas developed so far only help us to solve those that are infrared convergent. In the case of the integrals that are infrared divergent the introduction of an IR-regulator is needed and the study of the obtention of their regularized values becomes much more delicate, requiring a completely different approach from the one followed up to now. One of these integrals that require special attention is:

\begin{equation}
    I(z,\vec{y}) = \int d^{d+1}x\sqrt{g}\ G(x,z)\tilde{K}^\Delta(x,\vec{y})
\end{equation}\

where both propagators are of the same scaling dimension $\Delta$. This integral is found throughout our study of correlators in the reducible diagrams coming from both $\Phi^3$ and $\Phi^4$ theories on AdS, where the irreducible parts of the diagrams are connected by a single bulk-bulk propagator. The IR-divergence of this integral can be seen directly by studying how its integrand behaves as it approaches the boundary of AdS. In this case, using the explicit form of the metric and the known expansions of both propagators we obtain that:

\begin{align}
    \sqrt{g}\ G(x,z)\tilde{K}^\Delta(x,\vec{y}) \underset{x_0\rightarrow0}{\sim} &x_0^{-d-1}x_0^\Delta x_0^{d-\Delta}\nonumber\\
    =\ &x_0^{-1}
\end{align}\

behavior which clearly shows the divergence (logarithmic in this case) of the integral, for any value of $\Delta$, as it gets integrated closer and closer to the conformal boundary of the AdS space at $x_0=0$. As it has been discussed throughout this work, this particular type of divergence is to be expected due to the IR/UV duality that the AdS/CFT correspondence implies, motivating the introduction of the $\varepsilon$-regulator not only for the on-shell contributions coming from the AdS path integral through the study of the holographic renormalization procedure, but also for the off-shell contributions coming from it, where the regularization scheme used in the former naturally translates into the latter by simply performing the loops integral up to this same small distance $\varepsilon$. This brief discussion is to motivate the regularized version of the integral which is the quantity that one actually encounters:

\begin{equation}\label{reggkint}
    I(z,\vec{y}) = \int_{x_0=\varepsilon}d^{d+1}x\sqrt{g}\ G(x,z)\tilde{K}^\Delta(x,\vec{y})
\end{equation}\

There are many ways in which we can solve for the value of this integral not only as a function of the external points $z$ and $\vec{y}$, but also as a function of the IR-regulator $\varepsilon$. For instance, we could follow the same strategy used for the integral $\int KK$ studied in Appendix A, where simply using AdS isometry transformations we extract all the dependence on the external points of the integrand, remaining dependence only in the limits of integration for then solving the differential equation followed by the resulting integral. However, for the sake of diversity and completeness regarding the different approaches that one can take for these divergent integrals, here we will present an alternative approach. It is of course satisfactory that the method we will use next delivers exactly the same result as the one described above. Its motivation comes from noticing the easy differential equation that the integral satisfies when acting with the wave operator:

\begin{align}
    (-\Box_z+m^2)I(z,\vec{y}) &= \int_{x_0=\varepsilon}d^{d+1}x\sqrt{g}\ (-\Box_z+m^2)G(x,z)\tilde{K}^\Delta(x,\vec{y})\nonumber\\
    &= \int_{x_0=\varepsilon}d^{d+1}x\ \delta^{d+1}(x-z)\tilde{K}^\Delta(x,\vec{y})\nonumber\\
    &= \tilde{K}^\Delta(z,\vec{y})
\end{align}\

where we used that the bulk-bulk propagator is the Green's function of the wave operator and in the last line the limit $\varepsilon\rightarrow0$ is understood. In other words, the integral we are trying to compute satisfies the equation:

\begin{align}
    (-\Box_z+m^2)I(z,\vec{y}) = \tilde{K}^\Delta(z,\vec{y})
\end{align}\

But the solution to this equation is already known! It is given by $-\frac{1}{2\nu}\tilde{K}^\Delta(z,\vec{y})\ln{\bigl(\tilde{K}(z,\vec{y})\bigl)}$. Indeed:

\begin{align}
    (-\Box_z+m^2)\Bigl[-\frac{1}{2\nu}\tilde{K}^\Delta(z,\vec{y})\ln{\bigl(\tilde{K}(z,\vec{y})\bigl)}\Bigl] = &(-\Box_{z'}+m^2)\Bigl[-\frac{1}{2\nu}\tilde{K}^\Delta(z',\vec{0})\ln{\bigl(\tilde{K}(z',\vec{0})\bigl)}\Bigl]\nonumber\\
    = &(-\Box_{z''}+m^2)\Bigl[-\frac{1}{2\nu}z''^\Delta_0\ln{(z''_0)}\Bigl]\nonumber\\
    = &\bigl[-z''^2_0\partial^2_{0''}-(1-d)z''_0\partial_{0''}-z''^2_0\partial^2_{i''}+m^2\bigl]\Bigl[-\frac{1}{2\nu}z''^\Delta_0\ln{(z''_0)}\Bigl]\nonumber\\
    = &\frac{1}{2\nu}\Delta(\Delta-1)z''^\Delta_0\ln{(z''_0)}+\frac{1}{2\nu}(2\Delta-1)z''^\Delta_0\nonumber\\
    &+\frac{1}{2\nu}(1-d)\Delta z''^\Delta_0\ln{(z''_0)}+\frac{1}{2\nu}(1-d)z''^\Delta_0\nonumber\\
    &-\frac{1}{2\nu}m^2z''^\Delta_0\ln(z''_0)\nonumber\\
    = &\frac{1}{2\nu}\bigl[\Delta(\Delta-1)+(1-d)\Delta-m^2\bigl]z''^\Delta_0\ln(z''_0)\nonumber\\
    &+\frac{1}{2\nu}\bigl[(2\Delta-1)+(1-d)\bigl]z''^\Delta_0\nonumber\\
    = &\frac{1}{2\nu}\bigl[\Delta(\Delta-d)-m^2\bigl]z''^\Delta_0\ln(z''_0)+\frac{1}{2\nu}(2\Delta-d)z''^\Delta_0\nonumber\\
    = &z''^\Delta_0\nonumber\\
    = &\tilde{K}^\Delta(z,\vec{y})\nonumber\\
\end{align}\

where, using the invariance of the wave operator under isometries and the known transformation rules of the bulk-boundary propagator, in the first and second equalities we defined the translated point $z'=z-\vec{y}$ followed by the inverted point $z''^\mu=\frac{z'^\mu}{z'^2}$, in the third equality using eq. (\ref{eom1}) we wrote the wave equation explicitly, in the fourth equality we computed every derivative, in the fifth equality we conveniently factorized the resulting terms, in the sixth equality we simplified terms, in the seventh equality we used that $\Delta(\Delta-d)=m^2$ and $2\Delta-d=2\nu$, and in the final equality we wrote the result in terms of the original coordinates. By arguments of unicity and uniqueness of the solution then, this implies that the general solution of $I(z,\vec{y})$ is given by this quantity up to homogeneous solutions to the wave equation. But we already solved the homogeneous case, its solution are given by eq. (\ref{generalk}). Thus, the value of the integral eq. (\ref{reggkint}) can be written as:

\begin{equation}\label{intgkmedstep}
    \int_{x_0=\varepsilon}d^{d+1}x\sqrt{g}\ G(z,x)\tilde{K}^\Delta(x,\vec{y}) = -\frac{1}{2\nu}\tilde{K}^\Delta(z,\vec{y})\ln{\bigl(\tilde{K}(z,\vec{y})\bigl)}+c_1\tilde{K}^\Delta(z,\vec{y})+c_2\tilde{K}^{d-\Delta}(z,\vec{y})
\end{equation}\

for some constants $c_1$ and $c_2$, determined by the boundary conditions of the integral. As $z_0\rightarrow\infty$ it is direct to see that the LHS goes to 0, yet the last term of the RHS diverges. This implies the boundary condition $c_2=0$. Similarly, we can completely determine the coefficient $c_1$ studying the limit $z_0\rightarrow0$. However as we have already seen repeatedly for quantities being evaluated at the conformal boundary of AdS, the correct physics of their infrared behavior comes from studying them not at the boundary itself but at some small distance $\varepsilon$ from it, where of course this regulator is understood to be a small positive number. As expected, it turns out that the present case under study is no exception. Therefore, with the intention to study the boundary behavior of eq. (\ref{intgkmedstep}) in the limit $z_0\rightarrow0$, let us call $z_0=\varepsilon$ where of course this is the same IR-regulator as before:

\begin{equation}
    \int_{x_0=\varepsilon}d^{d+1}x\sqrt{g}\ G\bigl((\varepsilon,\vec{z}),x\bigl)\tilde{K}^\Delta(x,\vec{y}) = -\frac{1}{2\nu}\tilde{K}^\Delta\bigl((\varepsilon,\vec{z}),\vec{y}\bigl)\ln{\Bigl(\tilde{K}\bigl((\varepsilon,\vec{z}),\vec{y}\bigl)\Bigl)}+c_1\tilde{K}^\Delta\bigl((\varepsilon,\vec{z}),\vec{y}\bigl)
\end{equation}\

In the limit $\varepsilon\rightarrow0$, using the known expansion of both propagator and keeping the terms of order $\varepsilon^\Delta$, the resulting equation for $c_1$ is:

\begin{equation}
    \frac{c_\Delta}{2\nu}\varepsilon^{\Delta}\int_{x_0=\varepsilon}d^{d+1}x\sqrt{g}\ \tilde{K}^\Delta(x,\vec{z})\tilde{K}^\Delta(x,\vec{y}) = -\frac{1}{2\nu}\frac{\varepsilon^\Delta}{\lvert\vec{z}-\vec{y}\rvert^{2\Delta}}\ln{\Bigl(\frac{\varepsilon}{\lvert\vec{z}-\vec{y}\rvert^2}\Bigl)}+c_1\frac{\varepsilon^\Delta}{\lvert\vec{z}-\vec{y}\rvert^{2\Delta}}
\end{equation}\

In other words, we can determine the coefficient $c_1$ using the fact that in the appropriate limit the integral $\int GK$ must reduce to the value of the integral $\int KK$! Indeed, we already found the value of this integral in Appendix A, it is given by eq. (\ref{d2funcreg2}). Therefore, replacing its value we find the condition for $c_1$:

\begin{equation}
    -\frac{1}{\nu}\frac{\varepsilon^{\Delta}}{\lvert\vec{z}-\vec{y}\rvert^{2\Delta}}\ln{\Bigl(\frac{\varepsilon}{\lvert\vec{z}-\vec{y}\rvert}\Bigl)} = -\frac{1}{2\nu}\frac{\varepsilon^\Delta}{\lvert\vec{z}-\vec{y}\rvert^{2\Delta}}\ln{\Bigl(\frac{\varepsilon}{\lvert\vec{z}-\vec{y}\rvert^2}\Bigl)}+c_1\frac{\varepsilon^\Delta}{\lvert\vec{z}-\vec{y}\rvert^{2\Delta}}
\end{equation}\

Multiplying both sides by $\frac{\lvert\vec{z}-\vec{y}\rvert^{2\Delta}}{\varepsilon^\Delta}$ and solving for $c_1$:

\begin{align}
    c_1 &= \frac{1}{2\nu}\ln{\Bigl(\frac{\varepsilon}{\lvert\vec{z}-\vec{y}\rvert^2}\Bigl)}-\frac{1}{\nu}\ln{\Bigl(\frac{\varepsilon}{\lvert\vec{z}-\vec{y}\rvert}\Bigl)}\nonumber\\
    &= \frac{1}{2\nu}\ln{(\varepsilon)}-\frac{1}{2\nu}\ln{\bigl(\lvert\vec{z}-\vec{y}\rvert^{2}\bigl)}-\frac{1}{\nu}\ln{(\varepsilon)}+\frac{1}{\nu}\ln{\bigl(\lvert\vec{z}-\vec{y}\rvert\bigl)}\nonumber\\
    &=-\frac{1}{2\nu}\ln{(\varepsilon)}
\end{align}\

This result is noteworthy. It is not just a number as of course it should, but it is a divergent one. All the logarithmic infrared divergence of the integral $\int GK$ is correctly captured by this coefficient. Finally then, replacing it in the general form of the solution we find that the value of the regularized $\int GK$ integral is given by:

\begin{align}\label{scintgk}
    \int_{x_0=\varepsilon}d^{d+1}x\sqrt{g}\ G(z,x)\tilde{K}^\Delta(x,\vec{y}) &= -\frac{1}{2\nu}\tilde{K}^\Delta(z,\vec{y})\ln{\bigl(\tilde{K}(z,\vec{y})\bigl)}-\frac{1}{2\nu}\tilde{K}^\Delta(z,\vec{y})\ln{(\varepsilon)}\nonumber\\
    &= -\frac{1}{2\nu}\tilde{K}^\Delta(z,\vec{y})\ln{\bigl(\varepsilon\tilde{K}(z,\vec{y})\bigl)}
\end{align}\

This result not only satisfies the correct wave equation and shows the explicit logarithmic divergence of the integral, but also in the appropriate limit it reduces to the known value of the $\int KK$ integral. Moreover, as it is discussed in the main text, the form of this result allow us to factorize exactly the reducible diagrams present on both $\Phi^3$ and $\Phi^4$ theories, leading not only to a clear picture of the role of the quantum corrections to the holographic correlators, but also to a clear picture in their renormalization scheme.

\section{Special Case \texorpdfstring{$\int K\int GK$}{}}

Another infrared divergent integral involving the bulk-bulk propagator that is encountered in the computation of holographic correlators, whose study becomes delicate due to the need of introducing an IR-regulator and that therefore requires special attention is:

\begin{equation}
    I(\vec{y_1},\vec{y_2}) = \int d^{d+1}x_1\sqrt{g}\ \tilde{K}^\Delta(x_1,\vec{y_1})\int d^{d+1}x_2\sqrt{g}\ G(x_1,x_2)\tilde{K}^\Delta(x_2,\vec{y_2})
\end{equation}\

where all propagators are of the same scaling dimension $\Delta$. This integral is found throughout our study of correlators in the "double head" diagram present in the holographic 2-point function dual to a $\Phi^4$ theory on AdS. The IR-divergence of this quantity can be seen directly by analyzing first one of the integrals, say the $x_2$ integral, and then the remaining integral in $x_1$. Indeed, as we just saw in the previous section, the integral in $x_2$ is nothing but the special case $\int GK$ whose value was found to be IR-divergent as it gets integrated closer and closer to the conformal boundary of the AdS space. Moreover, since its value is proportional to $K$, the resulting integral in $x_1$ will be proportional to $KK$ and as we discussed in section A.4 of appendix A, such integral is also IR-divergent in the same region of integration. This fact suggests the regularization of both integrals in the lower limit of their radial coordinate. As it has been discussed repeatedly, this particular type of divergence is to be expected due to the IR/UV duality that the AdS/CFT correspondence implies, motivating the introduction of the $\varepsilon$-regulator not only for the on-shell contributions coming from the AdS path integral through the study of the holographic renormalization procedure, but also for the off-shell contributions coming from it, where the regularization scheme used in the former naturally translates into the latter by simply performing the loops integral up to this same small distance $\varepsilon$. This brief discussion is to motivate the regularized version of the integral which is the quantity that one actually encounters:

\begin{equation}\label{irregintkgk0}
    I(\vec{y_1},\vec{y_2}) = \int_{x_{1,0}=\varepsilon}d^{d+1}x_1\sqrt{g}\ \tilde{K}^\Delta(x_1,\vec{y_1})\int_{x_{2,0}=\varepsilon}d^{d+1}x_2\sqrt{g}\ G(x_1,x_2)\tilde{K}^\Delta(x_2,\vec{y_2})
\end{equation}\

The solving strategy for this quantity will be the same as the one used for the $\int KK$ integral, that is, extracting all the dependence on the external points of the integrand through AdS isometry transformations, remaining dependence only in the limits of integration, for then solving the differential equation followed by the resulting integral instead of solving the integral directly, process which will turn out to be much simpler to do. With these goals in mind then, we will start computing eq. (\ref{irregintkgk1}) by first performing the translations $x_i\rightarrow x_i+\vec{y_2}$:

\begin{equation}\label{irregintkgk1}
    I(\vec{y_1},\vec{y_2}) = \int_{x_{1,0}=\varepsilon}d^{d+1}x_1\sqrt{g}\ \tilde{K}^\Delta(x_1,\vec{y_{12}})\int_{x_{2,0}=\varepsilon}d^{d+1}x_2\sqrt{g}\ G(x_1,x_2)\tilde{K}^\Delta(x_2,\vec{0})
\end{equation}\

where we used that under AdS isometry both the AdS measure and the bulk-bulk propagator are invariant and that the bulk-boundary propagators transform according to eq. (\ref{bubotr}). The leftover dependency of the integrand on the external points can be removed by doing the rescaling $x_i\rightarrow\lvert\vec{y_1}-\vec{y_2}\rvert x_i$:

\begin{align}\label{irregintkgk2}
    I(\vec{y_1},\vec{y_2}) = &\int_{x_{1,0}=\frac{\varepsilon}{\lvert\vec{y_1}-\vec{y_2}\rvert}}d^{d+1}x_1\sqrt{g}\ \frac{1}{\lvert\vec{y_1}-\vec{y_2}\rvert^\Delta}\tilde{K}^\Delta\Bigl(x_1,\frac{\vec{y_{12}}}{\lvert\vec{y_1}-\vec{y_2}\rvert}\Bigl)\nonumber\\
    &\hspace{2cm}\times\int_{x_{2,0}=\frac{\varepsilon}{\lvert\vec{y_1}-\vec{y_2}\rvert}}d^{d+1}x_2\sqrt{g}\ G(x_1,x_2)\frac{1}{\lvert\vec{y_1}-\vec{y_2}\rvert^\Delta}\tilde{K}^\Delta(x_2,\vec{0})\nonumber\\
    = &\frac{1}{\lvert\vec{y_1}-\vec{y_2}\rvert^{2\Delta}}\int_{x_{1,0}=\sigma}d^{d+1}x_1\sqrt{g}\
    \tilde{K}^\Delta(x_1,\hat{n})\int_{x_{2,0}=\sigma}d^{d+1}x_2\sqrt{g}\ G(x_1,x_2)\tilde{K}^\Delta(x_2,\vec{0})
\end{align}\

where, again, we used the transformation rules of the measure and propagators, and defined the quantities $\sigma\equiv\frac{\varepsilon}{\lvert\vec{y_1}-\vec{y_2}\rvert}$ and $\hat{n}=\frac{\vec{y_{12}}}{\lvert\vec{y_1}-\vec{y_2}\rvert}$, $\hat{n}$ being a unit vector pointing in the direction of $\vec{y_{12}}$. Notice how under these transformations we managed to extract all the dependence of the external points of the integrand, remaining only in the lower limits of integration of the $x_1$ and $x_2$ integrals in the form of $\sigma$. It remains to compute the value of this last integral in the limit $\varepsilon\rightarrow0$, which in terms of $\sigma$ translates to $\sigma\rightarrow0$. The key realization here is that in this limit the differential equation in $\sigma$ satisfied by the integral is much easier to solve than the integral itself thanks to the convenient presence of Dirac deltas in the expansion of the integrand coming from the bulk-boundary propagators. Indeed, differentiating the integral in eq. (\ref{irregintkgk2}) with respect to $\sigma$ in the limit $\sigma\rightarrow0$:

\begin{align}
    &\frac{d}{d\sigma}\Bigl[\int_{x_{1,0}=\sigma}d^{d+1}x_1\sqrt{g}\
    \tilde{K}^\Delta(x_1,\hat{n})\int_{x_{2,0}=\sigma}d^{d+1}x_2\sqrt{g}\ G(x_1,x_2)\tilde{K}^\Delta(x_2,\vec{0})\Bigl]\nonumber\\
    &\hspace{3cm}=-\int d^dx_1\sqrt{g}\
    \tilde{K}^\Delta(x_1,\hat{n})\int_{x_{2,0}=\sigma}d^{d+1}x_2\sqrt{g}\ G(x_1,x_2)\tilde{K}^\Delta(x_2,\vec{0})\Bigl\rvert_{x_{1,0}=\sigma}\nonumber\\
    &\hspace{3.5cm}-\int_{x_{1,0}=\sigma}d^{d+1}x_1\sqrt{g}\
    \tilde{K}^\Delta(x_1,\hat{n})\int d^dx_2\sqrt{g}\ G(x_1,x_2)\tilde{K}^\Delta(x_2,\vec{0})\Bigl\rvert_{x_{2,0}=\sigma}\nonumber\\
    &\hspace{2.8cm}\underset{\sigma\rightarrow0}{=}-\frac{1}{2\nu\sigma}\int d^dx_1\ \delta^d(\vec{x_1}-\hat{n})\int_{x_{2,0}=\sigma}d^{d+1}x_2\sqrt{g}\ \tilde{K}^\Delta(x_2,\vec{x_1})\tilde{K}^\Delta(x_2,\vec{0})\nonumber\\
    &\hspace{3.6cm}-\frac{1}{2\nu\sigma}\int_{x_{1,0}=\sigma}d^{d+1}x_1\sqrt{g}\ \tilde{K}^\Delta(x_1,\hat{n})\int d^dx_2\ \tilde{K}^\Delta(x_1,\vec{x_2})\delta^d(\vec{x_2})+\mathcal{O}(\sigma^{-1<})\nonumber\\
    &\hspace{3cm}=-\frac{1}{\nu\sigma}\int_{x_0=\sigma}d^{d+1}x\sqrt{g}\ \tilde{K}^\Delta(x,\hat{n})\tilde{K}^\Delta(x,\vec{0})+\mathcal{O}(\sigma^{-1<})
\end{align}\

where in the first line we used the fundamental theorem of calculus, in the second line since $\sigma$ is understood to be small we used the known expansions of the propagators, and in the last line we trivially computed the respective integrals using the Dirac deltas coming from the bulk-boundary propagators. We can proceed with the calculation of this integral by multiplying both sides with $\sigma$ and then differentiating again with respect to $\sigma$:

\begin{align}
    &\frac{d}{d\sigma}\Bigl\{\sigma\frac{d}{d\sigma}\Bigl[\int_{x_{1,0}=\sigma}d^{d+1}x_1\sqrt{g}\
    \tilde{K}^\Delta(x_1,\hat{n})\int_{x_{2,0}=\sigma}d^{d+1}x_2\sqrt{g}\ G(x_1,x_2)\tilde{K}^\Delta(x_2,\vec{0})\Bigl]\Bigl\}\nonumber\\
    &\hspace{3cm}=\frac{d}{d\sigma}\Bigl[-\frac{1}{\nu}\int_{x_0=\sigma}d^{d+1}x\sqrt{g}\ \tilde{K}^\Delta(x,\hat{n})\tilde{K}^\Delta(x,\vec{0})+\mathcal{O}(\sigma^{0<})\Bigl]\nonumber\\
    &\hspace{3cm}=\frac{1}{\nu}\int d^dx\sqrt{g}\ \tilde{K}^\Delta(x,\hat{n})\tilde{K}^\Delta(x,\vec{0})\Bigl\rvert_{x_0=\sigma}+\mathcal{O}(\sigma^{-1<})\nonumber\\
    &\hspace{2.8cm}\underset{\sigma\rightarrow0}{=}(\text{contact terms})+\frac{1}{\nu c_\Delta\sigma}\int d^dx\ \Bigl[\frac{\delta^d(\vec{x}-\hat{n})}{\lvert\vec{x}\rvert^{2\Delta}}+\frac{\delta^d(\vec{x})}{\lvert\vec{x}-\hat{n}\rvert^{2\Delta}}\Bigl]+\mathcal{O}(\sigma^{-1<})\nonumber\\
    &\hspace{3cm}=(\text{contact terms})+\frac{2}{\nu c_\Delta\sigma}+\mathcal{O}(\sigma^{-1<})
\end{align}\

where in the second line we used the fundamental theorem of calculus, in the third line since $\sigma$ is understood to be small we used the known expansions of the propagators, and in the last line we trivially computed the respective integrals using the Dirac deltas coming from the bulk-boundary propagators, using the fact that the vector $\hat{n}$ is unitary. Notice how the resulting equation for the integral is very easy to solve! Simply integrating both sides with respect to $\sigma$ twice we find, up to integration constants, that the value of the integral is given by:

\begin{align}
    &\int_{x_{1,0}=\sigma}d^{d+1}x_1\sqrt{g}\
    \tilde{K}^\Delta(x_1,\hat{n})\int_{x_{2,0}=\sigma}d^{d+1}x_2\sqrt{g}\ G(x_1,x_2)\tilde{K}^\Delta(x_2,\vec{0})\nonumber\\
    &\hspace{6cm}=(\text{contact terms})+\frac{1}{\nu c_\Delta}\ln^2{(\sigma)}+\mathcal{O}(\sigma^{0<})
\end{align}\

Notice how the subleading terms of order $\mathcal{O}(\sigma^{0<})$ simply go to 0 in the limit $\sigma\rightarrow0$. Therefore, plugging this result for the integral back into eq. (\ref{irregintkgk2}) and remembering that $\sigma\equiv\frac{\varepsilon}{\lvert\vec{y_1}-\vec{y_2}\rvert}$, the value we find for the $\int K\int GK$ integral eq. (\ref{irregintkgk0}) (up to contact terms which can always be renormalized with appropriate local counterterms) is given by:

\begin{equation}\label{scintkgk}
    \int_{x_{1,0}=\varepsilon}d^{d+1}x_1\sqrt{g}\ \tilde{K}^\Delta(x_1,\vec{y_1})\int_{x_{2,0}=\varepsilon}d^{d+1}x_2\sqrt{g}\ G(x_1,x_2)\tilde{K}^\Delta(x_2,\vec{y_2}) = \frac{\nu^{-1}c_\Delta^{-1}}{\lvert\vec{y_1}-\vec{y_2}\rvert^{2\Delta}}\ln^2{\Bigl(\frac{\varepsilon}{\lvert\vec{y_1}-\vec{y_2}\rvert}\Bigl)}
\end{equation}\

This result seems to break the conformal structure expected for contributions to the 2-point function of a CFT, however as it is discussed in the main text, its form corresponds exactly to the expansion of a conformal anomaly, realization which will lead not only to a clear picture of the role of the quantum corrections to the holographic correlators, but also to a clear picture in their renormalization scheme.

\newpage


\printbibliography
\addcontentsline{toc}{chapter}{Bibliography}

\end{document}